\documentclass[11pt]{article}

\usepackage[T1]{fontenc}
\usepackage{lmodern}
\usepackage{amsmath,amssymb,amsthm,mathtools}
\usepackage{aliascnt}
\usepackage{booktabs,array,tabularx}
\usepackage{enumitem}
\usepackage[ruled,vlined,linesnumbered]{algorithm2e}
\DontPrintSemicolon
\SetAlFnt{\small}
\SetAlCapFnt{\small}
\SetAlCapNameFnt{\small\normalfont}
\SetArgSty{textnormal}
\usepackage[margin=1in]{geometry}
\usepackage{microtype}
\usepackage{needspace}
\usepackage{placeins}
\usepackage{tikz}
\usetikzlibrary{arrows.meta,positioning,calc}
\usepackage[colorlinks=true,linkcolor=blue,citecolor=blue,urlcolor=blue,bookmarksdepth=2]{hyperref}
\usepackage[nameinlink,capitalise,noabbrev]{cleveref}
\hypersetup{
 pdftitle={Gaussian Difference Dynamics: Breaking the 2 to the n/2 Barrier for the Shortest Vector Problem and Discrete Gaussian Sampling},
 pdfauthor={}
}

\newtheorem{theorem}{Theorem}[section]
\newaliascnt{lemma}{theorem}
\newtheorem{lemma}[lemma]{Lemma}
\aliascntresetthe{lemma}
\newaliascnt{proposition}{theorem}
\newtheorem{proposition}[proposition]{Proposition}
\aliascntresetthe{proposition}
\newaliascnt{corollary}{theorem}
\newtheorem{corollary}[corollary]{Corollary}
\aliascntresetthe{corollary}
\theoremstyle{definition}
\newaliascnt{definition}{theorem}
\newtheorem{definition}[definition]{Definition}
\aliascntresetthe{definition}
\newaliascnt{remark}{theorem}

\aliascntresetthe{remark}
\newtheorem*{remark*}{Remark}
\crefname{lemma}{Lemma}{Lemmas}
\Crefname{lemma}{Lemma}{Lemmas}
\crefname{proposition}{Proposition}{Propositions}
\Crefname{proposition}{Proposition}{Propositions}
\crefname{corollary}{Corollary}{Corollaries}
\Crefname{corollary}{Corollary}{Corollaries}
\crefname{definition}{Definition}{Definitions}
\Crefname{definition}{Definition}{Definitions}
\crefname{remark}{Remark}{Remarks}
\Crefname{remark}{Remark}{Remarks}
\crefname{algocf}{Algorithm}{Algorithms}
\Crefname{algocf}{Algorithm}{Algorithms}

\newcommand{\R}{\mathbb R}
\newcommand{\Z}{\mathbb Z}
\newcommand{\E}{\mathbb E}
\newcommand{\Prb}{\mathbb P}
\newcommand{\poly}{\operatorname{poly}}
\newcommand{\polylog}{\operatorname{polylog}}

\newcommand{\lab}{\operatorname{lab}}
\newcommand{\key}{\operatorname{key}}
\newcommand{\TV}{d_{\mathrm{TV}}}
\newcommand{\en}{\mathcal E}
\newcommand{\ind}{\mathbf 1}

\newcommand{\angles}[1]{\langle #1\rangle}
\newcommand{\pad}{b_n}
\newcommand{\errwin}{\ell_n}
\newcommand{\softB}{\mathsf B}
\newcommand{\amp}{\Upsilon}
\newcommand{\degpad}{b_{\mathrm{deg}}}
\newcommand{\logbudget}{L_{\mathrm{log}}}
\newcommand{\svptime}{\alpha_{\mathrm{SVP}}}
\newcommand{\svpspace}{\beta_{\mathrm{SVP}}}
\newcommand{\covrep}{r_{\mathrm{cov}}}
\newcommand{\Deltap}{\Delta_*}

\newcommand{\Var}{\operatorname{Var}}
\newcommand{\diag}{\operatorname{diag}}

\newcommand{\dgstime}{\alpha_{\mathrm G}}
\newcommand{\dgsspace}{\beta_{\mathrm G}}

\allowdisplaybreaks

\title{Gaussian Difference Dynamics: Breaking the $2^{n/2}$ Barrier for the Shortest Vector Problem and Discrete Gaussian Sampling}
\author{
Yiming Gao\textsuperscript{1} \quad
Xuyuan Han\textsuperscript{1} \quad
Honggang Hu\textsuperscript{1,2}\\[0.8em]
\small \textsuperscript{1}School of Cyber Science and Technology,\\
University of Science and Technology of China\\
\small \texttt{qw1234567@mail.ustc.edu.cn, hanxuyuan@mail.ustc.edu.cn}\\[0.15em]
\small \textsuperscript{2}Hefei National Laboratory, Hefei, China\\
\small \texttt{hghu2005@ustc.edu.cn}
}
\date{}

\begin{document}
\maketitle

\begin{abstract}
We give classical randomized algorithms in a word RAM model that break
the $2^{n/2}$ barrier for both discrete Gaussian sampling (DGS) and the
Shortest Vector Problem (SVP). For every fixed $\alpha>1$, shifted DGS
on arbitrary lattices at width $\sigma\ge\alpha\eta_{1/2}(L)$ takes
\[
 (3/2)^{n/2+o(n)}\text{ time}
 \qquad\text{and}\qquad
 (4/3)^{n/2+o(n)}\text{ space}.
\]
These bounds match the Becker--Ducas--Gama--Laarhoven (BDGL)
heuristic sieving exponents. At widths at least $3\eta_{1/2}(L)/2$,
the same total cost suffices for an entire batch of
$\lfloor(4/3)^{n/2}\rfloor$ samples jointly close to independent
target Gaussians.

For SVP on arbitrary lattices, our algorithm returns a shortest nonzero
vector with high probability in
\[
 2^{0.4130n+o(n)}\text{ time}
 \qquad\text{and}\qquad
 2^{0.3138n+o(n)}\text{ space}.
\]
As a corollary, we obtain centered DGS at every width with the same
exponents, answering a question of Stephens-Davidowitz (SODA 2016).

We introduce Gaussian difference dynamics (GDD), a framework for
tracking expected counts and pairwise clustering of vectors through
repeated differencing. A closed recursion for the first two counting
moments controls these quantities relative to a Gaussian reference,
without requiring independence within intermediate lists. Accurate
moment bounds let us extract a Gaussian sample by weighted selection,
while the weaker bounds needed to establish a shortest vector
representation allow us to retain more differences. We perform these
transitions on thin spherical shells, where the Gaussian rejection
weights can be renormalized at subexponential cost. For batch sampling,
we combine this geometry with the double rejection technique of
Doulgerakis and Laarhoven to control the joint distribution of outputs
while reusing vectors.
\end{abstract}

\clearpage
\begingroup
\setcounter{tocdepth}{2}
\normalsize
\tableofcontents
\endgroup
\clearpage

\section{Introduction}\label{sec:intro}

The Shortest Vector Problem (SVP) and discrete Gaussian sampling (DGS)
are basic problems in lattice algorithms, complexity, and cryptography.
For a full rank lattice $L\subset\R^n$, SVP asks for a shortest nonzero
vector, while DGS samples lattice points with probabilities proportional
to $e^{-\pi\|x\|^2/r^2}$ at a prescribed width $r$.
Discrete Gaussians connect sampling with the search for short vectors,
and Gaussian identities give quantitative control over the distributions
produced by repeatedly combining vectors. Geometric sieving exploits the
fact that nearby vectors have a shorter difference, suggesting
algorithms that combine Gaussian distributional control with the
efficiency of geometric search.

\paragraph{The $2^{n/2}$ barrier in provable algorithms.}
Aggarwal, Dadush, Regev, and Stephens-Davidowitz (ADRS) gave an SVP
algorithm using $2^{n+o(n)}$ time and space via Gaussian sampling at
arbitrary width~\cite{ADRSD15}, and Aggarwal and Stephens-Davidowitz
later simplified the algorithm by pairing and averaging~\cite{AS18}.
ADRS also give a faster sampler above smoothing that generates
$2^{n/2}$ independent samples in $2^{n/2+o(n)}$ time and space.
Building on sampling above smoothing, our earlier work uses random
superlattices to reach the shortest vector scale, first by direct
sampling in time $2^{0.7314n+o(n)}$~\cite{GFH26a} and then by geometric
pair search in time $2^{0.5596n+o(n)}$~\cite{GFH26b}.
More recently, Hhan's $q$-ary coset difference tree brings arbitrary
lattice SVP to $2^{n/2+o(n)}$ time and space~\cite{Hhan26}, while Kim
obtains one centered Gaussian sample at any width in expected
$2^{n/2+o(n)}$ time~\cite{Kim26}.

The fast ADRS sampler combines pairs $x,y$ by averaging, $(x+y)/2$.
Before rescaling, it pairs vectors from the same coset of
$L_{i+1}\subset L_i$ and retains their sum, with $2L_i\subseteq L_{i+1}$.
Its lattice tower satisfies
\[
 L_{i+2}\subseteq2L_i
 \quad\Longrightarrow\quad
 [L_i:L_{i+1}]\,[L_{i+1}:L_{i+2}]
 =[L_i:L_{i+2}]\ge2^n.
\]
Balancing the adjacent indices gives $2^{n/2+o(n)}$ cosets per level,
which sets the sampler's list scale~\cite[Proposition~5.7,
Definition~5.8, and Theorem~5.11]{ADRSD15}.
Algorithms that use this batch interface inherit its generation cost,
while Hhan's half exponent arises from a balance between Gaussian
signal, estimation error, and coset table size~\cite[Section~1.1 and
Remark~1.2]{Hhan26}. Improving these bounds calls for revisiting how
intermediate vectors are generated and how much supply the final task
needs. Stephens-Davidowitz asks for a faster centered sampler,
possibly producing only one sample, and writes that
\textquotedblleft entirely new techniques would be needed\textquotedblright{}
to go below $2^{n/2}$~\cite[Section~1.4, p.~7]{SD16}.

\paragraph{Making geometric sieving provable on arbitrary lattices.}
Sieving began with the algorithm of Ajtai, Kumar, and
Sivakumar~\cite{AKS01} and developed through both provable and heuristic
analyses~\cite{NV08,PS09,BGJ15}. Under heuristic assumptions on the
successive lists, the batched BDGL sieve takes
$(3/2)^{n/2+o(n)}$ time and $(4/3)^{n/2+o(n)}$ space~\cite{BDGL16}.
Recent work of Doulgerakis and Laarhoven~\cite{DL26} and Zhou, Wang,
and Wang~\cite{ZWW26} proves these SVP exponents for Haar-random
lattices, using distributional estimates for random lattice geometry
and the generated lists. Doulgerakis and Laarhoven also construct
Gaussian batches at every width.

Extending this efficiency to arbitrary input lattices requires control
of both the geometry and the dependence of the generated vectors.
Repeated differences share inputs and ancestors, while local lattice
geometry governs the numbers of partners of an input and
representations of an output, especially at widths below smoothing.
A rigorous iteration must therefore follow the supply at the intended
locations together with the correlations that affect its availability.

We combine Gaussian distributional control with geometric sieving to
obtain algorithms on every lattice. Above smoothing, Gaussian sampling
attains the BDGL time and space exponents, while SVP and centered
sampling at arbitrary width take $2^{0.4130n+o(n)}$ time and
$2^{0.3138n+o(n)}$ space. We first describe the counting framework behind the single sample
and shortest vector constructions, then state their guarantees and the
stronger batch result.

\subsection{Gaussian difference dynamics}

We start with wide Gaussian samples and repeatedly form weighted
short differences to reach a smaller Gaussian width. Once the
vectors reach a suitable scale, we use the resulting lists to
produce the desired output. For DGS above smoothing, one weighted
selection gives a Gaussian sample. For SVP, we first build additional
representations and then use BDGL filters to search for short
differences between two terminal lists. The final task thus
determines what guarantees the intermediate lists must provide.

The main difficulty is to control the lists produced by this
iteration. Differences sharing an input are correlated, and later
rounds inherit dependence through common ancestors. The heuristic
BDGL sieve analyzes successive lists under assumptions on their
angular distribution~\cite{BDGL16}. The ADRS Gaussian combiner
instead maintains a joint distribution close to independent
Gaussian samples through successive rounds~\cite{ADRSD15}.
Retaining many differences reuses the same inputs, so an analysis
of this iteration must account for the resulting dependence.

We track two counting moments relative to a Gaussian reference:
the expected number of records in a region and the expected number
of ordered pairs of distinct records in two regions. The first
describes where records appear, while the second controls their
clustering. At each merge, conditional independence between the
two separately generated child lists gives a closed recursion
for these moments, even though either list may be internally
correlated. These bounds connect expected counts to the probability
of producing useful vectors and to the work needed to find them.
We call this evolution \emph{Gaussian difference dynamics}.\footnote{Dynamical systems have
also been used in the analysis of lattice basis
reduction~\cite{HPS11,LN25,Wang24}.}

Thin spherical shells let us renormalize Gaussian rejection weights
at subexponential cost. On a fixed superlattice smooth at all midpoint
widths, the Gaussian midpoint identity gives the reference counts
for each cooling round. Geometric search finds the pairs needed
for the next list.

The final task determines the required moment accuracy. For Gaussian
sampling, weighted selection divides by the list's random total
weight. Accurate moments control this normalization when the
reference total weight is sufficiently large. Sparse activation
followed by isolation preserves this precision, retaining a pair
only when no other activated pair shares either input.

For SVP, it suffices that a shortest vector is represented with
probability $e^{-o(n)}$, since repetition then costs only a
subexponential factor. This weaker requirement permits a fixed
number of full differencing layers that retain every eligible pair,
building more representations before the final search. These layers
keep the reference width fixed and lower the shell energy. We compare
two terminal lists in a common random affine coset and select
differences that belong to the original lattice.

An exponential batch additionally requires the joint output law
to be close to a product of target Gaussians. We combine the same
shell geometry with the two rejection corrections of Doulgerakis
and Laarhoven~\cite{DL26}. At each merge, these corrections make
the complete trial law, including failure, independent of the
reused partner list. Uniform estimates for these lists and mixing
across radial bins then give the batch guarantee.
\Cref{sec:overview} develops these constructions.

The counting framework has also been applied to $\ell_\infty$-SVP
with uniform reference counts on cubes~\cite{GDDInfinity}.

\subsection{Main results}

For a rational lattice $L=B\Z^n$, write
\[
 \rho_\sigma(x)=e^{-\pi\|x\|^2/\sigma^2},\qquad
 D_{a+L,\sigma}(x)=\frac{\rho_\sigma(x)}{\rho_\sigma(a+L)}.
\]
The smoothing parameter is defined by
$\rho_{1/\eta_\epsilon(L)}(L^*\setminus\{0\})=\epsilon$.
We write $\mathcal L(X)$ for the law of $X$ and $\TV$ for total
variation distance.
Write
\[
 \dgstime=\tfrac12\log_2(3/2),\qquad
 \dgsspace=\tfrac12\log_2(4/3),\qquad
 J(z)=\frac{z-1-\log z}{2\log2}.
\]
Our algorithms use a randomized word RAM, and their time and bit space
bounds hold on every execution, including random bit generation and
finite arithmetic. We suppress polynomial factors in the binary input
length $\mathcal B$ and the accuracy parameter $f$, when present.
The computational model and finite implementation are given in
\cref{app:computational-model,prop:finite-realization}.

Our first application of GDD is a Gaussian sampler. Controlling the
first two counting moments of the correlated list suffices to obtain
one Gaussian sample by weighted selection, at the BDGL time and space
exponents.

\Needspace{11\baselineskip}
\begin{theorem}[Single sample DGS above smoothing]\label{thm:single-main}
Fix $\alpha>1$ and $\xi>0$. There is a classical randomized algorithm
that, given a full rank rational lattice $L$, a rational shift $a$,
and a positive rational squared width $\sigma^2$ with
$\sigma\ge\alpha\eta_{1/2}(L)$, outputs $X\in a+L$ with
\[
 \TV(\mathcal L(X),D_{a+L,\sigma})
 \le\exp[-\Omega(n/\log^{2+\xi}(n+2))].
\]
Its time is $2^{\dgstime n+o(n)}$ and its bit space is
$2^{\dgsspace n+o(n)}$.
\end{theorem}

\Cref{sec:single-dgs} proves this result. The constants may depend on
$\alpha$ and $\xi$.

To obtain an exponential batch within the time and space bounds of
\cref{thm:single-main}, we combine the Gaussian geometry above with the double
rejection technique of Doulgerakis and Laarhoven~\cite{DL26}.
This gives the stronger joint sampling guarantee below.

\begin{theorem}[Batch DGS above smoothing]\label{thm:batch-main}
Fix $\alpha>1$ and $0<b<\min\{\dgsspace,J(\alpha^{-2})\}$.
Given a full rank rational lattice $L=B_L\Z^n$, a rational shift $a$,
and a positive rational squared width $\sigma^2$ satisfying
$\sigma\ge\alpha\eta_{1/2}(L)$, an algorithm outputs
$B=\lfloor2^{bn}\rfloor$ points in $a+L$ with
\[
 \TV\bigl(\mathcal L(X_1,\ldots,X_B),
                D_{a+L,\sigma}^{\otimes B}\bigr)
 \le\exp[-\Omega(n/\log^2(n+2))].
\]
Its total time is $2^{\dgstime n+o(n)}$ and its bit space, including
the output, is $2^{\dgsspace n+o(n)}$.
If $\alpha\ge3/2$, one may take the full batch
$B=\lfloor2^{\dgsspace n}\rfloor$.
\end{theorem}

The single sample construction gives a direct application of the counting
moment framework, while the batch guarantee requires additional control
of the joint output law. Retaining the first batch output also gives a single sample with this
stronger error bound and the same BDGL exponents
$\dgstime\approx0.2925$ and $\dgsspace\approx0.2075$.

In the full batch case, shared preprocessing gives average time
$2^{\frac12\log_2(9/8)n+o(n)}$ per output.
\Cref{sec:batch-dgs} proves the joint law and total resource bounds.

\Needspace{10\baselineskip}
\begin{theorem}[SVP]\label{thm:svp-main}
Given a full rank rational lattice $L=B\Z^n$, there is a classical
randomized algorithm that returns a shortest nonzero vector with
probability at least $1-2^{-\Omega(n\log n)}$, in
\[
 T=2^{0.4130n+o(n)}\quad\text{time},\qquad
 S=2^{0.3138n+o(n)}\quad\text{space}.
\]
Every output is a nonzero vector of $L$.
\end{theorem}

Stephens-Davidowitz~\cite[Theorem~4.6]{SD16} gave a reduction from
centered DGS at arbitrary width to SVP, with polynomial expected
overhead and no increase in dimension. Applying this reduction to
\cref{thm:svp-main} gives a single Gaussian sample with the same
time and space exponents, answering his question about sampling
below $2^{n/2}$.

We use his notion of accuracy: a law is
$(\gamma,\varepsilon)$-close to the target if it
is within total variation distance $\varepsilon$ of a law whose
pointwise probabilities differ from the target by at most a factor
$\gamma$ (\cref{sec:cdgs-reduction}).

\Needspace{9\baselineskip}
\begin{corollary}[Centered Gaussian sampling at arbitrary width]
\label{cor:cdgs-main}
Let $f:\mathbb N\to\mathbb Z_{\ge1}$ be efficiently computable, with
$1\le f(n)\le\poly(n)$. For a full rank rational lattice $L=B\Z^n$ and
any positive rational width $\sigma$, one can sample $X\in L$ whose law
is $(1+1/f(n),2^{-f(n)})$-close to $D_{L,\sigma}$. The algorithm uses
\[
 T=2^{0.4130n+o(n)}\quad\text{time},\qquad
 S=2^{0.3138n+o(n)}\quad\text{space}
\]
per sample.
\end{corollary}

The total variation error is $O(1/f(n))+2^{-f(n)}$, permitting any
prescribed inverse polynomial accuracy without changing the leading
time or space exponent.

The comparison of direction weights in the proof of~\cite[Theorem~4.6]{SD16}
does not go through as stated for the additive grid of squared radii.
We obtain the required control using a multiplicative grid and
complete the proof within the original reduction framework.
\Cref{app:cdgs-oracle} gives the counterexample and the repaired argument.

\begin{table}[!ht]
\centering
\small
\setlength{\tabcolsep}{4pt}
\begin{tabularx}{\textwidth}{@{}Xlll@{}}
\toprule
Problem & Guarantee & Time exponent & Space exponent\\
\midrule
Single Gaussian sample & $\alpha>1$ & $\dgstime$ & $\dgsspace$\\
Full Gaussian batch & $\alpha\ge3/2$ & $\dgstime$ & $\dgsspace$\\
Reusable BDD & fixed $r<0.3157071$ & $P:\dgstime,\ D:\dgsspace$ & $\dgsspace$\\
Exact $3$-uSVP & $\lambda_2\ge3\lambda_1$ & $\dgstime$ & $\dgsspace$\\
$4.15$-GapSVP & promise decision & $\dgstime$ & $\dgsspace$\\
SVP & arbitrary lattice & $0.4130$ & $0.3138$\\
Centered DGS & every width & $0.4130$ & $0.3138$\\
$4$-CVP, SMP, SIVP & constant approximation & $0.4611$ & $0.3138$\\
$\gamma$-CVP, SMP, SIVP & $\gamma(n)=\omega(1)$ & $0.4130$ & $0.3138$\\
$O(\sqrt{n\log n})$-SIVP & near smoothing sampling & $\dgstime$ & $\dgsspace$\\
\bottomrule
\end{tabularx}
\caption{Algorithmic consequences on arbitrary rational lattices.
Time and bit space entries mean $2^{cn+o(n)}$, with polynomial
input factors suppressed. $\dgstime\approx0.2925$ and
$\dgsspace\approx0.2075$. BDD has preprocessing $P$ and per-target work
$D$.}
\label{tab:algorithmic-results}
\end{table}

\begin{table}[!ht]
\centering
\small
\setlength{\tabcolsep}{4pt}
\begin{tabularx}{\textwidth}{@{}Xllrr@{}}
\toprule
Reference & Input model & Guarantee & Time & Space\\
\midrule
Ajtai--Kumar--Sivakumar~\cite{AKS01} & Arbitrary & Randomized & $O(1)$ & $O(1)$\\
Pujol--Stehl\'e~\cite{PS09} & Arbitrary & Randomized & $2.465$ & $1.233$\\
Micciancio--Voulgaris~\cite{MV13} & Arbitrary & Deterministic & $2$ & $1$\\
Aggarwal--Dadush--Regev--Stephens-Davidowitz~\cite{ADRSD15} & Arbitrary & Randomized & $1$ & $1$\\
Gao--Feng--Hu~\cite{GFH26a} & Arbitrary & Randomized & $0.7314$ & $0.5$\\
Hhan (Hessian)~\cite{Hhan26a} & Arbitrary & Randomized & $0.6039$ & $0.5$\\
Gao--Feng--Hu~\cite{GFH26b} & Arbitrary & Randomized & $0.5596$ & $0.5$\\
Hhan (coset tree)~\cite{Hhan26} & Arbitrary & Randomized & $0.5$ & $0.5$\\
This work & Arbitrary & Randomized & $0.4130$ & $0.3138$\\
\midrule
Pouly--Shen~\cite{PS26} & Haar-random & Randomized & $0.6327$ & $0.5$\\
Doulgerakis--Laarhoven~\cite{DL26} & Haar-random & Randomized & $0.2925$ & $0.2075$\\
Zhou--Wang--Wang~\cite{ZWW26} & Haar-random & Randomized & $0.2925$ & $0.2075$\\
\midrule
Nguyen--Vidick~\cite{NV08} & Arbitrary & Heuristic & $0.4151$ & $0.2075$\\
Becker--Ducas--Gama--Laarhoven~\cite{BDGL16} & Arbitrary & Heuristic & $0.2925$ & $0.2075$\\
\bottomrule
\end{tabularx}
\caption{Classical SVP bounds, with time and space exponents in
$2^{\alpha n+o(n)}$.}
\label{tab:svp-comparison}
\end{table}

\FloatBarrier

Gaussian batches on the dual lattice give a reusable BDD data
structure whose advice is correct simultaneously for every promised
target. Accounting separately for its preprocessing and query costs
gives exact $3$-uSVP and $4.15$-GapSVP within the BDGL time and space,
with the latter using the smaller batch region of
\cref{thm:batch-main}.

Centered sampling at every width gives a second route. Gaussian small
ball estimates improve the explicit sampling-call exponent in the
Euclidean reduction of Aggarwal, Chen, Kumar, Li, and
Stephens-Davidowitz~\cite{ACKLS21}, cover every fixed approximation
factor greater than one, and give direct SMP and SIVP algorithms in
dimension $n$. Here $\gamma$-SMP returns independent vectors
$v_i$ with $\|v_i\|\le\gamma\lambda_i(L)$, while $\gamma$-SIVP bounds
their maximum length by $\gamma\lambda_n(L)$.
\Cref{sec:reductions} develops each reduction together with its
algorithmic consequences.

\paragraph{Organization.}
\Cref{sec:overview} develops the algorithms. After the preliminaries,
\cref{sec:replicas,sec:lattice} establish the counting framework and
Gaussian cooling, and \cref{sec:search} supplies geometric pair search.
\Cref{sec:single-dgs} completes the single-sample application.
\Cref{sec:bridges,sec:affine-root,sec:svp} then develop full differencing,
terminal coset search, and SVP. \Cref{sec:batch} establishes independent
Gaussian batches, followed by the reductions in \cref{sec:reductions}.
The appendices give the geometric estimates, finite implementations,
and proofs of the reductions.

We used Codex to assist with algorithm design, mathematical analysis,
and manuscript preparation. The
\hyperref[sec:ai-methodology]{AI methodology statement} at the end
of the paper describes its role.

\section{Technique overview}\label{sec:overview}

\subsection{A weighted Gaussian difference}
Write $\rho_r(x)=e^{-\pi\|x\|^2/r^2}$, and fix a desired contraction
$0<\gamma<1$. For two input vectors $x,y$, set $z=x-y$ and $m=(x+y)/2$.
Completion of squares separates their Gaussian weights:
\begin{align}
 \rho_r(x)\rho_r(y)
 &=\rho_{r/\sqrt2}(m)\rho_{\sqrt2r}(z),
 \label{eq:overview-completion}\\
 \rho_r(x)\rho_r(y)
 e^{-\pi(\gamma^{-2}-1/2)\|z\|^2/r^2}
 &=\rho_{r/\sqrt2}(m)\rho_{\gamma r}(z).
 \label{eq:overview-fiber}
\end{align}
The exponential factor in the second identity gives an acceptance rule
that changes the weight of the difference from width $\sqrt2r$ to
width $\gamma r$.

To see the operation directly, fix a lattice $\Gamma$ that is smooth
at the midpoint width $r/\sqrt2$. Consider the following trial:
\begin{enumerate}[leftmargin=*,itemsep=2pt,topsep=4pt]
\item Draw $x,y$ independently from $D_{\Gamma,r}$.
\item Compute $z=x-y$.
\item Accept $z$ with probability
$p_r(z)=e^{-\pi(\gamma^{-2}-1/2)\|z\|^2/r^2}$.
\end{enumerate}
For a fixed $z\in\Gamma$, we multiply each input pair's probability
mass by $p_r(z)$ and sum over all pairs with difference $z$.
Parametrizing these pairs by their midpoints gives
\[
 \Pr[\text{trial accepts }z]
 =\frac{\rho_{\gamma r}(z)}{\rho_r(\Gamma)^2}
   \sum_{m\in z/2+\Gamma}\rho_{r/\sqrt2}(m).
\]
Smoothing makes the midpoint sum nearly independent of its shift.
Consequently, conditional on acceptance, the output is close to
$D_{\Gamma,\gamma r}$. The midpoint records the different ways to
produce $z$.

To turn this single pair operation into an algorithm, we would like
to maintain a list of vectors and repeatedly contract its width,
using the output list at width $\gamma r$ as the input to the next
round. This suggests a layered sieve in the spirit of
BDGL~\cite{BDGL16}. We next choose the list sizes and acceptance
rule so that each round supplies enough vectors for this iteration.

\subsection{Supplying the next round via restriction to thin shells}
\label{sec:overview-shells}
Before rejection, the difference has Gaussian width $\sqrt2r$,
while the accepted difference has width $\gamma r$. At our smoothing
scales, Gaussian mass is proportional to the $n$th power of the
width, so the acceptance probability is
\[
 p_{\rm raw}\approx
 \frac{\rho_{\gamma r}(\Gamma)}{\rho_{\sqrt2r}(\Gamma)}
 \approx\left(\frac\gamma{\sqrt2}\right)^n.
\]
Among $N$ vectors there are order $N^2$ candidate pairs. Supplying
another list of size $N$ therefore calls for
\[
 N\gtrsim p_{\rm raw}^{-1}
 \approx\left(\frac{\sqrt2}\gamma\right)^n
 \ge 2^{n/2}.
\]
Thus simple Gaussian rejection cannot break the $2^{n/2}$ barrier.

Our earlier work~\cite[Sections~2.3 and~3.3]{GFH26b} restricts
Gaussian samples to thin shells before a single BDGL decoding
layer, giving $2^{0.5596n+o(n)}$ SVP time. Nearly equal endpoint
lengths turn the search for short differences into an angular
problem, while Gaussian shell estimates control the retained list
size. To eliminate this exponential rejection loss, we restrict
the inputs and outputs to thin norm shells on which the Gaussian
weight factor is nearly constant and can be normalized away.
We choose these shells so that the resulting differences supply
the next round at a smaller Gaussian width.

Measure squared lengths relative to $nr^2/(2\pi)$ by the normalized
energy
\[
 \en_r(x)=\frac{2\pi\|x\|^2}{nr^2},\qquad
 \mathcal S_r(a)=\{x:\en_r(x)\in[a(1-w_n),a(1+w_n)]\},
\]
where $w_n\to0$ is the relative shell width. A Gaussian of width $r$
has typical energy near $1$ at our smoothing scales, so $a$ specifies
the shell's squared radius relative to that typical value.
Let $R_-$ be the inner radius of the desired output shell and write
\[
 R_-^2=\frac{n(\gamma r)^2a(1-w_n)}{2\pi},\qquad
 A_r=\frac{\pi(\gamma^{-2}-1/2)}{r^2}.
\]
Consider the following trial:
\begin{enumerate}[leftmargin=*,itemsep=2pt,topsep=4pt]
\item Draw $x,y$ independently from
$D_{\Gamma,r}\mid\mathcal S_r(a)$.
\item Compute $z=x-y$. If $z\notin\mathcal S_{\gamma r}(a)$,
return failure.
\item Output $z$ with probability
\[
 \omega_r(z)=\frac{p_r(z)}{p_r(R_-)}
            =e^{-A_r(\|z\|^2-R_-^2)},
\]
and return failure otherwise.
\end{enumerate}
Here $p_r(R_-)=e^{-A_rR_-^2}$ is the value of the radial acceptance
weight at the inner boundary. It is the largest value on the output
shell, so $0<\omega_r(z)\le1$ there. Dividing by it preserves the
relative Gaussian weights and makes
$\omega_r(z)=2^{-O(nw_n)}=2^{-o(n)}$ throughout that shell.

We can now read off the leading acceptance rate from the shell
geometry. Write $u=1-\gamma^2/2$. The input lengths are nearly equal,
and the output shell forces
\[
 \frac{\angles{x,y}}{\|x\|\|y\|}
 =1-\frac{\gamma^2}{2}+O(w_n)=u+O(w_n).
\]
For a fixed endpoint direction, the partner therefore lies in a thin
spherical band of correlation $u$. Its transverse radius is
$\sqrt{1-u^2}$, giving angular measure
$(1-u^2)^{n/2}2^{o(n)}$. At our smoothing scales the shell counts
have this same leading rate, and the normalized Gaussian correction
adds only a subexponential factor. Thus
\[
 p_{\rm shell}=(1-u^2)^{n/2}\,2^{o(n)}.
\]
Two lists of size $N$ offer expected accepted supply of scale
$N^2p_{\rm shell}$. Supplying another list with the same leading
size calls for $N$ slightly above $1/p_{\rm shell}$. Since contraction
requires $u>1/2$, the smallest supply exponent is approached as
$u\downarrow1/2$. This gives the sampling list size
$(4/3)^{n/2+o(n)}$, with exponent about $0.2075$.
This is exactly the space exponent of the heuristic BDGL sieve!

\begin{figure}[!ht]
\centering
\begin{tikzpicture}[>=Latex,font=\small,line cap=round,line join=round]
 \definecolor{shellblue}{RGB}{48,103,151}
 \definecolor{diffgold}{RGB}{174,105,24}
 \definecolor{midgreen}{RGB}{33,122,103}
 \pgfmathsetmacro{\shellR}{1.8}
 \pgfmathsetmacro{\shellInner}{\shellR*sqrt(1-.065)}
 \pgfmathsetmacro{\shellOuter}{\shellR*sqrt(1+.065)}
 \pgfmathsetmacro{\halfZ}{\shellR/sqrt(5)}
 \pgfmathsetmacro{\midH}{2*\shellR/sqrt(5)}
 \pgfmathsetmacro{\outInner}{2*\shellInner/sqrt(5)}
 \pgfmathsetmacro{\outOuter}{2*\shellOuter/sqrt(5)}
 \path[use as bounding box] (-3.5,-2.65) rectangle (11.8,2.7);

 \node at (0,2.48) {(a) One input pair};
 \fill[shellblue!12,even odd rule]
   (0,0) circle (\shellOuter) (0,0) circle (\shellInner);
 \draw[shellblue!65,line width=.45pt] (0,0) circle (\shellInner);
 \draw[shellblue!65,line width=.45pt] (0,0) circle (\shellOuter);
 \fill[diffgold!13,even odd rule]
   (0,0) circle (\outOuter) (0,0) circle (\outInner);
 \draw[diffgold!65,line width=.45pt] (0,0) circle (\outInner);
 \draw[diffgold!65,line width=.45pt] (0,0) circle (\outOuter);
 \coordinate (O) at (0,0);
 \coordinate (X) at (\halfZ,\midH);
 \coordinate (Y) at (-\halfZ,\midH);
 \coordinate (M) at (0,\midH);
 \draw[->,shellblue,line width=.9pt] (O) -- (X);
 \draw[->,shellblue,line width=.9pt] (O) -- (Y);
 \draw[midgreen,densely dashed,line width=.8pt] (O) -- (M);
 \draw[->,diffgold,line width=1pt] (Y) -- (X)
   node[pos=.73,below=2pt] {$z$};
 \draw[->,diffgold,line width=1pt] (O) -- (\midH,0)
   node[midway,below=3pt] {$z=x-y$};
 \draw[black!60] (63.435:.53) arc[start angle=63.435,end angle=116.565,radius=.53];
 \node[fill=white,inner sep=1pt] at (0,.72) {$\theta$};
 \fill (O) circle (1.4pt) node[below left=2pt] {$0$};
 \fill[shellblue] (X) circle (1.7pt) node[above right=1pt] {$x$};
 \fill[shellblue] (Y) circle (1.7pt) node[above left=1pt] {$y$};
 \fill[midgreen] (M) circle (1.7pt) node[below left=3pt] {$m$};
 \node[shellblue,anchor=east] at (-1.62,1.14) {$\mathcal S_r(a_0)$};
 \draw[shellblue!65] (-1.61,1.02) -- (-1.38,1.17);
 \node[diffgold,anchor=west] at (1.62,-1.14) {$\mathcal S_{\gamma r}(a_0)$};
 \draw[diffgold!65] (1.60,-1.04) -- (1.26,-.98);
 \node at (0,-2.34) {$\cos\theta\approx u=1-\gamma^2/2$};

 \begin{scope}[xshift=8.05cm]
  \node at (0,2.48) {(b) Midpoints for a fixed $z$};
  \foreach \side in {-1,1}{
   \fill[shellblue!8,even odd rule]
     (\side*\halfZ,0) circle (\shellOuter)
     (\side*\halfZ,0) circle (\shellInner);
   \draw[shellblue!55,line width=.45pt]
     (\side*\halfZ,0) circle (\shellOuter);
   \draw[shellblue!55,line width=.45pt]
     (\side*\halfZ,0) circle (\shellInner);
  }
  \begin{scope}[even odd rule]
   \clip
     (-\halfZ,0) circle (\shellOuter)
     (-\halfZ,0) circle (\shellInner);
   \fill[midgreen!65,even odd rule]
     (\halfZ,0) circle (\shellOuter)
     (\halfZ,0) circle (\shellInner);
  \end{scope}
  \draw[black!35,densely dotted] (-2.82,0) -- (2.82,0);
  \draw[midgreen,densely dashed,line width=.8pt] (0,0) -- (0,\midH);
  \draw[->,shellblue,line width=.85pt]
    (-\halfZ,0) -- (0,\midH) node[pos=.50,left=3pt] {$x$};
  \draw[->,shellblue,line width=.85pt]
    (\halfZ,0) -- (0,\midH) node[pos=.50,right=3pt] {$y$};
  \fill (0,0) circle (1.1pt);
  \fill[shellblue] (-\halfZ,0) circle (1.5pt) node[below=4pt] {$-z/2$};
  \fill[shellblue] (\halfZ,0) circle (1.5pt) node[below=4pt] {$z/2$};
  \fill[midgreen] (0,\midH) circle (1.8pt) node[above right=2pt] {$m$};
  \node[midgreen] at (-1.55,2.02) {$\en_r(m)\approx 1/2$};
  \node[midgreen,anchor=west] at (.85,2.02) {$\mathcal W_r(z)$};
  \draw[midgreen,line width=.6pt] (.84,1.92) -- (.09,1.68);
  \node[shellblue] at (-1.66,-1.97) {$\mathcal S_r(a_0)-z/2$};
  \node[shellblue] at (1.66,-1.97) {$\mathcal S_r(a_0)+z/2$};
  \node at (0,-2.43) {$x=m+z/2,\quad y=m-z/2$};
 \end{scope}
\end{tikzpicture}
\caption{Shell geometry in a planar section, with thickness enlarged for visibility.
Left: two input vectors and their shorter difference.
Right: the shaded region $\mathcal W_r(z)$ contains the admissible midpoints
for a fixed difference $z$.}
\label{fig:shell-midpoints}
\end{figure}
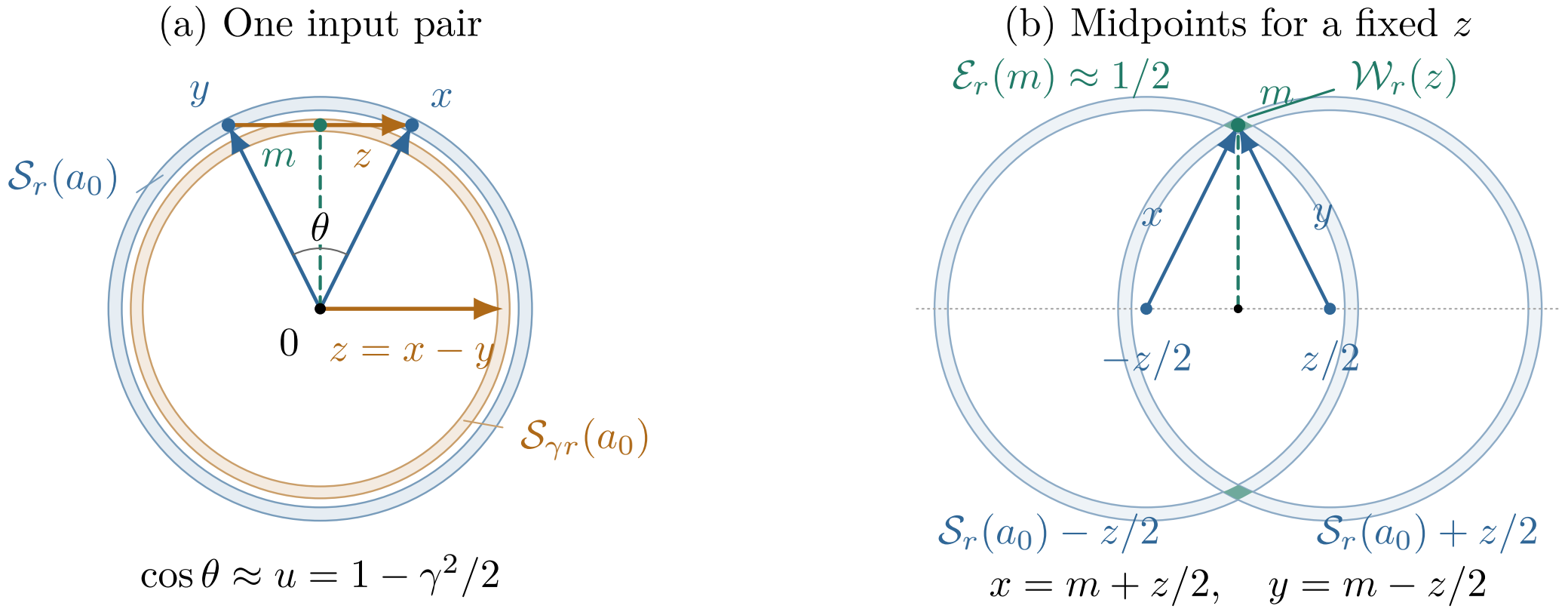

To iterate the trial, we also need its accepted distribution to have
the next Gaussian profile. For $z$ in the output shell,
\eqref{eq:overview-fiber} gives
\[
 \Pr[\text{trial returns }z]\propto
 \rho_{\gamma r}(z)
 \sum_{\substack{m\in z/2+\Gamma\\
                 m+z/2,m-z/2\in\mathcal S_r(a)}}
       \rho_{r/\sqrt2}(m).
\]
The endpoint and output shells put these midpoints at energy
$\en_r(m)\approx a(1+u)/2$. The midpoint Gaussian in the sum has
typical energy $1/2$, which suggests the choice
\[
 a=a_0=\frac1{1+u}.
\]
\Cref{fig:shell-midpoints} illustrates the input pair and its midpoint
region $\mathcal W_r(z)=\{m:m+z/2,m-z/2\in\mathcal S_r(a_0)\}$.
With this choice of $a_0$, the region contains almost all
of the midpoint Gaussian mass. Smoothing then makes the sum nearly
independent of $z$, so the accepted law is close to
$D_{\Gamma,\gamma r}\mid\mathcal S_{\gamma r}(a_0)$.
The same shell parameter can therefore be used in the next round.
\Cref{sec:soft} proves this Gaussian closure using only the input
and output shells.

\FloatBarrier

\subsection{Iterating the Gaussian transition}
\label{sec:overview-dynamics}\label{sec:binary-merge-tree}
The preceding calculation starts from two independent inputs. A list
of differences soon loses this independence: $x-y$ and $x-y'$ share
$x$, and later rounds can share entire chains of ancestors. Even if
each output has the desired Gaussian law, pairing these outputs does
not automatically repeat the single trial calculation.

We therefore organize the contractions as a binary cooling tree.
The two child lists at each merge use disjoint leaf pools and
independent private randomness. Conditional on the common lattice
and prescribed child cosets, the lists are therefore independent
and their expected counts can be multiplied. Each list may still
be internally correlated, which the counting argument below will control.

Generating fresh child lists has an affordable cost because cooling
needs only a shallow tree. If the leaves start at width $R$ and the
root has width $r_0$, then
\[
 t\approx\frac{\log(R/r_0)}{-\log\gamma},\qquad
 \text{number of leaf pools}=2^t.
\]
Basis preprocessing makes $R/r_0$ polynomial in $n$, giving
$t=O(\log n)$ for fixed $\gamma$, as used for SVP. Near the BDGL
endpoint, the sampling schedule lets $\gamma$ approach $1$ slowly
enough that $t=O(\log n\log\log n)$
(\cref{lem:endpoint-schedule}). Thus all the independent leaf pools
together add only a factor $2^{o(n)}$ to the cost of one pool.

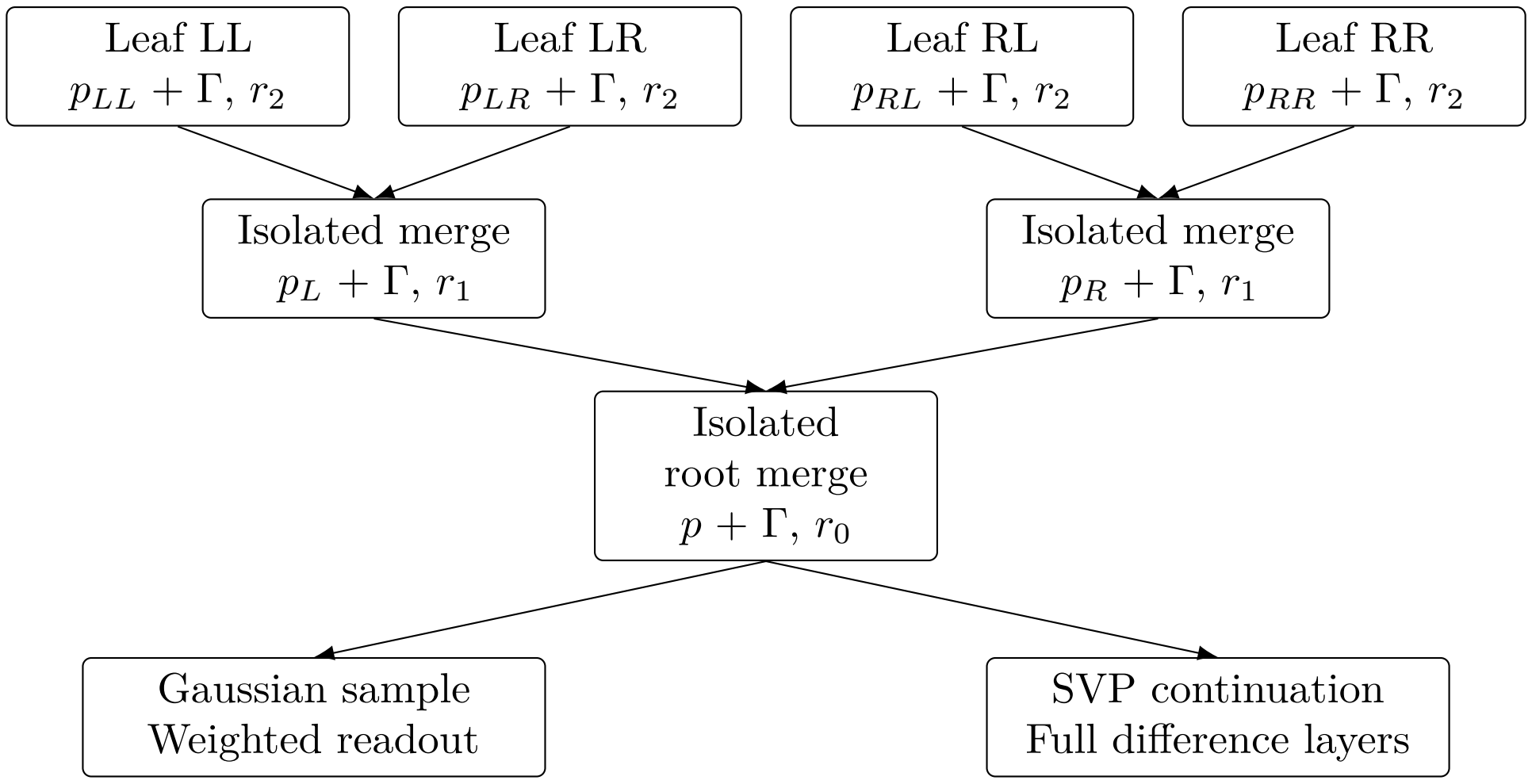
\begin{figure}[!ht]
\centering
\begin{tikzpicture}[
 >=Latex,
 treebox/.style={draw,rounded corners=2pt,align=center,text width=2.6cm,
                 minimum height=0.78cm,font=\small},
 node distance=6mm and 4mm]
 \node[treebox] (ll) {Leaf LL\\$p_{LL}+\Gamma$, $r_2$};
 \node[treebox,right=of ll] (lr) {Leaf LR\\$p_{LR}+\Gamma$, $r_2$};
 \node[treebox,right=of lr] (rl) {Leaf RL\\$p_{RL}+\Gamma$, $r_2$};
 \node[treebox,right=of rl] (rr) {Leaf RR\\$p_{RR}+\Gamma$, $r_2$};
 \node[treebox,below=of $(ll.south)!0.5!(lr.south)$] (ml)
       {Isolated merge\\$p_L+\Gamma$, $r_1$};
 \node[treebox,below=of $(rl.south)!0.5!(rr.south)$] (mr)
       {Isolated merge\\$p_R+\Gamma$, $r_1$};
 \node[treebox,below=of $(ml.south)!0.5!(mr.south)$] (root)
       {Isolated root merge\\$p+\Gamma$, $r_0$};
 \node[treebox,below left=8mm and 4mm of root,text width=3.6cm] (readout)
       {Gaussian sample\\Weighted readout};
 \node[treebox,below right=8mm and 4mm of root,text width=3.6cm] (svp)
       {SVP continuation\\Full difference layers};
 \draw[->] (ll.south) -- (ml.north);
 \draw[->] (lr.south) -- (ml.north);
 \draw[->] (rl.south) -- (mr.north);
 \draw[->] (rr.south) -- (mr.north);
 \draw[->] (ml.south) -- (root.north);
 \draw[->] (mr.south) -- (root.north);
 \draw[->] (root.south) -- (readout.north);
 \draw[->] (root.south) -- (svp.north);
\end{tikzpicture}
\caption{A depth-two isolated cooling tree, with $r_0=\gamma r_1=\gamma^2r_2$.
Under the respective parameter schedules, the root supplies weighted
Gaussian readout or the full differencing layers used for SVP.}
\label{fig:cooling-merge-tree}
\end{figure}

The random superlattice construction of~\cite{GFH26a} supplies
$\Gamma\supset L$ smooth at the smallest midpoint width and hence
at every wider level. Independent Klein/GPV proposals, which generate
lattice vectors at a sufficiently wide Gaussian scale, are restricted
to the leaf shells to initialize the tree~\cite{Klein00,GPV08}.
Random child cosets preserve the prescribed parent difference
coset while averaging midpoint locations for the later representation
and search estimates (\cref{sec:replicas,sec:soft}).
In \cref{fig:cooling-merge-tree}, offsets are assigned from the
root toward the leaves, and records are generated in the reverse direction.

We now have independent left and right supply at every merge without
changing the leading exponent. To make the resulting list useful
at the next merge, we must also control how its records cluster.

\subsection{Controlling correlations by isolated pairing}
The tree makes the two input lists independent, so their expected
counts determine the expected supply of differences. A large expected
supply can still come from very few executions. If a region $A$ has
Gaussian probability $p$, both $N$ independent samples and $N$ copies
of one sample give $\E[N_A]=Np$. Yet their probabilities of hitting
$A$ are $1-(1-p)^N$ and $p$, respectively. A large expected count
can therefore come entirely from rare lists containing many copies.
A second count distinguishes these behaviors.

Let $G_j$ be the reference Gaussian law on the shell and coset at
round $j$, and let $M_j$ be its reference list size. Write $N_A$ for
the number of records in $A$ and $N^{(2)}_{A,B}$ for the number of
ordered pairs of distinct records in $A\times B$. Different
occurrences of the same vector remain distinct records. With the
lattice and prescribed affine data fixed, we seek
\begin{equation}\label{eq:overview-factorial-invariant}
 \begin{aligned}
 e^{-\varepsilon_j}M_jG_j(A)
 &\le\E[N_A]\le e^{\varepsilon_j}M_jG_j(A),\\
 \E[N^{(2)}_{A,B}]
 &\le e^{\varepsilon_j}M_j^2G_j(A)G_j(B).
 \end{aligned}
\end{equation}
The first bound describes where the supply should appear. The second
limits how often two records appear together, relative to two
independent reference records, giving a quantitative bound on clustering.

The ancestry of two output records explains why these two counts
are enough. Their input pairs either use four distinct child records,
share a left record, or share a right record. In the first case,
independence of the child lists multiplies their two factorial
counts. A shared left record instead uses a first count on the left
and a factorial count on the right, with the roles reversed for a
shared right record. All three contributions thus depend only on
the first two child moments. Repeated differencing therefore admits
a closed recursion for just these two counting moments. This is what
lets us follow an entire tree of correlated lists without describing
their joint laws. We call this evolution of the reference counts and
their moment bounds \emph{Gaussian difference dynamics} (GDD).

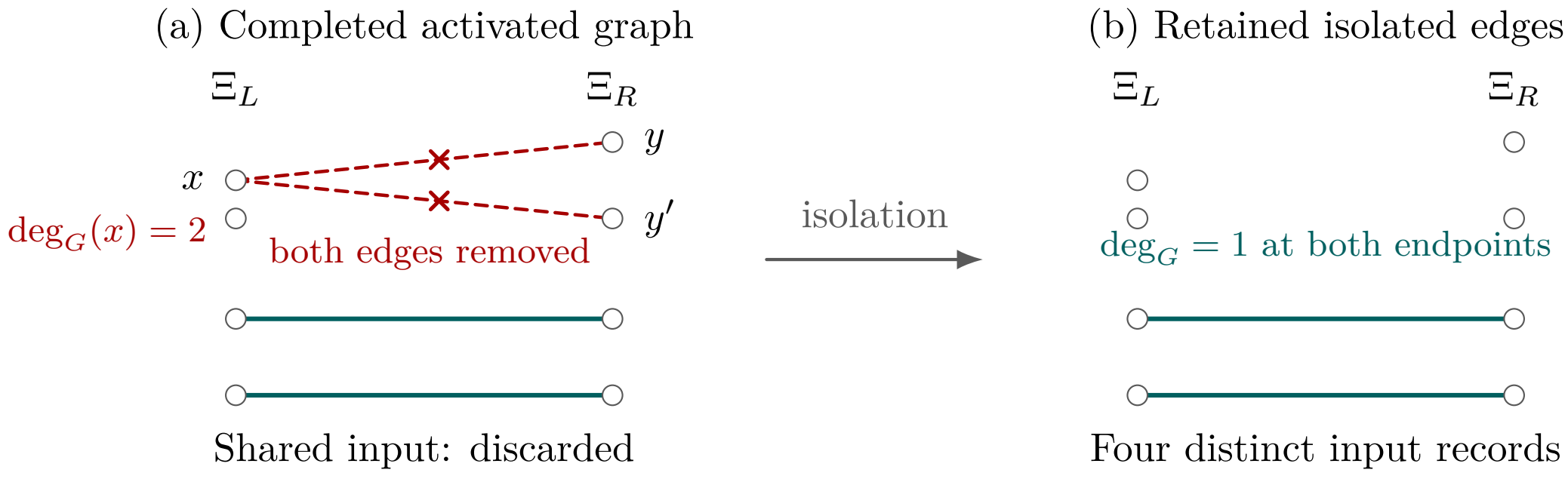
\begin{figure}[!ht]
\centering
\begin{tikzpicture}[>=Latex,font=\small,line cap=round,
 record/.style={circle,draw=black!65,fill=white,inner sep=0pt,minimum size=5pt},
 kept/.style={draw=teal!75!black,line width=1.15pt},
 removed/.style={draw=red!65!black,densely dashed,line width=.85pt}]
 \path[use as bounding box] (-1.3,-3.03) rectangle (12.5,1.05);
 \node at (1.65,.85) {(a) Completed activated graph};
 \node at (9.55,.85) {(b) Retained isolated edges};
 \foreach \shift/\prefix in {0/a,7.9/b}{
  \begin{scope}[xshift=\shift cm]
   \node at (0,.32) {$\Xi_L$};
   \node at (3.3,.32) {$\Xi_R$};
   \foreach \i/\height in {1/-.15,2/-.82,3/-1.70,4/-2.37}{
    \coordinate (\prefix L\i) at (0,\height);
    \coordinate (\prefix R\i) at (3.3,\height);
   }
   \coordinate (\prefix L1) at (0,-.485);
  \end{scope}
 }
 \draw[removed] (aL1) -- (aR1);
 \draw[removed] (aL1) -- (aR2);
 \foreach \target in {aR1,aR2}{
  \coordinate (cross) at ($(aL1)!.54!(\target)$);
  \draw[red!65!black,line width=.95pt]
    ($(cross)+(-.075,-.075)$) -- ($(cross)+(.075,.075)$)
    ($(cross)+(-.075,.075)$) -- ($(cross)+(.075,-.075)$);
 }
 \foreach \prefix in {a,b}{
  \draw[kept] (\prefix L3) -- (\prefix R3);
  \draw[kept] (\prefix L4) -- (\prefix R4);
  \foreach \i in {1,2,3,4}{
   \node[record] at (\prefix L\i) {};
   \node[record] at (\prefix R\i) {};
  }
 }
 \node[left=4pt] at (aL1) {$x$};
 \node[right=4pt] at (aR1) {$y$};
 \node[right=4pt] at (aR2) {$y'$};
 \node[red!65!black,anchor=east,align=right,font=\footnotesize]
   at (-.12,-.94) {$\deg_G(x)=2$};
 \node[red!65!black,font=\footnotesize] at (1.70,-1.13) {both edges removed};
 \draw[->,black!65,line width=.8pt] (4.65,-1.18) -- (6.55,-1.18)
   node[midway,above=4pt] {isolation};
 \node[teal!75!black,font=\footnotesize] at (9.55,-1.08)
   {$\deg_G=1$ at both endpoints};
 \node at (1.65,-2.86) {Shared input: discarded};
 \node at (9.55,-2.86) {Four distinct input records};
\end{tikzpicture}
\caption{Isolation in \textsc{CoolingMerge}. Keep exactly the edges whose
endpoints both have degree one in the completed activated graph.
Any two surviving edges use four distinct input records, eliminating
the shared input terms in the second factorial moment at this merge.}
\label{fig:isolated-edges}
\end{figure}

For the small errors needed through many cooling rounds, we can
simplify the recursion by removing the two shared input cases.
View eligible pairs as edges of a bipartite graph on the child
records. Keeping only edges that share neither endpoint with another
edge leaves exactly the four distinct input case. To keep this
operation from discarding most of the supply, we first activate
edges sparsely. Multiplying the Gaussian pair weight by a common
factor $c_r$ reduces the density without changing its reference
profile. These observations give the following ideal
\textsc{CoolingMerge} at input width $r$:
\begin{enumerate}[leftmargin=*,itemsep=2pt,topsep=4pt]
\item List all input record pairs $(x,y)$ whose difference lies in
$\mathcal S_{\gamma r}(a_0)$.
\item Activate each pair with probability $c_r\omega_r(x-y)$,
using the weight from \cref{sec:overview-shells}.
\item In the completed activated graph, keep only edges whose two
endpoints both have degree one, and output their differences.
\end{enumerate}
As \cref{fig:isolated-edges} illustrates, the surviving pair count
is bounded by the child factorial moments, including dependence
inherited from earlier ancestors.

The first moment survives for a different reason. Choose $c_r$ so
that the activated weighted reference degrees have a small budget
$\theta$. Every deleted edge has an activated competitor, and the
same child moment bounds count such competing pairs. Charging lost
edges to these pairs bounds the expected isolation loss by an
$O(\theta)$ fraction of the activated supply. Thus isolation gives
a sharper second moment while changing the Gaussian first moment
by only a small relative error.

\Needspace{9\baselineskip}
The choice of $\theta$ balances moment accuracy against the number
of retained records. Moment errors roughly double at each merge,
while smaller activation probabilities reduce the expected output
count. The main requirements are
\[
 2^t\theta=o(1),\qquad t\log_2(1/\theta)=o(n).
\]
The first keeps the accumulated isolation error small. The second
allows us to compensate for thinning by increasing the initial
list size by only a subexponential factor. Both requirements can
be met at depth $t=O(\log n\log\log n)$.
\Cref{sec:lattice} combines this choice with the geometric error
budget.

We find eligible pairs with BDGL product codes~\cite{BDGL16}, a structured
family of spherical cap filters, using the geometric search interface
of~\cite{GFH26b}. A cap of threshold
$u$ has measure $(1-u^2)^{n/2+o(n)}$, so a list of size
$N=(1-u^2)^{-n/2+o(n)}$ has subexponential expected cap occupancy.
A target pair of correlation $u$ meets in a common cap with probability
$\bigl(1-2u^2/(1+u)\bigr)^{n/2+o(n)}$, whose reciprocal gives
\[
 T_{\rm search}(u)
 =\left(1-\frac{2u^2}{1+u}\right)^{-n/2+o(n)}.
\]
Inserting the lists into the filters also fits this rate, and
\cref{sec:search} bounds the
candidate reports for our correlated lists at the same rate.
The search recovers the activated competitors needed for isolation.
As $u\downarrow1/2$, the time approaches $(3/2)^{n/2+o(n)}$.

\subsection{One Gaussian sample by weighted readout}
Cooling gives a correlated list whose counting moments follow a
Gaussian reference. To extract one sample, with a possible coset
restriction or width change, assign each record a weight
$b(x)\in[0,1]$. For root reference measure $\nu$, the target law is
\[
 \pi=\frac{b\nu}{M},\qquad M=\nu(b)>0.
\]
We call $M$ the \emph{effective mass}. Given the completed list
$\Xi=(X_i)$, let $W=\sum_i b(X_i)$ and choose record $i$ with
probability $b(X_i)/W$, returning a prescribed fallback point if $W=0$.
The difficulty is this random normalization. For a region $A$, write
\[
 Z_A=\sum_i b(X_i)\ind_{X_i\in A}.
\]
The output probability involves $\E[Z_A/W;\,W>0]$, rather than
the ratio of expected counts. The numerator and denominator come
from the same correlated list, so we use the two maintained moments
to control their ratio directly.

Writing $\varepsilon$ for the root moment error, those bounds give
\[
 \E[Z_A]\ge e^{-\varepsilon}M\pi(A),\qquad
 \E[Z_AW]\le e^\varepsilon M\pi(A)(M+1).
\]
The factorial bound controls distinct records in $Z_AW$, while
$b^2\le b$ controls its diagonal terms, producing the extra $1$.
Cauchy--Schwarz then gives
\[
 \E[Z_A/W;\,W>0]
 \ge\frac{(\E[Z_A])^2}{\E[Z_AW]}
 \ge e^{-3\varepsilon}\frac{M}{M+1}\,\pi(A).
\]
Since this holds for every event $A$ and the fallback adds only
nonnegative mass, \cref{thm:weighted-readout} gives
\[
 \TV\bigl(\mathcal L(X_{\mathrm{out}}),\pi\bigr)
 \le3\varepsilon+\frac1{M+1}.
\]
Thus moment accuracy and effective mass determine the readout error,
without requiring independence within the list.

For the requested Gaussian on $a+L$ at width $\sigma$, choose
terminal cooling width $r=\sigma/\sqrt{a_0}$, so the root shell
is centered at the typical target radius. On a root record
$x\in a+\Gamma$, use
\[
 b_{\rm out}(x)=\ind_{x-a\in L}
 e^{-\pi(\sigma^{-2}-r^{-2})(\|x\|^2-R_-^2)},
\]
where $R_-^2$ is the shell's inner squared radius. This weight
restricts to $a+L$ and changes width $r$ to $\sigma$. Anchoring at
$R_-^2$ keeps it in $[0,1]$, with normalized law equal to the target
Gaussian conditioned on the shell.

It remains to check that these corrections leave enough effective
mass. The radial weight varies by only $2^{o(n)}$ on the thin shell,
while restriction to $L$ costs a factor of $[\Gamma:L]$. With cooling
list exponent $P(u)=-\tfrac12\log_2(1-u^2)$, this gives
\[
 \nu(b_{\rm out})
 =\frac{\nu(\R^n)}{[\Gamma:L]}\,2^{o(n)}
 =2^{(P(u)-\iota)n+o(n)}.
\]
For the width promise $\sigma\ge\alpha\eta_{1/2}(L)$, our extension
has index rate
$\iota=\max\{0,\tfrac12\log_2(2a_0/\alpha^2)\}$, and
\[
 P(u)-\iota\ge\min\{\dgsspace,\log_2\alpha\}>0.
\]
This is the role of $\alpha>1$: after paying for the smoothing
extension, an exponential reserve remains for the weighted readout.
Gaussian concentration controls the omitted shell tail. Letting
$u\downarrow1/2$ gives time $(3/2)^{n/2+o(n)}$ and space
$(4/3)^{n/2+o(n)}$ (\cref{sec:single-dgs}). We next use the weaker
accuracy requirement of target existence to build more representations
for SVP.

\subsection{Representations of a shortest vector}
At the shortest vector scale, restricting a cooled list back to $L$
can leave too few records. We instead search for a shortest vector
among differences of two terminal lists. For the analysis at the
correct scale, fix a shortest vector $v$. Success depends on both
the number of records at each endpoint location and the number
of admissible positions $(x,x-v)$ at which the target can be
represented.

Let $A$ be the reference expected number of records at each endpoint
location and $\mu$ the average number of admissible positions.
The two lists are generated independently conditional on their
common lattice and affine data, so the reference target count is
\[
 \Lambda=A^2\mu.
\]
This factorization guides the construction. Full differencing
builds the endpoint intensity $A$, while the random affine geometry
controls the number of positions counted by $\mu$ and its second
moment.

Place both terminal lists on a common random affine coset $C$
of the superlattice $\Gamma$. If $N_\pm(x)$ count occurrences
in the two lists, the actual target count is
\[
 Z_v=\sum_xN_+(x)N_-(x-v).
\]
Quotient labels restrict differences to the original lattice:
$\lab(x)=\lab(y)$ exactly when $x-y\in L$.
One ideal search trial therefore has the following form:
\begin{enumerate}[leftmargin=*,itemsep=2pt,topsep=4pt]
\item Draw the superlattice $\Gamma$ and common affine coset $C$.
\item Conditional on this environment, independently generate two
terminal lists on the prescribed shell in $C$.
\item Return the shortest nonzero difference between equally labeled
records, or failure if there is none.
\end{enumerate}

To increase $A$, we use the weaker accuracy needed for target
existence. Sampling required logarithmic moment error $o(1)$,
whereas target existence allows error $o(n)$ while retaining
success probability $e^{-o(n)}$, provided the reference target
count is large enough. On a thin shell, Gaussian weights vary
by only $e^{O(nw_n)}=e^{o(n)}$, so replacing the Gaussian reference
by a constant intensity fits this error allowance. After isolated
cooling, a fixed number of full differencing layers retain every
eligible pair, with no sparse activation or isolation. Each layer
keeps the width $r$ fixed and lowers the shell energy. If $B$ is
the reference number of midpoint representations for one output,
then
\[
 A_{\rm new}=A_{\rm old}^2B.
\]
The shared input contributions from \cref{sec:replicas} have
relative size of order $1/(A_{\rm old}B)$. Thus sufficiently large
input counts both increase the endpoint intensity and control
the additional moment error caused by reuse.

The geometric factor $\mu$ comes from the same midpoint picture
as in \cref{sec:overview-shells}. For endpoints of energy $a$ and
a difference of energy $b$, the parallelogram identity gives
\[
 \en_r(m)\approx a-\frac b4=\frac E4,\qquad E=4a-b.
\]
The midpoint region therefore has volume
$r^n(eE/4)^{n/2+o(n)}$. In an internal full layer, two such counts
share the same random half grid shift, and propagating their moment
bound requires $E>1$ (\cref{sec:bridges}).

For the terminal target, write $c=\en_r(v)$ and $E=4a-c$.
Here a random common affine coset lets us average over a denser
ambient lattice, extending the affine slicing method
of~\cite[Section~6]{GFH26b}. This average counts the midpoint volume
at density $1/\det\Gamma$. Our chosen scale has
$r^n/\det\Gamma=2^{n/2+o(n)}$, so
\[
 \mu=\frac{r^n}{\det\Gamma}(eE/4)^{n/2+o(n)}
     =(eE/2)^{n/2+o(n)}.
\]
Thus $E>2/e$ already supplies exponentially many positions on
average. The terminal step can use smaller endpoint shells than
an internal layer, reducing the final search cost.

To make this average useful, we must also control collisions of
representation positions modulo $L$. The Kabatiansky--Levenshtein
bound gives at most $2^{\kappa_{\rm pack}n+o(n)}R^n$ lattice points
within radius $R\lambda_1(L)$ for fixed $R\ge1$, where
$\kappa_{\rm pack}=0.401$ (\cref{lem:kl-count}). Applied to these
collisions, it gives the leading factor
$2^{(\kappa_{\rm pack}-1/2)n}(ec)^{-n/2}$.
Choosing
\[
 c=\frac{2^{2\kappa_{\rm pack}-1}}e
\]
balances these factors. The resulting geometric second moment,
combined with the list moments, gives $Z_v>0$ with probability
$e^{-o(n)}$ whenever $\Lambda\ge e^{-o(n)}$
(\cref{sec:affine-root}).

The required target count now determines the terminal list size.
Write $2^{L_{\rm term}n+o(n)}$ for that size and
$u=1-c/(2a)$ for the target correlation. Expressing $A$ in terms
of the list size and substituting the geometric count $\mu$ gives
\[
 \frac1n\log_2\Lambda
 =2L_{\rm term}-P(u)-\kappa_{\rm pack}+o(1).
\]
We set the leading rate to zero and enlarge the lists by a
subexponential factor to absorb the moment and geometric losses.
This fixes the list size needed to represent a shortest vector.

Geometric filters then replace the exhaustive terminal comparison.
We choose their cap threshold $\tau$ by $P(\tau)=L_{\rm term}$,
so each filter has subexponential expected occupancy. As in the
cooling search, the reciprocal common cap probability gives time
\[
 T_{\rm term}
 =\left(1-\frac{2\tau^2}{1+u}\right)^{-n/2+o(n)}.
\]
Thus the target count determines the list size, and the list size
determines the filter threshold. With these two balances, letting
$E$ approach $2/e$ gives time exponent
\[
 -\frac12\log_2\!\left(1-2^{2\kappa_{\rm pack}-2}\right)
 =0.412961878\ldots.
\]
\Cref{prop:terminal-infimum} derives the terminal curve. Taking $E$
just above $2/e$ puts the time exponent below $0.4130$.

Earlier list sizes follow backward from this terminal requirement.
The pair supply calculation of \cref{sec:overview-shells} gives
\[
 L_j=2L_{j-1}-P(u_j),\qquad
 L_{j-1}=\frac{L_j+P(u_j)}2,\qquad
 u_j=1-\frac{a_j}{2a_{j-1}}.
\]
This is the same balance that sustained supply through cooling,
now used to work backward from a target. A constant number of full
layers connects the cooling shell to the terminal shell, with all
internal searches below the terminal cost. The largest intermediate
list determines space. \Cref{sec:svp} gives parameters realizing
$2^{0.4130n+o(n)}$ time and $2^{0.3138n+o(n)}$ space, including
subexponential repetition. Finally, the radius mixture in
\cref{sec:cdgs-reduction} converts SVP into centered Gaussian sampling
at every width with the same exponents.

\subsection{Independent Gaussian batches}
The single sample construction reads once from a correlated list,
whereas a batch requires joint control of all outputs. We retain
the Gaussian shell geometry and replace isolated pairing by the
fixed bank corrections of Doulgerakis and
Laarhoven~\cite[Sections~3 and~5]{DL26}. At each merge, reuse the
right child stream as a \emph{bank} $C=(y_i)$ and use fresh left
trials, independent of the bank and of one another.

Write $q(x)$ for the probability that a left trial returns $x$,
and $K(x,y)=\mathbf1_{\mathrm{Eligible}}\omega_r(x-y)$ for the
pair weight. For the fixed bank, put
\[
 A_C(x)=\sum_iK(x,y_i),\qquad
 B_C(z)=\sum_iq(z+y_i)K(z+y_i,y_i).
\]
These are the total partner weight for an input $x$ and the total
representation weight for a difference $z$. The target weight is
$h(z)=\rho_{\gamma r}(z)$ on the prescribed output shell and coset,
and zero elsewhere. Choose common constants $D,c>0$ so that every
good bank satisfies $A_C(x)\le D$ and $B_C(z)\ge ch(z)$ on the
relevant supports. For a successful left trial $x$, continue with
probability $A_C(x)/D$, choose bank record $i$ proportionally to
$K(x,y_i)$, and accept $z=x-y_i$ with probability $ch(z)/B_C(z)$.
Any failed trial or rejection returns failure.
For one input pair with $z=x-y_i$, the returned probability mass is
\[
 q(x)\frac{A_C(x)}D\frac{K(x,y_i)}{A_C(x)}
       \frac{ch(z)}{B_C(z)}
 =\frac{q(x)K(x,y_i)}D\frac{ch(z)}{B_C(z)}.
\]
The first rejection cancels the partner normalization $A_C(x)$.
Summing over all representations of the same $z$ then cancels
$B_C(z)$ in the second rejection:
\[
 \Pr[\text{trial returns }z\mid C]
 =\frac{ch(z)}{DB_C(z)}\sum_iq(z+y_i)K(z+y_i,y_i)
 =\frac cD h(z).
\]
The remaining failure probability, $1-(c/D)\sum_z h(z)$, is also
bank independent. Thus fresh left trials and private choices give
conditionally independent outputs with the same complete law on
every good bank. Averaging over good banks preserves this product
law (\cref{alg:fixed-bank-trial}).

Smoothing and uniform concentration establish the bank bounds with
common $D,c$. An iid reference tree uses this trial law at every
merge, and coupling until the first bad bank transfers its product
law to the construction (\cref{sec:bank-correction,sec:batch-dgs}).

For $B$ outputs, a per-sample tail $\delta$ can cost $B\delta$ in
joint error, so we cover a wider annulus with polynomially many
thin bins. Each bin uses its own smoothing extension, followed by
restriction to $a+L$ and correction to width $\sigma$.
Mixing with probabilities inversely proportional to the known
coefficients gives the Gaussian weight throughout the annulus.
An additional terminal merge supplies the inner bins for a full
batch at $\sigma\ge3\eta_{1/2}(L)/2$.

The list size, correction, and search schedules give total time
$(3/2)^{n/2+o(n)}$ and space $(4/3)^{n/2+o(n)}$. At the $3/2$
width threshold they supply $\lfloor(4/3)^{n/2}\rfloor$ outputs
jointly close to independent target Gaussians. Every fixed width
factor $\alpha>1$ supports the batch rate in
\cref{thm:batch-main}.

\FloatBarrier

\section{Preliminaries and lattice tools}\label{sec:prelim}
Gaussian identities propagate mass through the differences, random
superlattices supply smoothing, and lattice point bounds control both
terminal representations and dual Gaussian mass. Basis preprocessing
provides the initial samples.
All logarithms without a subscript are natural. Constants in asymptotic
notation are uniform over the compact parameter ranges specified below.
\subsection{Ideal analysis and finite realization}\label{sec:ideal-model}
We analyze ideal sampling primitives, exact geometric tests, and exact
probability weights. Lists retain multiplicities, so different records
may store the same vector. The ideal lists are defined independently
of the search codes, allowing their counting moments to control both
target representations and the work needed to find them.

The finite realization result in \cref{sec:finite-realization} transfers
these list size and search estimates to bounded word RAM executions,
with the stated errors and unchanged leading exponents. Record encodings,
data structures, and exceptional execution rules are collected in
\cref{app:conventions}.
We first introduce the counting measures and Gaussian tools used by
the analysis.

\subsection{Counting measures and conditioning}\label{sec:measure-tools}
\label{sec:occurrence-measures}
For nonnegative measures on a countable set, write
$\mu(f)=\sum_xf(x)\mu(x)$, $(b\mu)(x)=b(x)\mu(x)$, and
$(\mu\otimes\nu)(x,y)=\mu(x)\nu(y)$.
Measure inequalities are pointwise, and $T_*$ denotes pushforward.
For $\Delta(x)=(x,x)$, write $\Deltap\mu$ for the diagonal measure.
A reference $\nu=NG$ has intended list size $N$ and probability
profile $G$. We use $\mathcal L(X)$ for the law of $X$ and
$\TV$ for total variation distance.

A record consists of an ID and a vector value. For a random finite ID set
$I$ and values $(X_i)_{i\in I}$, the list's counting measure is
$\Xi=\sum_{i\in I}\delta_{X_i}$.
Thus $\Xi(A)$ counts records in $A$, including multiplicities:
IDs remain distinct even when their values coincide.
Its \emph{intensity} and \emph{second factorial measure} are
\begin{equation}\label{eq:moments}
 m_\Xi=\E[\Xi],\qquad
 S_\Xi=\E\!\left[\sum_{\substack{i,j\in I\\i\ne j}}
                         \delta_{(X_i,X_j)}\right].
\end{equation}
The exclusion is by ID, so for $N_x=\Xi(\{x\})$,
$S_\Xi(x,x)=\E[N_x(N_x-1)]$ can be positive.
The ordinary second moment restores the diagonal:
\begin{equation}\label{eq:ordinary-factorial-moment}
 \E[\Xi^{\otimes2}]=S_\Xi+\Deltap m_\Xi.
\end{equation}
Consequently, for nonnegative weights,
\begin{equation}\label{eq:weighted-factorial-moment}
 \E[\Xi(f)\Xi(g)]=S_\Xi(f\otimes g)+m_\Xi(fg).
\end{equation}
This correction enters both target counts and weighted selection.

\paragraph{Conditioning.}\label{sec:measure-conditioning}
Conditional moments replace the expectations in \eqref{eq:moments}
by $\E[\cdot\mid\mathcal H]$, where $\mathcal H$ is the retained
information. We form products of child moments with their common
environment fixed, then average that environment. This order preserves
the dependence introduced by shared shifts.
\Cref{tab:conditioning} specifies the retained information for each use.

\subsection{Gaussian scales and complexity parameters}
For an affine lattice $C$, let $\rho_t(C)=\sum_{x\in C}\rho_t(x)$.
It is convenient to measure squared lengths in units of the Gaussian scale.
We use the normalized energy and shells
\begin{equation}\label{eq:bridge-normalization}
 V_r=\frac{r^n}{\det\Gamma},\qquad
 \en_r(x)=\frac{2\pi\|x\|^2}{nr^2},\qquad
 \mathcal S_r(a)=\{x:\en_r(x)\in a(1\pm w_n)\}.
\end{equation}
The notation $a(1\pm w_n)$ denotes the closed interval
$[a(1-w_n),a(1+w_n)]$.

Three functions describe the leading costs along the sequence of shells:
\begin{equation}\label{eq:rates}
 J(z)=\frac{z-1-\log z}{2\log2},\qquad
 P(u)=-\frac12\log_2(1-u^2),\qquad
 W(z)=\frac12\log_2(2ez).
\end{equation}
Here $J$ is the Gaussian shell cost, $P$ the angular pair and cap cost,
and $W$ the shell volume exponent when $V_r=2^{n/2+o(n)}$.
The rate for two directions to enter a common cap of threshold $\tau$ is
\begin{equation}\label{eq:wedge-rate}
 K_\tau(u)=-\frac12\log_2\left(1-\frac{2\tau^2}{1+u}\right)
\end{equation}
whenever the logarithm has positive argument. If $b=2a(1-u)$ and $E=4a-b$,
then
\begin{equation}\label{eq:entropy-identity}
 \frac12\log_2(eE/2)=2W(a)-P(u)-W(b).
\end{equation}
The identity counts midpoint representations from the input and output
list rates. The shell width and logarithmic error in shell counts are
\[
 w_n=\lceil\log_2(n+2)\rceil^{-4},\qquad
 \errwin=nw_n+\lceil\log_2(n+2)\rceil.
\]
These are the default widths used for SVP and batches. The direct
single sample algorithm uses the wider shells of
\cref{app:single-schedules}. In either case the local Gaussian estimates
use $w_n\to0$, $nw_n^2\to\infty$, and smoothing and list size allowances
larger than the logarithmic shell errors.
The default error and list size allowances of \cref{eq:budgets} satisfy
\[
 \errwin\ll h_n\ll\chi_n\ll\pad\ll n.
\]
At the scheduled cooling depths, $t(h_n+\errwin)=o(\pad)$, so the
extra initial records can absorb the losses from all rounds. The Gaussian
tail scale is $nw_n^2$. \Cref{tab:notation} collects the quantities
used across sections. Local normalization constants are introduced
with their respective kernels. Explicit budget choices and their
compatibility are proved in \cref{app:budgets}.

\begin{table}[htbp]
\centering
\small
\begin{tabularx}{\textwidth}{@{}>{\raggedright\arraybackslash}p{0.20\textwidth}X@{}}
\toprule
Quantity & Meaning and use\\
\midrule
$\Xi,\ m_\Xi,\ S_\Xi$ & Counting measure of the record list, expected record counts, and expected ordered counts of distinct records (\cref{eq:moments}).\\
$\nu=NG$ & Reference expected counts. $N$ is its total mass and $G$ its normalized shape.\\
$V_r,\ \en_r,\ \mathcal S_r(a)$ & Gaussian volume factor, normalized squared length, and thin shell (\cref{eq:bridge-normalization}).\\
$J(z),\ P(u),\ W(z)$ & Gaussian shell cost, angular pair cost, and shell volume rate (\cref{eq:rates}).\\
$K_\tau(u)$ & Common cap cost for threshold $\tau$ and correlation $u$ (\cref{eq:wedge-rate}).\\
$w_n,\ \ell_n$ & Relative shell width and logarithmic error in shell counts.\\
$h_n,\ \chi_n,\ b_n$ & Smoothing and thinning budget, SVP repetition budget, and extra initial list exponent.\\
$\Upsilon_j$ & Coefficient of the Gaussian reference in \cref{eq:soft-reference}. The shell mass converts it to an expected list size (\cref{eq:soft-total}).\\
$A_j,\ L_j$ & Expected record count per location in a full layer, up to moment error, and its list size exponent (\cref{eq:nominal-recursion,eq:svp-backward-rates}).\\
$a_j,\ E_j$ & Shell energy and midpoint bridge parameter $E_j=4a_{j-1}-a_j$ (\cref{eq:lens-energy}).\\
\bottomrule
\end{tabularx}
\caption{Notation shared by the counting, geometry, and algorithm sections.
The number of sampling trials, the number of stored records, and the
number of representations of a target are counted separately.}
\label{tab:notation}
\end{table}

\subsection{Theta estimates}
Fix a full rank lattice $\Gamma\subset\R^n$ and a scale $s>0$ with
\begin{equation}\label{eq:smoothing}
 \rho_{1/s}(\Gamma^*\setminus\{0\})\le\varepsilon_n<1.
\end{equation}
This is the usual smoothing condition $s\ge\eta_{\varepsilon_n}(\Gamma)$
of~\cite{MR07}.
We use the Euclidean norm and the dual defined by integer pairings,
$\Gamma^*=\{y\in\R^n:\langle y,x\rangle\in\Z\text{ for all }x\in\Gamma\}$.
\begin{lemma}\label{lem:theta}
For every $t\ge s$ and every shift $b\in\R^n$,
\begin{equation}\label{eq:theta-flat}
 (1-\varepsilon_n)\frac{t^n}{\det\Gamma}
 \le\rho_t(\Gamma+b)
 \le(1+\varepsilon_n)\frac{t^n}{\det\Gamma}.
\end{equation}
For $\Gamma/t_0$, with positive integer $t_0$, the same estimate holds at
parameters at least $s/t_0$. For any lattice $K$ and any $t>0$,
$\rho_t(K+b)\le\rho_t(K)$.
\end{lemma}
These standard bounds follow from Poisson summation \cite{MR07}.

\subsection{Basis reduction and initial Gaussian samples}
The following proposition relates the block dimension to the cost
and quality of the initial basis.

\begin{proposition}\label{ing:reduction}
Given a full rank rational basis $B_0$ of a rank-$n$ lattice $\Gamma$
and an integer $2\le k<n$, put
\[
 H(n,k)=2k^{(n-1)/(2(k-1))}.
\]
One can deterministically compute a size-reduced basis
$B=(b_1,\ldots,b_n)$ satisfying
\[
 \max_i\|b_i^*\|\le H(n,k)\lambda_n(\Gamma).
\]
The algorithm makes polynomially many exact Euclidean SVP calls of
rank at most $k$. Its time and workspace are
$4^{k+o(k)}\poly(\operatorname{size}(B_0))$.
All projected bases, lifted vectors, and integer coordinates have
polynomial encoding length in the original input.
\end{proposition}

For fixed $0<\vartheta<1$, the choice $k=\lfloor\vartheta n\rfloor$
gives $H(n,k)=O_\vartheta(n^{1/(2\vartheta)})$.
Both Gaussian applications use $k=\lfloor n/10\rfloor$, giving
preprocessing exponent $0.2$ and a Gram--Schmidt factor $O(n^5)$.
\Cref{app:suffix-proof} proves these claims by applying the DBKZ theorem
of~\cite[Theorem~2.3]{ALNS20} in successive projections, with exact
SVP supplied by~\cite{MV13}. Fixed small dimensions use exact SVP.

\paragraph{Estimating the shortest length.}\label{sec:short-scale}
For the SVP scale grid, \cite[Corollary~4.4]{ALNS20} with an exact
oracle gives, for every fixed integer $c\ge1$, a nonzero $v\in L$ with
\begin{equation}\label{eq:short-scale}
 \lambda_1(L)\le\|v\|\le D_c(n)\lambda_1(L),
 \qquad D_c(n)=K_c(n+1)^c,
\end{equation}
where $K_c$ is an effective constant. With the oracle of \cite{MV13},
the cost is $4^{n/(c+1)+o(n)}\poly(\operatorname{size}(B_0))$,
using polynomially many calls of rank at most $\lfloor n/(c+1)\rfloor$
for large $n$ and polynomial encoding length.
We use $c=9$, so this estimate also has preprocessing exponent $0.2$.

The ideal leaf pools use independent proposals followed by a correction
whose accepted mass has a known Gaussian coefficient.
The following transference estimate relates this starting width to
the terminal smoothing scale.

A shortest dual vector and its negative give
$2e^{-\pi s^2\lambda_1(\Gamma^*)^2}\le\varepsilon_n$.
Euclidean transference
$\lambda_n(\Gamma)\lambda_1(\Gamma^*)\le n$~\cite[Theorem~2.1]{Ban93}
therefore yields
\begin{equation}\label{eq:smoothing-successive}
 \lambda_n(\Gamma)
 \le n\sqrt{\frac{\pi}{\log(2/\varepsilon_n)}}\,s
 \le n\sqrt{\frac{\pi}{\log2}}\,s.
\end{equation}
For $G=\max_i\|b_i^*\|$, this gives
$G\le H(n,k)n\sqrt{\pi/\log2}\,s$, uniformly for
$\varepsilon_n<1$, which supplies a polynomially wide initial sampler.

\begin{lemma}\label{lem:gaussian-proposal}
Let $B$ be a basis of $\Gamma$, let $p+\Gamma$ be an affine coset,
and let $\mathcal S$ be a prescribed shell. If
\[
 R\ge G\sqrt{\log(8n)/\pi},\qquad G=\max_i\|b_i^*\|,
\]
one independent ideal trial returns a point or a failure symbol
$\dagger$, with probability
\begin{equation}\label{eq:gaussian-proposal-law}
 q_{R,p}(x)=\frac{\rho_R(x)}{2V_R}
       \ind_{x\in(p+\Gamma)\cap\mathcal S}
\end{equation}
of returning each vector $x$, and the remaining probability at $\dagger$.
The trial uses a sequential Klein/GPV proposal and a rejection
probability strictly below one. With rational input and squared width,
it has a finite implementation of any prescribed TV accuracy $\delta$
in polynomial time and space in the input length and $\log(1/\delta)$.
\end{lemma}
The correction cancels the proposal's scalar normalizers, so the
coefficient in \eqref{eq:gaussian-proposal-law} is known without computing
the total Gaussian mass on the coset.
\Cref{app:gaussian-leaves} proves the law and its counting moments.

\subsection{Lattice point counts and random superlattices}
For SVP, a lattice point count and a random superlattice supply smoothing
at the scale of a shortest vector. Put
\begin{equation}\label{eq:svp-scale-constants}
 \kappa_{\rm pack}=\frac{401}{1000},\qquad
 A_{\rm G}=\frac{e}{2^{2\kappa_{\rm pack}}},\qquad
 c=\frac{1}{2A_{\rm G}}=\frac{2^{2\kappa_{\rm pack}-1}}e.
\end{equation}
In particular, $W(c)=\kappa_{\rm pack}$.

\begin{lemma}\label{lem:kl-count}
Write $\beta_{\rm KL}=2^{\kappa_{\rm pack}}=2^{0.401}$.
For every fixed $R_0\ge1$, there are constants $C_{R_0}$ and $n_{R_0}$
such that, for every $n\ge n_{R_0}$ and every full rank lattice
$L\subset\R^n$, simultaneously for all $1\le R\le R_0$,
\[
 \#\{x\in L:\|x\|\le R\lambda_1(L)\}
 \le C_{R_0}(n+1)\beta_{\rm KL}^{\,n}R^n.
\]
\end{lemma}
The underlying spherical-code bound is due to
Kabatiansky--Levenshtein~\cite{KL78}. The bounded radius lattice form
used here is \cite[Lemma~A.1]{GFH26a}, also \cite[Lemma~3.8]{GFH26b}.
Its polynomial factor contributes $O(\log n)$ to the rates.
We will also use the following weighted form, proved by integration
in \cref{app:kl-weighted}.

\begin{lemma}\label{lem:kl-weighted}
For fixed $R_0\ge1$, set $\xi_w=\|w\|^2/\lambda_1(L)^2$.
Uniformly for positive $H,c$ with $H/c\le R_0^2$,
\[
 \sum_{w\in L\setminus\{0\}}(H-c\xi_w)_+^{n/2}
 \le\poly(n)\,2^{\kappa_{\rm pack}n}
       \left(\frac{H^2}{4c}\right)^{n/2}.
\]
The polynomial depends only on $R_0$.
\end{lemma}

We combine it with the GFH random superlattice
\cite[Sections 3--5]{GFH26a}, \cite[Proposition 3.10]{GFH26b}.

\begin{proposition}\label{ing:gfh}
Suppose a squared scale $Q$ satisfies
$Q\le\lambda_1(L)^2<(1+1/n)Q$. Choose a rational approximation to
\[
 s^2=\frac{2\pi A_{\rm G}}nQ\,(1+O(1/n)).
\]
There is a prime
$p=\Theta(2^{h_n}\det L/s^n)$ of polynomial bit length such that, for
$K=p^{-1}L$ and a uniformly random one-dimensional subspace
$U\subset K/L\simeq\mathbb F_p^n$, its inverse image $\Gamma$ has
\begin{equation}\label{eq:gfh}
 [\Gamma:L]=p,\qquad
 \frac{s^n}{\det\Gamma}=2^{h_n+O(1)},\qquad
 \Prb\left[\rho_{1/s}(\Gamma^*\setminus\{0\})>2^{-h_n/3}\right]
 \le2^{-2h_n/3+O(\log n)}.
\end{equation}
For every permitted scale and prime, the ambient lattice $K$ satisfies the
following unconditional bound, before the random line is chosen. For every
fixed $a_*>0$, uniformly over shifts $b$ and $t\ge a_*s$,
\begin{equation}\label{eq:dense-ambient}
 \rho_t(b+K)=\frac{t^n}{\det K}(1+O(2^{-\Omega(h_n)})).
\end{equation}
\end{proposition}

The line representation is $\Gamma=L+\Z Bv/p$ for uniform nonzero
$v\in\mathbb F_p^n$, with scale and prime fixed before the draw.
Call $\Gamma$ good when
$\rho_{1/s}(\Gamma^*\setminus\{0\})\le2^{-h_n/3}$, and on this
event put $r_0=\sqrt2s$.
For any shortest vector $v$, the actual normalized energy satisfies
\begin{equation}\label{eq:actual-c}
 \en_{r_0}(v)=c_n=c+O(1/n),\qquad
 V_{r_0}=2^{n/2+h_n+O(1)}.
\end{equation}
The estimates below are uniform in this perturbation. At the affine terminal
step we first use the unconditional estimate on the dense ambient lattice,
then restrict to good superlattices.

\section{Counting moments under differencing}\label{sec:replicas}
Two differences $x-y$ and $x-y'$ share an input. To control the resulting
dependence, we classify two distinct output records by their ancestry
at the current merge: they use four distinct child records, share a
left record, or share a right record. Conditional independence of the
two child lists expresses each case through their first two factorial
moments. The classification therefore closes without tracking the
joint law of an entire list.

Full differencing keeps all three contributions and increases the
number of target representations. Isolation removes the shared endpoint
cases at the current merge, giving the sharper control needed through
many cooling rounds. We first establish these counting identities for
a general pair map and activation rule. \Cref{sec:lattice,sec:bridges}
supply the Gaussian estimates that make each operation useful.

\subsection{The first two moments of one merge}
Consider a merge of two record lists under fixed information that
makes them independent, using the counting measures and conditional
moments of \cref{sec:occurrence-measures}. For each ordered pair of
child IDs
$(i,j)$, let $B_{ij}\in\{0,1\}$ indicate whether the pair is activated.
The pair produces the value $T(X_i,Y_j)$ when $B_{ij}=1$, with
\[
 \E[B_{ij}\mid\Xi_L,\Xi_R]=q(X_i,Y_j)\in[0,1].
\]
The fixed information is suppressed here. Distinct pair coins are
conditionally pairwise independent given the completed child lists.
The full merge retains one record for every activated pair.
Different input record pairs remain distinct even when they produce
the same vector. This convention makes the following ancestry
classification exact.

The positive maps
\begin{align}
 \Phi_q(f,g)(z)
 &=\sum_{T(x,y)=z}q(x,y)f(x)g(y),\label{eq:Phi}\\
 \Psi_q(U,V)(z,z')
 &=\sum_{\substack{T(x,y)=z\\T(x',y')=z'}}
 q(x,y)q(x',y')U(x,x')V(y,y')\label{eq:Psi}
\end{align}
transport one and two copies of the input counts.

\begin{lemma}[Exact factorial recursion]\label{lem:exact-recursion}
Under the preceding conditional independence and activation law, the
full merge satisfies
\begin{align}
 m_a&=\Phi_q(m_L,m_R),\label{eq:exact-first}\\
 S_a^{\rm act}
 &=\Psi_q(S_L^{\rm act},S_R^{\rm act})
  +\Psi_q(\Deltap m_L,S_R^{\rm act})
  +\Psi_q(S_L^{\rm act},\Deltap m_R).\label{eq:exact-second}
\end{align}
All moments are conditional on the fixed information of the merge.
\end{lemma}
\begin{proof}
Linearity and independence of the children give the first identity.
For the second, classify two distinct parent IDs according to whether
their child IDs are distinct on both sides, shared only on the left, or
shared only on the right. Conditional pairwise independence of their
coins gives the three terms. Sharing both IDs would give the same parent
record, which the factorial measure excludes. Equal vector values can
still belong to different IDs and remain included.
\end{proof}

We compare the actual moments to a reference measure $\nu$, whose
value at each point specifies the expected number of records there.
Its total mass is the intended list size, and one error parameter
controls both these expected counts and pairs of distinct records.

\begin{definition}\label{def:leaf-cone}
\label{def:factorial-moment-bounds}
Let $\mathcal H$ be retained information, and let the nonnegative measure
$\nu$ and number $\varepsilon\ge0$ be $\mathcal H$-measurable.
Write $\Xi\in\mathcal C_{\mathcal H}(\nu,\varepsilon)$ if, almost surely,
its conditional moments satisfy
\begin{equation}\label{eq:leaf-cone}
 e^{-\varepsilon}\nu\le m_\Xi\le e^{\varepsilon}\nu,
 \qquad S_\Xi\le e^{\varepsilon}\nu^{\otimes2}.
\end{equation}
The comparisons are pointwise on the countable underlying space.
We omit $\mathcal H$ when conditioning on a fixed complete external
environment $\Omega$.
\end{definition}

A reference on a prescribed coset and shell is extended by zero outside
that support. The first upper bound then ensures that all records lie
in the same support almost surely. Nonnegative summation also gives the
corresponding bounds for weighted counts and pairs. The parameter
$\varepsilon$ measures logarithmic distortion: sampling will require
$\varepsilon=o(1)$, whereas target extraction can allow
$\varepsilon=o(n)$.

\subsection{Affine shifts and full differencing}
The two children share an affine shift, so their references vary
together. For a parent coset $p+\Gamma$, the local shift $d_a$ gives
child cosets $p/2+d_a+\Gamma$ and $-p/2+d_a+\Gamma$. Once $d_a$
is fixed, the children use independent private shifts, leaf pools,
and coin seeds. We can therefore multiply their conditional moments,
then average this product over $d_a$. Averaging the references
separately would lose their shared geometric dependence.

Before drawing $d_a$, let $\mathcal H_a$ retain the lattice, scales,
the node's prescribed affine offset,
and ancestor assignments. After the draw,
$\mathcal A_a=\mathcal H_a\vee\sigma(d_a)$ also fixes the child
offsets. The child moments are conditional on $\mathcal A_a$,
and averaging $d_a$ gives the parent moments conditional on
$\mathcal H_a$. Bounds uniform over child offsets let us repeat
this calculation throughout the tree.

\Cref{tab:conditioning} records what is fixed and what is averaged
in each application. Cooling bounds hold with all external shifts
fixed. Full merges average the local shift, while search work also
averages the ancestor shifts that determine the support density.
\begin{table}[htbp]
\centering
\small
\begin{tabularx}{\textwidth}{@{}>{\raggedright\arraybackslash}p{0.18\textwidth}>{\raggedright\arraybackslash}p{0.34\textwidth}X@{}}
\toprule
Use & Retained information & Order of averaging\\
\midrule
Cooling & $\Omega$: lattice, scales, and all external affine shifts &
Average the samples and pair coins within each independent subtree.\\
\addlinespace[3pt]
Full merge & $\mathcal H_a$: lattice, scales, prescribed node offset, and ancestor assignments &
Reveal $d_a$. Form products of child moments conditional on
$\mathcal A_a$. Then average $d_a$.\\
\addlinespace[3pt]
Search work & $\mathcal B_v$: lattice, scales, and prescribed affine data independent of the shifts being averaged &
Reveal local and ancestor shifts. Multiply conditional child intensities.
Then average those shifts (\cref{sec:search}).\\
\bottomrule
\end{tabularx}
\caption{Conditioning in the moment and work estimates. Fix the shared
shifts, multiply the child moments, then average the indicated shifts.}
\label{tab:conditioning}
\end{table}

For references $\nu_{L,d},\nu_{R,d}$ at the induced child offsets, put
\begin{align}
 \lambda_d&=\Phi_{q_d}(\nu_{L,d},\nu_{R,d}),
       \label{eq:full-reference-intensity}\\
 Q_{L,d}&=\Psi_{q_d}(\Deltap\nu_{L,d},\nu_{R,d}^{\otimes2}),&
 Q_{R,d}&=\Psi_{q_d}(\nu_{L,d}^{\otimes2},\Deltap\nu_{R,d}).
       \label{eq:full-shared-reference}
\end{align}
The two $Q$ terms bound the contribution from reusing a child record.
The next theorem controls the merge error using the child moment
bounds and two geometric estimates for the parent reference.

\begin{theorem}[Full differencing]
\label{thm:full-moment-step}
Consider a full merge under the preceding activation law and affine offset
construction.

\emph{Child bounds.} Suppose the two children satisfy
$\mathcal C_{\mathcal A_a}(\nu_{L,d},\varepsilon_L)$ and
$\mathcal C_{\mathcal A_a}(\nu_{R,d},\varepsilon_R)$, where
$\varepsilon_L,\varepsilon_R$ are $\mathcal H_a$-measurable bounds
uniform over $d$.
Set $D=\varepsilon_L+\varepsilon_R$.

\emph{Reference bounds.} For a reference $\nu'$ and numbers $\eta,\rho\ge0$, all
$\mathcal H_a$-measurable, assume
\begin{align}
 e^{-\eta}\nu'\le\E_d[\lambda_d]
       &\le e^\eta\nu',\label{eq:full-first-closure}\\
 \E_d[\lambda_d^{\otimes2}+Q_{L,d}+Q_{R,d}]
       &\le e^\eta(1+\rho)(\nu')^{\otimes2}.
       \label{eq:full-second-closure}
\end{align}
Here and below $\E_d[\cdot]$ keeps $\mathcal H_a$ fixed.

\emph{Parent bound.} Then the full output
satisfies
\begin{equation}\label{eq:full-moment-update}
 \Xi_a\in\mathcal C_{\mathcal H_a}
       \bigl(\nu',D+\eta+\log(1+\rho)\bigr).
\end{equation}
\end{theorem}
\begin{proof}
First condition on $\mathcal A_a$. Positivity and
\cref{lem:exact-recursion} give
\begin{align*}
 e^{-D}\lambda_d&\le m_a\le e^D\lambda_d,\\
 S_a^{\rm act}&\le
 e^D(\lambda_d^{\otimes2}+Q_{L,d}+Q_{R,d}).
\end{align*}
Each term uses one moment from each child, including the terms with a
shared record, so all have the same factor $e^D$.
Take conditional expectations over $d$ and apply the two closure bounds.
\end{proof}

On a finite binary tree, the update adds the leaf errors
$\varepsilon_i$ and the local losses $c_a=\eta_a+\log(1+\rho_a)$:
\begin{equation}\label{eq:full-tree-budget}
 \varepsilon_{\rm root}
 =\sum_{i\text{ leaf}}\varepsilon_i
  +\sum_{a\text{ internal}}c_a.
\end{equation}
Here the bounds must be uniform over the offsets being averaged
and determined by the retained root information.

The theorem separates the counting argument from the geometry of a
particular kernel. To use it, one supplies a reference intensity,
bounds its variation under the shared shift, and bounds the two
shared input terms.

\subsection{Isolation for accurate cooling}
Sampling needs closer control than target existence. Retain an activated
pair only if no other activated pair uses either endpoint. Distinct
surviving pairs then use four distinct child IDs. The lost intensity is
bounded by the expected number of activated competitors. Isolation
removes the two shared input contributions in \eqref{eq:exact-second},
leaving a product bound whose small error can pass through the growing
cooling depth. Full differencing retains those terms and controls them
with larger input lists when the geometry permits it.

\begin{proposition}\label{prop:isolated}
\label{lem:shadowing-budget}
Fix information under which the input lists are independent and satisfy
factorial moment bounds with references $\nu_L,\nu_R$ and errors
$\varepsilon_L,\varepsilon_R$. Set $D=\varepsilon_L+\varepsilon_R$.
Use the preceding activation law with $q=cK\le1$, and keep only isolated
edges.
Put
\begin{gather*}
 \lambda=\Phi_q(\nu_L,\nu_R),\qquad \theta=c(D_++D_-),\\
 D_+=\sup_x\sum_yK(x,y)\nu_R(y),\qquad
 D_-=\sup_y\sum_xK(x,y)\nu_L(x).
\end{gather*}
Then the conditional moments of the output obey
\begin{align}
 (e^{-D}-e^D\theta)\lambda
 &\le m_{\rm iso}\le e^D\lambda,\label{eq:iso-first}\\
 S_{\rm iso}^{\rm act}&\le e^D\lambda^{\otimes2}.
       \label{eq:iso-second}
\end{align}
\end{proposition}
\begin{proof}
Before isolation, the intensity is between $e^{-D}\lambda$ and
$e^D\lambda$. Charge a deleted edge to an activated competitor.
A shared left ID uses the first moment on the left and the second
factorial measure on the right, contributing at most
$e^DcD_+\lambda$. A shared right ID contributes at most
$e^DcD_-\lambda$. A union bound proves \eqref{eq:iso-first}.
Two surviving edges have distinct IDs on both sides. Dropping their
isolation indicators and using the first term of
\eqref{eq:exact-second} proves \eqref{eq:iso-second}.
\end{proof}

\begin{corollary}
\label{cor:isolated-environment}
Assume the input hypotheses of \cref{prop:isolated} conditional on
$\mathcal A_a$, with $\mathcal H_a$-measurable $D$ and a degree bound
$\theta$ uniform over $d$.
Let $\nu'$ and $0\le\eta<1$ also be $\mathcal H_a$-measurable, and
suppose
\begin{equation}\label{eq:iso-environment}
 (1-\eta)\nu'\le\E_d[\lambda_d]\le(1+\eta)\nu',\qquad
 \E_d[\lambda_d^{\otimes2}]\le(1+\eta)^2(\nu')^{\otimes2}.
\end{equation}
If $D\le1/10$ and $\eta,\theta\le1/4$, the output satisfies
$\mathcal C_{\mathcal H_a}(\nu',\varepsilon')$ with
\begin{equation}\label{eq:isolated-linear}
 \varepsilon'=D+4(\eta+\theta).
\end{equation}
Pointwise closure $(1-\eta)\nu'\le\lambda_d\le(1+\eta)\nu'$
gives the same conclusion in each complete environment in which
the input moment, independence, activation, and degree assumptions hold.
\end{corollary}
Averaging \eqref{eq:iso-first}--\eqref{eq:iso-second} gives factors
$e^{-D}(1-e^{2D}\theta)(1-\eta)$ and $e^D(1+\eta)^2$.
Elementary logarithmic bounds yield \eqref{eq:isolated-linear}.

\begin{corollary}\label{cor:isolated-iteration}
Consider a balanced isolated tree of depth $t$, with leaves at level $t$
and root at level $0$. Suppose its activation probabilities are valid,
its conditioning follows the preceding construction, and every level
satisfies the closure and degree assumptions of
\cref{cor:isolated-environment} with uniform
$\eta_j,\theta_j\in[0,1/4]$. If the leaf error is at most
$\varepsilon_t$ and
\begin{equation}\label{eq:finite-shadowing}
 B_t=2^t\varepsilon_t+
       4\sum_{j=1}^t2^{j-1}(\eta_j+\theta_j)\le1/20,
\end{equation}
then the root satisfies the factorial moment bounds with error at most
$B_t$. One may condition on the prescribed node offset or on the complete
environment, using the same convention throughout the tree.
\end{corollary}
\begin{proof}
Iterating $\bar\varepsilon_{j-1}
=2\bar\varepsilon_j+4(\eta_j+\theta_j)$ gives $\bar\varepsilon_0=B_t$.
Every intermediate envelope is at most $B_t\le1/20$, so
\eqref{eq:isolated-linear} applies throughout the induction.
\end{proof}

\subsection{From moment bounds to outputs}\label{sec:weighted-readout}
Weighted selection normalizes by the actual list's random total
weight. To control this normalization, we compare the weighted count
in an event with its mixed second moment against the total weight.
Cauchy--Schwarz turns these two bounds into a lower bound for the
selection probability of every event. The resulting comparison with
the weighted reference has error linear in the moment distortion
and inverse effective mass.

\begin{theorem}[Weighted readout]
\label{thm:weighted-readout}
Let $\Xi\in\mathcal C_{\mathcal H}(\nu,\varepsilon)$ be finite.
For an $\mathcal H$-measurable weight $b:\mathcal X\to[0,1]$, put
\[
 0<M=\nu(b)<\infty,\qquad W=\sum_i b(X_i),\qquad
 \pi_{\mathcal H}=\frac{b\nu}{M}.
\]
Choose record $i$ with probability $b(X_i)/W$ when $W>0$, and
return a fixed fallback point when $W=0$. Then, almost surely,
\begin{align}
 \TV\bigl(\mathcal L(X_{\rm out}\mid\mathcal H),\pi_{\mathcal H}\bigr)
 &\le 1-e^{-3\varepsilon}\frac{M}{M+1}
 \le 3\varepsilon+\frac1{M+1},
       \label{eq:weighted-tv}\\
 \Prb[W=0\mid\mathcal H]
 &\le 1-e^{-3\varepsilon}\frac{M}{M+1}.
       \label{eq:weighted-empty}
\end{align}
\end{theorem}
\begin{proof}
Fix $\mathcal H$ and write $\pi=b\nu/M$. For an event $A$, put
$Z_A=\Xi(b\ind_A)$ and $\nu_A=\nu(b\ind_A)$. The moment bounds
and $b^2\le b$ imply
\[
 \E[Z_A]\ge e^{-\varepsilon}\nu_A,\qquad
 \E[Z_AW]
 =S_\Xi((b\ind_A)\otimes b)+m_\Xi(b^2\ind_A)
 \le e^\varepsilon\nu_A(M+1).
\]
On $W=0$ we have $Z_A=0$, and define $Z_A/W=0$ there.
If $\nu_A>0$, Cauchy--Schwarz applied to
$\sqrt{Z_A/W}$ and $\sqrt{Z_AW}$ gives
\[
 \E[Z_A/W;W>0]
 \ge\frac{(\E[Z_A])^2}{\E[Z_AW]}
 \ge e^{-3\varepsilon}\frac{M}{M+1}\pi(A).
\]
For $\nu_A=0$ this lower bound is immediate. The fallback contributes
nonnegative event mass. Thus the output law dominates $\alpha\pi$,
where $\alpha=e^{-3\varepsilon}M/(M+1)<1$. It can be written as
$\alpha\pi+(1-\alpha)R$ for a probability law $R$, proving the
first inequality in \eqref{eq:weighted-tv}. The second follows from
$1-e^{-3\varepsilon}\le3\varepsilon$.

Finally $\E[W]\ge e^{-\varepsilon}M$ and
$\E[W^2]\le e^\varepsilon(M^2+M)$. The second moment inequality
gives $\Prb[W>0]\ge\alpha$, proving \eqref{eq:weighted-empty}.
\end{proof}

\begin{corollary}
\label{cor:root-extraction}
If $\Xi\in\mathcal C_{\mathcal H}(\nu,\varepsilon)$ and
$F=\nu(v)>0$, then
\begin{equation}\label{eq:full-extraction}
 \Prb[\Xi(\{v\})>0\mid\mathcal H]
 \ge\frac{e^{-3\varepsilon}}{1+1/F}.
\end{equation}
In particular, $\varepsilon=o(n)$ and $F\ge e^{-o(n)}$ give probability
$e^{-o(n)}$.
\end{corollary}
This is \eqref{eq:weighted-empty} with $b=\ind_{\{v\}}$.

\paragraph{Averaging good environments.}\label{sec:weighted-environments}
Suppose each good environment has the same normalized target $\pi$,
moment error at most $\varepsilon_*$, and effective mass at least
$M_*>0$. Convexity of TV then gives the unconditional bound
\[
 \varepsilon_{\rm env}+\varepsilon_{\rm impl}
   +1-e^{-3\varepsilon_*}\frac{M_*}{M_*+1},
\]
where the first two terms pay for bad environments and implementation
disagreement. A weight randomized independently of the list conditional
on retained information is handled by additionally conditioning on that
weight, provided these uniform hypotheses still hold.

The weighted choice can be performed by one scan using constant
additional record space. Its finite realization is described in
\cref{app:weighted-implementation}.

\section{Gaussian cooling by isolated pairing}\label{sec:lattice}
A Gaussian difference naturally has a larger width than either input.
To obtain a narrower profile, we select pairs at a suitable angle and
correct the weight of their difference. Thin shells make this correction
nearly constant on the exponential scale, while sparse activation and
isolation preserve enough moment precision for another round.
This section specifies the shell, \textsc{Eligible}, and pair weight
interfaces of \cref{alg:ideal-cooling-tree}, then proves its levelwise
moment guarantee.

\subsection{Shells and pair weights for repeated cooling}\label{sec:soft}
We use the shell geometry of \cref{sec:overview-shells}.
At input width $r$, write $z=x-y$, $M=(x+y)/2$, and
$A_r=\pi(\gamma^{-2}-1/2)/r^2$. The correction
$e^{-A_r\|z\|^2}$ in \eqref{eq:overview-fiber} gives the desired
Gaussian factor $\rho_{\gamma r}(z)$.

For $u=u_n\in(1/2,31/50]$, fix
\begin{equation}\label{eq:soft-family}
 a_0=\frac1{1+u},\qquad a_1=4a_0-2,\qquad u_1=u,\qquad
 \gamma^2=2(1-u)=\frac{a_1}{a_0},\qquad
 \beta=a_0(1-u^2)=1-u=\frac{\gamma^2}{2}.
\end{equation}
The identities $a_0=1/2+a_1/4$ and $a_1=\gamma^2a_0$ put the
midpoint at its typical energy and preserve the endpoint shell
parameter after contraction.
SVP uses $u=47/80$. For DGS, $u$ approaches $1/2$ according to the
schedule in \cref{lem:endpoint-schedule}.

Fix $\Gamma,s$ with
$\epsilon_\Gamma=\rho_{1/s}(\Gamma^*\setminus\{0\})\le2^{-h_n/3}$,
and terminal width $r_0=\sqrt2s$. This smoothing condition makes the midpoint masses almost
independent of their affine shifts. The DGS application sets $s=s_0$ from the requested width.
At scale $r$, use the reference
\begin{equation}\label{eq:soft-reference}
 \nu_{p,r,\amp}(x)=\frac{\amp}{V_r}\rho_r(x)
 \ind_{x\in(p+\Gamma)\cap\mathcal S_r(a_0)}.
\end{equation}
The coefficient controls the expected number of records, while the
remaining factors fix their Gaussian profile.
Uniformly over the scheduled scales,
\begin{equation}\label{eq:soft-total}
 \nu_{p,r,\amp}(\R^n)
 =\amp\,2^{-J(a_0)n+O(\errwin)}.
\end{equation}
Indeed, tilting to width $r\sqrt{a_0}>s$ concentrates the mass in this
shell and gives the rate $J(a_0)$. The quantitative calculation appears in
\cref{app:gaussian-details}.

For output coset $p+\Gamma$, draw a uniform
$d\in(\Gamma/2)/\Gamma$ and use child cosets
$p/2+d+\Gamma$ and $-p/2+d+\Gamma$.
The child lists are supported on $\mathcal S_r(a_0)$.
For $z=x-y$, the eligibility test is
\begin{equation}\label{eq:soft-tests}
 \textsc{Eligible}_r(z)
 \quad\Longleftrightarrow\quad
 \en_r(z)\in a_1(1\pm w_n).
\end{equation}
Thus the output lies in $\mathcal S_{\gamma r}(a_0)$ and has the
same shell parameter at the next width. For a fixed output $z$, its
admissible midpoint region is
\begin{equation}\label{eq:soft-midpoint-region}
 \mathcal W_r(z)=\{M:M+z/2,M-z/2\in\mathcal S_r(a_0)\}.
\end{equation}
The endpoint identities
\[
 \en_r(M)=\frac{\en_r(x)+\en_r(y)}2-\frac{\en_r(z)}4,
 \qquad
 \frac{2\pi\angles{M,z}}{nr^2}
 =\frac{\en_r(x)-\en_r(y)}2
\]
place the midpoint near energy $1/2$ and bound its normalized inner
product with $z$ by $O(w_n)$. The region $\mathcal W_r(z)$ contains
almost all midpoint Gaussian mass, as proved below.

Reweight eligible pairs by
\[
 R_-^2=\frac{nr^2a_1(1-w_n)}{2\pi},\qquad
 \omega_r(z)=e^{-A_r(\|z\|^2-R_-^2)}.
\]
On the output shell, $2^{-O(nw_n)}\le\omega_r\le1$.

\subsection{Sparse activation controls the moment error}
The Gaussian weight sets the new profile, while a common thinning
factor controls how often activated pairs share an endpoint. Normalize
by an
upper bound on the weighted degree, so a typical activated edge has
only a small chance of acquiring a competitor.

For input coefficient $\amp_j=2^{\varrho n+\sigma_j}$, set
\begin{equation}\label{eq:soft-normalization}
 \softB_j=e^{A_rR_-^2}\left(\frac\gamma{\sqrt2}\right)^n,
 \quad \degpad=\lceil 10\errwin\rceil,
 \quad D_j^+=\amp_j\softB_j2^{\degpad}.
\end{equation}
Activate a pair with probability
\begin{equation}\label{eq:soft-coin}
 q_j(x,y)=\ind_{x,y\in\mathcal S_r(a_0)}
          \ind_{\en_r(x-y)\in a_1(1\pm w_n)}
          \frac{2^{-h_n}}{D_j^+}\omega_r(x-y),
\end{equation}
and retain edges isolated at both endpoints. Thus the algorithm uses
$c_j=2^{-h_n}/D_j^+$ when its children are at level $j$, producing
level $j-1$ at width $\gamma r_j=r_{j-1}$. With this normalization,
the degree budget is $\theta\le2^{1-h_n}$, controlling the isolation
loss in \cref{prop:isolated}. The next lemma supplies the closure and degree
estimates behind these choices.

\begin{lemma}[Gaussian closure and weighted degrees]\label{lem:soft-midpoint}
\label{lem:soft-degree}
Uniformly over \eqref{eq:soft-family} with $\gamma^2\le1-w_n/4$,
for all sufficiently large $n$,
\begin{equation}\label{eq:soft-B-rate}
 \softB_j=2^{-(J(a_0)+P(u_1))n+O(nw_n)}.
\end{equation}
For every scheduled input $r\ge r_0/\gamma$, shift, and eligible
output $z=p+g$,
\begin{equation}\label{eq:midpoint-flat}
 \sum_{M\in d-g/2+\Gamma}\rho_{r/\sqrt2}(M)
          \ind_{M\in\mathcal W_r(z)}
 =V_{r/\sqrt2}(1\pm\eta_{\rm mid}),
 \quad
 \eta_{\rm mid}=O(\epsilon_\Gamma)+e^{-\Omega(nw_n^2)}.
\end{equation}
Every eligible endpoint has weighted reference degree at most $D_j^+$
before the common factor $2^{-h_n}/D_j^+$.
\end{lemma}
\begin{proof}
The coefficient follows from
$A_rR_-^2=n(1-a_0)(1-w_n)/2$.
Radial and linear Gaussian concentration give a lower bound on an
inner region of $\mathcal W_r(z)$, while the full shifted Gaussian
mass gives the upper bound. For degrees, the endpoint and output
shells bound the exponent after completion of squares. A full coset
sum then gives the bound.
The uniform estimates, including the validity of every tilt, are proved
in \cref{app:gaussian-profile}.
\end{proof}

\subsection{Initializing and repeating the cooling step}
The corrected proposals of \cref{lem:gaussian-proposal} initialize
the desired intensity exactly. Basis preprocessing gives
$G\le\poly(n)r_0$, so a schedule $R=r_0\gamma^{-t}$ with
$t=O(\log n)$ for fixed $u$ reaches
$R\ge G\sqrt{\log(8n)/\pi}$.

For fixed $C_a>0$ and a bounded raw exponent
$\varrho\ge P(u_1)+J(a_0)$, set
\begin{equation}\label{eq:raw-leaves}
 M_{\rm leaf}=\left\lceil2^{\varrho n+C_a\pad}\right\rceil
\end{equation}
and run $2M_{\rm leaf}$ independent corrected trials at width $R$,
with shell $\mathcal S_R(a_0)$. Store only nonfailure records.
By \eqref{eq:fixed-count-leaves}, their reference coefficient is
$M_{\rm leaf}$ and their initial moment error is zero.

Number levels from the root, with leaves at level $t$, and set
$r_j=r_0\gamma^{-j}$. Fix $\amp_t=M_{\rm leaf}$ and
$\amp_{j-1}=\amp_j2^{-h_n-\degpad}$ before the construction.
The complete ideal procedure is \cref{alg:ideal-cooling-tree},
called as \textsc{BuildIdeal}$(p,0)$.
Conditional on the completed children, distinct pair coins
are pairwise independent, with the probabilities in \eqref{eq:soft-coin}.
The all-pairs loop defines the ideal law. Pair search in
\cref{sec:search} recovers this graph with the stated coverage and
incidence bounds. \Cref{fig:cooling-merge-tree} illustrates the
affine routing and Gaussian scales in a tree of depth two.

\begin{algorithm}[htbp]
\caption{Ideal cooling tree: \textsc{BuildIdeal}$(p,j)$}
\label{alg:ideal-cooling-tree}
\KwIn{Offset $p$, level $j$, and the fixed cooling schedule.}
\KwOut{A record list in $(p+\Gamma)\cap\mathcal S_{r_j}(a_0)$.}
\If{$j=t$}{
  $\Xi\gets$ empty record list\;
  \For{$i=1,\ldots,2M_{\rm leaf}$}{
    Run a corrected trial at $(p,r_t,\mathcal S_{r_t}(a_0))$ using \cref{lem:gaussian-proposal}\;
    \If{the trial returns $X\ne\dagger$}{Append a record with value $X$ to $\Xi$\;}
  }
  \Return $\Xi$\;
}
Draw $d\sim\operatorname{Unif}((\Gamma/2)/\Gamma)$\;
$\Xi_L\gets\textsc{BuildIdeal}(p/2+d,j+1)$\;
$\Xi_R\gets\textsc{BuildIdeal}(-p/2+d,j+1)$ using independent private randomness\;
$r\gets r_{j+1}$\;
$G\gets$ the empty bipartite graph on the child records\;
\For{each record pair $(x,y)\in\Xi_L\times\Xi_R$}{
  $z\gets x-y$\;
  \If{$\textsc{Eligible}_r(z)$}{
    Activate edge $(x,y)$ in $G$ with probability $c_{j+1}\omega_r(z)$, using pairwise independent coins\;
  }
}
Keep the edges whose two endpoints both have degree one in the completed graph $G$\;
\Return one difference record per retained edge\;
\end{algorithm}
\FloatBarrier

Each round reduces the coefficient by $2^{-h_n-\degpad}$.
The extra initial records absorb these losses, while isolation
keeps the accumulated moment error small. Isolation uses the completed
graph so that every activated competitor enters the degree test.
\Cref{app:search-canonical} implements this test within the stated space.
The following theorem states
both recursions, including the growing depth case used at the DGS endpoint.

\begin{theorem}[Gaussian cooling]\label{thm:soft}
\emph{Parameters.}
Choose $u=u_n$ from \eqref{eq:soft-family}, a bounded raw sampling rate
$\varrho=\varrho_n\ge P(u_1)+J(a_0)$, and depth $t=t_n$ satisfying
\[
 \gamma^2\le1-w_n/4,\qquad t(h_n+\errwin)=o(\pad).
\]
For either schedule in \cref{app:budgets,app:single-schedules}, we also
have $t=o(nw_n^2)$ and $nw_n^2=o(h_n)$, which ensure that the Gaussian
tail error remains small through the entire tree.
\emph{Construction.} Fix $\Gamma,s$ as above, $C_a>0$, and the external offsets. Construct
the ideal isolated forest at starting width $R=r_0\gamma^{-t}$ satisfying
\cref{lem:gaussian-proposal}, with disjoint pools of $2M_{\rm leaf}$
corrected trials and $M_{\rm leaf}$ given by \eqref{eq:raw-leaves}.

\emph{Root guarantee.} For all sufficiently large $n$, every root satisfies
\[
 \Xi_0\in\mathcal C(\nu_{p,r_0,\amp_0},\varepsilon_0),
 \qquad \varepsilon_0\le e^{-\Omega(nw_n^2)}.
\]
Its reference coefficient and expected list size obey
\begin{equation}\label{eq:soft-roots}
 \amp_0=2^{\varrho n+\sigma_0},\qquad
 \sigma_0=C_a\pad-\sum_{j=1}^t(h_n+\degpad)+O(1)=\Theta(\pad).
\end{equation}
In particular, the expected size is $2^{[\varrho-J(a_0)]n+\Theta(\pad)}$.
All pair probabilities are at most one. Independently constructed roots
are independent conditional on their common public parameters and
external offsets.
\end{theorem}
\begin{proof}
Set $\amp_t=M_{\rm leaf}$ and
$\amp_{j-1}=\amp_j2^{-h_n-\degpad}$. We first verify that these
coefficients define valid pair probabilities. At every level,
\eqref{eq:soft-B-rate} gives
\[
 \log_2(\amp_j\softB_j)
 =[\varrho-J(a_0)-P(u_1)]n+C_a\pad
   -\sum_{k=j+1}^t(h_n+\degpad)+O(nw_n+1)
 \ge C_a\pad/2.
\]
Thus $D_j^+\ge2^{-h_n}$, making \eqref{eq:soft-coin} a valid probability.

Next, completion of squares and \eqref{eq:midpoint-flat} identify the
activated reference in each complete environment as
$\kappa_d=(1\pm\eta_{\rm mid})\nu_{p,\gamma r,\amp_{j-1}}$.
Indeed, writing $c_j=2^{-h_n}/D_j^+$, its coefficient is
\[
 c_j\amp_j^2e^{A_rR_-^2}
 \frac{V_{r/\sqrt2}V_{\gamma r}}{V_r^2}
 =c_j\amp_j^2\softB_j
 =\amp_j2^{-h_n-\degpad}.
\]
The weighted degree bound gives $\theta_j\le2^{1-h_n}$.
By \eqref{eq:fixed-count-leaves}, the initial moment error is zero.

The error budget in \cref{cor:isolated-iteration} is therefore
\begin{equation}\label{eq:cooling-error-budget}
 \bar\varepsilon_0
 =4(2^t-1)(\eta_{\rm mid}+2^{1-h_n})
 =e^{-\Omega(nw_n^2)}.
\end{equation}
Here $t=o(nw_n^2)$ and $nw_n^2=o(h_n)$ make the budget small enough
for the local estimates at every level. The bounds hold with all
external offsets fixed, and averaging private offsets preserves them
because the root reference depends only on its own offset.

Summing the coefficient losses proves \eqref{eq:soft-roots}.
Equation~\eqref{eq:soft-total} gives the expected size. Independence of the
private random sources gives the final conditional independence claim.
\end{proof}

\section{Finding pairs with BDGL product codes}\label{sec:search}
The cooling construction specifies useful differences through its
counting moments. Finding them efficiently requires a search that
preserves the transitions just analyzed. Spherical filters make
correlated pairs meet in common buckets. Their cost separates into
point incidences for insertion and traversal, candidate reports for
pair testing, and true outputs for dictionary storage.

We first bound these counts on fixed lists and then average using
the intensities and shell geometry of the lists generated by the
algorithm. This order connects the search cost to the same lists
that supply the output. Query and bank sizes are kept separate when
they differ, as in the terminal batch merge.

\subsection{Counting filter insertions and pair reports}
Write $C_{d_{\rm pc}}(\tau)$ for spherical cap measure and
$W_{d_{\rm pc}}(\tau,u)$ for common cap measure at correlation $u$.
The product code construction uses dimension $d_{\rm pc}=n+O(\log n)$.
The spherical cap and wedge estimates of~\cite[Lemmas~2.1--2.2]{BDGL16}
give the following rates, in the notation of~\cite[Section~4.1]{GFH26b}.
For $\tau\in(0,1)$ and $u\in(-1,1)$ in fixed compact ranges with
$1-2\tau^2/(1+u)$ bounded below by a positive constant,
\begin{equation}\label{eq:cap-rates}
 \log_2 C_{d_{\rm pc}}(\tau)=-P(\tau)n+O(\log n),\qquad
 \log_2 W_{d_{\rm pc}}(\tau,u)=-K_\tau(u)n+O(\log n).
\end{equation}
The product code and decoder originate
in~\cite[Lemma~5.1 and Theorem~5.1]{BDGL16}. We use the following
uniform coverage and restricted traversal interface
from~\cite[Theorem~4.2, Proposition~4.3, and Appendix~D]{GFH26b},
which avoids storing all filters explicitly.

\begin{proposition}\label{ing:bdgl}
Let $A,B$ be fixed finite, internally deduplicated lists of nonzero
vectors with equality keys, as in~\cite[Algorithm~1]{GFH26b}.
Assume $\log_2M_f=O(n)$.

\emph{Expected counts.} For a fresh code of $M_f$ filters,
\begin{align}
 \E[Z_A]&=|A|M_f C_{d_{\rm pc}}(\tau),&
 \E[Z_B]&=|B|M_f C_{d_{\rm pc}}(\tau),\label{eq:point-work}\\
 \E[P_{\rm bucket}]
 &=M_f\sum_{\substack{x\in A,y\in B\\\key(x)=\key(y)}}
 W_{d_{\rm pc}}(\tau,\angles{\widehat x,\widehat y}).
 \label{eq:pair-work}
\end{align}
Here $Z_A,Z_B$ count point and filter incidences, and $P_{\rm bucket}$
counts reports of equally keyed pairs in common filters.

\emph{Coverage.} Fix $\delta>0$ and suppose
\[
 \delta\le\tau\le1-\delta,\qquad |u|\le1-\delta,\qquad
 \tau(1-u)\ge\delta,\qquad
 1-\frac{2\tau^2}{1+u}\ge\delta.
\]
The last two inequalities keep both cap constraints active and their
intersection nondegenerate. Uniformly on this range, a fixed target pair
is missed with probability $\exp[-\Omega(\log^2 n)]$ if
$M_fW_{d_{\rm pc}}(\tau,u)\ge2^{H_{d_{\rm pc}}}$, where
$H_{d_{\rm pc}}=O(\sqrt n\,\polylog n)$.

\emph{Resources.} For live insertion budget $S\ge|A|$,
restricted Cartesian traversal uses space
\begin{equation}\label{eq:search-space}
 O(S+|A|+|B|+\poly(n)B_{\rm pc})
\end{equation}
and, up to polynomial factors, time at most
\begin{equation}\label{eq:search-time}
 Z_A+Z_B+P_{\rm bucket}
 +\left(1+\frac{nZ_A}{S}\right)(|A|+|B|)B_{\rm pc},
 \qquad \log_2B_{\rm pc}=O(n/\log n).
\end{equation}
Here $B_{\rm pc}$ is the block code size. Reports can be streamed,
so candidate enumeration does not require their simultaneous storage.
\end{proposition}

For target correlation $u_0$, choose the filter count
\begin{equation}\label{eq:filter-counts}
 \log_2 M_f=K_\tau(u_0)n+C_{\rm code}\pad+O(\log n).
\end{equation}
The padding is chosen to meet the product code coverage threshold in
\cref{ing:bdgl}. Independent repetitions then reduce the probability of
missing a target pair. Internal calls give every record the same key. The SVP
root instead uses quotient labels, so equally keyed pairs have differences
in the input lattice.

The interface extends to record lists, counting each record and
pair with its multiplicity (\cref{app:search-interface}).
During cooling, the search must recover every eligible activated edge
to identify all competitors and decide isolation correctly. Complete
coverage couples the implemented merge to the ideal transition analyzed
by the moments. At the SVP root, it suffices to recover one target pair:
the analysis fixes a witness from the completed ideal lists before
drawing the terminal codes.

\subsection{Expected search costs for the generated lists}
The affine offsets determine which lattice points can occur in each
input list. Averaging these offsets gives the support density used in
the candidate count.

Each routing step chooses a uniform half grid coset. Thus a
height-$j$ subtree below a fixed root of depth $d$ has offset uniform
over $\pm p_{\rm root}/t+\Gamma/t$ modulo $\Gamma$, where
$t=2^{d-j}$ and the path fixes the sign. For every nonnegative $f$,
\begin{equation}\label{eq:ancestor-average}
 \E_{\rm ancestors}\!\left[\sum_{z\in p_j+\Gamma}f(z)\right]
 =t^{-n}\sum_{z\in\pm p_{\rm root}/t+\Gamma/t}f(z).
\end{equation}

For a call at scale $r$, put $g=r^2/s^2$. A pair of endpoint energy $a$
and correlation $u$ has difference energy $b=2a(1-u)$ and midpoint
bridge energy $E=2a(1+u)$. Counting its midpoint representations uses
a Gaussian of width $r\sqrt E/2$ on $\Gamma/2$, and hence requires only
\begin{equation}\label{eq:one-copy-margin}
 gE>1.
\end{equation}
At $r=r_0$, this is $2E>1$. The stronger internal requirement $E>1$
comes from the second moment transfer, rather than this work estimate.

To bound the search work, let $N_{I_n(u)}$ count candidate record
pairs whose correlation lies in a bin
$I_n(u)\subseteq[u-1/n,u+1/n]$, before the activation coin is applied.
The next estimate separates the angular pair cost from the additional
cost of counting differences below smoothing.

The work estimate follows the same order as the moment recursion:
fix the shifts to multiply child intensities, then average the
resulting geometric count. Here we average both the local shift
$d_v$ and the ancestor shifts $\mathbf d_{\uparrow v}$, whose
support density is given by \eqref{eq:ancestor-average}.
At node $v$, let $\mathcal B_v$ retain the lattice, scale, and
prescribed affine data independent of these shifts, as in
\cref{tab:conditioning}. Revealing the shifts gives
$\mathcal A_v=\mathcal B_v\vee\sigma(d_v,\mathbf d_{\uparrow v})$,
which fixes the node and child offsets and leaves the two child
subtrees independent. When $t=1$, only the local shift is averaged,
so the prescribed node offset can remain in $\mathcal B_v$.

\begin{lemma}[Candidate pairs in a correlation interval]\label{lem:correlation}
At input energy $a$ and scale $r$, suppose the two child lists are
independent conditional on $\mathcal A_v$, with intensities at most
$Ae^{O(\pad)}$ on their input shells, uniformly in the shifts being
averaged. Assume the local half grid average and
\eqref{eq:ancestor-average} apply with denominator $t$, and write
\[
 A V_r(ea)^{n/2}=2^{\beta n+O(\pad)},\qquad c_0=2t^2ga.
\]
For $u$ in a fixed compact correlation interval with $gE>1$ by a fixed
margin, the conditional candidate count satisfies
\begin{equation}\label{eq:correlation-spectrum}
 \E[N_{I_n(u)}\mid\mathcal B_v]
 \le 2^{[2\beta-P(u)]n+O(\pad)}
 \max\{1,[c_0(1-u)]^{-n/2}\}.
\end{equation}
\end{lemma}
\begin{proof}
With the shifts fixed, independence gives a product of child
intensities inside the tower identity
\[
 \E[N_{I_n(u)}\mid\mathcal B_v]
 =\E_{d_v,\mathbf d_{\uparrow v}}
 \!\left[\E_{\rm private}[N_{I_n(u)}\mid\mathcal A_v]
 \,\middle|\,\mathcal B_v\right].
\]
For a fixed output, averaging the local half grid shift contributes
$V_r(eE/4)^{n/2}e^{O(\errwin)}$ midpoint representations.
The free output shell contributes the bound
\eqref{eq:subcritical-shell} proved in \cref{app:search-geometry}.
The support density cancels its covolume, and
$bE/4=a^2(1-u^2)$ gives the stated rate. The last factor is the
penalty when the output Gaussian lies below smoothing.
Shell perturbations and bin width contribute $O(\errwin)$.
The moment bounds for cooling and full differencing supply the required
conditional intensity estimate in each application.
\end{proof}

A lower bound $g_0\le r^2/s^2$ may replace $g$ in the smoothing factors
and in $c_0$, while $V_r$ retains the actual scale. For cooling, we use
$g_0=2/\gamma^2$ to make the correlation cutoff independent of the level.

\subsection{Time and space bounds}
Let $\log_2M_f=K_{\rm code}n+O(\pad)$. Multiplying the spectrum by
common cap probabilities controls the expected bucket reports.
In the region $c_0(1-u)\ge1$, the identity
\begin{equation}\label{eq:angle-identity}
 P(u)+K_\tau(u)\ge2P(\tau),\qquad
 (1-\tau^2)^2-(1-u)(1+u-2\tau^2)=(u-\tau^2)^2
\end{equation}
gives rate $K_{\rm code}+2(\beta-P(\tau))$.
In the subcritical region, the rate is
$K_{\rm code}+2\beta+\frac12\log_2((1+u-2\tau^2)/c_0)$,
which increases with $u$.

For the calls used here, $\beta<1/2$. With the strict smoothing margins
of \cref{lem:correlation} and the cutoff conditions in
\cref{sec:terminal-cap}, the expected pair work has rate
\begin{equation}\label{eq:dummy-rate}
 T_{\rm pair}\le\max\left\{
 K_{\rm code}+2(\beta-P(\tau)),\quad
 K_{\rm code}+2\beta+\frac12\log_2\frac{2(1-\tau^2)}{c_0}
 \right\}.
\end{equation}
The same rate extends across the remaining correlations by the boundary
estimates in \cref{sec:terminal-cap}.

For input lists of size at most $2^{\beta n+O(\pad)}$, choose
a live insertion budget $S=2^{\beta n+C_S\pad}$.
The point incidence rate is $K_{\rm code}+\beta-P(\tau)$.
Equation~\eqref{eq:search-time} shows that traversal adds at most this
rate and the list rate $\beta$. Its live storage is
$2^{\beta n+O(\pad)}$, including the implicit block codes.

Depth first evaluation stores the current lists and traversal state.
The resulting list size and incidence bounds are the inputs to
\cref{prop:finite-realization}. Its application preserves the leading
rates while accounting for incomplete search and exceptional resource use.

\subsection{From ideal operations to finite algorithms}
\label{sec:finite-realization}
The incidence estimates specify what the search must recover and how
much work it may use. Fixed resource limits and finite arithmetic turn
these ideal constructions into algorithms whose costs are bounded on
every execution. The following result collects the transfer guarantee.
The common cap rule is given in \cref{app:application-guards}, with
execution schedules in
\cref{app:single-execution,app:batch-execution,app:svp-execution}.
\Cref{app:realization-proof} proves the transfer guarantee.

\begin{proposition}\label{prop:finite-realization}
\label{prop:word-ram-transfer}
Fix one of the single sample, batch, or SVP constructions below with
these resource limits. Let $T_0$ be its total work cap and let $S_0$
be its peak storage cap, counting ideal operations and stored data items
before the finite arithmetic overhead. These budgets include preprocessing,
search workspace, and all prescribed repetitions. They have leading
time and space exponents $\dgstime,\dgsspace$ for the sampling
constructions and $0.4130,0.3138$ for SVP.
For every requested $\lambda\ge1$, the chosen construction admits
a finite word RAM implementation with
\[
 T_{\rm word}\le T_0\,2^{o(n)}
       \poly(\mathcal B,n,\lambda),\qquad
 S_{\rm bits}\le S_0\,2^{o(n)}
       \poly(\mathcal B,n,\lambda)
\]
on every execution. Once the search and resource limits have been
fixed, the finite and exact-arithmetic executions can be coupled
so that their discrete transcripts of record lists, pair decisions,
and sampling decisions differ with probability at most $2^{-\lambda}$.
Let $\mathcal G$ be the good environment event used in the application.
On the ideal construction, let $\delta_{\rm search}$ bound the probability
that $\mathcal G$ holds but the required search coverage fails, and let
$\delta_{\rm resource}$ bound the probability that $\mathcal G$ holds
but a resource cap is exceeded.
For sampling, the ideal and finite outputs admit a coupling with
\[
 \Prb[\mathcal G\ \text{and}\ X_{\rm finite}\ne X_{\rm ideal}]
 \le\delta_{\rm search}+\delta_{\rm resource}+2^{-\lambda}.
\]
For SVP, fix a shortest vector $v$ and let $E_v$ be the event that
the ideal terminal lists contain a representation of $v$. Then
\[
 \begin{aligned}
 \Prb[\text{the implemented trial finds a shortest vector}]
 &\ge\Prb[\mathcal G\cap E_v]\\
 &\quad-\delta_{\rm search}-\delta_{\rm resource}-2^{-\lambda}.
 \end{aligned}
\]
The search and resource allowances are specified in
\cref{app:realization-proof}. For sampling they fit within the stated
statistical errors. For SVP their per-trial sum is smaller than the
ideal target probability by a factor tending to zero.
Bad environment probabilities are accounted for in the respective
applications. The batch allowance applies to its complete output law.
\end{proposition}

\section{Gaussian sampling by weighted readout}\label{sec:dgs}
\label{sec:single-dgs}
Cooling gives a correlated Gaussian list on a smoothing superlattice.
To sample from the requested coset of the original lattice, we choose
one record with a weight that both restricts its coset and adjusts its
Gaussian width. The weighted sampling theorem reduces the analysis
to showing that the expected total selection weight remains large
after these two corrections.

\subsection{Choosing the readout weight}
\label{sec:gaussian-readout-example}
Fix $\alpha>1$ and rational input $(L,a,\sigma^2)$ with
\begin{equation}\label{eq:dgs-promise-new}
 \sigma\ge\alpha\eta_{1/2}(L),
\end{equation}
and write $V_r(H)=r^n/\det H$. For $1/2<u\le3/5$, use the shell
parameters in \eqref{eq:soft-family}, so $a_0=1/(1+u)$.
Choose the terminal cooling width
\[
 r=\sigma/\sqrt{a_0},\qquad \mathcal S=\mathcal S_r(a_0).
\]
Then $\mathcal S$ is centered at the typical target radius
$\|x\|^2=n\sigma^2/(2\pi)$.

Suppose cooling on $\Gamma\supset L$ supplies a root
$\Xi\in\mathcal C(\nu,\varepsilon_{\rm cool})$, where
\[
 \nu(x)=\frac{\amp}{V_r(\Gamma)}\rho_r(x)
          \ind_{x\in(a+\Gamma)\cap\mathcal S}.
\]
The profile is wider than the target Gaussian, so restricting to
$a+L$ and multiplying by
$\exp[-\pi(\sigma^{-2}-r^{-2})\|x\|^2]$ gives the desired relative
weights. To normalize this factor on the shell, anchor it at the
inner squared radius $R_-^2$ and set
\begin{equation}\label{eq:dgs-weight-new}
 b_{\rm out}(x)=\ind_{x-a\in L}
 e^{-\pi(\sigma^{-2}-r^{-2})(\|x\|^2-R_-^2)}.
\end{equation}
Thus $b_{\rm out}\in[0,1]$ on the root support, and multiplication
into $\nu$ gives
\begin{equation}\label{eq:dgs-readout-law-new}
 \frac{b_{\rm out}\nu}{\nu(b_{\rm out})}
 =D_{a+L,\sigma}(\,\cdot\mid\mathcal S).
\end{equation}
One weighted selection therefore has distance at most
\[
 3\varepsilon_{\rm cool}+\frac1{M_{\rm eff}+1},
 \qquad M_{\rm eff}=\nu(b_{\rm out}),
\]
from this conditional target by \cref{thm:weighted-readout}.
The Gaussian tail outside $\mathcal S$ is
$e^{-\Omega(nw_n^2)}$ by \cref{app:dgs-tail}.
It remains to choose the smoothing extension and the cooling list size
so that $M_{\rm eff}$ is large.

\subsection{Paying for the smoothing extension}
At terminal width $r$, the cooling estimates use smoothing at
$s=r/\sqrt2$. A denser superlattice provides it, while restriction back to $L$
costs a factor equal to its index. The following construction
quantifies this tradeoff and will also be used for radial bins in
\cref{sec:batch}.

\begin{lemma}\label{lem:dgs-index-wrapper}
Let $s/\sigma$ be bounded below by a fixed positive constant and put
$\iota=\max\{0,\log_2(\sigma/(\alpha s))\}$.
Choose a prime with $\log_2p=\iota n+3h_n+O(1)$.
For a uniform nonzero $v\in\mathbb F_p^n$, set
\begin{equation}\label{eq:dgs-extension-new}
 \Gamma=L+\Z Bv/p,\qquad L=B\Z^n.
\end{equation}
Then $[\Gamma:L]=p$, and
\begin{equation}\label{eq:dgs-wrapper-good}
 \Prb[\rho_{1/s}(\Gamma^*\setminus\{0\})>2^{-h_n}]
 \le2^{-2h_n+O(1)}.
\end{equation}
\end{lemma}
\begin{proof}
The input promise and theta monotonicity bound the dual mass at $s$
by $(3/2)2^{\iota n}$. A random modular constraint retains each
unforced dual vector with probability at most $1/p$, while the mass
on $pL^*$ is $2^{-\Omega(p^2)}$. Markov's inequality gives the claim.
\Cref{app:dgs-environment} supplies the mass calculation.
\end{proof}

Call the extension good when
$\rho_{1/s}(\Gamma^*\setminus\{0\})\le2^{-h_n}$.
At our chosen width $s=\sigma/\sqrt{2a_0}$, the index rate is
\[
 \iota=\max\left\{0,\frac12\log_2\frac{2a_0}{\alpha^2}\right\}.
\]
On a good extension, the effective mass can be computed directly
from \eqref{eq:dgs-weight-new}:
\begin{equation}\label{eq:dgs-mass-exact-new}
 M_{\rm eff}=
 \frac{\amp}{p}a_0^{n/2}e^{n(1-a_0)(1-w_n)/2}
 \frac{\rho_\sigma((a+L)\cap\mathcal S)}{V_\sigma(L)}.
\end{equation}
The last ratio lies between positive constants for large $n$ by
\cref{app:dgs-tail}, and the middle factors equal
$2^{-J(a_0)n+O(nw_n)}$. Hence the readout spends the shell rate
$J(a_0)$ on width correction and the index rate $\iota$ on
returning to the original lattice.

To pay these costs, we choose the following list, raw sampling, and
forward search rates:
\begin{align}
 \ell(u)&=P(u),\qquad m(u)=P(u)+J(a_0),\label{eq:dgs-rates-new}\\
 K(u)&=K_u(u)=\frac12\log_2\frac{1+u}{(1-u)(1+2u)}.\nonumber
\end{align}
With raw trial rate $m(u)$, \cref{thm:soft} gives root coefficient
$\log_2\amp=m(u)n+\Theta(\pad)$ and expected list size
$2^{P(u)n+\Theta(\pad)}$, provided the depth and padding absorb
the accumulated cooling loss. We choose such a schedule below.
Substituting this coefficient into \eqref{eq:dgs-mass-exact-new},
using $nw_n+h_n=o(\pad)$, cancels the shell cost:
\begin{equation}\label{eq:single-effective-rate}
 \log_2M_{\rm eff}=(P(u)-\iota)n+\Theta(\pad).
\end{equation}
If $\iota=0$, then $P(u)\ge\dgsspace$. Otherwise,
$2a_0(1-u^2)=\gamma^2<1$ gives
\[
 P(u)-\iota=\log_2\alpha-\frac12\log_2\gamma^2
 \ge\log_2\alpha.
\]
In either case,
\[
 P(u)-\iota\ge\min\{\dgsspace,\log_2\alpha\}>0.
\]
Thus exponentially large effective mass survives the return to $L$.
This positive margin is what makes every fixed $\alpha>1$ sufficient:
the smoothing extension enables the cooling estimates while leaving
enough total selection weight for one Gaussian readout.

\subsection{Reaching the BDGL exponents}
The forward search rate $K(u)$ determines time, while the list rate
$P(u)$ determines space. To approach the BDGL exponents, let $u$
decrease toward $1/2$. This slows the contraction, so we choose the
schedule below to keep the overhead from the additional cooling levels
subexponential.

\begin{lemma}\label{lem:endpoint-schedule}
Set
\[
 k_n=\max\{10,\lceil\log_2\log_2(n+4)\rceil\},\qquad
 u_n=\frac12+\frac1{k_n},\qquad \gamma_n^2=1-\frac2{k_n},
\]
and use $a_0,a_1$ from \eqref{eq:soft-family}.
Starting from a basis with $G\le\poly(n)s$ and an endpoint width
$r=\Theta(s)$, a depth
\begin{equation}\label{eq:dgs-endpoint-depth}
 t=O(k_n\log n)=O(\log n\log\log n)
\end{equation}
reaches the proposal threshold of \cref{lem:gaussian-proposal}.
For either schedule in \cref{app:budgets,app:single-schedules},
\[
 t(h_n+\degpad+2)=o(\pad),\qquad
 K(u_n)\to\dgstime,\quad P(u_n)\to\dgsspace,\quad
 m(u_n)\to\frac12-\frac1{6\log2}<\dgstime.
\]
\end{lemma}
\begin{proof}
Since $-\log\gamma_n\ge1/k_n$, the scheduled width
$r\gamma_n^{-t}$ reaches any prescribed polynomial multiple of $s$
within the stated depth. The default budgets \eqref{eq:budgets}
give total loss $O(n\log\log n/\log n)=o(\pad)$, and
\eqref{eq:single-budget-relations} gives the same conclusion for
the single sample schedule. Substitution yields the rate limits.
The energies remain in compact positive ranges and $k_nw_n\to0$,
so the shell and search estimates hold uniformly along the schedule.
\end{proof}

For the accuracy exponent $\xi>0$ in \cref{thm:single-main}, choose
a fixed rational $0<\delta\le\min\{1,\xi\}$ and shell width
$w_n\asymp(\log n)^{-1-\delta/2}$. The budgets of
\cref{app:single-schedules} absorb the cooling loss above and give
$nw_n^2=\Theta(n/\log^{2+\delta}n)$.
\Cref{alg:single-dgs} combines the construction with one weighted
selection. Its bounded implementation is given in
\cref{app:single-execution}.

\begin{algorithm}[H]
\caption{One Gaussian sample: \textsc{GaussianReadout}}
\label{alg:single-dgs}
\KwIn{A rational lattice $L$, shift $a$, squared width $\sigma^2$,
and fixed $\alpha>1$, $\xi>0$ as in \cref{thm:single-main}.}
\KwOut{One point of $a+L$.}
Choose the endpoint parameters of \cref{lem:endpoint-schedule} and set
$r=\sigma/\sqrt{a_0}$, $s=r/\sqrt2$\;
Construct the smoothing extension $\Gamma$ at scale $s$ and preprocess
its basis\;
Generate a cooling root $\Xi$ in $a+\Gamma$ at width $r$ using
\textsc{BuildIdeal}$(a,0)$\;
Select one record from $\Xi$ with weights $b_{\rm out}(x)$ in
\eqref{eq:dgs-weight-new}, using the readout convention of
\cref{thm:weighted-readout}\;
\Return its value\;
\end{algorithm}

\begin{proof}[Proof of \cref{thm:single-main}]
On each good extension, cooling and the effective mass estimate
\eqref{eq:single-effective-rate} give
$\varepsilon_{\rm cool}\le e^{-\Omega(nw_n^2)}$ and
$M_{\rm eff}=2^{\Omega_\alpha(n)}$.
The weighted sampling theorem therefore bounds the distance from
\eqref{eq:dgs-readout-law-new} by
$3\varepsilon_{\rm cool}+1/(M_{\rm eff}+1)$.
Because the normalized target is the same on every good extension,
this comparison survives averaging over extensions. Adding the
shell tail and bad extension probability gives
\[
 \TV(\mathcal L(X_{\rm ideal}),D_{a+L,\sigma})
 \le e^{-\Omega(nw_n^2)}+2^{-2h_n+O(1)}.
\]
The implementation in \cref{app:single-execution} adds search,
resource, and numerical loss at most
$2^{-\Omega(n)}+2^{-\Omega(\pad)}+2^{-Cn}$.
These errors and the bad extension term are smaller than the
shell and moment error under \eqref{eq:single-budget-relations},
so the final distance is at most
\[
 e^{-\Omega(nw_n^2)}
 \le\exp[-\Omega(n/\log^{2+\xi}(n+2))].
\]

For time, forward pair search is the dominant operation. Its expected
work at a node is $2^{K(u_n)n+O(\pad)}$ by \cref{sec:dgs-work}, and the
$2^{O(t)}$ nodes multiply this by only $2^{o(n)}$.
The other operations fit this rate: raw sampling has rate
$m(u_n)\to1/2-1/(6\log2)<\dgstime$, basis preprocessing has rate
$1/5$, and the final weighted scan has the smaller list rate $P(u_n)$.
The work caps in \cref{app:single-execution} therefore give time
$2^{K(u_n)n+o(n)}
=2^{\dgstime n+o(n)}$.

For space, depth first evaluation retains the current lists and
search workspace, while leaf proposals and candidate reports are
streamed. Each list has rate $P(u_n)$, and the weighted scan needs
only one additional record. With the caps of
\cref{app:single-execution}, the bit space on every execution is
$2^{P(u_n)n+o(n)}=2^{\dgsspace n+o(n)}$.
\end{proof}

\begin{remark*}
Our recent work~\cite{GH26} identifies a route to improving the BDGL
sieving exponent through faster matrix multiplication. We find that
this approach also applies to our provable single sample DGS algorithm.
Specifically, let $\omega$ denote the matrix multiplication exponent.
If $\omega<9/4$, then for every fixed $\alpha>1$, a single sample can
be obtained with the same width range and accuracy as in
\cref{thm:single-main} in $(3/2-\delta)^{n/2+o(n)}$ time for some
constant $\delta=\delta(\alpha,\omega)>0$. The result can be obtained
by replacing pair enumeration within the Gaussian cooling buckets
with compressed correlation search and enlarging the buckets to
reduce the number of filters. The surplus supply available for a
single output allows us to strengthen smoothing, which controls the
recovery cost of highly correlated pairs.
The fixed-list coverage guarantee preserves the ideal cooling
transitions, while GDD correlation estimates bound the search work
after averaging the generated lists.
\end{remark*}

Polynomially many independent executions give a joint Gaussian
approximation with the same leading resource exponents. To obtain
an exponential number of outputs within the same total cost,
\cref{sec:batch} reuses partner lists and corrects each trial's
complete output law.
\FloatBarrier

\section{Gaussian shells and full differencing}\label{sec:bridges}
At the shortest vector scale, restricting to the input lattice can
consume many records in the cooled list. To create more representations
before this restriction, the SVP algorithm keeps every eligible
difference through a fixed number of layers. Each layer squares the
expected count at a location and multiplies it by the number of
possible midpoints. This gain permits the larger moment error allowed
when the goal is to find one target vector.

For each output $z$, its \emph{midpoint lens} consists of the
midpoints whose two endpoints lie in the input shells. We must control
both the size of this lens under a shared affine shift and the effect
of reusing an input record. The former requires bridge energy $E>1$,
while a sufficiently large input list controls the latter.
The terminal coset argument in \cref{sec:affine-root} will allow the
smaller threshold $E>2/e$ because no further moment propagation is
needed there.

Here we keep the width $r=\sqrt2s$ fixed and vary the shell energy from
one layer to the next. Thus smaller energies give shorter vectors at
the same reference width. Assume \eqref{eq:smoothing} with
$\varepsilon_n\to0$ and set $w=w_n$.

\subsection{Counting representations by their midpoints}
For a transition from child energy $a$ to output energy $b$, assume
\begin{equation}\label{eq:lens-energy}
 E=4a-b\ge1+\kappa
\end{equation}
with fixed $\kappa>0$. At parent offset $p$, draw
$d\in(\Gamma/2)/\Gamma$ uniformly and use
\[
 C_+=p/2+d+\Gamma,\qquad C_-=-p/2+d+\Gamma.
\]
For a point $z=p+g$ in the prescribed output coset and shell, with
$e_z=\en_r(z)\in b(1\pm w)$,
let $J_d(z)$ count pairs $z=x-y$ with
$x\in C_+\cap\mathcal S_r(a)$ and $y\in C_-\cap\mathcal S_r(a)$.
Put
\begin{equation}\label{eq:lens-volume}
 \beta_z=a-e_z/4,\quad E_z=4\beta_z,\quad
 \sigma_z=r\sqrt{\beta_z},\quad
 \mathcal B_z=V_r(e\beta_z)^{n/2},\quad
 \mathcal B=V_r(eE/4)^{n/2}.
\end{equation}
In midpoint coordinates
the normalized energy is approximately $E/4$. The parameter $E$
makes the second moment smoothing condition take the simple form $E>1$.
Writing
$M=(x+y)/2\in d-g/2+\Gamma$, the shell conditions are exactly
\begin{equation}\label{eq:lens-window}
 |\en_r(M)-\beta_z|
 +\left|\frac{2\pi\angles{M,z}}{nr^2}\right|\le aw.
\end{equation}
The inner rectangle has both summands at most $aw/2$, and the outer
radial window has $|\en_r(M)-\beta_z|\le aw$. Averaging the local shift
counts on a finer lattice:
\begin{equation}\label{eq:halfgrid-average}
 \E_d\!\left[\sum_{M\in d-g/2+\Gamma}f(M)\right]
 =2^{-n}\sum_{M\in\Gamma/2}f(M),\qquad f\ge0.
\end{equation}

\Needspace{13\baselineskip}
\begin{theorem}[Midpoint lens moments]\label{thm:lens}
Let $(a,b)$ range over a fixed compact subset of $(0,\infty)^2$
satisfying \eqref{eq:lens-energy} for a fixed $\kappa>0$.
Under \eqref{eq:smoothing} with $\varepsilon_n\to0$, the following
estimates hold uniformly over this range for all sufficiently large $n$.
There is a constant $C$ depending only on the range and $\kappa$ such that, with
$\eta_n=C(nw+\log(n+2))$,
\begin{align}
 e^{-\eta_n}\mathcal B\le\E_d[J_d(z)]
 &\le e^{\eta_n}\mathcal B,\label{eq:lens-first}\\
 \E_d[J_d(z)J_d(z')]
 &\le e^{\eta_n}\mathcal B^2.\label{eq:lens-second}
\end{align}
The second estimate uses the same $d$ in both copies and is uniform over
$z,z'$ in the same parent coset and output shell, including $z=z'$.
\end{theorem}
\begin{proof}
Apply the window count of \cref{lem:window-count} on $\Gamma/2$
at width $\sigma_z$, using the inner rectangle and outer radial
window from \eqref{eq:lens-window}. Its radial tilts have a fixed
smoothing margin since $4\sigma_z^2/s^2=2E_z>1$.
The factor $2^{-n}$ in \eqref{eq:halfgrid-average} cancels the finer
lattice's covolume, and $\beta_z=E/4+O(w)$ gives
\eqref{eq:lens-first}.

For the second moment, write $G_d(t)=\rho_t(d+\Gamma)$. The half grid
has a useful square-sum identity:
\begin{equation}\label{eq:halfgrid-square}
 \E_d[G_d(t)^2]=2^{-n}\rho_{\sqrt2t}(\Gamma)^2.
\end{equation}
Indeed, the expanded sum runs over $x,y\in\Gamma/2$ with $x-y\in\Gamma$.
The map $(x,y)\mapsto(x+y,x-y)$ is a bijection onto $\Gamma^2$, with
inverse $(u,v)\mapsto((u+v)/2,(u-v)/2)$. The parallelogram identity
separates its Gaussian weight into
$\rho_{\sqrt2t}(u)\rho_{\sqrt2t}(v)$.

The outer window gives
$J_d(z)\le e^{n/2+O(nw)}G_{d-g/2}(\sigma_z)$.
Since $g\in\Gamma$, translating $d$ by $g/2$ permutes its possible
values. Also $2\sigma_z^2=E_zs^2>s^2$, so \cref{lem:theta} yields
\[
 \E_d[J_d(z)^2]
 \le e^{n+O(nw)}2^{-n}
       \left((1+\varepsilon_n)
       \frac{(\sqrt2\sigma_z)^n}{\det\Gamma}\right)^2
 \le e^{O(nw)}\mathcal B_z^2.
\]
Thus the threshold $E>1$ comes from smoothing at width
$\sqrt2\sigma_z$. Cauchy--Schwarz now bounds
$\E_d[J_d(z)J_d(z')]$ by the geometric mean of the two square
moments. Using $E_z,E_{z'}=E+O(w)$ proves \eqref{eq:lens-second}
for the same local shift in both copies.
\end{proof}

\subsection{Moment bounds for repeated full differencing}
Consider a balanced tree of fixed depth $d$.
At height $j$, retain pairs with kernel
\[
 q_j(x,y)=\ind_{x,y\in\mathcal S_r(a_{j-1})}
                   \ind_{x-y\in\mathcal S_r(a_j)},
\]
using the preceding local environment at every node. Cooling supplies
each leaf at offset $p$ with Gaussian reference
\begin{equation}\label{eq:shell-leaf}
 \nu_{0,p}(x)=\frac{\amp_0}{V_r}\rho_r(x)
 \ind_{x\in(p+\Gamma)\cap\mathcal S_r(a_0)}.
\end{equation}
The recursion is made explicit in \cref{alg:full-difference-tree}.
Here height counts upward from the cooled leaves, whereas the level
argument of \textsc{BuildIdeal} counts downward from a cooling root.
Each call at height zero starts a fresh cooling tree with the prescribed
offset. Consequently a root of height four uses sixteen cooling pools.

\begin{algorithm}[htbp]
\caption{Ideal full difference tree: \textsc{BuildFullTree}$(p,j)$}
\label{alg:full-difference-tree}
\KwIn{Offset $p$, height $j$, fixed width $r$, energies $(a_i)$,
and a cooling schedule ending at width $r$.}
\KwOut{A record list in $(p+\Gamma)\cap\mathcal S_r(a_j)$.}
\If{$j=0$}{
  \Return $\textsc{BuildIdeal}(p,0)$ with a fresh leaf pool and private randomness\;
}
Draw $d\sim\operatorname{Unif}((\Gamma/2)/\Gamma)$\;
$\Xi_L\gets\textsc{BuildFullTree}(p/2+d,j-1)$\;
$\Xi_R\gets\textsc{BuildFullTree}(-p/2+d,j-1)$ with independent private randomness\;
$\Xi\gets$ empty record list\;
\For{each record pair $(x,y)\in\Xi_L\times\Xi_R$}{
  \If{$x-y\in\mathcal S_r(a_j)$}{
    Append a record with value $x-y$ to $\Xi$\;
  }
}
\Return $\Xi$\;
\end{algorithm}

The all-pairs loop specifies the ideal transition. The searches of
\cref{sec:search} recover these eligible pairs with the
coverage and work guarantees used below. Every eligible record pair
contributes a difference, preserving the multiplicities needed for
the intensity recursion. The shared input terms remain in its second
moment and will be bounded using the available number of records.

The nominal intensity per atom is
\begin{equation}\label{eq:nominal-recursion}
 A_0=\frac{\amp_0}{V_r}e^{-na_0/2},\qquad
 E_j=4a_{j-1}-a_j,\qquad
 \mathcal B_j=V_r(eE_j/4)^{n/2},\qquad
 A_j=A_{j-1}^2\mathcal B_j.
\end{equation}
The list size also depends on the number of output atoms.
Three quantities govern the iteration. Bridge energy controls the
common environment, $A_{j-1}\mathcal B_j$ controls the relative error
from reusing an input, and $A_d$ sets the final target count.
We express the needed
moment bounds directly relative to the constant intensity $A_j$ on each
shell, reducing their propagation to two scalar recursions.

At a leaf, the Gaussian reference differs from $A_0$ by at most
$e^{\pm b_0^*}$, where $b_0^*=na_0w_n/2$. Replacing this reference by
$A_0$ costs $b_0^*$ in the first moment and $2b_0^*$ in the second.
Thus cooling error $\varepsilon_{\rm cool}$ gives leaf error
\begin{equation}\label{eq:shell-leaf-error}
 \varepsilon_0=\varepsilon_{\rm cool}+2b_0^*
 =O(nw_n)+e^{-\Omega(nw_n^2)}=o(n).
\end{equation}

\begin{theorem}[Full difference trees]\label{thm:shell-tree}
\label{cor:shell-extraction}
\emph{Geometry and expected counts.}
Let $d$ be fixed, with energies in a fixed compact positive range, under
the smoothing assumption of \cref{thm:lens}. Suppose constants
$\kappa,\zeta>0$ satisfy, at every height,
\begin{equation}\label{eq:shell-tree-hyp}
 E_j\ge1+\kappa,\qquad A_{j-1}\mathcal B_j\ge2^{\zeta n}.
\end{equation}

\emph{Leaf bounds.} Assume each leaf satisfies
$\mathcal C(A_0\ind_{(p+\Gamma)\cap\mathcal S_r(a_0)},\varepsilon_0)$
uniformly in its prescribed offset, with $\varepsilon_0=o(n)$. Use independent
private randomness in disjoint subtrees.

\emph{Propagation.} For sufficiently large $n$, each height-$j$ list, conditional on its
prescribed offset and ancestor assignments, satisfies
\begin{equation}\label{eq:atom-envelope}
 e^{-\varepsilon_j}A_j\le m_{j,p}(z)\le e^{\varepsilon_j}A_j,
 \qquad S^{\rm act}_{j,p}(z,z')\le e^{\varepsilon_j}A_j^2
\end{equation}
for $z,z'$ in its prescribed coset and shell, where
\begin{equation}\label{eq:shell-moment-update}
 \varepsilon_j=2\varepsilon_{j-1}+\eta_n+
 \log\!\left(1+\frac{2}{A_{j-1}\mathcal B_j}\right),
 \qquad \eta_n=C(nw_n+\log(n+2)).
\end{equation}
The root error obeys
\begin{equation}\label{eq:shell-tree-alpha}
 \varepsilon_d\le2^d\varepsilon_0+(2^d-1)\eta_n
                +2(2^d-1)2^{-\zeta n}=o(n).
\end{equation}
\end{theorem}
\begin{proof}
Fix the prescribed node offset and reveal its local shift $d_a$. Conditional on
this information the children are independent. Write
$A=A_{j-1}$, $B=\mathcal B_j$ and take
$\nu'=A_j\ind_{\mathrm{support}}$. The reference intensity is
$\lambda_{d_a}(z)=A^2J_{d_a}(z)$. By \cref{thm:lens},
\[
 e^{-\eta_n}A_j\le\E_{d_a}[\lambda_{d_a}(z)]\le e^{\eta_n}A_j,
 \qquad
 \E_{d_a}[\lambda_{d_a}(z)\lambda_{d_a}(z')]
 \le e^{\eta_n}A_j^2.
\]
For a shared left value $x$, the right values are determined by $z,z'$.
The admissible triples are a subset of those counted by $J_{d_a}(z)$.
Consequently
\[
 \E_{d_a}[Q_{L,d_a}(z,z')]
 \le A^3\E_{d_a}[J_{d_a}(z)]
 \le e^{\eta_n}\frac{A_j^2}{AB}.
\]
The right shared input term has the same bound. Thus
\cref{thm:full-moment-step} applies with
$\eta=\eta_n$ and $\rho=2/(AB)$, proving
\eqref{eq:shell-moment-update}.
Now apply \eqref{eq:full-tree-budget} and
$\log(1+2/(AB))\le2^{1-\zeta n}$ to obtain
\eqref{eq:shell-tree-alpha}.
\end{proof}

For a specified root point $v$ in the output support,
\cref{cor:root-extraction} gives occurrence probability $e^{-o(n)}$
whenever $A_d\ge e^{-o(n)}$. For SVP we instead apply the target bound
for two lists in \cref{lem:dependent-root}. In both cases, merging equal
values at the terminal search preserves every represented target.

\section{Shortest vector search in a random coset}\label{sec:affine-root}
An internal layer must support further propagation. The root only needs
one shortest difference. A random affine coset lets us use this simpler
requirement to lower the bridge parameter from the internal threshold
$E>1$ to $E>2/e$. We extend the affine slicing argument
of~\cite[Section~6, especially Proposition~6.4]{GFH26b} from one pairing
layer to the lists generated by our dynamics.

Fix a scale $Q$ satisfying $Q\le\lambda_1(L)^2<(1+1/n)Q$, and take
$p,s$ from \cref{ing:gfh}. Choose a uniform one-dimensional subspace
$U\subset p^{-1}L/L\simeq\mathbb F_p^n$ and let $\Gamma$ be its inverse
image. For a fixed shortest vector $v$ used in the analysis, write
\begin{equation}\label{eq:affine-root-scales}
 K=p^{-1}L,\qquad r=\sqrt2s,\qquad
 V=\frac{r^n}{\det\Gamma}=2^{n/2+h_n+O(1)},\qquad
 c_n=\en_r(v)=c+O(1/n).
\end{equation}
Here $\kappa_{\rm pack}=401/1000$,
$ec=2^{2\kappa_{\rm pack}-1}$, and $W(c)=\kappa_{\rm pack}$.
Recall $\ell_n=nw_n+\lceil\log_2(n+2)\rceil=o(h_n)$. We also have $nw_n^2\to\infty$.
The dense lattice clause of \cref{ing:gfh} gives shifted theta flatness
on $K$ at every fixed positive multiple of $s$.
The scale grid in \cref{sec:svp-environment} ensures that the algorithm
has a branch with the required $Q$.

\subsection{A random common coset}
Independently of the random line, choose $z$ uniformly in $K/L$ and
$\delta$ uniformly in $(L/2)/L$. Both root lists are constructed in
\begin{equation}\label{eq:affine-root-coset}
 C=z+\delta+\Gamma.
\end{equation}
Sampling $z$ uses $n$ field elements. Since $p$ is odd,
$L\cap2\Gamma=2L$, so
$L/2L\longrightarrow\Gamma/2\Gamma$ is an isomorphism.
Thus $\delta$ is also uniform modulo $\Gamma$ on the half grid, supplying
the ancestor averages for internal work.

Conditional on $(\Gamma,z,\delta)$, use disjoint raw pools, independent
internal shifts, and independent pair coins for the two subtrees. Their
output lists are independent, and dependence within each list is
controlled by the factorial moment bounds.
Store $x=z+\delta+g_x$, with $g_x\in\Gamma$,
and label it by $g_x+L$. Then
\begin{equation}\label{eq:affine-root-label}
 \lab(x)=\lab(y)\quad\Longleftrightarrow\quad x-y\in L.
\end{equation}

\subsection{Counting representations in the coset}
For positive child energy $a$, define
\begin{equation}\label{eq:affine-lens-def}
 E=4a-c,\qquad E_n=4a-c_n,\qquad
 f_v(x)=\ind_{\mathcal S_r(a)}(x)\ind_{\mathcal S_r(a)}(x-v),\qquad
 J_C(v)=\sum_{x\in C}f_v(x).
\end{equation}
Throughout, $a$ ranges over a fixed compact set with $E-2/e$ bounded
below by a positive constant. All estimates are uniform on this range.

\begin{theorem}[Affine target representations]\label{thm:affine-lens}
Suppose $E>2/e$ with a fixed positive margin. For each fixed $\delta$,
let $\mu_\delta=\E_{\Gamma,z}[J_C(v)]$. Uniformly in $\delta$,
\begin{align}
 \mu_\delta
 &=V(eE/4)^{n/2}\exp\bigl(\pm O(\ell_n)\bigr),
 \label{eq:affine-lens-mean}\\
 \frac{\Var_{\Gamma,z}J_C(v)}{\mu_\delta^2}
 &\le\frac1{\mu_\delta}+2^{-h_n+O(\ell_n)}.
 \label{eq:affine-lens-variance}
\end{align}
For every fixed $\Gamma$, the conditional mean
$\E_z[J_C(v)\mid\Gamma]$ is also $\mu_\delta$.
\end{theorem}
\begin{proof}
Flatness on the dense ambient lattice counts the lens even when its
midpoint width is below $s$. The Gaussian estimate
\eqref{eq:affine-ambient-lens}, proved in \cref{app:affine-lens-details},
gives its ambient mass.

For $t\in K/L$, put $F_t=\sum_{x\in\delta+t+L}f_v(x)$ and
$\theta=p^{1-n}$. A uniform affine line contains each residue with
probability $\theta$. Distinct residues have nonpositive covariance
because their joint probability is
$\theta(p-1)/(p^n-1)\le\theta^2$. Hence
\begin{equation}\label{eq:affine-line-two-moments}
 \E[J_C(v)]=\theta\sum_tF_t,\qquad
 \Var J_C(v)\le\theta\sum_tF_t^2.
\end{equation}
Since $\theta r^n/\det K=V$, the ambient mass proves the mean estimate.
The second sum counts collisions modulo the original lattice:
\begin{equation}\label{eq:affine-residue-collision}
 \sum_t F_t^2=
 \sum_{w\in L}\ \sum_{x\in\delta+K}f_v(x)f_v(x+w).
\end{equation}
Its diagonal contributes $\mu_\delta$ after multiplication by $\theta$.
For nonzero $w$, put $\xi=\|w\|^2/\|v\|^2$.
The two midpoint balls intersect inside a ball of normalized squared
radius at most $(E-c\xi)/4+O(w_n)$.
The KL bound counts the possible $w$, while this shrinking intersection
bounds the midpoint locations for each one. Maximizing
$\xi(E-c\xi)\le E^2/(4c)$ and dividing by the squared mean gives
\[
 V^{-1}2^{\kappa_{\rm pack}n}(ec)^{-n/2}e^{O(\ell_n)}
 =2^{-h_n+O(\ell_n)}.
\]
The choice $ec=2^{2\kappa_{\rm pack}-1}$ makes the packing and
volume factors cancel. \Cref{app:affine-lens-details} supplies the
weighted radial estimate and its uniform boundary bound, proving the variance
claim.

For every fixed direction $\Gamma$, averaging $z$ includes each
ambient residue with the same probability $\theta$, proving the last
claim. Also
$\log_2\mu_\delta=\frac n2\log_2(eE/2)+h_n+O(\ell_n)$,
so $E>2/e$ gives exponentially many representations on average.
\end{proof}

The lower threshold comes from averaging on the dense lattice $K$ and
controlling collisions in $L$. Internal layers instead require uniform
transfer over their shared half grid environments, which is where
the stronger condition $E>1$ enters.

\subsection{A target difference in the two lists}
The geometric count supplies possible endpoint locations. The following
lemma turns these locations into a target occurrence using only the
first two factorial measures.

\begin{lemma}[Target success]\label{lem:dependent-root}
Fix the scale, prime, and $\delta$ above. Let $\mathcal G$ depend only
on $\Gamma$, with $g=\Prb[\mathcal G]>0$, and put
$\mu=\mu_\delta$ and $\xi=2^{-h_n+O(\ell_n)}$ as in
\cref{thm:affine-lens}, with $\mu\ge1$ and $\xi\le1$.
Conditional on each environment
$(\Gamma,z,\delta)$ in $\mathcal G$, suppose the two record lists
are independent, supported on $C\cap\mathcal S_r(a)$, and satisfy
the following bounds with uniform $A>0$ and $D\ge0$:
\begin{equation}\label{eq:dependent-root-envelope}
 e^{-D}A\le m_\pm(x)\le e^D A,\qquad
 S_\pm(x,x')\le e^D A^2
 \quad(x,x'\in C\cap\mathcal S_r(a)).
\end{equation}
Let $Z_v$ count record pairs with difference $v$, and put
$Y=\mathbf1_{\mathcal G}Z_v$ and $\Lambda=A^2\mu$. Then
\begin{align}
 \E[Y]&\ge ge^{-2D}\Lambda,\label{eq:dependent-root-first}\\
 \E[Y^2]&\le e^{2D}(4\Lambda^2+2\Lambda).
 \label{eq:dependent-root-second}
\end{align}
Consequently
\begin{equation}\label{eq:dependent-root-success}
 \Prb[\mathcal G,\ Z_v>0]\ge
 \frac{g^2}{6}e^{-6D}\min\{1,\Lambda\}.
\end{equation}
If $g=1-o(1)$, $\Lambda\ge e^{-o(n)}$, and $D=o(n)$,
this is $e^{-o(n)}$.
\end{lemma}
\begin{proof}
First fix a good environment. Let
$\mathcal R_v=\{x\in C:x,x-v\in\mathcal S_r(a)\}$, $J=|\mathcal R_v|$, and write
$Z_v=\sum_{x\in\mathcal R_v}N_x^+N_{x-v}^-$.
Independence gives conditional mean at least $e^{-2D}A^2J$. For
$x\ne x'$, the product of counts within either list has expectation at most $e^DA^2$.
For $x=x'$, the diagonal correction gives at most $e^D(A^2+A)$.
Consequently,
\[
 \E[Z_v^2\mid\Gamma,z,\delta]\le e^{2D}\bigl(J(J-1)A^4+J(A^2+A)^2\bigr)
 =e^{2D}(A^4J^2+2A^3J+A^2J).
\]
The coset average in \cref{thm:affine-lens} equals $\mu$ for every
fixed $\Gamma$. Since $\mathcal G$ depends only on $\Gamma$, it
follows that $\E[\mathbf1_{\mathcal G}J]=g\mu$, proving the first bound.
For the second, extend the nonnegative terms to all environments and
use $\E[J^2]\le(1+\xi)\mu^2+\mu$. The inequalities
$\mu\ge1$, $\xi\le1$, and $2A\le A^2\mu+1$ give
\eqref{eq:dependent-root-second}.
The second moment method now yields
$\Prb[Y>0]\ge g^2e^{-6D}\Lambda/(4\Lambda+2)$, which implies
\eqref{eq:dependent-root-success}.
\end{proof}

The success estimate averages the root environment, with $\delta$ fixed,
and is uniform in $\delta$. The algorithm samples every environment
without testing $\mathcal G$. Keeping one representative of each vector
value preserves target existence, so the final search can use distinct
values while this analysis counts occurrences.

\subsection{Cost of the final pair search}
The affine average also bounds all other reports from the decoder.
If equally labeled endpoints differ by $w\in L$, membership of one
endpoint in $C$ already ensures membership of the other. Such a pair
therefore has one affine membership factor $\theta$, rather than two.

The search parameters balance the terminal list size against the
number of filters. We choose the list rate so that the expected number
of target representations has leading exponent zero, then choose the
cap threshold so that point incidence work has the same rate as the
filter count. For an energy $a$ satisfying \cref{thm:affine-lens}, set
\begin{equation}\label{eq:affine-root-balance}
 u=1-\frac{c}{2a},\qquad
 \mathsf L=\frac{P(u)+\kappa_{\rm pack}}2,\qquad
 P(\tau)=\mathsf L.
\end{equation}
We merge equal vector values within each list and use quotient labels
to restrict the search to differences in $L$. The filter count is
\eqref{eq:filter-counts}, so
$\log_2M_f=K_\tau(u)n+O(\pad)$.

\begin{proposition}\label{prop:terminal-infimum}
Fix $0<\kappa<1/2$ and $c=2^{2\kappa-1}/e$. Under the balance
\eqref{eq:affine-root-balance} with $\kappa_{\rm pack}=\kappa$ and
$E=4a-c>2/e$, the terminal work rate is
\begin{equation}\label{eq:terminal-closed-form}
 K_{\rm term}(E)
 =-\frac12\log_2\!\left(2^{1-\kappa}\sqrt{c/E}-c/E\right).
\end{equation}
Its infimum on this curve is
\begin{equation}\label{eq:terminal-infimum}
 \inf_{E>2/e}K_{\rm term}(E)
 =-\frac12\log_2(1-2^{2\kappa-2}).
\end{equation}
\end{proposition}
\begin{proof}
Put $x=\sqrt{c/E}$. The identities
$1-u^2=cE/(4a^2)$ and $1-\tau^2=2^{-2\mathsf L}$ give
$1-2\tau^2/(1+u)=2^{1-\kappa}x-x^2$.
The energy condition is equivalent to $0<x<2^{\kappa-1}$.
Since $\kappa<1/2$, the quadratic is increasing throughout this
interval. Letting $x\uparrow2^{\kappa-1}$ proves the infimum.
\end{proof}

At $\kappa_{\rm pack}=0.401$, the infimum is
$0.412961878\ldots$. Our parameters approach this value while
keeping a fixed positive geometric margin. Earlier energies can
reduce space, whereas a substantial time improvement requires a
different terminal balance or geometric estimate.
\Cref{tab:svp-bottlenecks} records the costs and the geometric margin.

\begin{proposition}
\label{prop:affine-root-work}
Use the preceding search parameters, with
$1-2\tau^2/(1+u)>0$ by a fixed margin. Conditional on each root
environment $(\Gamma,z,\delta)$ with $\Gamma$ good, suppose the two
lists are independent, supported on
$C\cap\mathcal S_r(a)$, and satisfy $m_\pm(x)\le e^{D_n}A_n$ uniformly,
where
\begin{equation}\label{eq:affine-root-work-amplitude}
 D_n=o(n),\qquad \log_2A_n=(\mathsf L-W(a))n+o(n).
\end{equation}
Then the following hold for the search of \cref{ing:bdgl}.
\begin{enumerate}[label=(\roman*),leftmargin=*]
\item Averaged over the root environment conditional on a good
superlattice, the expected list sizes are at most
$2^{\mathsf L n+o(n)}$, and the expected point incidences and bucket
reports are at most
\begin{equation}\label{eq:affine-root-work-bound}
 2^{K_\tau(u)n+o(n)}.
\end{equation}
If $D_n$, the amplitude remainder, and code padding are $O(\pad)$,
the logarithmic remainders in both bounds are $O(\pad)$.
\item The code padding can be chosen so that every fixed shortest vector
representation has the coverage guarantee of \cref{ing:bdgl}.
\end{enumerate}
\end{proposition}
\begin{proof}
First apply the uniform intensity bounds on the good event. The remaining
geometric sums are nonnegative, so extend them to all environments and
divide by the good event probability $1-o(1)$. The affine average then
has its original uniform direction.

The average list size is at most
$A_nV(ea)^{n/2}e^{D_n+O(\ell_n)}(1+o(1))$.
Its rate is $\mathsf L$, and the point incidence rate is
$\mathsf L+K_\tau(u)-P(\tau)=K_\tau(u)$.
For nonzero differences, the conditional intensities multiply, followed
by the single affine membership factor. \Cref{app:affine-work-details}
combines the midpoint volume with common cap probability, uniformly
up to the cap boundary.

The choice of parameters balances this work against the target count.
Put $\mathcal B_E=V(eE/4)^{n/2}$ and $Q=4a(1-\tau^2)$. Then
\begin{equation}\label{eq:affine-root-target-balance}
 A_n^2\mathcal B_E=2^{o(n)},\qquad
 \frac{Q^2}{4cE}=2^{-2\kappa_{\rm pack}}.
\end{equation}
These follow from $1-u^2=cE/(4a^2)$ and
$1-\tau^2=2^{-2\mathsf L}$. The first sets the target mass.
The second cancels the factor from \cref{lem:kl-count} when summing
the filtered difference counts, giving rate $K_\tau(u)$.
The appendix supplies the uniform bin estimates.

After deduplication, each zero pair in a filter is charged to one
point incidence. A shortest vector representation has correlation
$u+O(w_n)$ with the stated active constraint margin, giving coverage.
All other logarithmic losses are $O(\ell_n+\log n)$, which also proves
the quantitative $O(\pad)$ assertion.
\end{proof}

\section{Solving the shortest vector problem}\label{sec:svp}

The SVP algorithm cools thirty-two pools, combines them through four full
difference layers, and searches the resulting two lists on a common affine
coset. We choose the list sizes backward from the mass needed for one
shortest difference. This makes the target condition, rather than the
initial sampler, determine the required list sizes.

Each cooled pool is a tree of the form in
\cref{fig:cooling-merge-tree,alg:ideal-cooling-tree}. The thirty-two
pools form the leaves of two full difference trees of depth four, followed
by the terminal affine pair search. Cooling retains only isolated
activated edges, whereas the four full layers retain every eligible
pair. Their
analysis uses the conditioning on prescribed offsets in \cref{tab:conditioning}:
average private child randomness, form the conditional products, and
only then average the shared local shift. This order lets the same
two moment bounds pass through the four full layers.

\subsection{Choosing the list sizes from the target}\label{sec:svp-orbit}

Use the rates in \eqref{eq:rates} and the scale constants in
\eqref{eq:svp-scale-constants}, so $W(c)=\kappa_{\rm pack}=401/1000$.
Let $a_0,\ldots,a_4$ be the energies of the generated lists and
$a_5=c$ the target energy. For each transition, write
\begin{equation}\label{eq:svp-hard-correlations}
 u_j=1-\frac{a_j}{2a_{j-1}},\qquad E_j=4a_{j-1}-a_j.
\end{equation}

A terminal list of rate $L_4$ has nominal intensity rate
$L_4-W(a_4)$ per location. Combining two such intensities with the
midpoint count, \eqref{eq:entropy-identity} gives target mass rate
\[
 2L_4-P(u_5)-W(c).
\]
We set this rate to zero and add sublinear padding to make the actual
mass large. Earlier full layers obey the list size balance
$L_j=2L_{j-1}-P(u_j)$, so all required list rates follow backward:
\begin{equation}\label{eq:svp-backward-rates}
 L_4=\frac{P(u_5)+W(c)}2,\qquad
 L_{j-1}=\frac{L_j+P(u_j)}2\quad(1\le j\le4).
\end{equation}

The terminal geometry anchors the energy choice.
\Cref{prop:terminal-infimum} shows that choosing
$E_5=4a_4-c$ just above $2/e$ puts the terminal work rate near
$0.412962$. It also fixes $a_4\approx0.264120$ and
$L_4\approx0.260971$. The four internal layers connect this endpoint
to the cooling shell $a_0=1/(1+u_{\rm s})$. Their energies must
meet three goals:
\begin{enumerate}[label=(\roman*),leftmargin=*,nosep]
\item leave a positive cooling list margin, internal bridge margins
$E_j>1$, and abundance margins for the moment transfer,
\item keep cooling and internal search work below the terminal rate,
\item reduce the largest intermediate list, which determines space.
\end{enumerate}
The backward recurrence \eqref{eq:svp-backward-rates} couples these
choices: changing an internal energy changes both adjacent correlations
and the list size required at every earlier level.
Scheme A below keeps cooling and the largest internal search close
to the terminal rate while reducing peak storage.
The following rational parameters realize this tradeoff with fixed
positive margins.
\begin{equation}\label{eq:svp-soft-parameters}
 u_{\rm s}=\frac{47}{80},\qquad
 z_{\rm s}=\frac{80}{127},\qquad
 \gamma_{\rm s}^2=\frac{33}{40},\qquad
 q_{\rm s}=\frac{66}{127}.
\end{equation}
\begin{equation}\label{eq:svp-energy-orbit}
 (a_0,a_1,a_2,a_3,a_4,a_5)
 =\left(\frac{80}{127},\frac{5201}{10000},\frac{851}{2000},
          \frac{179}{500},\frac{6603}{25000},c\right).
\end{equation}
Here $a_0=z_{\rm s}$, $z_{\rm s}(1+u_{\rm s})=1$, and
$q_{\rm s}=4z_{\rm s}-2=\gamma_{\rm s}^2z_{\rm s}$, as required by
\cref{thm:soft}. The initial padding pays for cooling and leaves the target mass needed
by the final second moment bound.
Cooling needs raw sampling rate
\begin{equation}\label{eq:svp-raw-rate}
 m_{\rm raw}=L_0+J(z_{\rm s}).
\end{equation}
The resulting parameters are
\begin{center}
\begin{tabular}{@{}ccc@{}}
\toprule
Height & Energy & List rate\\
\midrule
0 & $80/127$ & $0.306032\ldots$\\
1 & $5201/10000$ & $0.307101\ldots$\\
2 & $851/2000$ & $0.304330\ldots$\\
3 & $179/500$ & $0.313711\ldots$\\
4 & $6603/25000$ & $0.260970\ldots$\\
\bottomrule
\end{tabular}
\end{center}
In particular, $L_0-P(u_{\rm s})>0.000643$ and
$m_{\rm raw}<0.372456$, so the cooling list exceeds its critical size.

\subsection{Constructing the two final lists}\label{sec:svp-environment}

The length estimate \eqref{eq:short-scale} gives a polynomial-factor
estimate of $\lambda_1(L)$. The multiplicative scale grid in
\cref{app:svp-environment} has $O(n\log n)$ branches, including one with
$Q_i\le\lambda_1(L)^2<(1+1/n)Q_i$. Thus one branch places every shortest
vector at the target energy $c+O(1/n)$.

On each branch, choose the scale and prime by the bounded initialization
in \cref{app:svp-environment}.
A complete trial samples a fresh uniform one dimensional subspace
$U=\Gamma/L$ in
$K/L\simeq\mathbb F_p^n$, where $K=p^{-1}L$.
Put $r=\sqrt2s$ and $V=r^n/\det\Gamma$. On the correct branch,
the good smoothing event $\mathcal G$ has probability
$1-2^{-\Omega(h_n)}$, and
\begin{equation}\label{eq:svp-scale-volume}
 \log_2V=n/2+v_n,\qquad v_n=h_n+O(1),\qquad
 \frac{2\pi\|v\|^2}{nr^2}=c_n=c+O(1/n)
\end{equation}
for every fixed shortest vector $v$. The index is always $p$, so
$V$ is fixed before the line is drawn.

Draw the common coset as in \cref{eq:affine-root-coset}, independently
of the line. The two terminal lists have support
\begin{equation}\label{eq:svp-root-coset}
 C=z+\delta+\Gamma.
\end{equation}
The half grid property established there supplies the ancestor averages.
Build two trees of depth four with root offset $z+\delta$, using
\textsc{BuildFullTree} from \cref{alg:full-difference-tree}.
At their thirty-two leaves, use independent cooling forests with the
parameters above, terminal scale $r$, and raw rate $m_{\rm raw}$.

The wide starting scale is polynomially larger than $r$.
A fixed logarithmic cooling depth, with raw rate $m_{\rm raw}$,
therefore supplies each initial pool. The schedule is given in \cref{app:svp-environment}.
The reference from \cref{thm:soft} is
\begin{equation}\label{eq:svp-soft-reference}
 \nu_{0,b}(x)=\frac{\amp_0}{V}\rho_r(x)
       \ind_{x\in(b+\Gamma)\cap\mathcal S_r(a_0)},\qquad
 \log_2\amp_0=m_{\rm raw}n+\sigma_0,
\end{equation}
and factorial error $\varepsilon_{\rm cool}\le\exp[-\Omega(nw_n^2)]$, where
\[
 \sigma_0=C_{\rm leaf}\pad-
 O\bigl((h_n+\errwin)\log n\bigr)=\Theta(\pad).
\]
The amplitudes are fixed before the affine environment is chosen.
\Cref{app:svp-environment} verifies these scale and list size choices.

\subsection{The target representation count}\label{sec:svp-hard-dynamics}

At height $j=1,\ldots,4$, retain every record pair whose difference
lies in $\mathcal S_r(a_j)$. Apply \eqref{eq:nominal-recursion} with
$d=4$ and $V_r=V$. Writing $k_j=L_j-W(a_j)$ gives
\begin{equation}\label{eq:svp-amplitude-rates}
 \log_2A_j=k_jn+2^j\sigma_0-v_n,\qquad 0\le j\le4.
\end{equation}
Substitution of \eqref{eq:svp-energy-orbit} gives
$\min_{1\le j\le4}E_j>1.16$ and an abundance exponent
greater than $0.13$ for the four internal layers.
Thus \cref{thm:shell-tree} applies with bridge margin $0.16$ and
abundance parameter $0.12$.

Retain $\mathcal H_{\rm term}=\sigma(\Gamma,z,\delta)$ and average the
private environments, samples, and coins of each tree. The two trees
remain independent given this retained information.
On the input shell, replace the Gaussian reference by its central
intensity $A_0$. By \eqref{eq:shell-leaf-error}, the resulting leaf
error is
\[
 \varepsilon_0=\varepsilon_{\rm cool}+na_0w_n=O(nw_n).
\]
For target extraction we only need logarithmic distortion $o(n)$.
Applying \cref{thm:shell-tree} now gives
\[
 \varepsilon_4\le16\varepsilon_0+15\eta_n+30\cdot2^{-0.12n}
 =O(\errwin)+e^{-\Omega(nw_n^2)}.
\]
Choose the uniform upper bound $D_n=C_D(h_n+\errwin)$ with $C_D$
sufficiently large. The terminal moments then satisfy
\begin{equation}\label{eq:svp-terminal-envelopes}
 e^{-D_n}A_4\le m_\pm(x)\le e^{D_n}A_4,\qquad
 S_\pm(x,x')\le e^{D_n}A_4^2,\qquad
 D_n\le C_D(h_n+\errwin),
\end{equation}
on $C\cap\mathcal S_r(a_4)$ for every good $\Gamma$.
The second factorial measure counts distinct IDs, including equal values.

The terminal bridge has energy
\begin{equation}\label{eq:svp-affine-energy}
 E_5=4a_4-c=0.735778066\ldots
       >2/e+10^{-5},
\end{equation}
so \cref{thm:affine-lens} applies to
$J_C(v)=\#\{x\in C:x,x-v\in\mathcal S_r(a_4)\}$.
For every fixed $\delta$, its mean satisfies
\[
 \mu_\delta=V(eE_5/4)^{n/2}e^{\pm O(\errwin)}.
\]
The actual energy $c_n=c+O(1/n)$ changes its logarithm by $O(1)$.
The terminal list size balance now gives
\begin{equation}\label{eq:svp-target-padding}
 A_4^2\mu_\delta\ge2^{32\sigma_0-v_n-O(\errwin)}
                 =2^{\Omega(\pad)}.
\end{equation}
Apply \cref{lem:dependent-root} with $A=A_4$ and $D=D_n$.
The good superlattice probability tends to one, and the estimate is
uniform in $\delta$. Averaging $\delta$ gives
\begin{equation}\label{eq:svp-codefree-success}
 \Prb[\mathcal G\text{ and some terminal pair has difference }v]
       \ge2^{-B_n},\qquad
 B_n=C_B(h_n+\errwin)=o(\chi_n).
\end{equation}
This proves target existence. Deduplicating terminal values preserves it.

\subsection{The cost of finding the target}\label{sec:svp-work}

The moments also control the search work. Averaging the extra half grid
shift and the ancestor shifts gives
\begin{equation}\label{eq:svp-list-first}
 \E[|\Xi_j|\mid\Gamma\text{ good},z]
 \le2^{L_jn+O(\pad)},\qquad 1\le j\le4.
\end{equation}
The cooling lists satisfy the corresponding bound at rate $L_0$.
\Cref{app:svp-search} derives these bounds and the candidate counts by
first fixing the common environment and then multiplying the conditional
first intensities of the independent children.

Cooling uses $\tau_{\rm s}=u_{\rm s}$ and
$K_{\rm s}=K_{u_{\rm s}}(u_{\rm s})$. At heights $j=1,\ldots,5$, use
\begin{equation}\label{eq:svp-code-parameters}
 P(\tau_j)=L_{j-1},\qquad K_j=K_{\tau_j}(u_j).
\end{equation}
The cooling pair work has rate
\begin{equation}\label{eq:svp-soft-work}
 T_{\rm s}=K_{\rm s}+2(L_0-P(u_{\rm s}))
          =0.412924647\ldots.
\end{equation}
For each internal full layer, the ordinary correlation term in
\eqref{eq:dummy-rate} gives rate $K_j$ and dominates the contribution
of nearly parallel pairs. The strict smoothing and wedge margins needed
for this comparison are checked in \cref{app:svp-search}.

At the terminal call, write $x=z+\delta+g_x$ and assign
$\mathsf{lab}(x)=g_x+L$. Then
\begin{equation}\label{eq:svp-label-exactness}
 \mathsf{lab}(x)=\mathsf{lab}(y)\quad\Longleftrightarrow\quad x-y\in L.
\end{equation}
The affine search bound \cref{prop:affine-root-work}, applied to
\eqref{eq:svp-terminal-envelopes}, gives rate $K_5$ after deduplication.
Its expectation uses the same unconditioned coset average as the target bound.
The chosen parameters give
\[
 \max\{0.2,m_{\rm raw},T_{\rm s},K_1,\ldots,K_4\}
 =K_3<0.412960<K_5<0.412965<0.413.
\]
Thus the terminal search determines time and is already close to the
minimum of its balance curve (\cref{prop:terminal-infimum}).
The earlier energies mainly determine peak storage.

The largest retained list determines space. The input rate of the
last full layer reflects its angular pair loss:
\[
 u_4=0.631117\ldots,\qquad P(u_4)=0.366452\ldots,\qquad
 L_3=\frac{L_4+P(u_4)}2.
\]
This layer needs a larger input list to supply the smaller terminal
list after its angular restriction. Changing $a_3$ changes this loss
and also the preceding correlation $u_3$, so reducing storage must
respect the earlier search budget. For the chosen energies, set
\begin{equation}\label{eq:svp-live-space}
 L_{\max}=L_3<0.313712,
 \qquad s_{\rm live}=0.31375,
 \qquad S=2^{s_{\rm live}n+C_S\pad},
\end{equation}
with $C_S$ sufficiently large. Every input rate is at most
$L_{\max}<s_{\rm live}$, so the restricted traversal of
\cref{ing:bdgl} fits this storage with the same leading time rate.
Raw samples are streamed, and depth first evaluation keeps only the
current lists and traversal state. These estimates supply the work
and storage inputs to \cref{prop:finite-realization}.

\Cref{tab:svp-bottlenecks} collects the costs and the terminal geometric
margin for this parameter choice.
\begin{table}[htbp]
\centering
\small
\begin{tabularx}{\textwidth}{@{}>{\raggedright\arraybackslash}p{0.23\textwidth}>{\raggedright\arraybackslash}p{0.40\textwidth}X@{}}
\toprule
Quantity & Value or bound & Role\\
\midrule
Cooling work $T_{\rm s}$ & $0.4129246\ldots$ & Initial list generation\\
\addlinespace[3pt]
Internal work $K_3$ & $0.4129598\ldots$ & Largest nonterminal cost\\
\addlinespace[3pt]
Terminal work $K_5$ & $0.412964<K_5<0.412965$ & Time bottleneck\\
\addlinespace[3pt]
List rate $L_3$ & $<0.313712$ & Space bottleneck\\
\addlinespace[3pt]
Terminal margin & $19\cdot10^{-6}<E_5-2/e<20\cdot10^{-6}$ & Strict affine lens margin\\
\bottomrule
\end{tabularx}
\caption{The main costs and terminal margin of the SVP construction.}
\label{tab:svp-bottlenecks}
\end{table}

\subsection{The algorithm and success amplification}\label{sec:svp-completion}
\Cref{alg:svp} assembles the construction. The scale grid has a branch
at the shortest length, and independent trials on that branch turn
the target probability $2^{-B_n}$ into constant success. A final
polynomial number of complete executions gives the stated high
probability. Keeping the shortest valid vector makes it unnecessary
to identify which branch or trial succeeded.

The recursive calls in the algorithm use the finite implementation
of \cref{prop:finite-realization}. Initialization checks, resource
limits, and
exceptional execution rules are specified in \cref{app:svp-execution}.

\begin{algorithm}[!ht]
\caption{Shortest vectors: \textsc{SVP}}
\label{alg:svp}
\KwIn{A rational basis $B$ of $L$ and the fixed parameters of \cref{sec:svp-orbit}.}
\KwOut{A nonzero vector of $L$, shortest with the probability in \cref{thm:svp-main}.}
$v_{\rm best}\gets$ a nonzero column of $B$\;
Construct the polynomial scale grid $(Q_i)$ of \cref{eq:svp-scale-grid}\;
\For{$e=1,\ldots,\lceil n\log_2(n+2)\rceil$}{
  \For{each scale $Q_i$}{
    Initialize its squared width and prime $p$ as in \cref{app:svp-environment}, skipping the scale on failure\;
    \For{$t=1,\ldots,\lceil2^{\chi_n}\rceil$}{
      Draw a fresh uniform line defining $\Gamma$ as in \cref{ing:gfh}\;
      Independently draw $z\sim\operatorname{Unif}((p^{-1}L)/L)$ and $\delta\sim\operatorname{Unif}((L/2)/L)$\;
      Preprocess $\Gamma$ for cooling\;
      $b\gets z+\delta$\;
      $\Xi_+\gets\textsc{BuildFullTree}(b,4)$\;
      $\Xi_-\gets\textsc{BuildFullTree}(b,4)$ with independent private randomness\;
      Merge equal values within each terminal list\;
      Label $x=b+g_x$ by $g_x+L$ and draw fresh terminal search codes\;
      \For{each equally labeled pair $(x,y)$ reported by the terminal search}{
        \If{$x\ne y$ and $\|x-y\|<\|v_{\rm best}\|$}{
          $v_{\rm best}\gets x-y$\;
        }
      }
    }
  }
}
\Return $v_{\rm best}$\;
\end{algorithm}

On the correct branch, \eqref{eq:svp-codefree-success} and
\cref{prop:finite-realization} give success at least $2^{-B_n-1}$
per trial, conditional on a valid public prime. Since $B_n=o(\chi_n)$,
the scheduled $2^{\chi_n}$ trials give constant success even after
the public prime and sampling errors.
Taking the shortest output of $\lceil n\log_2(n+2)\rceil$
independent complete executions reduces failure to
$2^{-\Omega(n\log n)}$, with only $2^{o(n)}$ extra time and the
same workspace. \Cref{app:svp-amplification} accounts for the
implementation error over this amplified execution, proving
\cref{thm:svp-main}.
\FloatBarrier

\subsection{Other choices of the four layer energies}\label{sec:svp-tradeoffs}
The terminal energy fixes the final search cost, while the earlier
energies determine the list sizes needed to reach it. Varying these
energies therefore changes peak storage and may also make an internal
search the dominant cost. We describe this tradeoff by the largest
list rate and the largest work rate, then state the conditions under
which the preceding analysis applies.

\Needspace{14\baselineskip}
Fix rational $u_{\rm s}\in(1/2,31/50]$ and positive rational
$a_0,\ldots,a_4$ with $a_0=1/(1+u_{\rm s})$.
Keep $\kappa_{\rm pack}=401/1000$, $a_5=c$, and define
$u_j,E_j,L_j,\tau_j,K_j$ by
\eqref{eq:svp-hard-correlations}, \eqref{eq:svp-backward-rates},
and \eqref{eq:svp-code-parameters}. Put
\[
\begin{aligned}
 \Delta_{\rm s}&=L_0-P(u_{\rm s}),\\
 s(\mathbf a)&=\max_{0\le j\le4}L_j,\\
 t(\mathbf a)&=\max\{1/5,L_0+J(a_0),
      K_{u_{\rm s}}(u_{\rm s})+2\Delta_{\rm s},K_1,\ldots,K_5\}.
\end{aligned}
\]
Here $\Delta_{\rm s}$ is the extra list exponent available for cooling,
$s(\mathbf a)$ sets peak storage, and $t(\mathbf a)$ counts
preprocessing, raw sampling, cooling search, and the five later
searches. Choose a rational storage rate $s_*>s(\mathbf a)$.

The construction works when the following conditions hold with
positive margins independent of $n$.
\begin{enumerate}[label=(\roman*),leftmargin=*]
\item Cooling has enough records: $\Delta_{\rm s}>0$.
\item Internal layers preserve the moment bounds. For $1\le j\le4$,
\[
 E_j>1,\qquad
 \mathfrak a_j:=L_{j-1}-W(a_{j-1})
       +\frac12+\frac12\log_2(eE_j/4)>0.
\]
The first inequality controls the shared midpoint shift, and the
second bounds the relative contribution from reusing an input.
\item Pair search satisfies the cap and smoothing estimates.
For $1\le j\le5$, require
$0<u_j<1$, $0<\tau_j^2<1$, and
\mbox{$1-2\tau_j^2/(1+u_j)>0$}.
For the remaining comparisons put
$t_j=2^{5-j}$, $c_{0,j}=4t_j^2a_{j-1}$ for $1\le j\le4$,
$c_{0,\rm s}=2/(1-u_{\rm s}^2)$, and $\Delta=1/2000$.
The smoothing and nearly parallel pair bounds require, for $1\le j\le4$,
\[
 \begin{gathered}
 2t_j^2a_j>1,\qquad 8a_{j-1}\tau_j^2>1,\qquad c_{0,j}>4,\\
 L_{j-1}-\tfrac12\log_2(c_{0,j}/2)<0,\\
 c_{0,j}\Delta<1,\qquad 4a_{j-1}(2-\Delta)>1.
 \end{gathered}
\]
For cooling the corresponding checks are
$4u_{\rm s}^2/(1-u_{\rm s}^2)>1$,
$c_{0,\rm s}\Delta<1$, and $c_{0,\rm s}(2-\Delta)>1$.
\item The terminal affine count and search apply: $E_5>2/e$.
With $\zeta=\tfrac12\log_2(eE_5/2)$,
$\omega=L_4-W(a_4)$, and $Q=4a_4(1-\tau_5^2)$, require
$\zeta>0$, $\omega+\zeta>0$, and $Q/(2c)>1$.
\item Storage permits restricted traversal:
$1/5<s(\mathbf a)<\min\{s_*,1\}$.
Use the depth and list size schedule of \cref{app:svp-environment}
with this storage rate.
\end{enumerate}

\begin{proposition}\label{prop:four-layer-parameters}
Every parameter choice satisfying conditions (i)--(v) above gives an
SVP algorithm with the success probability of \cref{thm:svp-main}
and, on every execution,
\[
 T\le2^{t(\mathbf a)n+o(n)},\qquad S\le2^{s_*n+o(n)}.
\]
\end{proposition}
\begin{proof}
Conditions (i) and (ii) give valid cooling coefficients through
\cref{thm:soft} and propagate the moment bounds through four full
layers by \cref{thm:shell-tree}. The backward recurrence sets the
leading target count exponent to zero, and padding makes that count
$2^{\Omega(\pad)}$. Condition (iv) permits the affine count and work
estimates, so \cref{lem:dependent-root} gives target probability
$e^{-o(n)}$ per trial.

The ancestor denominators remain $16,8,4,2$. Condition (iii) gives
the list smoothing, active cap, nearly parallel pair, and terminal
cap comparisons used in \cref{app:svp-search}. The bound $L_j<1$
also makes the contribution with common cap factor at most $1/16$
smaller than the filter rate. The cooling point rate
$K_{u_{\rm s}}(u_{\rm s})+\Delta_{\rm s}$ lies between $L_0$ and
the cooling pair rate, while $K_j>L_{j-1}$ at every later search.
Thus (v) fits restricted traversal into the stated work and storage.
The amplification in \cref{sec:svp-completion} and
\cref{prop:finite-realization} give the success and resource bounds.
\end{proof}

The main instance above is scheme A. A second feasible instance, scheme B,
keeps $u_{\rm s},a_0,a_4,c$ unchanged and takes
\[
 (a_1,a_2,a_3)
 =\left(\frac{6501}{12500},\frac{42549}{100000},
                         \frac{35777}{100000}\right).
\]
Its largest list rate is $L_3<0.313533$, so
$s_*=0.31355$ gives a displayed space bound $2^{0.3136n+o(n)}$.
Its largest work rate is the internal rate
$K_2<0.412980<0.4130$. Thus it offers less space at
a slightly higher time exponent than A before rounding. Both have the same
terminal geometry, and direct substitution verifies the admissibility
conditions for both. Schemes A and B give two operating points
in the four layer parameter region: increasing the overall work rate
by about $1.6\cdot10^{-5}$ lowers the displayed space exponent from
$0.3138$ to $0.3136$.

\section{Independent Gaussian batches}\label{sec:batch}
The single sample construction reads once from a correlated list.
To obtain an exponential number of outputs within the same total
cost, we instead reuse a fixed partner list, called a bank, at each merge. Two
rejections make every trial's complete output law independent of that
list, allowing conditional product laws to pass through the recursion.

We retain the Gaussian shell geometry, smoothing extensions, and
endpoint schedule of \cref{sec:single-dgs}, using the default budgets
\eqref{eq:budgets}. The new ingredients are uniform control of partner
and representation weights, a correction against each fixed bank,
and radial coverage sufficient for the entire batch.

\subsection{Two rejections against a fixed bank}
\label{sec:bank-correction}
Selecting a partner from a bank introduces a normalization that depends
on the left record. Even after this dependence is removed, a difference
with more representations is proposed more often. We correct these
two effects in order. Accepting the left record with probability
$A_C(x)/D$ cancels the denominator used to select its partner. A second
rejection divides by the total proposal weight of the difference and
inserts its desired target weight.

The correction has two consequences. First, the complete trial law,
including failure, is the same for every good bank, so independent
trials remain independent after averaging over banks. Second, a
matching upper bound on representation weights bounds the proposal
law before the last rejection by a constant times the output law
at each fixed bank. Applying this comparison
before averaging the bank-dependent query cost lets the bank's
original first moment control the reverse search. The lemma records
both properties.

\begin{lemma}[Correction against a fixed bank]\label{lem:bank-correction}
Fix public information $H$. Conditional on $H$, let the left trials
be iid on $\mathcal X\cup\{\dagger\}$, with their entire stream
independent of a finite bank $C=(y_i)$. For each vector $x\in\mathcal X$,
write $q(x)=\Prb[X=x\mid H]$. Thus $\sum_xq(x)\le1$, with the
remaining probability assigned to failure $\dagger$.
For $K\ge0$, define
\[
 A_C(x)=\sum_iK(x,y_i),\qquad
 B_C(z)=\sum_iq(z+y_i)K(z+y_i,y_i).
\]
Let $D,c>0$ and a summable target weight $h\ge0$ depend only on $H$.
Call a bank good if $A_C(x)\le D$ for every $x$ with $q(x)>0$,
and $B_C(z)\ge c h(z)$ for every $z$ with $h(z)>0$.
Apply \cref{alg:fixed-bank-trial} to each left trial, using independent
private randomness for its choices. On a good bank the output trials are
iid, with probability
\begin{equation}\label{eq:bank-complete-law}
 q_o(z)=c h(z)/D
\end{equation}
of returning each vector $z$ and the remaining probability at $\dagger$.
Both parts of this distribution are independent of $C$.
If also $B_C(z)\le C_0c h(z)$ for every $z$, the proposal law before the second
rejection is at most $C_0q_o(z)$.
\end{lemma}
\begin{proof}
Fix a good bank. For $h(z)>0$, summing over representations gives
\[
 \Prb[Z=z\mid H,C]
 =\sum_iq(z+y_i)\frac{K(z+y_i,y_i)}D
                 \frac{c h(z)}{B_C(z)}
 =\frac{c h(z)}D.
\]
Outputs with $h(z)=0$ are rejected, leaving failure mass
$1-(c/D)\sum_z h(z)$. Thus the complete trial law, including failure,
is the same for every good bank. Conditional independence of the
left trials and their private choices gives a product of this common
law, which remains a product after averaging the bank.
Before the second rejection the mass is $B_C(z)/D$, giving the
proposal bound.
\end{proof}

\begin{algorithm}[htbp]
\caption{Two rejections: \textsc{FixedBankTrial}$(X,C)$}
\label{alg:fixed-bank-trial}
\KwIn{A left trial $X\in\operatorname{supp}(q)\cup\{\dagger\}$, fixed bank $C=(y_i)$,
and $q,K,D,c,h$ from \cref{lem:bank-correction}.}
\KwOut{A difference $z$ or $\dagger$. The bank is retained for later trials.}
\If{$X=\dagger$}{\Return $\dagger$\;}
$x\gets X$, $A\gets\sum_iK(x,y_i)$\;
\If{a fresh Bernoulli trial of probability $\min\{1,A/D\}$ fails}{\Return $\dagger$\;}
Choose a bank record $i$ with probabilities $K(x,y_i)/A$\;
$z\gets x-y_i$\;
$B\gets\sum_kq(z+y_k)K(z+y_k,y_k)$\;
\If{a fresh Bernoulli trial of probability $\min\{1,ch(z)/B\}$ fails}{\Return $\dagger$\;}
\Return $z$\;
\end{algorithm}

The bank is retained across trials, so the complete law in
\eqref{eq:bank-complete-law} applies to every reuse of it. Clipping
defines the procedure on arbitrary banks. The implementation of the
two sums, including the positive selected contribution to the reverse
sum, is given in \cref{app:batch-execution}.

For a recursive construction, define a reference tree that replaces
the whole output of each bad bank by fresh iid trials of
\eqref{eq:bank-complete-law}. The common complete law proves the
product property by induction. Coupling the actual and reference trees
until their first bad bank transfers it to the algorithm.
\Cref{app:batch-search} applies the fixed bank proposal bound to the
reverse search cost.

\FloatBarrier
\subsection{Exponential batches with a fixed bank}\label{sec:batch-dgs}
We apply \cref{lem:bank-correction} recursively with the Gaussian
kernel. The construction has three parts: produce independent trials
on thin shells, combine radial bins using their known coefficients,
and add one terminal merge for the inner bins needed at width $3/2$.
We use the fixed bank strategy of~\cite{DL26}. On an arbitrary lattice,
the main task is to control both the partner weight of every possible
left record and the representation weight of every possible output.
Gaussian identities estimate their means on a smoothing superlattice,
and uniform concentration bounds make these estimates hold
simultaneously for the realized bank. The two rejections then remove
the bank from each trial's complete output law. Each radial bin uses
its own smoothing extension so that its index cost reflects the
lengths it must supply.

Use the endpoint parameters $u_n,\gamma,a_0,a_1$ of
\cref{lem:endpoint-schedule}. Put
\[
 \degpad=\lceil10\errwin\rceil,\quad
 M=\left\lceil2^{m(u)n+C_a\pad}\right\rceil,\quad
 R(u)=\tfrac12\log_2(2a_0).
\]
A stream consists of $M$ trials, each returning a vector or $\dagger$.
Including failed trials keeps the Gaussian coefficient explicit
throughout the recursion and radial mixing. Conditioning on success
would instead introduce the unknown Gaussian mass of the shell.
Only successful records occupy vector storage, as described in
\cref{app:batch-execution}.

\subsubsection{A smoothing extension for each radial bin}
A single shell of relative width $w_n$ discards Gaussian mass
$\exp[-\Omega(nw_n^2)]$. This error is small for one output, but
truncating an independent $B$-product can cost up to $B$ times that
mass. We therefore cover an annulus of constant radial width whose
large deviation tail is exponentially smaller than $B^{-1}$.
Product code search still uses thin shells, so we partition the annulus
into polynomially many bins, build a separate tree on each bin, and
balance the root streams. The choice below makes the endpoint tail
rates $J(z_-),J(z_+)$ exceed the batch rate.

Choose fixed rational $\alpha^{-2}<z_-<1<z_+$.
Partition $[z_-,z_+]$ into disjoint half-open bins $I_j$ of width at
most $c z_-w_n$, including the final endpoint in the last bin.
Let $z_j$ be their rational centers and set
\[
 A_j=\{x:2\pi\|x\|^2/(n\sigma^2)\in I_j\},\quad
 r_j^2=\sigma^2z_j/a_0,\quad
 s_j^2=\sigma^2z_j/(2a_0).
\]
The number $Q$ of bins is polynomial and
$A_j\subseteq\mathcal S_{r_j}(a_0)$ for small fixed $c$.
For each bin choose an independent $\Gamma_j\supset L$ using
\cref{lem:dgs-index-wrapper} at scale $s_j$, with
\[
 \iota_j=\max\{0,\tfrac12\log_2(2a_0/(\alpha^2z_j))\},\qquad
 \log_2p_j=\iota_jn+3h_n+O(1).
\]
Except with probability $2^{-2h_n+O(1)}$ its dual mass at $s_j$ is
at most $2^{-h_n}$. A union bound pays for the polynomially many public
choices. Basis preprocessing has rate $0.2$ and gives
$G_j\le\poly(n)s_j$. Within each tree suppress the bin index on its
lattice and write $V_r=r^n/\det\Gamma_j$.

Use one common depth $t$ for all bins so that each leaf width
$R_j=r_j\gamma^{-t}$ satisfies
$(R_j/G_j)^2\ge\log(8n)/\pi$.
Then $t=O(\log n\log\log n)$ and
$t(\degpad+2)=o(\pad)$. Generate each of the $M$ leaf labels by
\cref{lem:gaussian-proposal}. Its probability mass on vector outputs is
$\lambda_t\rho_{R_j}(x)/V_{R_j}$ on the prescribed shell and coset,
with $\lambda_t=1/2$.

\subsubsection{Two rejections and independent output trials}
At a node of input width $r$, both child streams assign probability
\[
 q(x)=\frac{\lambda}{V_r}\rho_r(x)
       \ind_{x\in\text{prescribed coset}\cap\mathcal S_r(a_0)}
\]
to each vector $x$, with the remaining mass at $\dagger$.
Route the half grid shift as in \cref{sec:soft}.
Keep the right child vectors as a fixed bank $C=(y_i)$.
Selecting a partner leaves every bank record available for later trials.
Use $K(x,y)=\ind_{\mathrm{Eligible}}\omega_r(x-y)$ on the source
shells, extended by zero elsewhere, in
\cref{lem:bank-correction}, with partner and representation weights
$A_C(x)$ and $B_C(z)$ defined there.
The deterministic normalizers are
\[
 \mathcal B_r=e^{A_rR_-^2}(\gamma/\sqrt2)^n,\quad
 A_{\max}=2M\lambda\mathcal B_r2^{\degpad},\quad
 b_*=\frac{M\lambda^2}{V_r^2}e^{A_rR_-^2}V_{r/\sqrt2}.
\]
\Cref{lem:uniform-bank} proves, with failure
$\exp[-2^{\Omega(\pad)}]$ uniformly in the input width and determinant,
the simultaneous inequalities
\begin{equation}\label{eq:batch-good-bank}
 A_C(x)\le A_{\max},\qquad
 \tfrac12b_*\rho_{\gamma r}(z)\le B_C(z)\le
                         2b_*\rho_{\gamma r}(z)
\end{equation}
for every admissible source or output query.

Apply \cref{lem:bank-correction} with
$D=A_{\max}$, $c=b_*/2$, and
$h(z)=\rho_{\gamma r}(z)$ on the prescribed output shell and coset,
and zero elsewhere. The bank bounds give $C_0=4$.
The output trial therefore assigns mass
\begin{equation}\label{eq:batch-output-trial}
 q_{\rm out}(z)=\frac{b_*/2}{A_{\max}}\rho_{\gamma r}(z)
 =\frac{\lambda2^{-(\degpad+2)}}{V_{\gamma r}}
       \rho_{\gamma r}(z)
       \ind_{z\in\text{output coset}\cap\mathcal S_{\gamma r}(a_0)}
\end{equation}
to each vector $z$, with the remaining probability at $\dagger$.
The actual algorithm clips probabilities on bad branches.
Use the reference replacement of \cref{sec:bank-correction} for the
analysis. Independent child streams satisfy the lemma's hypotheses
at each node, so induction gives iid root labels and
\[
 \lambda_{j-1}=\lambda_j2^{-(\degpad+2)},\qquad
 \lambda_0=2^{-1-t(\degpad+2)}=2^{-o(\pad)}.
\]
Use disjoint private randomness for different bins. Conditional on the
public environment their root streams are independent.
The coupling charges the first bad bank in the whole forest.

Thus the recursive procedure uses the same affine routing as cooling,
with the two rejections replacing activation and isolation.

\subsubsection{Balancing radial bins and extracting a product law}
\label{sec:batch-mixing}
Restrict root $j$ to $A_j\cap(a+L)$. Correct its width to $\sigma$ by
radial rejection, anchored at the bin's least squared norm if
$r_j\ge\sigma$ and at its greatest squared norm otherwise.
Let $\zeta_j$ be that bin endpoint in normalized target energy.
It is rational and $\zeta_j=z_j+O(w_n)$. The surviving probability mass on vector outputs is
\[
 \widetilde q_j(x)=\kappa_j\rho_\sigma(x)\ind_{A_j\cap(a+L)},
 \qquad
 \kappa_j=\frac{\lambda_0}{V_{r_j}}
       e^{n\zeta_j(1-a_0/z_j)/2}.
\]
Each coefficient uses its own determinant and is explicitly known.
Choose bin $j$ with probability proportional to $\kappa_j^{-1}$, and put
\[
 \kappa_{\rm mix}=\left(\sum_{j=1}^Q\kappa_j^{-1}\right)^{-1}.
\]
The bin probability $\kappa_{\rm mix}/\kappa_j$ cancels its coefficient,
giving the same Gaussian weight throughout the annulus.

In each of $M$ mixture trials, independently choose a bin, consume
its next unused label, and apply the radial correction.
Each bin supplies $M$ labels, and the conditional law of its unused
stream gives iid mixture trials with
\[
 q_{\rm mix}(x)=\kappa_{\rm mix}\rho_\sigma(x)
          \ind_{\{x\in a+L:\,\en_\sigma(x)\in[z_-,z_+]\}}.
\]
Return the first $B$ successful values, or the fallback tuple if
fewer occur. For iid trials the success indicators are independent
of the successful values, so conditioning on enough successful trials
leaves exactly the truncated Gaussian product law.
This law is independent of the public extensions and therefore
survives their average.

The identity
\[
 \kappa_j\frac{\sigma^n}{\det L}
 =\frac{\lambda_0}{p_j}
   2^{-J(a_0)n+J(z_j)n+O(nw_n)}
\]
becomes, after multiplying by the number $M$ of trial labels,
\[
 M\kappa_j V_\sigma(L)
 =2^{[P(u)-\iota_j+J(z_j)]n+\Theta(\pad)}.
\]
This expresses the expected successful count after normalization by
the target Gaussian volume. Its limiting exponent is
\begin{equation}\label{eq:batch-bin-supply}
 F_E(z)=\min\left\{\dgsspace+J(z),\,
                \log_2\alpha+\frac{z-1}{2\log2}\right\}.
\end{equation}
Since $\kappa_{\rm mix}\ge Q^{-1}\min_j\kappa_j$ and $Q$ is polynomial,
the smallest count exponent among bins also bounds the mixture count.
The radial rarity and the expansion loss are therefore charged within
the same bin. The following theorem records the resulting range.

\begin{theorem}[Batches with independently chosen extensions]
\label{thm:batch-region}
Fix $\alpha>1$ and
\begin{equation}\label{eq:batch-exponent-range}
 0<b<\min\{\dgsspace,J(\alpha^{-2})\}.
\end{equation}
On the inputs of \cref{thm:batch-main}, there is an algorithm producing
$\lfloor2^{bn}\rfloor$ points with joint statistical distance at most
$\exp[-\Omega(n/\log^2(n+2))]$ from the target product distribution,
in time $2^{\dgstime n+o(n)}$ and bit space $2^{\dgsspace n+o(n)}$.
\end{theorem}
\begin{proof}
Choose $\alpha^{-2}<z_-<1<z_+$ with $J(z_-),J(z_+)>b$.
The minimum of \eqref{eq:batch-bin-supply} on this annulus exceeds $b$,
because
\[
 \log_2\alpha+\frac{z_--1}{2\log2}-J(z_-)
 =\log_2(\alpha\sqrt{z_-})>0.
\]
The expected mixture success count exceeds $2^{bn}$ by a fixed
exponential factor, so a Chernoff bound makes a shortfall doubly
exponentially unlikely. The tail estimate below and the search and
realization bounds at the end of this subsection complete the proof.
\end{proof}

The radial exponential moment of \cref{app:gaussian-tails}, optimized
at the two bin endpoints, gives
\[
 \delta_{\rm tail}\le3\,2^{-J(z_-)n}+3\,2^{-J(z_+)n}.
\]
The condition $z_->\alpha^{-2}$ keeps the smaller tilted width
$\sigma\sqrt{z_-}$ above $\eta_{1/2}(L)$.
Truncating a $B$-product costs at most $B\delta_{\rm tail}$.
\subsubsection{A terminal merge for full batches at width $3/2$}
At $\alpha=3/2$, the general theorem allows rates only below
$J(4/9)\approx0.1842$, whereas a full batch has rate
$\dgsspace\approx0.2075$. Reaching that rate requires more radial
coverage. We supply the extra inner bins with one additional merge
and estimate their lower tail using a tilt that stays at smoothing.
For $z\in[5/12,1/2]$ the merge uses
\[
 u_T=3/5,\qquad c_T=5/8,\qquad d_T=1/2,\qquad\gamma_T^2=4/5.
\]
Here $c_T$ and $d_T$ are the input and difference energies at the same
reference width. Since $d_T/\gamma_T^2=c_T$, rescaling the output
returns it to the input energy. For each inner bin, first generate two
independent streams using the ordinary batch construction above.
Choose their squared width $R^2=(c_T/a_0)r^2$, so that their
stored vectors already lie in the shell of energy $c_T$ at width $r$.
A radial rejection changes their Gaussian profile while preserving
this support. We then apply one additional fixed bank merge with
correlation $u_T$. The two input lists still have size exponent
$\dgsspace$. \Cref{eq:batch-terminal-widths,eq:batch-terminal-child}
give the width and coefficient calculations.

The forward residual Gaussian can lie below smoothing, introducing the
factor $H_n$ in the theta bound \eqref{eq:batch-terminal-degree}. We
absorb this factor into the degree normalizer to keep rejection valid.
The chosen parameters cancel its leading exponent against the
degree estimate, leaving a $2^{o(n)}$ normalizer and allowing the same
two rejections. The output count has rate
$q_T=2\dgsspace-P(3/5)=\log_2(16/15)$.
The inner bins have exponentially small target mass, so this smaller
count suffices for their share of a full batch. After restricting
to $a+L$ and correcting the width, the normalized output count exponent is
\begin{equation}\label{eq:batch-terminal-supply}
 F_T(z)=q_T-\max\{0,-\tfrac12\log_2(135z/64)\}+J(z).
\end{equation}
Its minimum on $[5/12,1/2]$ exceeds $\dgsspace$.
The outer bins $[1/2,2]$ use \eqref{eq:batch-bin-supply}.
Their limiting minimum is $\dgsspace$, and the finite endpoint parameters
and positive padding provide the extra subexponential number of records needed
to extract $\lfloor2^{\dgsspace n}\rfloor$ samples.

\begin{theorem}[Full batches at width $3/2$]\label{thm:batch-three-halves}
If $\sigma\ge3\eta_{1/2}(L)/2$, the algorithm outputs
$\lfloor2^{\dgsspace n}\rfloor$ points jointly within
$\exp[-\Omega(n/\log^2(n+2))]$ of
$D_{a+L,\sigma}^{\otimes\lfloor2^{\dgsspace n}\rfloor}$,
in time $2^{\dgstime n+o(n)}$ and bit space $2^{\dgsspace n+o(n)}$.
\end{theorem}
\begin{proof}
\Cref{app:batch-terminal} proves the terminal law, output count, and search
bounds. The ordinary tilt to $\sigma\sqrt{5/12}$ may be below smoothing.
Tilting only to $\sigma/\alpha$ instead gives the lower-tail rate
\[
 H_{3/2}(5/12)=\frac{\log(9/4)-25/48}{2\log2}>\dgsspace,
 \qquad J(2)>\dgsspace.
\]
The bounds $\log(9/5)>7/12$ and $\log(27/16)>25/48$ verify
these inequalities and $\min F_T>\dgsspace$.
Consequently the full product truncation costs $2^{-\Omega(n)}$.
Mixing the known coefficients gives the same normalized target on
every good public environment. The remaining joint errors are those
below and in \cref{eq:batch-terminal-joint-error}.
\end{proof}

The general region in \cref{thm:batch-region} remains useful:
smaller batches allow a smaller width and yield the GapSVP application.
\Cref{thm:batch-three-halves} is a separate strengthening at full size.
\Cref{alg:gaussian-batch} collects the two cases into one procedure.
In the full batch case we use the width promise with $\alpha=3/2$,
even if a stronger promise is available, and split bins at $z=1/2$.
Resource limits,
stream representation, and exceptional execution rules are given in
\cref{app:batch-execution} under \cref{prop:finite-realization}.

\begin{algorithm}[!ht]
\caption{Gaussian sampling by radial bins: \textsc{GaussianBatch}}
\label{alg:gaussian-batch}
\KwIn{$L,a,\sigma^2$ and feasible batch parameters from
\cref{thm:batch-region} or \cref{thm:batch-three-halves}.}
\KwOut{$B$ points in $a+L$, or failure.}
Fix the bins $(I_j)_{j=1}^Q$, trial count $M$, and depths of this section\;
\For{each bin $j$}{
  Draw its independent extension $\Gamma_j$ and preprocess its basis\;
  \eIf{full batch mode and $I_j\subseteq[5/12,1/2]$}{
    Generate two independent ordinary streams on the prescribed child cosets\;
    Correct their widths and apply the additional fixed bank merge of \cref{app:batch-terminal}\;
  }{
    Generate its root stream from corrected leaves, using \textsc{FixedBankTrial} at each merge\;
  }
  Compute $\kappa_j$, the known coefficient after restriction to $A_j\cap(a+L)$ and radial correction\;
}
Initialize an empty output list $\mathcal O$\;
\For{$m=1,\ldots,M$}{
  Choose bin $j$ with probabilities proportional to $\kappa_j^{-1}$ and consume its next unused trial $X$\;
  Restrict $X$ to $A_j\cap(a+L)$ and apply bin $j$'s radial rejection, preserving failures\;
  \If{$X\ne\dagger$}{
    Append $X$ to $\mathcal O$\;
    \If{$|\mathcal O|=B$}{\Return $\mathcal O$\;}
  }
}
\Return failure\;
\end{algorithm}

Every mixture trial consumes a new label, including when that label
fails. Each bin has $M$ labels, so none can be exhausted by the $M$
mixture trials. The reference sequence continues to $M$ trials for
the analysis, while execution stops when the output list fills.
The coefficient calculation uses determinants and Gaussian
weights, without computing the Gaussian mass of any bin.

\subsubsection{Search, space, and the joint error}
The sums used by \textsc{FixedBankTrial} require two searches.
Forward queries enumerate all eligible pairs of left and bank records,
giving the partner weights $A_C(x)$.
Reverse search uses fresh codes on the complete ideal forward proposal
list and computes $B_C(z)$, counting each bank record once.
The second search must recover all representations of each
proposed difference, because its rejection probability divides by
their total weight. To find those bank records, query
the bank with $-z$. Indeed, writing $x=z+y_i$, the norm identity gives
\[
 2\langle-z,y_i\rangle
 =\|z\|^2+\|y_i\|^2-\|x\|^2.
\]
Both endpoints have energy $a_0$, while $z$ has energy
$a_1=2a_0(1-u)$. Their thin shells therefore force
\[
 \frac{\langle-z,y_i\rangle}{\|z\|\,\|y_i\|}
 =\frac{\|z\|}{2\|y_i\|}+O(w_n)
 =\sqrt{\frac{a_1}{4a_0}}+O(w_n)
 =\sqrt{\frac{1-u}{2}}+O(w_n).
\]
Thus reverse search uses target correlation
$v_{\rm rev}=\sqrt{(1-u)/2}$ and code rate
\[
 K_{\rm rev}(u)=-\tfrac12\log_2
 \left(1-\frac{2u^2}{1+\sqrt{(1-u)/2}}\right).
\]
\Cref{app:batch-search} proves the reverse work bound while retaining
the bank dependence of these proposals.
The forward and reverse dictionaries contribute storage rates $P(u)$
and $2P(u)-R(u)$, respectively. Streams store only vector records,
while processing all $M$ trial labels contributes work at the raw
label rate.
Depth first evaluation and storing the polynomial number of root bins
give
\[
 K(u_n),K_{\rm rev}(u_n)\longrightarrow K_*,
 \qquad P(u_n),\,2P(u_n)-R(u_n)\longrightarrow\dgsspace,
\]
where $K_*=\frac12\log_2(3/2)$.
The exponential rate of the total number of trial labels tends to $1/2-1/(6\log2)<K_*$.
All forest sizes, code repetitions, and bin counts are $2^{o(n)}$.
These list size and work bounds allow \cref{prop:finite-realization}
to transfer the ideal batch law to a finite algorithm. Combining its
whole batch allowance with the public extension failure, bad bank
probabilities, and annulus tail gives
\begin{align*}
 \TV(\mathcal L(\text{batch}),D_{a+L,\sigma}^{\otimes B})
 &\le\epsilon_{\rm global}+B\delta_{\rm tail},\\
 \epsilon_{\rm global}
 &\le2^{-\Omega(h_n)}+2^{-\Omega(\pad)}+2^{-\Omega(n)}
                       +\exp[-2^{\Omega(\pad)}].
\end{align*}
In this combiner, midpoint error verifies the good bank inequalities.
The actual reverse count cancels it from \eqref{eq:batch-output-trial}.
Summing bad bank probabilities over the forest accounts for this error
jointly for the entire batch.
The public extension failure
is the leading term, yielding $\exp[-\Omega(n/\log^2(n+2))]$.
These rates prove \cref{thm:batch-region}. Together with the terminal
analysis they prove \cref{thm:batch-three-halves} and hence
\cref{thm:batch-main}.

\section{Reductions and algorithmic consequences}\label{sec:reductions}
\label{sec:consequences}
The sampling results lead to two kinds of applications. A Gaussian
batch on the dual lattice builds advice that can answer many decoding
queries. SVP instead supplies centered Gaussian samples at every
width, making rare useful vectors accessible by repeated sampling.
We develop each reduction together with its resulting algorithm.

\subsection{Accuracy and composition}\label{sec:reduction-accuracy}
We use the accuracy convention of~\cite[Definition~2.10]{SD16}.
A law $\mu$ is $(\gamma,\varepsilon)$-close to $D$ if some law
$\widetilde\mu$ satisfies
\[
 \TV(\mu,\widetilde\mu)\le\varepsilon,\qquad
 \gamma^{-1}D(x)\le\widetilde\mu(x)\le\gamma D(x)
 \quad\text{for every }x.
\]
Its total variation error is $O(\gamma-1)+\varepsilon$.
The pointwise comparison is essential for our sampling reductions:
it preserves useful events whose probability is exponentially
smaller than this total variation error.

If $\mu$ is $(1+\delta,\varepsilon)$-close to $D$, then every event
$E$ satisfies
\begin{equation}\label{eq:reduction-event}
 \frac{D(E)}{1+\delta}-\varepsilon
 \le\mu(E)\le(1+\delta)D(E)+\varepsilon.
\end{equation}
If $D(E)\ge p$, $\delta\le1$, and $\varepsilon\le p/4$, this
gives success probability at least $p/4$, which determines the
number of sampling calls in each application.

For a joint batch within TV distance $\varepsilon_B$ of $D^{\otimes B}$,
\[
 \Pr[\exists i:X_i\in E]\ge1-(1-D(E))^B-\varepsilon_B.
\]
If advice is simultaneously correct on all promised targets except
with probability $\varepsilon_A$, and the conditional search failure
on every good advice is at most $\zeta$, total failure is at most
$\varepsilon_A+\zeta$.

This advice event is charged once across all queries. For an event
of probability $n^{-O(1)}2^{-\kappa n}$, use
$Q=\poly(n)2^{\kappa n}$ queries and count time as
$\poly(n,\mathcal B)(P+QD)$, with one preprocessing cost $P$ per
lattice and query cost $D$. Sequential calls reuse the workspace.
Fixed call budgets and exact membership and length checks let us
retain the best valid candidate over all scales.

\subsection{Reusable decoding from Gaussian batches}\label{sec:reusable-bdd}
The preprocessing of a periodic Gaussian decoder consists of samples
on the dual lattice~\cite{DRS14}. Our batch interface supplies these
samples at a total cost much larger than a single decoding query.
The same advice can therefore serve an exponential number of attempts.

\begin{theorem}[Decoding from a feasible Gaussian batch]
\label{thm:batch-bdd}
Let $(\alpha,b)$ satisfy either \cref{eq:batch-exponent-range} or
$(\alpha,b)=(3/2,\dgsspace)$, and put $A_{\rm KL}=2^{0.401}$.
For every fixed
\begin{equation}\label{eq:batch-bdd-radius}
 0<r<r(\alpha,b):=\frac{\sqrt{e\,b\log2}}{\alpha A_{\rm KL}},
\end{equation}
there is an $r$-BDD data structure on every rational lattice, with
\[
 P=2^{\dgstime n+o(n)},\quad D=2^{bn+o(n)},\quad
 S=2^{\dgsspace n+o(n)}.
\]
With probability $1-o(1)$ the same advice determines a function that
returns the exact nearest lattice point for every rational target at
distance at most $r\lambda_1(L)$. All resource bounds hold on every
execution, with polynomial input and target encoding factors suppressed.
\end{theorem}
\begin{proof}[Proof outline]
Set $\varepsilon=2^{-2bn}$. The KL counting estimate gives a dual
sampling width just above
$\alpha A_{\rm KL}\sqrt{n/(2\pi e)}/\lambda_1(L)$.
The feasible parameters obey $\alpha>2^{2b}$, so this width both
satisfies the batch promise and makes
$\rho_{1/s}(L\setminus\{0\})\le\varepsilon$.
The periodic Gaussian decoder needs
$N=O(n\log(1/\varepsilon)/\sqrt\varepsilon)=\poly(n)2^{bn}$
samples and decodes to any fixed fraction below
$\sqrt{\log(1/\varepsilon)/\pi}/(2s)$.
Substitution gives \eqref{eq:batch-bdd-radius}.
\Cref{app:decoding-advice} proves the version using a known mass
upper bound, the uniform empirical estimates, and the unknown length
grid. Polynomially many independent batches contribute only $o(1)$
joint TV error to the advice. Its global good event is used once.
\end{proof}

\subsection{Shortest vectors and gap tests from decoding}
\label{sec:decoding-reductions}
Assume an $r$-BDD interface with $0<r<1/2$, preprocessing $P$,
query work $D$, and space $S$. With probability at least $9/10$ its
advice determines a function correct on all promised rational targets.
The function has a fixed resource bound even outside the promise.
The reductions below preserve dimension and polynomial encoding length.

A sphere perturbation reaches a specified decoding ball with a
probability determined by a spherical cap. If $\|v\|=d>a>0$ and
$U$ is uniform on $S^{n-1}$, then
\begin{equation}\label{eq:reduction-cap}
 \Pr[\|v+\sqrt{d^2-a^2}U\|\le a]
 =\Theta_{a/d}(n^{-1/2}(a/d)^n).
\end{equation}
Enumerating the unknown error length lets us use this optimal
perturbation radius for any fixed distance promise.

\begin{proposition}
\label{prop:bdd-radius-amplification}
For every fixed $R>r$, exact CVP under the promise
$\operatorname{dist}(t,L)\le R\lambda_1(L)$ can be solved with
constant success probability in time
\[
 \poly(n,\mathcal B_t)[P+(R/r)^nD]
\]
and space $S+\poly(n,\mathcal B_t)$. For $R<1/2$ this is
$R$-BDD. The statement also permits $R\ge1/2$ and multiple closest points.
\end{proposition}

\begin{theorem}[Direct unique-SVP tradeoff]\label{thm:usvp-bdd-tradeoff}
For every fixed $g\ge1$, exact $g$-uSVP, with promise
$\lambda_2(L)\ge g\lambda_1(L)$, is solvable with probability at
least $2/3$ in time and space
\begin{equation}\label{eq:usvp-bdd-tradeoff}
 T\le\poly(n,\mathcal B)[P+\max\{1,(gr)^{-n}\}D],\qquad
 S+\poly(n,\mathcal B).
\end{equation}
At $g=1$ the problem is ordinary SVP.
\end{theorem}
\begin{proof}[Proof outline]
A shortest vector $v$ is primitive. Choose a fixed prime $p>g$ and
a uniform nonzero linear function on basis coordinates modulo $p$.
Its kernel defines an index-$p$ sublattice $L'$. With probability
$p^{n-1}(p-1)/(p^n-1)>1-1/p$, it excludes $v$, and then
$\lambda_1(L')\ge g\|v\|$. Among its $p-1$ nonzero coset
representatives $x_j$ there is one with $x_j-v\in L'$.
Preprocess $L'$ once. Direct decoding suffices if $gr\ge1$.
Otherwise perturb $x_j$ and use \eqref{eq:reduction-cap} to hit
the decoding ball of $x_j-v$ in $\poly(n)(gr)^{-n}$ attempts.
For each legal reply $y\in L'$ keep the unperturbed residual
$x_j-y\in L\setminus\{0\}$, and return the shortest candidate.
Unknown scales, rational perturbations, and candidate checks are
proved in \cref{app:decoding-perturbations}.
\end{proof}

This uses the sublattice idea of Lyubashevsky and Micciancio~\cite{LM09}.
At $r=1/g$ it recovers the polynomial query relation. Below that
radius the cap probability gives an explicit exponential query cost,
while retaining a single preprocessing cost for each sublattice.

\begin{theorem}[GapSVP from reusable decoding]\label{thm:gapsvp-bdd-tradeoff}
Fix $\gamma>1$ with $r\gamma>1/2$, and put
\begin{equation}\label{eq:gapsvp-overlap-exponent}
 c(r,\gamma)=-\tfrac12\log_2\left(1-\frac1{4r^2\gamma^2}\right).
\end{equation}
There is a randomized algorithm distinguishing
$\lambda_1(L)\le d$ from $\lambda_1(L)\ge\gamma d$ with
success probability at least $2/3$, time
$\poly(n,\mathcal B)[P+2^{c(r,\gamma)n}D]$, and space
$S+\poly(n,\mathcal B)$.
\end{theorem}
\begin{proof}[Proof outline]
Fix the advice as a deterministic function $F$ on a half-open
fundamental parallelepiped. Draw $e$ from a ball of radius
$R=r\gamma d$, decompose $e=x+y$ with $y\in L$, and check
$F(x)=-y$. In the NO case good advice passes every check, since
$\|e\|\le r\lambda_1(L)$.

In the YES case take $0\ne v\in L$ with $\|v\|\le d$.
Whenever both $e$ and $e+v$ lie in the ball, they have the same
representative $x$ but demand different answers. Every fixed $F$
therefore fails on at least half the overlap, a fraction at least
\[
 n^{-O(1)}\left(1-\frac1{4r^2\gamma^2}\right)^{n/2}
 =n^{-O(1)}2^{-c(r,\gamma)n}.
\]
Repeated checks detect a conflict with constant probability.
\Cref{app:decoding-gap} proves this argument on a finite rational
grid containing $L$ and implements ball sampling with polynomial
overhead. The YES argument holds for every advice function, including
one obtained outside a sampling promise.
\end{proof}

The two geometries yield different query exponents. With
$P=2^{an+o(n)}$ and $D=2^{bn+o(n)}$, the total exponents are
\[
 \max\{a,b+[\log_2(1/(gr))]_+\},\qquad
 \max\{a,b+c(r,\gamma)\},
\]
respectively. The following choices fit both query costs into the
preprocessing time.

\paragraph{Algorithms at the BDGL exponents.}\label{sec:bdgl-consequences}
The full batch gives the radius supremum
\begin{equation}\label{eq:bdd-full-radius}
 r_*=\frac{\sqrt{e\dgsspace\log2}}{(3/2)2^{0.401}}
     =0.315707104\ldots.
\end{equation}
Every fixed $r<r_*$ has preprocessing exponent $\dgstime$, query
exponent $\dgsspace$, and space exponent $\dgsspace$ by
\cref{thm:batch-bdd}. The query savings leave room for
$2^{(\dgstime-\dgsspace)n}$ attempts during the preprocessing time.

\begin{corollary}
\label{cor:bdgl-promise-problems}
Exact $3$-uSVP and $4.15$-GapSVP admit classical randomized algorithms
with success probability at least $2/3$, time $2^{\dgstime n+o(n)}$,
and bit space $2^{\dgsspace n+o(n)}$ on every execution.
More generally, the same bounds hold for exact $g$-uSVP for every fixed
$g>2.986340909\ldots$.
\end{corollary}
\begin{proof}
For uSVP, the ratio of preprocessing time to query time is
$2^{(\dgstime-\dgsspace)n}=(3/\sqrt8)^n$. Hence
\cref{thm:usvp-bdd-tradeoff} fits the budget whenever
$g>1/(r_*2^{\dgstime-\dgsspace})=2.986340909\ldots$.

For the explicit factors, use the parameters in the table below.
For $3$-uSVP take prime $5$ and geometric hit radius $0.3155$.
Then $\log_2(1/(3\cdot0.3155))<0.084$.
For $4.15$-GapSVP take geometric radius $0.2635$.
The smaller batch is feasible since
$b<J(\alpha^{-2})$ and
$0.26355<r(1.382,0.1228)=0.263595555\ldots$.
Its call exponent is $c(0.2635,4.15)<0.1694$.
In each row, this exponent plus $b$ is below $\dgstime$.
The gaps between the base and geometric radii pay for finite
perturbations.
\end{proof}

\begin{center}
\small
\begin{tabular}{@{}lcccc@{}}
\toprule
Application & $\alpha$ & $b$ & Base radius & Query rate bound\\
\midrule
$3$-uSVP & $1.5$ & $\dgsspace$ & $0.3156$ & $0.084$\\
$4.15$-GapSVP & $1.382$ & $0.1228$ & $0.26355$ & $0.1694$\\
\bottomrule
\end{tabular}
\end{center}
The query rate column counts calls, while each call has work exponent
$b$. In both rows their sum is below $\dgstime$.
The full batch serves uSVP, while the smaller batch permits the
narrower width used by the GapSVP test.

\subsection{Centered Gaussian sampling from SVP}
\label{sec:cdgs-reduction}
The shortest vector algorithm also gives centered Gaussian samples at
every width. Following the primitive direction sparsification approach
of Stephens-Davidowitz~\cite[Section~4]{SD16}, we reduce sampling to
SVP calls in the same dimension.

Each primitive direction $y$ carries the Gaussian mass of all its
nonzero multiples,
\[
 h_\sigma(\|y\|)=2\sum_{k\ge1}e^{-\pi k^2\|y\|^2/\sigma^2}.
\]
Sparsification gives approximately uniform directions inside a ball.
If the ball of radius $r_i$ contains $N_i$ directions, select it with
weight $N_i(h_\sigma(r_i)-h_\sigma(r_{i+1}))$, then draw a direction
nearly uniformly inside it. The factor $N_i$ cancels, and summing over
balls telescopes to the desired direction weight. A one-dimensional
Gaussian then selects the multiple of that direction.

Near zero, $h_\sigma(r)$ varies like $1/r$, so a uniform additive grid
can miss a large change in direction weight. A sufficiently fine
multiplicative grid controls adjacent weights throughout the truncated
range and still has polynomially many radii.
\Cref{app:cdgs-oracle} proves this comparison and treats balls with too
few primitive directions.
\Cref{alg:centered-dgs} displays the complete mixture, including the
zero vector and the separate treatment of those sparse layers.
This yields the following oracle reduction.

\begin{theorem}[Centered sampling from SVP]
\label{thm:cdgs-reduction}
Let $L=B\Z^n$ be a full rank rational lattice, $\sigma^2>0$ rational,
and $f\ge1$ an integer. There is an oracle algorithm whose output law is
$(1+1/f,2^{-f})$-close to $D_{L,\sigma}$ in the convention of
\cref{sec:reduction-accuracy}. In the declared computational model, its
work, counting each SVP call as one operation, is bounded by a
polynomial in $\operatorname{size}(B,\sigma^2)$ and $f$ on every
execution. The same is true of its workspace and number of oracle calls.
Every query is a full rank sublattice of $L$ with
polynomial encoding length. The oracle may choose among shortest vectors
using any rule that depends on the current and previous queries.
\end{theorem}

\begin{proof}[Proof of \cref{cor:cdgs-main}]
Apply \cref{thm:cdgs-reduction} with parameter $f=f(n)$, replacing each
oracle query by amplified calls to \cref{thm:svp-main}.
The queries have rank $n$ and polynomial encoding length, so their
sequential execution and the reduction's polynomial overhead preserve
the time exponent $0.4130$ and space exponent $0.3138$.
The budget in \cref{cdgs:implementation} includes all oracle failures,
giving additive error below $2^{-f(n)}$ and multiplicative error
at most $1+1/f(n)$, with resources bounded on every execution.
\end{proof}

\subsection{Gaussian samples for CVP and successive minima}\label{sec:sampling-reductions}
For $\gamma>1$, $\gamma$-CVP asks, given a rational target $t$, for
$v\in L$ with $\|v-t\|\le\gamma\operatorname{dist}(t,L)$.
The problem $\gamma$-SMP asks for independent $v_1,\ldots,v_n\in L$
with $\|v_i\|\le\gamma\lambda_i(L)$ for every $i$. This also solves
$\gamma$-SIVP. Define
\begin{equation}\label{eq:approximation-phi}
 \Phi(\Delta)=
 \begin{cases}
  1+\log(1/\Delta),&0<\Delta\le1,\\
  1/\Delta,&\Delta\ge1.
 \end{cases}
\end{equation}
Here an effective constant admits polynomial-time rational approximation
to any requested precision. An effective factor depending on $n$ is
polynomial-time computable with polynomial encoding length.

Time--approximation tradeoffs from CVP to SVP were established
by Aggarwal, Chen, Kumar, Li, and Stephens-Davidowitz~\cite{ACKLS21}.
For the Euclidean norm and $\gamma\ge4$, their Theorem~5.1 gives
reduction cost $\widetilde O(\exp(16n/\gamma^2))$, using SVP
instances in dimension $n+1$. Their Theorem~5.2 isolates the sampling
step. We refine the explicit sampling-call exponent, cover every
fixed $\gamma>1$, and obtain direct SMP and SIVP bounds in dimension
$n$ from the same estimates. We first separate these geometric
arguments from the cost of our sampler.

\begin{proposition}
\label{prop:sampler-approximation}
\emph{Sampler assumption.}
Suppose a sampler accepts every full rank rational lattice
$K\subset\R^N$ and every positive rational width $s$, and permits
$(1+1/f(N),2^{-f(N)})$-closeness to $D_{K,s}$ for every efficiently
computable polynomially bounded precision parameter $f$.

\emph{Approximation algorithms.}
For every fixed effective $\gamma>1$, there are randomized algorithms
for $\gamma$-SMP, $\gamma$-SIVP, and $\gamma$-CVP with success
probability at least $2/3$, making at most
\begin{equation}\label{eq:sampler-approximation-calls}
 Q_\gamma(n,I)
 =\poly(|I|,n)\exp\!\left(\frac n2\Phi(\gamma^2-1)+O(1)\right)
\end{equation}
sampler calls. Here $I$ is the rational input. All calls for SMP and
SIVP have dimension $n$. All calls for CVP have dimension $n+1$.

\emph{Resources and growing factors.}
Their inputs have polynomial encoding length. The number of calls is
fixed before sampling, the additional work per call is polynomial,
and the additional storage is polynomial.

For every effective $\gamma(n)=\omega(1)$, the construction is uniform
and uses $2^{o(n)}\poly(|I|,n)$ calls. In the asymptotic range the
precision choice is $f(N)=N^2$.
\end{proposition}

\Cref{app:consequences} proves the proposition, including rational widths,
unknown scales, candidate verification, and predetermined call budgets.
Sequential calls reuse the sampler workspace.

The common geometric observation is a choice of sign. In each pair
$\{x,-x\}$, choose the sign with $\langle u,\pm x\rangle\le0$.
Then
\[
 \|u\pm x\|^2\le\|u\|^2+\|x\|^2,\qquad
 \rho_s(u\pm x)\ge\rho_s(u)\rho_s(x).
\]
A centered small ball estimate supplies the mass of the possible $x$.
For a closest displacement this yields \cref{lem:shifted-small-ball}.
For a vector $u$ outside a fixed subspace it yields
\cref{lem:small-ball-escape}. Both give, up to polynomial factors,
\[
 p_\gamma\ge
 \beta^{n/2}\exp[-n\beta/(2\Delta)-O(1)],\qquad
 \Delta=\gamma^2-1,\quad0<\beta\le1.
\]
Minimizing $\beta/\Delta+\log(1/\beta)$ gives
$\beta=\min\{1,\Delta\}$ and \eqref{eq:approximation-phi}.
The event guarantee \eqref{eq:reduction-event}, with $f(N)=N^2$,
preserves this exponentially small success probability.
For SMP, fix the current span and draw fresh samples until a short
vector outside it appears. For CVP, the rational embedding
$\{(v-kt,ks):v\in L,k\in\Z\}$ uses centered sampling in dimension $n+1$.
Its layers $k=\pm1$ contain the desired candidates.

In the Euclidean setting and the common range $\gamma\ge4$, the
sampling overhead expressed with natural exponentials decreases from $16/\gamma^2$
in~\cite[Theorems~5.1--5.2]{ACKLS21} to $1/[2(\gamma^2-1)]$.
At $\gamma=4$ the coefficients are $1$ and $1/30$.
The logarithmic branch of $\Phi$ also covers every fixed
$1<\gamma<\sqrt2$. SMP and SIVP retain dimension $n$.

\paragraph{Instantiating the sampler.}
Write $\svptime=0.4130$ and $\svpspace=0.3138$ for the displayed
time and space exponents of \cref{cor:cdgs-main}. Sequential calls add
the sampling-call exponent to $\svptime$ and reuse the same workspace.

\begin{corollary}
\label{cor:approximation-curve}
For every fixed effective $\gamma>1$, there are classical randomized
algorithms for $\gamma$-CVP, $\gamma$-SMP, and $\gamma$-SIVP with
success probability at least $2/3$ and resource bounds
\[
 T\le2^{\left(\svptime+\Phi(\gamma^2-1)/(2\log2)\right)n+o(n)},
 \qquad S\le2^{\svpspace n+o(n)}.
\]
For every effective $\gamma(n)>1$ with $\gamma(n)=\omega(1)$, these
problems admit time $2^{\svptime n+o(n)}$ and space $2^{\svpspace n+o(n)}$.
Both resource bounds hold on every execution.
\end{corollary}

For example, the time exponents are below $0.4968$ at $\gamma=3.1$
and $0.4611$ at $\gamma=4$, with space exponent $0.3138$ in both cases.

The extra time pays for rare useful samples. As $\gamma(n)\to\infty$,
their call exponent tends to zero.

\subsection{Short independent vectors near smoothing}
Polynomially many samples above smoothing contain $n$ linearly
independent vectors with high probability, and their lengths are
controlled by the last successive minimum.
The standard reduction of~\cite[Lemma~7.5 and Corollary~7.6]{ADRSD15}
therefore gives the following consequence.

\begin{corollary}\label{cor:fast-sivp}
Given a full rank rational lattice $L\subset\R^n$, there is a classical
randomized algorithm that returns $n$ linearly independent vectors
$v_1,\ldots,v_n\in L$ and, with probability at least $2/3$, satisfies
\[
 \max_i\|v_i\|\le C\sqrt{n\log(n+2)}\,\lambda_n(L)
\]
for an absolute constant $C$. Its time and space are
\[
 T=(3/2)^{n/2+o(n)},\qquad S=(4/3)^{n/2+o(n)}.
\]
\end{corollary}

The proof in \cref{sec:fast-sivp} uses independent executions of the
weighted readout sampler in \cref{thm:single-main}. Polynomially many
sequential calls preserve its leading time and space exponents.

\subsection{Polynomial-factor approximations by block reduction}
\label{sec:block-consequences}

Block reduction uses SVP in a smaller dimension at the cost of a larger
approximation factor. Applying \cref{thm:svp-main} in~\cite{ALNS20}
gives the bounds below. Hermite SVP (HSVP) measures length against
$\det(L)^{1/n}$, and SVP against $\lambda_1(L)$.

\Needspace{16\baselineskip}
\begin{corollary}
\label{cor:block-consequences}
For every fixed $c$ in the indicated range, there is a classical
randomized algorithm that, on input a full rank rational lattice
$L\subseteq\R^n$, returns a nonzero $v\in L$ with probability at least
$2/3$ of satisfying the following bound:
\begin{center}
\begin{tabular}{@{}llll@{}}
\toprule
Problem and range & Output length & Time & Space\\
\midrule
SVP, $c\ge1$
  & $\|v\|\le O_c(n^c)\lambda_1(L)$
  & $2^{\svptime n/(c+1)+o(n)}$ & $2^{\svpspace n/(c+1)+o(n)}$\\
SVP, $1/2<c<1$
  & $\|v\|\le O_c(n^c)\lambda_1(L)$
  & $2^{\svptime n/(2c)+o(n)}$ & $2^{\svpspace n/(2c)+o(n)}$\\
HSVP, $c\ge1/2$
  & $\|v\|\le O_c(n^c)\det(L)^{1/n}$
  & $2^{\svptime n/(2c)+o(n)}$ & $2^{\svpspace n/(2c)+o(n)}$\\
\bottomrule
\end{tabular}
\end{center}
\end{corollary}

For example, $O(n)$-SVP takes $2^{0.2065n+o(n)}$ time and
$2^{0.1569n+o(n)}$ space. The SVP rows use
\cite[Corollaries~4.4 and~3.4]{ALNS20} with block sizes
$\lfloor n/(c+1)\rfloor$ and $\lceil n/(2c)\rceil$, respectively.
For HSVP, use \cite[Theorem~2.3]{ALNS20} with
$k=\lceil n/(2c)\rceil$. At $c=1/2$, SVP and Minkowski's bound suffice.
Only constant-factor block SVP is needed, so rational Gram-matrix
approximation permits our oracle on projected lattices.
\Cref{app:block-consequences} gives the interface and resource analysis.

\section{Conclusion and further discussion}\label{sec:further-directions}

Gaussian difference dynamics controls the expected counts and dependence of
lattice vectors through their first two counting moments. Accurate
moment bounds yield a Gaussian sample, while weaker bounds allow more
differences to be retained for shortest vector search. This gives
provable DGS above smoothing at the BDGL heuristic exponents and exact
SVP on arbitrary lattices in $2^{0.4130n+o(n)}$ time.

\paragraph{SVP at the BDGL exponent.}
Can exact SVP on arbitrary lattices be solved in $(3/2)^{n/2+o(n)}$
time, matching the BDGL heuristic exponent? Our sampler attains this
exponent above smoothing, while the shortest vector construction
requires additional representations at the terminal scale.
Under the terminal balance of \cref{prop:terminal-infimum}, the
optimized time exponent is approximately $0.412962$. Changing the
intermediate shell energies may improve space usage, but does not
lower this terminal cost.

\paragraph{Exact CVP below $2^n$.}
Another central question is whether exact CVP on arbitrary lattices
admits time $2^{(1-\epsilon)n+o(n)}$ for some fixed $\epsilon>0$.
The equivalence between CVP and DGS at arbitrary shifts and widths
gives a sampling formulation of this question
\cite[Theorem~3.6 and Section~7]{SD16}.
An extension of fast sampling to arbitrary shifts below smoothing
would therefore also give faster exact CVP. Our two sampling routes
suggest a starting point: direct sampling handles arbitrary shifts
above smoothing, and the SVP reduction handles centered sampling at
every width.

\paragraph{Practical and heuristic algorithms.}
A further goal is to implement algorithms based on DGS and bring their
practical performance closer to that of heuristic sieving. For SVP,
can their time and memory costs be made comparable to those of heuristic
sieves in dimensions of practical interest? Alongside implementations
of the provable constructions, heuristic variants of DGS algorithms
are also worth exploring.

\paragraph{Extending GDD to other norms.}
The counting framework provides a common way to control dependence,
while each construction supplies its own reference profile and geometric
estimates. The cube construction of~\cite{GDDInfinity} illustrates this
flexibility beyond Gaussian profiles. Extending the approach to other
$\ell_p$ norms calls for suitable reference profiles, difference rules,
and efficient pair searches. Can these ingredients improve the provable
time and space bounds for SVP, and how do the resulting exponents depend
on $p$?

\appendix
\phantomsection
\addcontentsline{toc}{part}{Appendices}
\crefalias{section}{appendix}
\crefalias{subsection}{appendix}
\crefalias{subsubsection}{appendix}
\crefname{appendix}{Appendix}{Appendices}
\Crefname{appendix}{Appendix}{Appendices}
\addtocontents{toc}{\protect\setcounter{tocdepth}{1}}
\section{Lattice and Gaussian tools}\label{app:lattice-estimates}
\subsection{Centered Gaussian moments and scale comparison}
For every
full rank lattice $L\subset\R^n$, $s>0$, and $0<\beta\le1$,
\begin{equation}\label{eq:consequence-gaussian-facts}
 \E_{X\sim D_{L,s}}[\|X\|^2]\le\frac{ns^2}{2\pi},\qquad
 \rho_{\sqrt\beta s}(L)\ge\beta^{n/2}\rho_s(L),\qquad
 \rho_s(L-t)\le\rho_s(L).
\end{equation}
These standard consequences of Poisson summation follow by taking
a logarithmic derivative, comparing widths, and using positivity
of the Fourier coefficients, respectively.

\subsection{Weighted radial counts}\label{app:kl-weighted}
\begin{proof}[Proof of \cref{lem:kl-weighted}]
Put $a=n/2$ and $N(x)=\#\{w\ne0:\xi_w\le x\}$.
For $x<1$, $N(x)=0$, and \cref{lem:kl-count} bounds it by
$C_{R_0}(n+1)2^{\kappa_{\rm pack}n}x^a$ on the remaining
integration range. Stieltjes integration by parts gives
\[
 \sum_{w\ne0}(H-c\xi_w)_+^a
 =ac\int_0^{H/c}N(x)(H-cx)^{a-1}\,dx.
\]
With $x=(H/c)t$, the integral after inserting the count bound,
apart from its factor $C_{R_0}(n+1)2^{\kappa_{\rm pack}n}$, is
\[
 H^{2a}c^{-a}\frac{\Gamma(a+1)^2}{\Gamma(2a+1)}
 \le\poly(n)\left(\frac{H^2}{4c}\right)^a.
\]
The last estimate follows from Stirling's inequality. Boundary terms
vanish since $N(0)=0$ and the weight is
zero at $H/c$.
\end{proof}

\subsection{Basis preprocessing and length estimates}
\label{app:suffix-proof}\label{app:interfaces}
We derive \cref{ing:reduction} from DBKZ in successive projections.
For exact SVP on a rational projected lattice of rank $d$ in $\R^n$,
\cite[Theorem~5.4 and Corollary~5.5]{MV13} gives $4^{d+o(d)}$ time
and $2^{d+o(d)}$ space, with polynomial input overhead.

\paragraph{Projected calls.}
Fix $2\le k<n$.
Suppose a prefix $F=(b_1,\ldots,b_i)$ has been fixed as part of a basis
$(F,D)$ of $\Gamma$. Its orthogonal projection and remaining lattice are
\[
 \pi=I-F(F^{\mathsf T}F)^{-1}F^{\mathsf T},\qquad
 \Lambda=(\pi D)\Z^d=\pi\Gamma,\qquad d=n-i,
\]
with $\pi=I$ when $i=0$. Clear one common positive denominator in $\pi D$ to obtain the integer
input required by ALNS20. This preserves length ratios and coefficients:
an output with coefficient vector $z$ lifts to $Dz$.
Use the same conversion for rational queries to the exact oracle.

If $d>k$, apply \cite[Theorem~2.3]{ALNS20} with block dimension $k$,
exact oracle factor $\delta=1$, and slack $\varepsilon=1$. It returns
a nonzero projected vector $v$ with
\[
 \|v\|\le2\gamma_k^{(d-1)/(2(k-1))}\det(\Lambda)^{1/d}
       \le2k^{(d-1)/(2(k-1))}\det(\Lambda)^{1/d},
\]
using polynomially many SVP calls of rank $k$.
Here $\gamma_k\le k$ is the Euclidean Hermite bound, and the
determinant is taken within the span of $\Lambda$.
If $d\le k$, use exact SVP on $\Lambda$.

\paragraph{From projected vectors to a basis.}
Take the primitive core of the returned coefficient vector, complete
it unimodularly, and size reduce the lifted column against the fixed
prefix. LLL reduction of the remaining projected suffix, followed by
size reduction in the original space, prepares the next round while
preserving the chosen Gram--Schmidt vectors~\cite{LLL82,MV13}.

The projections of $n$ independent vectors of length at most
$\lambda_n(\Gamma)$ span the remaining space. Choose $d$ independent
projections. Their integer span is a sublattice of $\Lambda$, so
Hadamard's inequality gives
\[
 \det(\Lambda)^{1/d}\le\lambda_d(\Lambda)\le\lambda_n(\Gamma).
\]
For $d>k$, the selected column therefore satisfies
\[
 \|b_{i+1}^*\|\le2k^{(d-1)/(2(k-1))}\lambda_n(\Gamma)
                 \le H(n,k)\lambda_n(\Gamma).
\]
For $d\le k$, exact SVP gives the stronger bound
$\|b_{i+1}^*\|\le\lambda_n(\Gamma)$.
Later operations preserve each selected Gram--Schmidt vector, proving
the geometric conclusion of \cref{ing:reduction}.

\paragraph{Representation and resource bounds.}
The usual LLL bound gives Gram--Schmidt lengths at most
$2^{n/2}\lambda_n(\Gamma)$ in each remaining suffix. After size
reduction, every current column therefore has length at most
\[
 \sqrt{1+n/4}\,
 \max\{H(n,k),2^{n/2}\}\lambda_n(\Gamma).
\]
Its denominator divides the original common denominator, and
$\lambda_n(\Gamma)\le\max_j\|b_{0,j}\|$. Thus each round begins with
a polynomial description bound in terms of $B_0$, preventing
encoding growth across projections. Rational linear algebra and the
intermediate bounds of \cite[Section~2.3]{ALNS20} give the same
control within a round. The $n$ rounds make polynomially many calls
of rank at most $k$, proving the resource bound.
For $k=\lfloor\vartheta n\rfloor$, direct substitution gives the
stated factor $O_\vartheta(n^{1/(2\vartheta)})$ and exponent
$2\vartheta$. Fixed small ranks use exact SVP.

\subsection{Rational projected basis interfaces}\label{app:rational-projections}
The block reductions query projected lattices and their duals,
which have rational rectangular bases. The following construction
puts these queries in the square basis format of \cref{thm:svp-main}.

Let $A\in\mathbb Q^{m\times d}$ have full column rank. In polynomial
time in its rational input length, one can compute an invertible
$C\in\mathbb Q^{d\times d}$, of polynomial encoding length, such that
\begin{equation}\label{eq:rational-block-gram}
 \|Az\|^2\le\|Cz\|^2\le\tfrac32\|Az\|^2
 \qquad\text{for every }z\in\R^d.
\end{equation}
To obtain $C$, compute the rational factorization
$A^{\mathsf T}A=U^{\mathsf T}\diag(d_i)U$ with $d_i>0$.
Choose rational $r_i$ with $d_i\le r_i^2\le3d_i/2$ and set
$C=\diag(r_i)U$. The Gram matrix inequalities give
\eqref{eq:rational-block-gram}. A shortest vector $Cz$ maps back to
$Az$, whose length is at most $\sqrt{3/2}\lambda_1(A\Z^d)$.
This supplies the $2$-approximate SVP oracle needed by the block
reductions without changing the rank.

Orthogonal projections and dual bases within rational spans are
rational, so the construction applies to both types of block query.
Lift the returned integer coordinates in the original block,
taking their primitive core when needed. Clearing denominators
allows the use of the integer-lattice statements in \cite{ALNS20}
without changing approximation factors.

\subsection{Radial and linear tails}\label{app:gaussian-tails}
\label{app:gaussian-details}
The following estimates support Gaussian cooling and midpoint lenses,
uniformly on the compact parameter ranges in
\cref{sec:lattice,sec:bridges}.

Let $X$ be a discrete Gaussian of width $t$ on a shifted lattice.
Suppose the shifted theta estimates hold with error $\varepsilon$ at
the widths used below. Put
$U=\pi\|X\|^2/t^2$ and
$R_\varepsilon=(1+\varepsilon)/(1-\varepsilon)$.
Ratios of theta masses and completion of squares give
\[
 \E[e^{aU}]\le R_\varepsilon(1-a)^{-n/2},\qquad
 \E[e^{b\langle X,z\rangle}]
 \le R_\varepsilon e^{b^2t^2\|z\|^2/(4\pi)}.
\]
The first requires $a<1$ and theta control at $t/\sqrt{1-a}$.
The second holds for every real $b$, since the tilt translates the coset.
Optimizing gives, for $0<\delta\le1/2$ and, in the last bound,
$z\ne0$ and $h>0$,
\begin{align*}
 \Prb[U\ge(1+\delta)n/2]
 &\le R_\varepsilon e^{-n[\delta-\log(1+\delta)]/2},\\
 \Prb[U\le(1-\delta)n/2]
 &\le R_\varepsilon e^{-n[-\delta-\log(1-\delta)]/2},\\
 \Prb[|\langle X,z\rangle|\ge h]
 &\le2R_\varepsilon e^{-\pi h^2/(t^2\|z\|^2)}.
\end{align*}

For \eqref{eq:soft-total}, tilt the reference Gaussian to width
$r\sqrt{a_0}$. The shell density factor is
$e^{-n(a_0-1)/2+O(nw_n)}$, and the volume factor is $a_0^{n/2}$.
The tilted width has a fixed smoothing margin
$r\sqrt{a_0}/s\ge\sqrt{2a_0}>1$.
The radial bound puts $1-o(1)$ of the tilted mass in a narrower shell.
Multiplying the density and volume factors gives
$\amp2^{-J(a_0)n+O(\errwin)}$.

\subsection{Gaussian profile and pairing degree bounds}
\label{app:gaussian-profile}
For the midpoint in \cref{lem:soft-midpoint}, take
\[
 t=r/\sqrt2,\qquad U=n\en_r(M),\qquad
 L_z=\frac{2\pi\langle M,z\rangle}{nr^2},\qquad
 R_\epsilon=\frac{1+\epsilon_\Gamma}{1-\epsilon_\Gamma}.
\]
For each eligible $z$, the region $\mathcal W_r(z)$ contains
\[
 \mathcal T_r(z)=\{M:|\en_r(M)-1/2|\le w_n/8,
                         \ |L_z|\le w_n/8\}.
\]
Indeed, on this region,
\[
 |\en_r(M\pm z/2)-a_0|
 \le w_n/8+a_1w_n/4+w_n/8
 =(a_0-1/4)w_n<a_0w_n.
\]
The preceding bounds give
\begin{align*}
 \Prb[|U-n/2|>nw_n/8]
 &\le2R_\epsilon e^{-nw_n^2/128},\\
 \Prb[|L_z|>w_n/8]
 &\le2R_\epsilon\exp[-nw_n^2/(64\en_r(z))].
\end{align*}
Every radial tilt is valid because
$t\sqrt{1-w_n/4}\ge(s/\gamma)\sqrt{1-w_n/4}\ge s$.
The linear tilt uses shifted theta flatness at $t$.
Consequently the mass on $\mathcal T_r(z)$ is at least
$V_{r/\sqrt2}(1-O(\epsilon_\Gamma)-e^{-\Omega(nw_n^2)})$.
The mass on $\mathcal W_r(z)$ lies between this lower bound and
the full shifted mass, at most $(1+\epsilon_\Gamma)V_{r/\sqrt2}$.
This proves \eqref{eq:midpoint-flat}.

For the coefficient in \eqref{eq:soft-B-rate}, use
\[
 e^{n(1-a_0)/2}\beta^{n/2}
 =2^{-(J(a_0)+P(u_1))n},\qquad
 A_rR_-^2=n(1-a_0)(1-w_n)/2.
\]
To bound a weighted degree, fix an eligible endpoint $x$.
Completion of squares gives
\[
 \rho_r(y)=\rho_{r\sqrt\beta}(y-ux)
 \exp\left\{\frac n2
 \left[-u\en_r(x)+\frac{u}{1-u}\en_r(x-y)\right]\right\}.
\]
The bounds $\en_r(x)\ge a_0(1-w_n)$ and
$\en_r(x-y)\le a_1(1+w_n)$ put the bracket at most
$(1-a_0)(1+3w_n)$.
Since $r\sqrt\beta\ge s$ and $\omega_r\le1$, summing over the full
coset bounds the degree by
\[
 \amp_j\beta^{n/2}e^{n(1-a_0)(1+3w_n)/2}(1+\epsilon_\Gamma)
 =\amp_j\softB_j e^{2(1-a_0)nw_n}(1+\epsilon_\Gamma)
 \le D_j^+.
\]
The last inequality uses $\degpad=\lceil10\errwin\rceil$.
Fixing the right endpoint gives the same bound.

\subsection{A common Gaussian window count}
\label{app:gaussian-window}
\begin{lemma}\label{lem:window-count}
Let $K+b$ be a full rank affine lattice and let $\beta$ range over a
fixed compact positive interval. Put $t=r\sqrt\beta$ and suppose
shifted theta flatness holds with error at most $1/4$ at $t$ and its
relative $O(w_n)$ radial tilts. Let $\|z\|^2=O(nr^2)$.
Suppose a window $W$ contains
\[
 |\en_r(x)-\beta|\le c_1w_n,\qquad
 |2\pi\langle x,z\rangle/(nr^2)|\le c_1w_n
\]
and is contained in $|\en_r(x)-\beta|\le c_2w_n$, where
$0<c_1<c_2$ are fixed. Then
\[
 \#((K+b)\cap W)
 =\frac{r^n}{\det K}(e\beta)^{n/2}e^{\pm O(nw_n+\log n)}.
\]
\end{lemma}
\begin{proof}
The radial and linear exponential moments above put
$1-O(e^{-\Omega(nw_n^2)})$ of $D_{K+b,t}$ in the inner rectangle.
On either radial window,
$\rho_t(x)=e^{-n/2+O(nw_n)}$.
Multiply the Gaussian mass of the inner rectangle and of the outer
window by the corresponding inverse weight bounds. Theta flatness
gives $\rho_t(K+b)=(1\pm1/4)t^n/\det K$ and proves both inequalities.
\end{proof}

\subsection{Corrected Gaussian proposals and leaf moments}
\label{app:gaussian-leaves}
\begin{proof}[Proof of \cref{lem:gaussian-proposal}]
The Klein/GPV proposal~\cite{Klein00,GPV08} starts with $v_n=-p$.
For $i=n,\ldots,1$, set
\[
 h_i=\frac{\langle v_i,b_i^*\rangle}{\|b_i^*\|^2},\qquad
 s_i=\frac R{\|b_i^*\|},\qquad
 z_i\sim D_{\Z,s_i,h_i},\qquad v_{i-1}=v_i-z_i b_i.
\]
It proposes $x=p+\sum_i z_i b_i$. Orthogonal decomposition makes
the product of the Gaussian numerators equal to $\rho_R(x)$, hence
\[
 Q_R(x)=\frac{\rho_R(x)}{\prod_i\theta_{s_i}(h_i)},\qquad
 \theta_s(h)=\sum_{k\in\Z}e^{-\pi(k-h)^2/s^2}.
\]
Poisson summation and $e^{-\pi s_i^2}\le1/(8n)$ give
$|\theta_{s_i}(h_i)/s_i-1|\le2/(8n-1)\le1/(3n)$.
Since $\prod_i s_i=V_R$, accept with probability
\[
 a(x)=\frac1{2V_R}\prod_i\theta_{s_i}(h_i)
     =\frac12\prod_i\frac{\theta_{s_i}(h_i)}{s_i}
     \le\frac12e^{1/3}<1.
\]
Then reject points outside $\mathcal S$. The product $Q_R(x)a(x)$
is exactly \eqref{eq:gaussian-proposal-law}.

For finite implementation, use the uniform scalar sampler of
\cref{app:scalar-sampling} with error $\delta/(2n)$ at each coordinate.
After reducing $h_i$ modulo $\Z$, the Poisson series for
$\theta_{s_i}(h_i)/s_i$ converges geometrically and is bounded away
from zero. Certified intervals evaluate the product and its acceptance
coin to error $\delta/2$ using \cref{app:numerical-primitives}.
The scalar guarantee and sequential coupling give total error
at most $\delta$ within deterministic polynomial budgets.
\end{proof}

Make $2M$ independent trials of \cref{lem:gaussian-proposal}, where
$M\ge1$ is an integer, and retain their nonfailure records with their
trial IDs. For
\[
 \nu(x)=\frac M{V_R}\rho_R(x)
          \ind_{x\in(p+\Gamma)\cap\mathcal S},
\]
independence gives the leaf moments
\begin{equation}\label{eq:fixed-count-leaves}
 m=\nu,\qquad
 S^{\rm act}=\left(1-\frac1{2M}\right)\nu^{\otimes2}
 \le\nu^{\otimes2}.
\end{equation}
Thus the retained list belongs to $\mathcal C(\nu,0)$.

Cooling uses these $2M$ trials to initialize coefficient $\amp_t=M$.
Batch streams instead keep $M$ trial labels and use the same single
trial law, with coefficient $\lambda_t=1/2$.
Thus the fixed trial counts determine known coefficients in both
constructions, while only nonfailure records occupy vector storage.

\section{Search and application estimates}\label{app:search-details}
\subsection{Midpoint and output counts}\label{app:search-geometry}
We justify the geometric factors in \cref{lem:correlation}. At bridge
energy $E$, the midpoint Gaussian has width $r\sqrt E/2$ on $\Gamma/2$.
The first moment argument of \cref{thm:lens} therefore applies whenever
$gE>1$ with a fixed margin, giving
$V_r(eE/4)^{n/2}e^{O(\errwin)}$ representations per output.

After the ancestor average, the output support is $p_*+\Gamma/t$ with
density $t^{-n}$. For an output shell of fixed positive energy $b$, put
$p_-=t^2gb$. If $p_-\ge1$, the Gaussian of width $r\sqrt b$ is above
smoothing on $\Gamma/t$. If $p_-<1$, the shifted maximum and monotonicity give
\[
 \rho_{r\sqrt b}(p_*+\Gamma/t)
 \le\rho_{s/t}(\Gamma/t)
 \le(1+\varepsilon_n)\frac{(r\sqrt b)^n}{\det(\Gamma/t)}p_-^{-n/2}.
\]
The usual Gaussian majorant on a thin shell consequently yields
\begin{equation}\label{eq:subcritical-shell}
 t^{-n}\#\bigl((p_*+\Gamma/t)\cap\mathcal S_r(b)\bigr)
 \le V_r(eb)^{n/2}e^{O(nw_n+\log n)}
             \max\{1,p_-^{-n/2}\}.
\end{equation}
All estimates are uniform on fixed compact positive energy ranges.
Multiplication with the midpoint count gives
$V_r^2(ea)^n(1-u^2)^{n/2}$ and
$p_-=c_0(1-u)$, proving the rate in \cref{lem:correlation}.
Endpoint perturbations and bins of width at most $2/n$ cost
$O(\errwin)$ in the logarithm. A fixed compact interval uses $O(n)$ bins.

\subsection{Nearly parallel pairs and equal values}\label{sec:terminal-cap}
Choose a fixed $\Delta\in(0,1)$ with $c_0\Delta<1$.
Put $u_*=1-\Delta>0$ and require the midpoint smoothing factor
$2ga(1+u_*)$ to exceed one by a fixed margin.
For correlations $u\ge u_*$, the difference energy is at most
$b_*=2a\Delta+O(w_n)$. Write $E_*=4a-b_*$ at the central parameters.
The applications choose fixed cutoffs satisfying these conditions.

Use Gaussian widths $r\sqrt{E_*}/2$ for the midpoint and $r\sqrt{b_*}$
for the output. On the endpoint shells,
$\en_r(M)=a-\en_r(z)/4+O(w_n)$, and
\[
 \frac{\en_r(M)}{E_*/4}+\frac{\en_r(z)}{b_*}
 =\frac{4a}{E_*}
 +\en_r(z)\left(\frac1{b_*}-\frac1{E_*}\right)+O(w_n)
 \le2+O(w_n).
\]
The coefficient of $\en_r(z)$ is positive because $u_*>0$.
Thus $e^{n+O(nw_n)}$ times this product Gaussian bounds the counting
weight throughout the region, including $z=0$. The local and ancestor
averages separate the midpoint and output sums. Their smoothing factors
are $gE_*>1$ and $t^2gb_*=c_0\Delta<1$, respectively.
The theta and monotonicity estimates above give candidate mass
\begin{equation}\label{eq:terminal-count}
 2^{[2\beta-P(1-\Delta)]n+O(\pad)}(c_0\Delta)^{-n/2}.
\end{equation}
A common cap is contained in either single cap. Multiplying by
$M_fC_{d_{\rm pc}}(\tau)$ therefore gives work rate
\begin{equation}\label{eq:terminal-rate}
 K_{\rm code}+2\beta+\frac12\log_2
              \frac{(2-\Delta)(1-\tau^2)}{c_0}.
\end{equation}
Replacing $2-\Delta$ by $2$ yields the second term of
\eqref{eq:dummy-rate}. Every occurrence is counted with its multiplicity.

\subsection{Pairs near the common cap threshold}
The exact geometric estimate from~\cite[Lemma~4.1]{GFH26b} is
\begin{equation}\label{eq:exact-wedge}
 W_{d_{\rm pc}}(\tau,u)
 \le\left(1-\frac{2\tau^2}{1+u}\right)^{(d_{\rm pc}-2)/2}
\end{equation}
for $-1<u<1$ and positive slack, with zero measure otherwise.
If the slack is at most $1/16$, input size limits
$|A|,|B|\le2^{\beta n+O(\pad)}$ give expected bucket work at most
$2^{(K_{\rm code}+2\beta-2)n+O(\pad)}$.
The forward calls have $\beta<1/2$ and $\beta\ge P(\tau)$, so this
is below the first rate in \eqref{eq:dummy-rate}.
The remaining correlation range has a fixed positive slack and is covered
by the compact bins and \eqref{eq:angle-identity}.
Nonnegativity permits the same full reference sums to bound capped sums.

The traversal bound also holds with these work estimates.
Choose $S=2^{\beta n+C_S\pad}$ after the list caps, with $C_S$ larger
than their insertion slack constants. In \eqref{eq:search-time}, the term
without $Z_A/S$ has rate $\beta+o(1)$, and the other has at most the point
incidence rate, up to $O(\pad)$ in the exponent. Polynomially many
sequential codes preserve the stated time and space rates.

\subsection{Representation counts in random cosets}\label{app:affine-details}
Use the notation of \cref{sec:affine-root}. Fix an upper bound
$a_{\max}$ for the compact energy range and choose $R_0\ge1$ with
$R_0^2>16a_{\max}/c+1$. For large $n$, $c_n\ge c/2$ and $w_n\le1$.
This radius bound covers all uses of \cref{lem:kl-count,lem:kl-weighted} below.

\subsubsection{The ambient lens and residue collisions}\label{app:affine-lens-details}
We supply the geometric estimates used in \cref{thm:affine-lens}.
Put $M=x-v/2$ and $\beta_n=E_n/4$. The endpoint shell conditions become
\begin{equation}\label{eq:affine-lens-window}
 \left|\en_r(M)-\beta_n\right|
 +\left|\frac{2\pi\langle M,v\rangle}{nr^2}\right|\le aw_n.
\end{equation}
On $\delta-v/2+K$, use a Gaussian of width $t=r\sqrt{\beta_n}$.
Dense ambient flatness remains uniform at $t$ and its relative
$O(w_n)$ radial tilts, including when $t<s$.
Applying \cref{lem:window-count} to the inner rectangle with both
window constants $aw_n/3$ and the outer radial window in
\eqref{eq:affine-lens-window} gives
\begin{equation}\label{eq:affine-ambient-lens}
 \sum_{x\in\delta+K}f_v(x)
 =\frac{r^n}{\det K}(eE_n/4)^{n/2}e^{\pm O(\ell_n)}.
\end{equation}
Replacing $E_n$ by $E$ changes this by $e^{O(1)}$.

For the collision sum \eqref{eq:affine-residue-collision}, take
$w\ne0$ and put $\xi=\|w\|^2/\|v\|^2\ge1$.
A nonzero summand puts both $M$ and $M+w$ in the midpoint ball.
The parallelogram identity gives
\begin{equation}\label{eq:affine-collision-ball}
 \en_r(M+w/2)\le\frac{E_n-c_n\xi}{4}+aw_n.
\end{equation}
It follows that $c_n\xi\le E_n+4aw_n$ and hence
$\xi\le16a_{\max}/c<R_0^2-1$, so all contributing difference radii
are below $R_0$.

Choose a fixed $0<b_0<E/32$ throughout the range.
When $(E_n-c_n\xi)/4\ge b_0$, the ball majorant on $K$ is uniform.
Put $H_n=E_n+4aw_n$. The contribution of each such difference to
the nonzero part of $\theta\sum_tF_t^2$ is at most
\[
 V\left(\frac{e(H_n-c_n\xi)}4\right)^{n/2}e^{O(\ell_n)}.
\]
Extend this nonnegative majorant to all nonzero differences.
\Cref{lem:kl-weighted} bounds their sum by
\[
 V\,2^{\kappa_{\rm pack}n}
 \left(\frac{eH_n^2}{16c_n}\right)^{n/2}e^{O(\ell_n)}.
\]
Since $H_n=E+O(w_n+1/n)$, dividing by the mean squared gives
\[
 V^{-1}2^{\kappa_{\rm pack}n}(ec_n)^{-n/2}e^{O(\ell_n)}
 =2^{-h_n+O(\ell_n)}.
\]
Here $ec=2^{2\kappa_{\rm pack}-1}$ and $c_n=c+O(1/n)$ give the
cancellation.

For the remaining boundary range, use one ball majorant of normalized
squared radius $b_0+aw_n$ and count all differences with
$\xi\le E_n/c_n+O(w_n)$. Before division by the mean squared, its
exponential base is at most $e b_0E_n/c_n+O(w_n)$.
The interior bound has base $eE_n^2/(16c_n)$, so the choice of $b_0$
leaves a fixed exponential margin. This proves the nonzero collision
bound used in \eqref{eq:affine-lens-variance}.

\subsubsection{Filtered differences and the affine work bound}
\label{app:affine-work-details}
Continue the proof of \cref{prop:affine-root-work} after applying the
uniform intensity bounds on the good event $\mathcal G$.
Write $g=\Prb[\mathcal G]=1-o(1)$.
Nonnegativity extends the remaining geometric sums to the original
uniform affine experiment at a factor $g^{-1}$.
For any $w\in L\setminus\{0\}$, the affine membership average is exactly
$\theta$ times the ambient sum over $\delta+K$.

Put $\xi=\|w\|^2/\|v\|^2$ and $d=c_n\xi$.
For endpoint energies $R,S\in a(1\pm w_n)$, their correlation is
$t=(R+S-d)/(2\sqrt{RS})$.
A positive common cap requires $1+t>2\tau^2$, hence
$d\le Q+O(w_n)$ and $4a-d\ge4a\tau^2-O(w_n)>0$.
The midpoint Gaussian majorant is therefore uniform on the whole
contributing range. Also $\xi\ge1$ bounds the correlation away from one,
so the Gram determinant of the two endpoint directions has a fixed
positive margin.

Let $B_\pm=4a(1\pm w_n)$. The common cap radius factor satisfies
\[
 1-\frac{2\tau^2}{1+t}
 =\frac{(\sqrt R+\sqrt S)^2-d-4\tau^2\sqrt{RS}}
         {(\sqrt R+\sqrt S)^2-d}
 \le\frac{Q+4a(1+\tau^2)w_n-d}{B_--d}.
\]
The midpoint ball has normalized squared radius at most $(B_+-d)/4$,
and throughout the contributing range,
\[
 B_--d\ge4a\tau^2-O(w_n)>0,\qquad
 \frac{B_+-d}{B_--d}=1+O(w_n).
\]
Choose $Q_n=Q+Cw_n+1/n$, with fixed
$C>4a(1+\tau^2)+1$ uniformly over the energy range.
The inflated cap numerator is at least $1/n$ on the support, and its
squared radius factor is uniformly below one. The spherical wedge
estimate in dimension $n+O(\log n)$ can therefore use exponent $n/2$
at cost $e^{O(\ell_n)}$.

Multiplying by the midpoint ball estimate cancels the denominator up
to $(1+O(w_n))^{n/2}=e^{O(nw_n)}$, and conditional independence of the
lists gives
\begin{equation}\label{eq:affine-root-filtered-difference}
 \E[\text{reports of difference }w\mid\mathcal G]
 \le g^{-1}e^{2D_n}M_f A_n^2V
 \left(\frac{e(Q_n-c_n\xi)}4\right)^{n/2}e^{O(\ell_n)}.
\end{equation}
Its support has $Q_n-c_n\xi>0$. The endpoint shells also imply
$c_n\xi\le4a(1+w_n)\le8a_{\max}$, so $\xi\le16a_{\max}/c$
and the fixed $R_0$ covers the counting bins.
Flatness on $K$ is the smoothing input here, replacing the internal
condition $8a\tau^2>1$.

Apply \cref{lem:kl-weighted} with $H=Q_n$ and $c=c_n$ to
\eqref{eq:affine-root-filtered-difference}. Together with the
target normalization in \eqref{eq:affine-root-target-balance}, it gives
\[
 2^{\kappa_{\rm pack}n}M_f
 \left(\frac{Q_n^2}{4c_nE}\right)^{n/2}2^{o(n)}
\]
as an upper bound for the sum over all nonzero differences.
Since $Q_n=Q+O(w_n)$ and $c_n=c+O(1/n)$, the second identity in
\eqref{eq:affine-root-target-balance} cancels $2^{\kappa_{\rm pack}n}$.
The total rate is therefore $K_\tau(u)$.
If all amplitude and code remainders are $O(\pad)$, the same calculation
keeps every logarithmic remainder $O(\pad)$.

For target coverage, choose the code padding to dominate $O(\ell_n)$
and $H_{d_{\rm pc}}$ from \cref{ing:bdgl}. The target correlation
$u+O(w_n)$ stays inside the prescribed active constraint range.
Equal values have already been charged to point incidences in the main
proof, completing the work and coverage bounds.

\subsection{Uniform concentration for fixed banks}\label{app:batch-concentration}
The relative approximation theorem
\cite[Theorem~2.11(ii)]{HS11} controls all queries without enumerating
their lattice supports. Let $P_M$ be the empirical law of $M$ iid
draws from $P$, and let the positive superlevel sets of a class
$\mathcal F$ of $[0,1]$ functions have VC dimension $d$.
For $0<\epsilon\le1/2$, $0<p\le1$, and $0<\delta<1$, the sample size condition
\[
 M\ge\frac{C}{\epsilon^2p}
       \left(d\log\frac2p+\log\frac2\delta\right)
\]
gives $|P_MA-PA|\le\epsilon\max\{PA,p\}$ simultaneously on these
sets, with failure at most $\delta$.
The finite-set formulation extends to our finitely supported laws
by approximating their masses by rational frequencies and replicating
the atoms. As the replicated population grows, uniform sampling
without replacement converges to the required iid law.
Integrating over $\{f>t\}$, $0<t<1$, yields
\begin{equation}\label{eq:weighted-relative}
 |P_Mf-Pf|\le\epsilon(Pf+p)\qquad(f\in\mathcal F).
\end{equation}
The failure symbol belongs to no positive superlevel set.

\paragraph{A common test for the two bank queries.}\label{app:bank-test}
For a forward family $0\le f_x\le1$ with $Pf_x\le H$ and a reverse
family $0\le g_z\le1$ with
$Pg_z=\bar\mu(1\pm\eta_{\rm mid})$, $\eta_{\rm mid}\le1/4$,
apply \eqref{eq:weighted-relative} at error $1/16$ and floor
$p_0=\min\{1,H,3\bar\mu/4\}$. In both applications,
$\log(1/p_0)=O(n)$ and the shatter coefficients satisfy
$S(N)\le(CN)^{Cn^2}$, giving $d=O(n^2\log(n+2))$.
Thus $Mp_0\ge2^{\Omega(\pad)}$ permits
$\delta=\exp[-cMp_0]=\exp[-2^{\Omega(\pad)}]$ for a small
absolute $c>0$.
The forward sum is at most $2MH$.
If $B_C(z)=h(z)U_r\sum_i g_z(y_i)$, the reverse sum lies between
$\tfrac12MU_r\bar\mu h(z)$ and $2MU_r\bar\mu h(z)$.
The two kernels below supply $H,U_r,\bar\mu$ and verify these conditions.

\begin{lemma}[Simultaneously good banks]\label{lem:uniform-bank}
For the node parameters in \cref{sec:batch-dgs}, the bank satisfies
\eqref{eq:batch-good-bank} for all admissible queries except with
probability $\exp[-2^{\Omega(\pad)}]$. This holds uniformly in the
input width, lattice determinant, and affine shift.
\end{lemma}
\begin{proof}
Condition on the node's lattice, scales, and routing shift.
The full right trial law is $q(y)=\lambda\rho_r(y)/V_r$ on its
source shell, with the remaining mass at $\dagger$.
For forward queries use $f_x(y)=K(x,y)$ and the mean envelope
$Pf_x\le H_r=\lambda\mathcal B_r2^{\degpad}$ from
\cref{lem:soft-degree}. For reverse queries let
\[
 t_z(y)=\frac{q(z+y)K(z+y,y)}{\rho_{\gamma r}(z)},\qquad
 g_z(y)=t_z(y)/U_r,\quad U_r=(\lambda/V_r)e^{nw_n}.
\]
Completion of squares gives
\[
 t_z(y)=\frac{\lambda}{V_r}
 e^{A_rR_-^2-n\en_r(z+y)/2+n\en_r(z)/4}.
\]
The source and output shells put the exponent after $\lambda/V_r$
at most $na_1w_n/2\le nw_n$, so $g_z\in[0,1]$.
The midpoint identity gives
\[
 Pg_z=\bar\mu_r(1\pm\eta_{\rm mid}),\qquad
 \bar\mu_r=\lambda e^{A_rR_-^2-nw_n}2^{-n/2}
 =\lambda2^{-[J(a_0)+R(u)]n-O(nw_n)}.
\]

For either family, a positive superlevel set is given by a constant
number of quadratic inequalities in $y$ (take logarithms of the
positive weight). Lifting to degree-two monomials gives
$S(N)\le(CN)^{Cn^2}$. Queries are restricted to prescribed cosets:
then $z+y$ automatically lies in the source coset. Varying coset
membership is not included among the quadratic predicates.
The failure symbol contributes to no positive superlevel.

Use \cref{app:bank-test} with $H=H_r$ and the displayed reverse
mean and envelope. The recursion gives $\lambda\ge2^{-o(\pad)}$,
and
\[
 P(u)-R(u)=-\tfrac12\log_2\gamma^2\ge0,\qquad
 M\min\{H_r,3\bar\mu_r/4\}\ge2^{C_a\pad-o(\pad)}.
\]
Since $MU_r\bar\mu_r=b_*$, the common test proves
\eqref{eq:batch-good-bank}.
\end{proof}

\subsection{Reverse-query work with a shared bank}\label{app:batch-search}
Conditional on bank $C$, the proposal before the second rejection
has mass $B_C(z)/A_{\max}$. On a good bank this is at most four times
the prescribed output subprobability in \eqref{eq:batch-output-trial}.
Applying this domination at fixed $C$ removes the bank-dependent
proposal weight before averaging the query cost $g(z)+C(H_z)$.
We can then drop the good bank indicator and bound the bank contribution
by its original first moment. The sum retains terms involving the
selected record itself, so it also pays for inserting that record
before reverse search.

Write $q=-z$, so query and bank have energies $a_1,a_0$.
After averaging the local half grid shift, sum first over the bank
and then over the queries. On a correlation bin near $v$, the
triangular change
\[
 (q,y)\longmapsto
 \left(q,y-v\sqrt{a_0/a_1}\,q\right)
\]
admits Gaussian widths
$r\sqrt{a_1}$ on $\Gamma$ and
$r\sqrt{a_0(1-v^2)}$ on $\Gamma/2$.
Their squared normalized radii add to $2+O(w_n)$ on the shells.
Shifted theta upper bounds apply for each fixed $q$ even though this
change need not preserve the lattice.
Writing $g=r^2/s^2$, we have
\[
 ga_1\ge2a_0>1,\qquad
 C_{\rm rev}=4ga_0\ge C_0:=4/(1-u^2).
\]
The half grid factor $2^{-n}$ cancels its covolume.
Converting Gaussian densities to query and bank list sizes
$N_Q,N_C$ uses
$e^{(1-a_0)n}a_0^n=2^{-2J(a_0)n}$.
Consequently the angular spectrum is
\begin{equation}\label{eq:batch-reverse-spectrum}
 N_QN_C\,2^{-P(v)n+O(nw_n+\log n)}
 \max\{1,[C_{\rm rev}(1-v^2)]^{-n/2}\}.
\end{equation}

For uniform control near parallel directions, fix $v_0<1$ with
$C_0(1-v_0^2)<1$. Throughout $v\ge v_0$, use the single residual
$y-v_0\sqrt{a_0/a_1}\,q$, of energy at most
$a_0(1-v_0^2)+O(w_n)$.
Sequential Gaussian sums bound that region by the angular factor
\[
 \max\{(1-v_0^2)^{n/2},C_{\rm rev}^{-n/2}\}
 \le C_0^{-n/2}
\]
times $2^{O(nw_n+\log n)}$. This includes parallel directions and
repeated values. It uses the fixed lower bound $C_0$, rather than
requiring a cutoff subcritical for arbitrarily large actual
$C_{\rm rev}$. Away from this region, replace the penalty in
\eqref{eq:batch-reverse-spectrum} by the weaker $C_0$ if needed.

Set $\tau=u$ and use code rate $K_{\rm rev}(u)$ from
\cref{sec:batch-dgs}. The identity
$P(v)+K_\tau(v)\ge2P(\tau)$ gives a report rate in the ordinary region
at most $K_{\rm rev}$, since both list size exponents are $P(u)$.
In the subcritical and near-parallel regions the single-cap bound gives
\[
 K_{\rm rev}+2P(u)-\tfrac12\log_2C_0-P(u)
 =K_{\rm rev}-1.
\]
For common cap slack at most $1/16$, \eqref{eq:exact-wedge} instead
gives $K_{\rm rev}+2P(u)-2$, also smaller.
The other correlations lie in fixed compact bins.
Point incidences and restricted traversal have rate at most
$K_{\rm rev}$ by \cref{ing:bdgl}.
Independent fresh reverse codes are introduced only after ideal
proposals have been fixed, so complete coverage and
\cref{lem:finite-code-coupling} apply.

True forward dictionaries have rate $P(u)$.
For reverse dictionaries evaluate the spectrum at
$v_{\rm rev}$, where $P(v_{\rm rev})=R(u)$, giving
$2P(u)-R(u)$. The full bank charging argument above includes
contributions from repeated IDs. Sort by canonical pair IDs before adding
weights, and use independent coins for left trials. Depth-first list storage,
sparse labels, dictionaries, and code descriptors therefore fit the
limits in \cref{sec:batch-dgs}.
\subsection{Independent extensions and the terminal batch merge}
\label{app:batch-terminal}
We prove the additional statements used in
\cref{thm:batch-region,thm:batch-three-halves}. Throughout, write
$u_E=1/2+1/k_n$, $c_E=(1+u_E)^{-1}$,
$\ell_E=P(u_E)$, and $R_E=\tfrac12\log_2(2c_E)$, with $k_n$ as in
\cref{lem:endpoint-schedule}. Thus $c_E\to2/3$ and
$\ell_E,R_E\to\dgsspace$. The existing endpoint forest gives $M$ iid
reference labels with law on vector outputs
\begin{equation}\label{eq:batch-endpoint-stream}
 q_E(x)=\frac{\lambda_E}{V_R(\Gamma)}\rho_R(x)
 \ind_{\en_R(x)\in c_E(1\pm w_n)},\qquad
 M=\left\lceil2^{[\ell_E+J(c_E)]n+C_M\pad}\right\rceil,
 \quad\lambda_E=2^{-1-t(\degpad+2)}.
\end{equation}
The affine support is the prescribed coset. The forest has common
depth $t=O(k_n\log n)$, and $-\log_2\lambda_E=o(\pad)$.
The reference coupling is the one in \cref{sec:bank-correction}.

\subsubsection{Extensions and exact mixing coefficients}
Use the independent extensions of \cref{lem:dgs-index-wrapper}, with
$\iota_j=\max\{0,\log_2(\sigma/(\alpha s_j))\}$ and exact index
$p_j$. Here $s_j/\sigma$ stays in a fixed compact positive range,
so the lemma's failure bound applies to every bin.
\Cref{alg:gaussian-batch} specifies the balancing and label consumption.
The remaining calculation keeps the finite rates needed at full size.

For ordinary bins take $R_j^2=\sigma^2z_j/c_E$ and $s_j^2=R_j^2/2$.
Radial correction anchored at endpoint $\zeta_j=z_j+O(w_n)$ gives
\[
 \kappa_j=\frac{\lambda_E}{V_{R_j}(\Gamma_j)}
       e^{n\zeta_j(1-c_E/z_j)/2},\qquad
 M\kappa_jV_\sigma(L)
 =2^{[\ell_E-\iota_j+J(z_j)]n+C_M\pad-o(\pad)+O(nw_n)-3h_n}.
\]
Taking the endpoint limit gives \eqref{eq:batch-bin-supply}.
In particular, using the innermost index for every bin would lose the
gain. On $[1/2,2]$ with $\alpha=3/2$, the second branch of that
minimum exceeds $\dgsspace$, since $\log(27/16)>1/2$.
The first branch attains its minimum at $z=1$, where the exact
finite rate $\ell_E>\dgsspace$ and padding give the required surplus.

\subsubsection{The new terminal kernel and its degree}
Fix $u=3/5$, $c=5/8$, $d=1/2$, and $\gamma^2=4/5$.
For an inner bin $z_j\in[5/12,1/2]$, set
\begin{equation}\label{eq:batch-terminal-widths}
 r_o^2=\sigma^2z_j/c,\qquad r^2=r_o^2/\gamma^2=2\sigma^2z_j,
 \qquad R^2=(c/c_E)r^2,\qquad s_j^2=R^2/2.
\end{equation}
For output coset $a+\Gamma_j$, route a uniform
$e\in(\Gamma_j/2)/\Gamma_j$ to the child cosets
$a/2+e+\Gamma_j$ and $-a/2+e+\Gamma_j$.
Their endpoint shells are precisely $\en_r(x)\in c(1\pm w_n)$.
For sufficiently large $n$, $R<r$. If $U$ is the maximum squared
norm on this shell, a radial acceptance
$\exp[\pi(\|x\|^2-U)(R^{-2}-r^{-2})]$ converts each child to
\begin{equation}\label{eq:batch-terminal-child}
 q(x)=\frac{\bar\lambda}{V_r(\Gamma_j)}\rho_r(x)
       \ind_{\en_r(x)\in c(1\pm w_n)},\qquad
 \bar\lambda=\lambda_E(r/R)^n e^{-\pi U(R^{-2}-r^{-2})}.
\end{equation}
Thus $\log_2\bar\lambda=\log_2\lambda_E+[J(c)-J(c_E)]n+O(nw_n)$.
The coefficient can exceed one, since it multiplies a restricted
Gaussian density. The acceptance probability above already guarantees
a subprobability. With $A=M\bar\lambda$,
$\log_2A=[\ell_E+J(c)]n+O(\pad)$, and the stored child list
still has rate $\dgsspace$.

For $z=x-y$, the eligibility test is
\[
 \en_r(z)\in d(1\pm w_n).
\]
Put $a_r=\pi(\gamma^{-2}-1/2)/r^2$,
$L_-^2=nr^2d(1-w_n)/(2\pi)$, and
$K(x,y)=\ind_{\rm Eligible}e^{-a_r(\|x-y\|^2-L_-^2)}$.
Writing $m=(x+y)/2$, completion of squares gives
\[
 \rho_r(x)\rho_r(y)K(x,y)
 =e^{a_rL_-^2}\rho_{r/\sqrt2}(m)\rho_{\gamma r}(z)
       \ind_{\rm Eligible}.
\]
For the fixed right bank define $A_C,B_C$ as in
\cref{sec:batch-dgs}, using the child law
\eqref{eq:batch-terminal-child}. The forward residual width
$r\sqrt{1-u}$ can be smaller than $s_j$. Its cost is
\begin{equation}\label{eq:batch-terminal-degree}
 H_n=\max\{1,(s_j/(r\sqrt{1-u}))^n\},\quad
 \mathcal B_r=e^{a_rL_-^2}(\gamma/\sqrt2)^n,\quad
 D=8A\mathcal B_rH_n2^{C_D(nw_n+\log(n+2))}.
\end{equation}
Indeed for $t<s_j$, shift maximality and monotonicity give
\[
 \rho_t(b+\Gamma_j)\le\rho_t(\Gamma_j)
 \le\rho_{s_j}(\Gamma_j)
 \le(1+2^{-h_n})V_t(\Gamma_j)(s_j/t)^n.
\]
This yields $\E[A_C(x)]\le2A\mathcal B_re^{O(nw_n)}H_n$.
Now $s_j^2/r^2\to15/32$, so
\[
 n^{-1}\log_2H_n\to\tfrac12\log_2(75/64)=P(3/5)-\dgsspace,
 \qquad \mathcal B_r=2^{-[J(c)+P(3/5)]n+O(nw_n)}.
\]
The fixed degree exponent cancels. At finite $n$ the remaining term is
\begin{equation}\label{eq:batch-terminal-finite-degree}
 \log_2D=(\ell_E-R_E)n+O(\pad)=o(n).
\end{equation}
In particular $O(n/k_n)$ need not be $O(\pad)$ and is retained in
the work budgets.

\subsubsection{Uniform banks and the complete output law}
Set
\[
 b_0=\frac{M\bar\lambda^2}{V_r^2}e^{a_rL_-^2}V_{r/\sqrt2}.
\]
We claim that, except with probability $\exp[-2^{\Omega(\pad)}]$,
\begin{equation}\label{eq:batch-terminal-bank}
 A_C(x)\le D,\qquad
 \tfrac12b_0\rho_{\gamma r}(z)\le B_C(z)
                  \le2b_0\rho_{\gamma r}(z)
\end{equation}
simultaneously for all admissible queries. For the forward family take $H=D/(4M)$ in
\cref{app:bank-test}. The expected sum is at most $D/4$, and
positive padding gives $MH=D/4\ge2^{\Omega(\pad)}$.
For the reverse class use
\[
 t_z(y)=\frac{q(z+y)K(z+y,y)}{\rho_{\gamma r}(z)},\quad
 g_z=t_z/U_r,\quad U_r=(\bar\lambda/V_r)e^{Cnw_n}.
\]
On the eligible support,
\[
 t_z(y)=\frac{\bar\lambda}{V_r}
 e^{a_rL_-^2-2\pi\|y+z/2\|^2/r^2+\pi\|y\|^2/r^2}.
\]
Equivalently the exponent is
$a_rL_-^2-n\en_r(z+y)/2+n\en_r(z)/4$.
The source and output shells bound it by $ndw_n/2$, so
$0\le g_z\le1$ for fixed large $C$.
The midpoint width satisfies $(r^2/2)/s_j^2\to16/15>1$.
Since $c=1/2+d/4$, the endpoint region contains the same inner
midpoint window as in \cref{app:gaussian-profile}. Radial and
linear tilts retain smoothing, and comparison with the full
midpoint mass shows
\[
 Pg_z=\bar\mu_r(1\pm\eta_{\rm mid}),\quad
 \eta_{\rm mid}=O(2^{-h_n})+e^{-\Omega(nw_n^2)},\quad
 M\bar\mu_r=2^{(\ell_E-R_T)n+O(\pad)},\quad
 R_T=\tfrac12\log_2(5/4).
\]
Here $\dgsspace-R_T>0.0465$. Both positive superlevel classes use a constant number of quadratic
predicates, so $S(N)\le(CN)^{Cn^2}$.
The forward and reverse sample floors satisfy \cref{app:bank-test},
with $MU_r\bar\mu_r=b_0$, proving \eqref{eq:batch-terminal-bank}.

Apply \cref{lem:bank-correction} with the normalizer $D$ in
\eqref{eq:batch-terminal-degree}, $c=b_0/2$, and the output Gaussian
weight restricted to its shell and coset. The bank bounds give $C_0=4$
and the following probability mass on vector outputs
\begin{equation}\label{eq:batch-terminal-trial}
 q_o(z)=\frac{b_0}{2D}\rho_{\gamma r}(z)
       \ind_{z\in a+\Gamma_j,\ \en_r(z)\in d(1\pm w_n)}.
\end{equation}
The reference replacement in \cref{sec:bank-correction} gives iid labels.
The known output coefficient is
$\lambda_o=b_0V_{\gamma r}/(2D)$.

\subsubsection{Output counts at finite dimension}
The live count rate tends to
$q_T=2\dgsspace-P(3/5)=\log_2(16/15)$.
More precisely, putting $d_E=n^{-1}\log_2H_n=P(3/5)-R_E$ gives
\[
 \log_2(M\lambda_o)-J(c)n
 =[\ell_E+R_E-P(3/5)]n+\Theta(\pad).
\]
The padding is positive after choosing $C_M$ large enough.
The function $P(u)+\tfrac12\log_2(2/(1+u))$ is increasing for
$u\ge1/2$, so this finite rate is at least $q_T$.
The actual extension rate is
\[
 \iota_E(z)=\max\{0,\tfrac12\log_2(c_E/(\alpha^2cz))\}.
\]
At $\alpha=3/2$ it is no larger than its limit, since $c_E\le2/3$.
Restricting to $a+L$ and changing width to $\sigma$ consequently
gives \eqref{eq:batch-terminal-supply}, with errors
$O(1/k_n)+O(\pad/n)$. Its minimum is at an endpoint:
it increases up to $64/135$ and then decreases up to $1/2$.
At those endpoints,
\[
 F_T(5/12)=J(5/12)>\dgsspace,\qquad
 F_T(1/2)=\log_2(16/15)+J(1/2)>\dgsspace.
\]
These strict margins absorb all finite parameter changes.

\subsubsection{Forward and reverse search}
The two stored child lists have size exponent $\dgsspace$.
Write $g=r^2/s_j^2\to32/15$ and $c_0=2gc\to8/3$.
The angular spectrum from \cref{sec:search} is
\[
 2^{[2\dgsspace-P(v)]n+o(n)}
 \max\{1,[c_0(1-v)]^{-n/2}\}.
\]
Use forward filter threshold $\tau=1/2$, target correlation $3/5$,
and define $K_\tau(v)=-\tfrac12\log_2(1-2\tau^2/(1+v))$.
The point and the two report bounds all have rate at most
\[
 T_F=K_{1/2}(3/5)=\tfrac12\log_2(16/11)<0.2703<\dgstime.
\]
The forward true edge dictionary has rate $q_T<\dgsspace$.
The common cap compact region has midpoint smoothing margin
$g\,2c(1+v)>4/3+o(1)$ whenever $v>-1/2$.
The fixed cutoff and single-cap bounds in \cref{app:batch-search}
handle its parallel and degenerate limits.

The reverse query depends on the bank. Conditional on a good bank,
its proposal density is $B_C(z)/D\le4q_o(z)$.
For any nonnegative cost, apply this domination before averaging
over the bank and removing its good event indicator. The right side
does not depend on the bank, so its original first moment controls
the bank cost, including contributions from the same record.
Fresh reverse codes are sampled only after ideal proposals are fixed.

The reverse target correlation is $v_T=1/\sqrt5$.
Its query list has size exponent $q_T$ and the bank has size exponent
$\dgsspace$. The triangular Gaussian transformation in
\cref{app:batch-search}, with different query and bank sizes retained, gives
\[
 N_QN_C2^{-P(v)n+o(n)}
 \max\{1,[C_{\rm rev}(1-v^2)]^{-n/2}\},\qquad
 C_{\rm rev}=4gc\to16/3.
\]
The query width has squared ratio $gd\to16/15>1$ to smoothing.
The residual sum is on $\Gamma_j/2$, which explains the factor four.
Use reverse threshold $1/3$ and put $K_R=K_{1/3}(1/\sqrt5)$.
Point, ordinary report, and subcritical report rates are at most
\begin{align*}
 K_R+\dgsspace-P(1/3)&<0.2429,\\
 K_R+q_T+\dgsspace-2P(1/3)&<0.2510,\\
 K_R+q_T+\dgsspace-\tfrac12\log_2(16/3)-P(1/3)&<-0.8715.
\end{align*}
The ordinary bound uses $P(v)+K_\tau(v)\ge2P(\tau)$.
The single-cap bound controls the subcritical and parallel regions.
The exact wedge estimate at very small common cap slack gives the
smaller rate $K_R+q_T+\dgsspace-2$.
Thus every reverse rate is below $\dgstime$ with a fixed margin.
The reverse dictionary rate is
$q_T+\dgsspace-R_T<0.1397<\dgsspace$.
The raw label rate tends to
$\dgsspace+J(2/3)=1/2-1/(6\log2)<0.2596<\dgstime$.
Sparse labels, record IDs, both dictionaries, and depth first
forest storage fit $2^{\dgsspace n+o(n)}$ bits. Restricted Cartesian
traversal adds no larger rate. Polynomially many bins and
$2^{o(n)}$ forest nodes preserve these bounds.

\subsubsection{Tails, joint error, and finite realization}
The lower tail cannot be tilted to width $\sigma\sqrt{5/12}$,
which may be below smoothing. Instead,
\[
 \E_{D_{a+L,\sigma}}[e^{-(\alpha^2-1)n\en_\sigma(X)/2}]
 =\frac{\rho_{\sigma/\alpha}(a+L)}{\rho_\sigma(a+L)}
 \le3\alpha^{-n}.
\]
At $\alpha=3/2$ this gives lower-tail exponent
$H_{3/2}(5/12)=[\log(9/4)-25/48]/(2\log2)$.
Direct substitution gives
$J(5/12),H_{3/2}(5/12),J(2)>\dgsspace$ and
$F_T(1/2)>\dgsspace$, with fixed positive margins.
Thus the annulus tail $\delta_{\rm ann}$ obeys
$2^{\dgsspace n}\delta_{\rm ann}=2^{-\Omega(n)}$.
The expected mixture success count is at least
$2^{\dgsspace n+\Omega(\pad)}$ and its shortfall is doubly
exponentially unlikely. The full joint error is bounded by
\begin{equation}\label{eq:batch-terminal-joint-error}
 \poly(n)2^{-2h_n+O(1)}
 +2^{o(n)}e^{-2^{\Omega(\pad)}}
 +2^{-\Omega(\pad)}+2^{-\Omega(n)}
 +\lfloor2^{\dgsspace n}\rfloor\delta_{\rm ann}.
\end{equation}
Here $\log(\text{forest size})=O(k_n\log n)=o(\pad)$.
The leading error is $\exp[-\Omega(n/\log^2(n+2))]$.

For finite execution use rational squared radii and half-open bins,
with normalized endpoints within $2^{-Cn}$ of the stated values.
Use exact shell comparisons, including at boundary lattice points,
and the exact finite rates, in particular
$(\ell_E-R_E)n=O(n/k_n)$, before enlarging work and storage limits
by $2^{C\pad}$. First-overflow bounds cost $2^{-\Omega(\pad)}$.
The positive coefficient formulas and whole batch numerical budget
are implemented in \cref{app:batch-numerics}. Their loss is charged
once in \eqref{eq:batch-terminal-joint-error}. Fixed-dimensional
sampling handles the remaining small ranks.

\subsection{Sampling estimates}\label{app:application-details}
We verify the scale and search margins used in \cref{sec:dgs,sec:svp}.

\subsubsection{The smoothing extension for DGS}\label{app:dgs-environment}
Put $\eta=\eta_{1/2}(L)$ and
$A(t)=\sum_{y\in L^*}e^{-\pi t^2\|y\|^2}$.
For $t\ge\eta$, $A(t)\le3/2$. For $t<\eta$, Poisson summation and
$\rho_t(L)\le\rho_\eta(L)$ give
$A(t)\le(\eta/t)^nA(\eta)$. Hence
\begin{equation}\label{eq:dgs-dual-budget}
 A(s)\le\tfrac32\,2^{\iota n}.
\end{equation}
The extension has index $p$. Every dual vector outside $pL^*$
survives its random modular constraint with probability
$(p^{n-1}-1)/(p^n-1)\le1/p$.
For $a_y=e^{-\pi\eta^2\|y\|^2}$, $y\ne0$, one has
$\sum_y a_y\le1/2$, so the forced mass is at most
$\sum_y a_y^T\le2^{-T}$, where $T=(ps/\eta)^2$.
Since $s/\eta$ has a fixed positive lower bound,
\[
 \E_\Gamma[\rho_{1/s}(\Gamma^*\setminus\{0\})]
 \le A(s)/p+2^{-T}\le2^{-3h_n+O(1)}.
\]
Markov's inequality proves \cref{lem:dgs-index-wrapper}.

\subsubsection{The target shell tail}\label{app:dgs-tail}
For $Y\sim D_{a+L,\sigma}$, put $T=2\pi\|Y\|^2/\sigma^2$.
At $|\theta|=O(w_n)$, both widths below are above smoothing, and
\[
 \E[e^{\theta T}]
 =\frac{\rho_{\sigma/\sqrt{1-2\theta}}(a+L)}{\rho_\sigma(a+L)}
 \le3(1-2\theta)^{-n/2}.
\]
Chernoff bounds with $\theta=\pm\Theta(w_n)$ give
$\Prb[|T/n-1|>w_n]\le6e^{-c_\alpha nw_n^2}$.
The target shell mass divided by $V_\sigma(L)$ is consequently
between two positive constants for large $n$.

\subsubsection{Ordinary forward search}\label{sec:dgs-work}
For the ordinary shell geometry, the smallest
input scale has $r^2/s^2=2/\gamma^2$ and free output denominator one.
The two smoothing factors are
\begin{equation}\label{eq:dgs-search-factors-new}
 c_0(1-v),\quad c_0(1+v),\qquad c_0=\frac2{1-u^2}.
\end{equation}
A nonempty common cap has $v>2u^2-1$, hence midpoint factor at least
$4u^2/(1-u^2)\ge4/3$. Multiplying child intensities before averaging
the shift gives, by \cref{lem:correlation}, the candidate count
\begin{equation}\label{eq:dgs-correlation-new}
 2^{[2\ell(u)-P(v)]n+O(\pad)}
 \max\{1,[c_0(1-v)]^{-n/2}\}.
\end{equation}
Earlier scales improve these factors. Substituting $\ell(u)=P(u)$ and $\tau=u$ into
\eqref{eq:dummy-rate} makes both report rates equal to $K(u)$,
as is the point incidence rate. The endpoint bins
also fit: with $\Delta=1/100$, $c_0\Delta<1$ and the terminal cap rate
is $K(u)+\tfrac12\log_2(1-\Delta/2)<K(u)$, including equal values.
When $1-2u^2/(1+v)\le1/16$, \eqref{eq:exact-wedge} gives
$K(u)+2\ell(u)-2<K(u)$. The remaining $O(n)$ bins are compact.
Restricted traversal costs at most
$2N_{\rm cap}B_{\rm pc}+2nZ_AB_{\rm pc}$.
The bound $\log_2B_{\rm pc}=o(\pad)$ therefore gives, per call within the prescribed limits,
\begin{equation}\label{eq:dgs-work-bound-new}
 \E[\text{forward search work}]\le2^{K(u)n+O(\pad)}.
\end{equation}

\subsection{Shortest-vector scales and search margins}
\subsubsection{Scale selection and initial pools}\label{app:svp-environment}
Apply \eqref{eq:short-scale} to $L$ with $c=9$, and set
$\widehat Q=\|v\|^2$ and $\Delta_n=D_9(n)$. Then
$\widehat Q/\Delta_n^2\le\lambda_1(L)^2\le\widehat Q$.
The grid
\begin{equation}\label{eq:svp-scale-grid}
 Q_i=\frac{\widehat Q}{\Delta_n^2}(1+1/n)^i
\end{equation}
up to $\widehat Q$ has $O(n\log n)$ branches, one with
$Q_i\le\lambda_1(L)^2<(1+1/n)Q_i$.
On each branch choose the rational squared width of \cref{ing:gfh},
put $Y=\max\{100,\lceil2^{h_n}\det L/s^n\rceil\}$,
and search for a prime in $[Y,2Y]$
using \cref{app:numerical-primitives}. Skip the scale if the search
exhausts its budget. On the correct branch, the Gaussian mass bound
underlying \cref{ing:gfh} gives
$2^{h_n}\det L/s^n\ge2^{h_n}/\poly(n)>100$ for large $n$, so the chosen prime has
the required scale.

Within a trial, apply \cref{ing:reduction} to $\Gamma$ with
$k=\lfloor n/10\rfloor$, at cost $2^{0.2n+o(n)}$.
On $\mathcal G$, \eqref{eq:smoothing-successive} gives
$G\le\poly(n)s$. A fixed depth
$t_{\rm s}=\lceil C_t\log(n+2)\rceil$ therefore makes
$R=r\gamma_{\rm s}^{-t_{\rm s}}$ a valid starting width.
Each leaf uses $2M$ corrected trials, where
$M=\lceil2^{m_{\rm raw}n+C_{\rm leaf}\pad}\rceil$, and
$\degpad=\lceil10\errwin\rceil$. Its root padding is
\[
 \sigma_0=\log_2M-m_{\rm raw}n-t_{\rm s}(h_n+\degpad)
 =C_{\rm leaf}\pad-O((h_n+\errwin)\log n)=\Theta(\pad).
\]
These predetermined schedules supply the cooling calls in
\cref{alg:full-difference-tree,alg:svp}, with the conditional
independence specified in \cref{sec:svp-environment}.

\subsubsection{List sizes and pair search}\label{app:svp-search}
At height $j=1,\ldots,4$, put $t_j=2^{5-j}$.
The ancestor average \eqref{eq:ancestor-average} becomes
\begin{equation}\label{eq:svp-ancestor-average}
 \E_{\delta,\mathrm{ancestors}}\!\left[\sum_{x\in b_j+\Gamma}f(x)\right]
 =t_j^{-n}\sum_{x\in\varepsilon z/2^{4-j}+\Gamma/t_j}f(x),
 \qquad f\ge0,
\end{equation}
where $\varepsilon\in\{1,-1\}$ depends on the path.
Fixing $\Gamma,z$, the path, and scale leaves the local and ancestor
shifts to be averaged. Conditional on those shifts, child intensities
are each at most $A_{j-1}e^{O(\pad)}$ by \cref{thm:shell-tree},
and multiply by private independence. Cooling gives the analogous
Gaussian bounds. The shell majorants on $\Gamma/t_j$ have squared
smoothing factors $266.2912,54.464,11.456,2.11296$.
Their covolumes cancel $t_j^{-n}$, proving \eqref{eq:svp-list-first}.

For input rate $\beta$, threshold $\tau$, free denominator $t$,
and $g=r_{\rm in}^2/s^2$, let $c_0=2t^2ga$.
Equation~\eqref{eq:dummy-rate} gives
\begin{equation}\label{eq:svp-generic-work}
 K_{\rm code}+
 \max\left\{2(\beta-P(\tau)),
  2\beta+\tfrac12\log_2\frac{2(1-\tau^2)}{c_0}\right\}.
\end{equation}
In cooling, $g\ge\bar g:=2/\gamma_{\rm s}^2$ and $t=1$.
Use the fixed comparison factor
$c_{0,\rm s}=2\bar g a_0=2/(1-u_{\rm s}^2)$ throughout.
Equivalently, at input width $r_{\rm in}$ use the auxiliary smoothing
scale $s_{\rm eff}=r_{\rm in}/\sqrt{\bar g}\ge s$.
The lattice is smooth there by monotonicity, so this weaker comparison
gives uniform upper bounds at every cooling depth. Substitution of
$\tau=u_{\rm s}$ gives \eqref{eq:svp-soft-work}. Point work is smaller,
and the comparison midpoint factor exceeds $2.10$ wherever the common
cap is nonempty.

At full height $j$, use $g=2$, $a=a_{j-1}$, $t=t_j$, and
$\beta=L_{j-1}$. Substituting \eqref{eq:svp-energy-orbit} gives
$8a_{j-1}\tau_j^2>1.008$ and $c_0\ge716/125>4$.
Since $P(\tau_j)=L_{j-1}$, both point and pair rates are at most $K_j$.
For full layers use $c_{0,j}=4t_j^2a_{j-1}$, and for cooling use
$c_{0,\rm s}$ above. With these comparison factors,
$\Delta=1/2000$ gives $c_{0,j}\Delta,c_{0,\rm s}\Delta\le1024/3175<1$
and midpoint factor at correlation
$1-\Delta$ at least $715821/250000>1$.
The common cap region of measure factor at most $1/16$ costs
$K_{\rm code}+2\beta-2<K_{\rm code}$. The other $O(n)$ bins are
compact. Window perturbations cost $O(\pad)$.

The terminal bound is \cref{prop:affine-root-work} with
\cref{eq:svp-terminal-envelopes,eq:svp-label-exactness}, averaging
the common coset. Deduplicated equal values cost point work.
Restricted traversal has rate at most
$\max\{\beta,\beta+T_{\rm point}-s_{\rm live}\}$.
Every input has $\beta\le L_{\max}<s_{\rm live}$ with gap
greater than $38\cdot10^{-6}$, so traversal fits the point rate.
The factor $B_{\rm pc}$ is subexponential, and both preprocessing
stages have exponent $0.2<s_{\rm live}$.

\section{Finite realization}\label{app:conventions}
We first give the common data representations, random primitives,
and search coupling, then choose the budgets for single samples,
batches, and SVP. The application analyses supply their list and work
rates. Predetermined caps turn these estimates into bounds on every
execution, and one coupling accounts for all implementation errors.

\subsection{Finite data and random tools}
\subsubsection{Computational model}\label{app:computational-model}
We use a randomized word RAM with independent uniform bits and word
size $w=\Theta(n+\log(\mathcal B+f+2))$, where $\mathcal B$ is the
binary input length and $f$ is an accuracy parameter when present.
Time counts word operations, including multiword arithmetic and
numerical evaluation. Space counts bits. Polynomial input factors
are suppressed, and polynomial factors in $f(n)\le\poly(n)$ are
included in $2^{o(n)}$. The same leading exponents hold with bit
costs for random-access operations.

All resource bounds hold on every execution. Lists, work, and random
trials have predetermined caps. On failure, shifted DGS returns its
affine offset, centered DGS returns zero, and SVP retains its best
nonzero lattice vector. A failed batch returns copies of its fallback.

\subsubsection{Sublinear budgets}\label{app:budgets}
The default schedules for SVP and batches are
\begin{equation}\label{eq:budgets}
 \begin{aligned}
 \logbudget&=\lceil\log_2(n+2)\rceil,&w_n&=\logbudget^{-4},\\
 h_n&=\lceil n/\logbudget^2\rceil,&\errwin&=nw_n+\logbudget,\\
 \chi_n&=\lceil n/\logbudget^{3/4}\rceil,&\pad&=n/\sqrt{\logbudget}.
 \end{aligned}
\end{equation}
Round resource budgets upward to integers. These schedules satisfy
$\errwin\ll h_n\ll\chi_n\ll\pad\ll n$ and
$(h_n+\errwin)\log n=o(\chi_n)$.
Extra records and resource allowances use a fixed multiple of
$\pad$, while $\chi_n$ controls SVP repetitions and $nw_n^2$ is
the Gaussian tail exponent.

\subsubsection{Choosing the constants}\label{app:constant-order}
Choose the fixed padding constants in the order: reference list sizes
and depths, moment bounds, coverage, storage, then work and call caps.
Each choice may depend on earlier choices. In particular, the ideal
reference sequence is fixed before the implementation budgets.

\subsubsection{Fixed dimensions}\label{app:finite-dimensions}
For each fixed rank under the smoothing promise, an LLL basis has
$G\le C_n\sigma$ by transference. The GPV proposal can then be
corrected by rejection with probability
$\prod_i\theta_{s_i}(h_i)/\theta_{s_i}(0)$, where
$s_i=\sigma/\|b_i^*\|\ge1/C_n$. This produces the target Gaussian
and has acceptance bounded below by a positive rank-dependent
constant. Standard theta evaluation and \cref{app:scalar-sampling}
give polynomial bit cost, and logarithmically many attempts achieve
any prescribed error. Batch size is constant at fixed rank, and
fixed-dimensional SVP uses \cite{MV13}. Thus the finitely many ranks
excluded by the asymptotic estimates affect only constants.

\subsubsection{Finite representations}\label{app:implementation}
Bases, lattice coordinates, quotient labels, and affine offsets are
stored as exact rational data. Shell supports bound vector norms,
and each difference increases coordinate length by only $O(1)$ bits.
Together with the polynomial basis bounds in \cref{ing:reduction},
this gives polynomial encoding length throughout the forest.

Each record has a node ID and a local index. Leaves use their
trial indices, isolated merges use left child order, and full merges
use lexicographic order of the distinct child ID pairs. Equal vector
values from different pairs remain distinct records. These
choices make the output IDs independent of report order, so complete
coverage gives the same IDs and subsequent pair coins in the ideal
and implemented forests. Batch labels continue to denote positions
in the original trial stream.

For DGS bin $j$, the squared scale $r_j^2=\sigma^2z_j/a_0$ is rational,
as are all later scales obtained from the exact ratio $\gamma^2$.
For SVP, approximate the initial $s^2$ with relative error $O(n^{-2})$
and then use the same exact ratios. Thus the Gaussian identities
remain exact along each sequence of widths. Membership and squared norms use
rational comparisons, while the shell predicates and spherical
search are treated in \cref{app:finite-precision}.

\subsubsection{Numerical and random primitives}
\label{app:numerical-primitives}
Elementary functions and Bernoulli and categorical draws use standard
polynomial bit algorithms. After canceling common factors, positive
weights are evaluated to relative error $\epsilon<1$, giving normalized
TV error at most $\epsilon/(1-\epsilon)$. A known lower bound on total
mass also permits absolute weight error, as used in the radius mixture.

For rational $x\ge1$, write
$\operatorname{clog}_2x=\min\{k\in\Z_{\ge0}:2^k\ge x\}$.
Uniform integers use binary rejection. Primes in $[Y,2Y]$, $Y\ge100$,
use uniform candidates, primality testing~\cite{AKS04}, and the density
bound of~\cite[(3.8)]{RS62}. For at most $N$ requests, fixed polynomial
trial budgets give total failure at most $\delta$, with logarithmic
dependence on $N/\delta$ and polynomial dependence on the input length.

For adaptive calls we use sequential coupling: conditional errors
$\delta_i$ at each common preceding history give complete-execution
error at most $\sum_i\delta_i$. Resource caps and the fallbacks of
\cref{app:computational-model} keep the same budgets on every path.

\subsubsection{Scalar Gaussian sampling}\label{app:scalar-sampling}
Let $D_{S,a,c}$ have mass proportional to $e^{-\pi(k-c)^2/a^2}$
on $S\subseteq\Z$.
For rational $a^2>0$, $c$, and $0<\delta\le1$, bounded interval
rejection samples within TV distance $\delta$ of $D_{\Z,a,c}$ and
$D_{\Z\setminus\{0\},a,0}$. Normalize rejection by a point of maximum
weight. With $T=\Theta(1+\log(1/\delta))$, an interval of radius
$\Theta(\max\{1,a\}T)$ has discarded mass $e^{-\Omega(T^2)}$ and
acceptance probability $\Omega(1/T)$, uniformly in the width.
For the centered nonzero law, omit zero and normalize by the weight
at $1$. A cap of $O(T^2)$ trials and numerical error at most $\delta/2$
give polynomial work, space, and random bit use in the input length
and $\log(1/\delta)$ on every execution.

For GPV, apply the uniform guarantee to the rational squared widths
and centers supplied by Gram--Schmidt. Sequential coupling sums the
scalar errors over the prescribed call bound.

\subsubsection{Weighted selection}\label{app:weighted-implementation}
Standard weighted reservoir sampling implements
\cref{thm:weighted-readout} in one pass with one stored sample:
at each positive-weight record $k$, replace the current sample with probability
$b(X_k)/\sum_{i\le k}b(X_i)$. Return the prescribed fallback if
the total weight is zero. The finite positive-weight computations
use \cref{app:numerical-primitives} and share the execution's
numerical error allowance.

\subsection{Search and the execution coupling}
\subsubsection{Finite precision spherical search}
\label{app:finite-precision}
We prove a finite implementation of the fixed list product code
interface in \cref{ing:bdgl}. Lattice operations and eligibility tests
remain exact, so the implemented shell supports agree with those of
the ideal process throughout the coupling.

Shell predicates reduce to comparisons $A\pi\le B$ with rational
$A,B$. The irrationality measure bound $\mu(\pi)<8$ of \cite{ZZ20}
gives a separation of $2^{-O(L)}$ at input length $L$, so interval
evaluation decides them in polynomial bit time. For the initial
Gaussian proposal, the sufficient
condition $R^2/G^2\ge\operatorname{clog}_2(8n)$ is rational.

Write the ideal code as
$\mathcal C=Q(\mathcal C_1\times\cdots\times\mathcal C_m)$, with
Haar $Q$ and block vectors of norm $m^{-1/2}$.
For nonzero $x$ let
$s_{j,k}(x)=\langle(Q^T\widehat x)_j,c_{j,k}\rangle$.
Choose dyadic upper scores and a dyadic lower threshold satisfying
\[
 s_{j,k}\le\overline s_{j,k}\le s_{j,k}+\eta/m,\qquad
 \tau-\eta\le\underline\tau\le\tau.
\]
Decoding the dyadic scores with threshold $\underline\tau$ gives
\begin{equation}\label{eq:finite-incidence}
 \mathcal I_\tau(x)\subseteq\widetilde{\mathcal I}(x)
 \subseteq\mathcal I_{\tau-2\eta}(x),
\end{equation}
because the total score error is at most $\eta$. The Cartesian
decoder of \cite[Section~5]{BDGL16} and its restricted traversal
\cite[Appendix~D]{GFH26b} apply directly to these additive scores,
with the resource bounds of \cref{ing:bdgl}.

\begin{lemma}[Finite code coupling]
\label{lem:finite-code-coupling}
Consider finitely many calls whose input lists obey prescribed size
limits and are defined independently of their fresh codes. If call $a$
uses $R_a$ codes with $M_a$ implicit
filters and list caps $N_{a,L},N_{a,R}$, put
\[
 U_{\rm cap}=\sum_a R_aM_a(N_{a,L}+N_{a,R}).
\]
For requested failure $\delta$, finite bit code descriptions and
scores can be generated so that all incidence sets on these lists agree
with the ideal ones except with probability at most $\delta$.
Their precision and the cost per explicitly generated scalar are
polynomial in the rational input length, $n$, $\log(1/\delta)$,
and $\log(U_{\rm cap}+2)$.
\end{lemma}
\begin{proof}
For $d\ge3$, a coordinate of a uniform point on $S^{d-1}$ has density
\[
 \frac{\Gamma(d/2)}{\sqrt\pi\Gamma((d-1)/2)}
             (1-t^2)^{(d-3)/2}\le C\sqrt d.
\]
Each fixed product code word has this uniform marginal under the
outer Haar rotation. Therefore a width $2\eta$ boundary band has
probability $O(\sqrt n\,\eta)$. By \eqref{eq:finite-incidence},
a union bound over all potential point-filter incidences costs at most
$C\sqrt n\,U_{\rm cap}\eta$. Choose
\[
 \log_2(1/\eta)\ge\log_2(1/\delta)+\log_2(U_{\rm cap}+2)
                         +C\log_2(n+2)+C.
\]

Generate the rotation by Gaussian QR and the block points by
normalizing Gaussian vectors. Couple their finite approximations
in distance to the continuous code. For $D$ generated scalars,
Gaussian tails and the bound $\Pr[\chi_k<\zeta]=O(\zeta)$ for
$k\ge1$ exclude large coordinates and small normalization
denominators at total cost $\delta/4$. On the remaining event,
standard perturbation bounds show that
\[
 O\bigl(\log(1/\eta)+n\log(nD/\delta)\bigr)
\]
working bits, in addition to input encodings, suffice for score
error $\eta$. Each code uses one fixed description throughout its
traversal. Combining generation and boundary errors proves the claim.
\end{proof}

\paragraph{Finite canonical pair coins.}
Let $U_{\rm pair}=\sum_aN_{a,L}N_{a,R}$ bound potential tagged pairs.
Use a standard pairwise independent family of uniform $p$-bit integers
$J_e$, indexed by the ordered child ID pairs. Its seed has polynomial
length and is drawn after the children are completed, independently
of the search codes. Repeated reports reuse the same $J_e$.
For the coupling, add independent $V_e\sim U[0,1)$ and put
$U_e=(J_e+V_e)/2^p$. This gives the ideal pairwise independent
uniform family. Approximating the activation thresholds to $p$ bits
therefore costs $O(U_{\rm pair}2^{-p})$ over all potential pairs.

\subsubsection{Product code search on lists with repeated values}
\label{app:search-interface}
Write $C_{d_{\rm pc}}$ and $W_{d_{\rm pc}}$ as in \cref{sec:search}.
The source construction uses zero padding to dimension
$d_{\rm pc}=n+O(\log n)=m_{\rm pc}b_{\rm pc}$, where
$m_{\rm pc}=\lceil\log_2d_{\rm pc}\rceil$.
An integer block code size $B_{\rm pc}$ gives $M_f=B_{\rm pc}^{m_{\rm pc}}$
filters. The fixed list work, space, and coverage interface is
\cref{ing:bdgl}. We extend it to records with repeated values and explain
how complete coverage reconstructs the merges used in the dynamics.

Filter membership depends only on vector values, so the incidence
and resource bounds of \cref{ing:bdgl} count records and pairs with
their multiplicities. Each eligible tagged pair with unequal values
has the coverage probability of its values. Thus the same search
interface applies to record lists.

For target correlation $u_0$, choose the least integer
$B_{\rm pc}\ge2$ with
$m_{\rm pc}\log_2B_{\rm pc}\ge K_\tau(u_0)n+C_{\rm code}\pad$.
This gives \eqref{eq:filter-counts}.
The padding dominates the coverage threshold and shell perturbations.
All eligible correlations $u_0+O(w_n)$ have uniform active constraint
margins. In particular, one code misses any fixed eligible record
edge with probability at most $1/2$ for large $n$.

\subsubsection{Canonical pairs and complete coverage}\label{app:search-canonical}
A record has a node ID and a local integer index as in
\cref{app:implementation}. At a cooling node, conditional on the
completed children and their IDs, draw a fresh seed for a pairwise independent activation family indexed
by ordered child ID pairs. This seed is independent of all earlier nodes
and search codes. Repeated reports reuse the same coin.
Distinct coins have the pairwise independence required by the dynamics,
and the activated graph is independent of the search codes.
A compact seed recovers the same decision on every repeated report
without storing the rejected pairs. Resampling on each report would
make the activation probability depend on how often search finds the pair.

A cooling call records each endpoint's unique activated neighbor,
or marks it as having several distinct neighbors. Repeated reports
leave these states unchanged. A final scan in left ID order emits
the mutually unique pairs, using $O(N_L+N_R)$ space and work beyond
the reports. Complete coverage is essential: it recovers every
competitor and makes mutual uniqueness agree with isolation in the
ideal graph.

A full call deduplicates eligible pair IDs in a dictionary and emits
their differences in lexicographic ID order. The dictionary uses
space proportional to the output count. Distinct pairs with equal
output values remain distinct records.

Integer block sizes change $\log_2M_f$ by only $O(\log n)$,
which is absorbed by the padding. The active constraints, including
the unequal-threshold case of \cite[Appendix~D]{GFH26b}, hold
uniformly over the parameter ranges used here.

Run $\covrep=\lceil C_{\rm cov}n\rceil$ independent codes per intermediate
call with $C_{\rm cov}=10$. All codes share the canonical state and,
at a cooling node, the pair coin seed. Their implicit descriptors are generated and traversed
sequentially, using $\poly(n)B_{\rm pc}$ auxiliary space.
The miss probability of a fixed eligible activated edge is at most
$2^{-C_{\rm cov}n}$. Charge code generation, state updates, dictionaries,
and the final endpoint traversal to the operation cap. At the SVP root, deduplicate
values and stream reports without a pair dictionary. If a target exists,
choose one canonical witness from the ideal lists before drawing top codes.

\begin{lemma}\label{lem:first-failure}
Let an ideal forest on a prescribed finite tree and an event $\mathcal G$
be independent of all search codes. Couple it to an implementation with
list limits $N_v$, work caps $W_a^{\rm cap}$, and $R_a$ codes at call $a$.
Conditional on ideal inputs, shared pair coin seeds, and the environment
independent of the codes, these codes are independent with their original
law. For each fixed input within its list size limits, suppose one code
misses a fixed eligible edge with probability at most $q_a<1$.

Let $\epsilon_{\rm num}$ bound the additional finite arithmetic and
random bit coupling loss on histories within the resource limits, with
zero in the exact arithmetic case. Let $\Xi_v$ be the ideal lists and
$W_a$ the work of a complete search
on ideal inputs, set to zero when an input exceeds its limit. Work includes
code generation, the report and endpoint scans, canonicalization, state updates, and
dictionary operations. Then
\begingroup
\interdisplaylinepenalty=10000
\begin{align}\label{eq:general-coupling}
 \Prb[\mathcal G\text{ and disagreement}]
 \le{}&\epsilon_{\rm leaf}+\epsilon_{\rm num}
 +\sum_v\frac{\E[\mathbf1_{\mathcal G}|\Xi_v|]}{N_v}
 +\sum_a\frac{\E[\mathbf1_{\mathcal G}W_a]}{W_a^{\rm cap}}\notag\\
 &+\sum_{a\ {\rm internal}}N_{a,L}N_{a,R}q_a^{R_a}
 +\epsilon_{\rm top}.
\end{align}
\endgroup
Here disagreement means a mismatch of the reconstructed internal lists
or sampling outputs, or failure to report the chosen terminal witness
when it exists.
The first error bounds leaf coupling.
The numerical allowance includes code generation, boundary incidences,
finite coins, and readout for the word RAM implementation, as in
\cref{prop:word-ram-transfer}.
For a terminal witness fixed from the ideal lists before the top codes,
take $\epsilon_{\rm top}=q_{\rm top}^{R_{\rm top}}$, and include missing
that witness in disagreement. Otherwise set $\epsilon_{\rm top}=0$.
For at most $N_{\rm raw}$ adaptive raw calls of uniform statistical error
$\epsilon_{\rm raw}$, one may use
$\epsilon_{\rm leaf}=N_{\rm raw}\epsilon_{\rm raw}$.
\end{lemma}
\begin{proof}
Share leaves, shifts, and canonical pair coins, and expose calls from the
leaves upward. Until the first disagreement, the inputs coincide.
For inputs within the list size limits, a union bound gives incomplete coverage probability
$N_{a,L}N_{a,R}q_a^{R_a}$. Complete coverage reconstructs the competitor
graph and hence the ideal output. At the top, apply the bound to the
canonical witness alone.

Markov's inequality on ideal list sizes and complete search work,
with the indicator of $\mathcal G$, gives the two expectation sums.
Sequential coupling bounds the adaptive raw calls uniformly over their
histories. All moments are taken on the
ideal process before conditioning on coverage, resource success, or
existence of a terminal witness.
\end{proof}

\subsubsection{Proof of the finite realization interface}
\label{app:realization-proof}
The key comparison is with a forest defined independently of every
search code. Complete intermediate coverage recovers its activated
graphs, while terminal SVP needs only one fixed target witness.
By \cref{lem:first-failure} and the budgets in
\cref{app:application-guards}, search and resource failures cost
\[
 \delta_{\rm search}+\delta_{\rm resource}
 \le2^{-\Omega(n)}+2^{-\Omega(\pad)}
\]
per core execution. The single sample budgets are given in
\cref{app:single-execution}. For SVP, \cref{sec:svp-implementation}
makes this loss $o(2^{-B_n})$, where $2^{-B_n}$ is the ideal
success probability.

For the finite arithmetic, let $N$ bound all scalar decisions and
let $U_{\rm cap},U_{\rm pair}$ bound potential incidences and pairs.
Their logarithms are $O(n)+o(n)+O(\log(\mathcal B+2))$ per core
execution, with the logarithmic repetition count added for amplification.
Allocate total error $2^{-\lambda}$ over these decisions using
\cref{app:numerical-primitives,lem:finite-code-coupling}.
The positive-weight formulas in \cref{app:batch-numerics} and the
scalar sampler require polynomial precision in
$\mathcal B,n,\lambda,\log N$.
The stable record IDs then allow sequential coupling of the
record lists, pair decisions, and sampling decisions over the
forest. For sampling, this gives
\[
 \Prb[\mathcal G\text{ and }X_{\rm finite}\ne X_{\rm ideal}]
 \le\delta_{\rm search}+\delta_{\rm resource}+2^{-\lambda}.
\]
Here $\mathcal G$ is the application's good environment event.
For SVP, on $\mathcal G\cap E_v$ choose one representation of the
fixed shortest vector $v$ before drawing the terminal codes. Unless
one of the charged failures occurs, the implementation reports this
witness and hence finds a shortest vector. This gives the SVP
success bound in \cref{prop:finite-realization}.
Polynomial arithmetic costs and the decoder's
$\poly(n)B_{\rm pc}$ workspace preserve the leading exponents.

Take $\lambda=Cn$ for a single sample or an entire Gaussian batch.
Combining the implementation loss with the analytic error bounds
proves the corresponding claims of \cref{prop:finite-realization}.
For SVP, first subtract the per-trial loss from $2^{-B_n}$ and
amplify as in \cref{sec:svp-completion}, then couple the complete
execution with $\lambda=Cn\log_2(n+2)$.
The centered reduction uses the query and accuracy bounds of
\cref{cdgs:implementation}.

\subsection{Application budgets}
\label{app:application-guards}
The applications use one common cap rule. For an expected list size
or work bound $2^{rn+O(\pad)}$, choose a cap $2^{rn+C\pad}$
with a larger fixed $C$, following \cref{app:constant-order}.
Use the finite rates in these caps, including any $O(n/k_n)$
correction, before adding the padding. The table gives the rates
and storage needed by each construction.

\begin{center}
\small
\begin{tabularx}{\textwidth}{@{}lcl>{\raggedright\arraybackslash}X@{}}
\toprule
Construction & Leaf trials & Live and dictionary rates & Search rates\\
\midrule
Single-sample cooling & $2M_{\rm leaf}$ & $P(u_n)$ & $K(u_n)$\\
SVP cooling and full layers & $2M_{\rm leaf}$ &
$\max_jL_j<s_{\rm live}$ &
$T_{\rm s},K_1,\ldots,K_5$\\
Ordinary batch & $M$ &
$P(u),\,2P(u)-R(u)$ &
$K(u),K_{\rm rev}(u)$\\
Terminal batch & $M$ &
$\dgsspace,\,q_T+\dgsspace-R_T$ &
$T_F$ and the reverse rates of \cref{app:batch-terminal}\\
\bottomrule
\end{tabularx}
\end{center}
The SVP quantities are defined in \cref{sec:svp}, ordinary batch
rates are in \cref{sec:batch-dgs}, and terminal rates in
\cref{app:batch-terminal}. Full-layer dictionaries contain output
records. Batch dictionaries contain true eligible pairs.
Single-sample rates are defined in \cref{sec:single-dgs}. Its cooling
tree and final weighted scan use the separate schedule and coupling
in \cref{app:single-execution}.

For a prescribed forest, the number of raw trials is at most its
number of leaves times the largest trial count in the table.
The number of leaves is $2^{O(t)}$ per bin, with polynomially many
bins and $t=O(\log n\log\log n)$. Include all SVP repetitions and
scale branches in the global call bound $N_{\rm trial}$.
Each trial uses $n$ scalar draws and one acceptance coin.
Assign their total error from the requested whole-execution allowance.

Evaluate forests depth first, streaming leaf proposals and candidate
reports. Retain only the active lists, endpoint states, dictionaries,
and search workspace described in the table.
Use $\lceil10n\rceil$ independent codes per merge, processed sequentially.
The one-code miss probability is at most $1/2$ on inputs within the prescribed list limits,
and all list rates are below $1/2$. A union over the potential pairs
and forest nodes therefore gives probability of incomplete coverage
$2^{-\Omega(n)}$. For isolation, the final endpoint traversal
of \cref{app:search-canonical} recovers the output after the report scan.

Markov's inequality on ideal list sizes and complete search work
gives resource loss $2^{-\Omega(\pad)}$ over a core forest.
The coupling in \cref{lem:first-failure} combines this loss with
coverage and finite arithmetic. Failures use the returns specified in
\cref{app:computational-model}. Subsequent SVP trials retain their
fresh randomness and fixed budgets.

\subsubsection{Cooling schedules and their implementation}
\label{app:cooling-initialization}
\label{sec:implemented-cooling}
Apply \cref{ing:reduction} with $k=\lfloor n/10\rfloor$ to obtain
a basis with $G\le\poly(n)s$.
Choose the first scheduled width $R=r_0\gamma^{-t}$ satisfying
\cref{lem:gaussian-proposal}. The depth is $O(\log n)$ for fixed
$u$, and is bounded by \cref{lem:endpoint-schedule} otherwise.
Enforce the depth cap before constructing a forest.

Implement \cref{alg:ideal-cooling-tree,alg:full-difference-tree}
using the report scan and endpoint traversal of
\cref{app:search-canonical} for isolation, and its ordered pair
dictionary for full merges. These procedures preserve the ideal
local IDs on complete coverage and use the common call budgets of
\cref{app:application-guards}.

\subsubsection{Single-sample schedules and execution}
\label{app:single-execution}
\paragraph{Shell width and sublinear budgets.}\label{app:single-schedules}
Fix the rational $0<\delta\le\min\{1,\xi\}$ used in
\cref{sec:single-dgs}, and set $\logbudget=\lceil\log_2(n+2)\rceil$.
For this algorithm use
\begin{equation}\label{eq:single-budgets}
 w_n=\frac1{\lceil\logbudget^{1+\delta/2}\rceil},\qquad
 h_n=\left\lceil\frac n{\logbudget^{1+\delta/3}}\right\rceil,
 \qquad
 \pad=\left\lceil\frac n{\logbudget^{\delta/6}}\right\rceil,
 \qquad \errwin=nw_n+\logbudget.
\end{equation}
For $t=O(\logbudget\log\logbudget)$ and
$\degpad=\lceil10\errwin\rceil$, these choices give
\begin{equation}\label{eq:single-budget-relations}
 \errwin=o(h_n),\quad t(h_n+\degpad+2)=o(\pad),\quad
 t=o(nw_n^2),\quad nw_n^2=o(h_n),\quad
 \frac n{\log n}=o(\pad),\quad \pad=o(n).
\end{equation}
Also $k_nw_n\to0$, so the endpoint contraction and shell margins
hold uniformly. The Gaussian estimates use only shrinking shells,
diverging $nw_n^2$, and the fixed smoothing margins of the relevant
tilts. Their logarithmic shell losses are $O(\errwin)$ for these
widths as well. The cooling error is therefore
$e^{-\Omega(nw_n^2)}=\exp[-\Omega(n/\log^{2+\delta}n)]$.

\paragraph{Initialization and expected counts.}
Choose the prime and extension by \cref{lem:dgs-index-wrapper}, using
the bounded prime procedure of \cref{app:numerical-primitives}, with
total prime and uniform-integer failure allowance $2^{-Cn}$.
On a good extension, basis preprocessing with block size
$\lfloor n/10\rfloor$ gives $G\le\poly(n)s$ within rate $1/5$.
Choose a deterministic depth $t=O(k_n\log n)$ large enough for this
bound and check the resulting starting width before generating records.
The good-extension bound ensures that the check succeeds there.
Give every leaf $2M_{\rm leaf}$ corrected trials with
$\varrho=m(u_n)$ in \eqref{eq:raw-leaves}.
The total number of leaves is $2^t$, and
\eqref{eq:single-budget-relations} makes the root coefficient
$2^{m(u_n)n+\Theta(\pad)}$ as required in
\eqref{eq:single-effective-rate}.

\paragraph{Work, space, and coupling.}
Applying the common cap rule of \cref{app:application-guards},
choose at each cooling node
\[
 N_{\rm cap}=\left\lceil2^{P(u_n)n+C_{\rm list}\pad}\right\rceil,
 \qquad
 W_{\rm cap}=\left\lceil2^{K(u_n)n+C_{\rm work}\pad}\right\rceil.
\]
The list first moment and \eqref{eq:dgs-work-bound-new}, averaged
over shifts on each good extension, supply the expectations in
\cref{lem:first-failure}. The common schedules of
\cref{app:application-guards} give resource and coverage losses
$2^{-\Omega(\pad)}$ and $2^{-\Omega(n)}$.
Lists and endpoint states use rate $P(u_n)$, and code descriptors
fit because $\log_2B_{\rm pc}=O(n/\log n)=o(\pad)$.
Keep the exact finite rates $P(u_n)$, $m(u_n)$, and $K(u_n)$
in every cap before adding padding.

Finish with the reservoir scan of \cref{app:weighted-implementation}
using \eqref{eq:dgs-weight-new}.
On the retained shell, every positive weight lies in
$[e^{-O(nw_n)},1]$, so the relative weight computations of
\cref{app:numerical-primitives} use polynomial precision.
There are $2^{O(t)}2M_{\rm leaf}$ leaf proposals, giving logarithmic
decision count $O(n)+o(n)$ and total numerical error $2^{-Cn}$
under the common finite realization.

\subsubsection{Finite batch execution}\label{app:batch-execution}
Implement \cref{alg:fixed-bank-trial,alg:gaussian-batch} with the
forward and reverse searches of \cref{sec:batch-dgs}, using the
finite decoder of \cref{app:finite-precision}.
Define the complete ideal forward proposal list before drawing
the independent reverse codes.

Dictionaries store true eligible pairs keyed by record IDs,
counting each pair once and using ID order for accumulation and
partner selection. Each reverse query includes its selected
forward partner, which has positive weight. Thus its denominator
is positive even if search does not report that partner again.
The forward and reverse dictionary rates are $P(u)$ and
$2P(u)-R(u)$, with the smaller terminal rates in
\cref{app:batch-terminal}. Store trial streams as sparse lists of
live records and labels, so scanning all $M$ labels takes $O(M)$
work without an array of $M$ failure symbols.

With the common budgets of \cref{app:application-guards},
\cref{app:realization-proof} couples the entire batch with numerical
error $2^{-Cn}$. This is the implementation allowance used in the
joint error bound of \cref{sec:batch-dgs}.

\subsubsection{Evaluation of the batch weights}
\label{app:batch-numerics}
The leaf correction uses the Poisson series for $\theta_s(h)/s$,
whose widths have a fixed positive lower bound.
The forward weights $K$ lie in $[e^{-O(nw_n)},1]$ when nonzero.
For reverse sums, cancel the factors $\lambda/V_r$ and
$\rho_{\gamma r}(z)$ before evaluation:
\[
 \frac{B_C(z)}{(\lambda/V_r)\rho_{\gamma r}(z)}
 =\sum_{i:\,\mathrm{eligible}}
 e^{A_rR_-^2-2\pi\|y_i+z/2\|^2/r^2+\pi\|y_i\|^2/r^2}.
\]
Every nonzero summand lies in $e^{\pm O(nw_n)}$, and the selected
forward partner supplies a positive term. The same cancellation
applies to $b_*\rho_{\gamma r}/B_C$, so relative approximation of
positive sums suffices for both reverse operations.

For bin selection, put $\vartheta_j=\kappa_jV_\sigma(L)$,
using each bin's own determinant in $\kappa_j$.
The common volume cancels, leaving weights $\vartheta_j^{-1}$.
Since $\log\vartheta_j=O(n)+o(n)$, these weights also admit relative
approximation with polynomial precision. The terminal branch uses
the same cancellation and positive sums.
All these decisions share the complete execution allowance in
\cref{app:realization-proof}.

\subsubsection{Finite shortest vector execution}\label{app:svp-execution}
We implement the fixed trial schedule of \cref{alg:svp}.
On each scale branch use the bounded initialization of
\cref{app:svp-environment}. On the correct branch, prime search
exhausts its budget with probability at most $2^{-\Omega(n)}$.
Conditional on successful prime selection, the starting width and
depth checks pass on the good smoothing event $\mathcal G$.

\paragraph{Resource loss compared with target success.}\label{sec:svp-implementation}

The implementation loss must be smaller than the ideal target
probability $2^{-B_n}$. Use the list rates and storage of
\eqref{eq:svp-live-space} with the common caps in
\cref{app:application-guards}. The ideal list and search bounds in
\cref{sec:svp-work}, averaged over the common coset on good
superlattices, give the per-trial Markov bound
\begin{equation}\label{eq:svp-cap-failure}
 \Prb[\mathcal G\text{ and some list or work cap fails}]
       \le2^{-\Omega(\pad)}.
\end{equation}
The per-trial coverage and terminal witness losses are $2^{-\Omega(n)}$.

Preprocessing uses a $2^{0.2n+O(\pad)}$ cap.
Use the deterministic raw call budget
\[
 N_{\rm call}\le\poly(n)2^{\chi_n}2^{m_{\rm raw}n+C_N\pad}
\]
and error $2^{-3n}/N_{\rm call}$ per corrected trial. Sequential coupling
changes the complete transcript by at most $2^{-3n}$.
Consider first trials with ideal corrected leaves. Their resource
and coverage losses on good environments sum to
\[
 \delta_{\rm trial}\le2^{-\Omega(\pad)}+2^{-\Omega(n)}
 =o(2^{-B_n}).
\]
Subtracting this loss leaves success at least $2^{-B_n-1}$ per
trial, uniformly on the correct branch. Independent repetition
therefore amplifies success before the Gaussian call coupling adds
its $2^{-3n}$ error over the complete execution.
\paragraph{Complete execution.}\label{app:svp-amplification}
The uniform trial guarantee permits the amplification in
\cref{sec:svp-completion}. Prime exhaustion and corrected-leaf
coupling cost $2^{-\Omega(n)}+2^{-3n}$ per complete execution,
leaving success at least $2/3$ for large $n$.
After the final polynomial repetitions, finite realization with
$\lambda=Cn\log_2(n+2)$ preserves failure
$2^{-\Omega(n\log n)}$ and both resource exponents.

\section{Centered DGS by a mixture over radii}
\label{app:cdgs-oracle}

We follow the primitive direction sparsification approach of
Stephens-Davidowitz~\cite[Section~4]{SD16}. A multiplicative radius
grid controls the change in Gaussian direction weight throughout the
truncated range. Treating layers with few primitive directions separately
allows the same mixture to work at every width.

We first establish the direction weight comparison, then use the
sparsification facts of~\cite{SD16} to construct the radius mixture.
The proof separates the multiplicative comparison of direction
weights from the additive errors due to truncation and failed draws.
These steps prove \cref{thm:cdgs-reduction}. We then replace its
exact SVP oracle by \cref{thm:svp-main} to obtain \cref{cor:cdgs-main}.
All polynomial bounds include
the numerical accuracy parameters and depend on
$\sigma/\lambda_1(L)$ through its logarithm and the binary input
length. Prime and uniform integer draws are first treated as ideal
primitives, with their finite implementation and numerical allowances
given in \cref{app:cdgs-numerics,cdgs:implementation}.

\subsection{Gaussian direction weights and the radius grid}
\label{cdgs:grid}

Let $\mathcal P(L)$ contain one representative of each primitive
direction modulo sign. For $r>0$, put
\[
 h_\sigma(r)=2\sum_{k\ge1}e^{-\pi k^2r^2/\sigma^2}.
\]
Every nonzero vector is uniquely $zy$, with $y\in\mathcal P(L)$ and
$z\in\Z\setminus\{0\}$. Therefore
\begin{equation}\label{eq:cdgs-direction-mass}
 \rho_\sigma(L)=1+\sum_{y\in\mathcal P(L)}h_\sigma(\|y\|).
\end{equation}
After selecting a direction with this weight, the conditional
coefficient law is $D_{\Z\setminus\{0\},\sigma/\|y\|,0}$.

\paragraph{Why the grid must control relative changes.}
\label{cdgs:sd16-grid}
The mixture uses samplers for directions inside successive balls.
For exact counts and uniform sampling, weighting the sampler at $r_i$
by its count times $h_\sigma(r_i)-h_\sigma(r_{i+1})$ cancels the
count. Give the last layer its residual weight $h_\sigma(r_{\rm last})$.
A direction first included at $r_j$ then receives total weight
$h_\sigma(r_j)$ by telescoping. Thus the grid must make
$h_\sigma(r_j)$ close to the desired $h_\sigma(\|y\|)$
multiplicatively.

Normalize $\sigma=1$ and write $h=h_1$. The proof of
\cite[Theorem~4.6]{SD16} uses
$r_i^2=\lambda_1(L)^2+i\Delta$, where $\Delta=1/(100nf)$, and claims
\begin{equation}\label{eq:cdgs-sd16-comparison}
 h(r_i)\le h(\|y\|)\le(1+1/f)^{1/10}h(r_i)
 \quad(r_{i-1}\le\|y\|\le r_i).
\end{equation}
The difficulty is the behavior near zero. Poisson summation gives
$h(r)=r^{-1}(1+h(1/r))-1$, so $h(r)\sim r^{-1}$ as $r\downarrow0$.
An additive step in $r^2$ can therefore cause an arbitrarily large
relative change in direction weight. For example, if
$\lambda_1(L)=a$ and a primitive direction has length $2a$, then
for $a^2\ll\Delta$ it first appears at
$r_1=\sqrt{a^2+\Delta}$, while
$h(2a)/h(r_1)\to\infty$ as $a\to0$.
The required comparison concerns the mass of all nonzero multiples
of the direction. Short directions have many multiples with
appreciable Gaussian weight, which produces this reciprocal growth.

\paragraph{A numerical counterexample.}
Take $\sigma=1$, $n=f=20$, and
\[
 a=2^{-20},\qquad
 L=\diag(a,2a,32,\ldots,32)\Z^{20},\qquad y=2ae_2.
\]
Here $\lambda_1(L)=a$ and $y$ is primitive, with
$r_0=a<\|y\|=2a<r_1=\sqrt{a^2+1/40000}$.
Poisson summation and comparison with the Gaussian integral give
$r^{-1}-1\le h(r)\le r^{-1}$, hence
\[
 h(\|y\|)\ge2^{19}-1=524287,
 \qquad
 (1+1/f)^{1/10}h(r_1)
 <(21/20)^{1/10}\cdot200<201.
\]
Thus \eqref{eq:cdgs-sd16-comparison} fails by more than a factor
of $2600$, even for this rational diagonal lattice.
Even with exact counts and uniform direction sampling, the mixture
assigns this direction the telescoped weight $h(r_1)$.
The missing control therefore lies in the radius discretization.

\paragraph{The multiplicative grid.}
We use multiplicative steps in squared radius instead. They control
the ratio of adjacent radii near zero, while a step size chosen for
the largest radius also controls the Gaussian decay at the other
end of the range. \Cref{cdgs:grid-bound} proves the resulting weight
comparison. Sparsification then supplies the counts and direction
samplers, and a thin interval estimate bounds the contribution of
layers with too few directions. After selecting a direction, a scalar
Gaussian chooses its nonzero multiple.

Fix
\begin{equation}\label{eq:cdgs-parameters}
 \begin{gathered}
 F=\max\{f,2\},\qquad \varepsilon=2^{-f},\qquad
 \eta=\frac1{1000F},\qquad \kappa=\varepsilon/4,\\
 Q=n+F+20,\qquad G=\lceil10^4FQ\rceil,\qquad
 T=\sigma\sqrt Q,\qquad R_{\rm g}=\frac{T}{\sqrt{1-G^{-2}}}.
 \end{gathered}
\end{equation}
The radius $T$ separates the region with a multiplicative guarantee
from the Gaussian tail. The grid extends to $R_{\rm g}$, and its
entire range is available to the sampler.

The centered case of \cite[Lemma~2.5]{SD16}, at radius
$T=\sigma\sqrt Q$, gives
\begin{equation}\label{cdgs:tail}
 \Pr_{X\sim D_{L,\sigma}}[\|X\|\ge T]
 <(2\pi eQ/n)^{n/2}e^{-\pi Q}
 <e^{-Q}<\varepsilon/64.
\end{equation}
This also bounds the mass of points whose primitive direction is
longer than $T$.

For $n=1$ use \cref{app:scalar-sampling} on the input generator,
with error $\kappa/8$. Suppose henceforth $n\ge2$.
One SVP call gives a shortest direction $y_0$ and the rational
number $q_0=\|y_0\|^2/\sigma^2$. If $q_0\ge Q$, output zero.
Equation~\eqref{cdgs:tail} proves the required guarantee for this branch.
Otherwise define $D=16000FQ$ and
\begin{equation}\label{eq:cdgs-grid}
 q_{i+1}=\min\{(1+1/D)q_i,q_{\rm g}\},
 \qquad q_{\rm g}=\frac{Q}{1-G^{-2}}.
\end{equation}
Stop at the first $J$ with $q_J=q_{\rm g}$, and write
$r_i=\sigma\sqrt{q_i}$. The implementation stores only rational
squared radii. All length comparisons are squared comparisons.

\begin{lemma}[Adjacent direction weights]\label{cdgs:grid-bound}
The grid is strictly increasing and satisfies
\[
 1\le\frac{h_\sigma(r_i)}{h_\sigma(r_{i+1})}\le1+\eta,\qquad
 J\le1+(D+1)\log(q_{\rm g}/q_0).
\]
Its points have polynomial encoding length in the input and $f$.
\end{lemma}
\begin{proof}
For $H_0(t)=2\sum_{k\ge1}e^{-\pi k^2t^2}$,
\begin{equation}\label{cdgs:log-derivative}
 M(t):=-\frac{d\log H_0(t)}{d\log t}
 =2\pi t^2
 \frac{\sum_{k\ge1}k^2e^{-\pi k^2t^2}}
      {\sum_{k\ge1}e^{-\pi k^2t^2}}
 \le13\max\{1,t^2\}.
\end{equation}
Indeed, for $t\le1/2$, Poisson summation for $1+H_0$ gives
$-tH_0'(t)\le1+H_0(t)$ and $H_0(t)\ge t^{-1}-1\ge1$.
For $t\ge1/2$, divide the numerator by its first term and use
$k^2\le4^{k-1}$ and $k^2-1\ge3(k-1)$.
The resulting geometric series is at most
$(1-4e^{-3\pi/4})^{-1}<2$, proving the bound.
Throughout the grid, $q\le q_{\rm g}\le4Q/3$ and $q_{\rm g}>1$.
Integrating the derivative across any step, including the truncated
last step, gives
\[
 \log\frac{h_\sigma(r_i)}{h_\sigma(r_{i+1})}
 \le\frac{13q_{\rm g}}2\log(1+1/D)
 \le\frac{13}{24000F}<\log(1+\eta).
\]
The step count and polynomial encoding length follow directly
from the geometric progression \eqref{eq:cdgs-grid}.
\end{proof}

\subsection{Primitive directions and sparsification}
\label{cdgs:primitive-tools}

We use the primitive direction sparsification method of
Stephens-Davidowitz~\cite[Sections~2.6 and~4]{SD16}.
The statements below give the counting and sampling interfaces needed
for the radius mixture, including their behavior at equal lengths and
under arbitrary causal choices among shortest vectors.
For a basis $B$ of $L$, put
\[
 \mathcal P(L)=L_{\rm prim}/\{\pm1\},\qquad
 \mathcal Y_r=\{y\in\mathcal P(L):\|y\|\le r\},\qquad
 C_L(r)=|\mathcal Y_r|.
\]
Choose one canonical representative of each direction. Taking the
$L$-primitive core of a nonzero vector means dividing its coordinates
in $B$ by their greatest common divisor, and then choosing this sign.
It preserves the real direction and cannot increase length.

We use three sparsification facts from \cite{SD16}. Its counting
function $\xi(L,r)$ is exactly $C_L(r)$, with closed balls and
opposite vectors identified. First, \cite[Lemma~2.18]{SD16} gives
the following collision criterion: if $p\ge101$ is prime and
distinct $x,y\in\mathcal Y_r$ satisfy
$B^{-1}x\equiv aB^{-1}y\pmod p$ for some
$a\in\mathbb F_p^\times$, then
\begin{equation}\label{cdgs:collision}
 C_L(r)>\frac{p}{20\log p}.
\end{equation}

For uniform $z\in\mathbb F_p^n$, define the full rank sublattice
\begin{equation}\label{cdgs:kernel-lattice}
 L_z=\{Bx:x\in\Z^n,\ \langle z,x\rangle=0\pmod p\}.
\end{equation}
For $z=0$ use $B$. Otherwise choose $j$ with $z_j\ne0$, let
$c_k\equiv-z_kz_j^{-1}\pmod p$ with $0\le c_k<p$, and use
\begin{equation}\label{cdgs:kernel-basis}
 b_k+c_kb_j\quad(k\ne j),\qquad pb_j.
\end{equation}
These columns generate the kernel, with determinant $p$ in
$B$-coordinates. Constructing each query afresh from $B$ gives encoding
length polynomial in $\operatorname{size}(B)$ and $\log p$.

\begin{lemma}\label{cdgs:isolation}
Suppose $r\ge\lambda_1(L)$, $p\ge101$ is prime, and, with
$C=C_L(r)$,
\begin{equation}\label{cdgs:isolation-conditions}
 p\lambda_1(L)>r,\qquad C\le\frac{p}{20\log p}.
\end{equation}
For every SVP oracle, including any choice among tied shortest
vectors, the one trial probability $a_y$ of returning direction
$y\in\mathcal Y_r$ satisfies
\begin{equation}\label{cdgs:one-direction}
 \frac1p-\frac{C-1}{p^2}\le a_y\le\frac1p.
\end{equation}
Consequently,
\begin{equation}\label{cdgs:hit-probability}
 \frac Cp-\frac{C(C-1)}{p^2}
 \le\Pr[\lambda_1(L_z)\le r]\le\frac Cp.
\end{equation}
\end{lemma}
\begin{proof}
By \cite[Theorem~4.1 and Proposition~4.2]{SD16}, the event that $y$
survives and all competing directions of length at most $\|y\|$
do not survive has probability at least $1/p-(C-1)/p^2$.
The condition $p\lambda_1(L)>r$ excludes a short nonprimitive
multiple whose primitive core does not survive. Hence this event
forces every correct oracle to return $\pm y$, regardless of ties
or previous replies. The upper bound is the survival probability
$1/p$, and summing the direction bounds gives
\eqref{cdgs:hit-probability}. Fresh $z$ makes the same argument
valid conditional on any prior oracle history.
\end{proof}

\subsection{A counting and sampling interface}
\label{cdgs:counting}

At each radius we need an approximate number of primitive directions
and a procedure that samples them nearly uniformly. The next lemma
either supplies these two objects or identifies a radius with few
directions, whose contribution is treated separately in the mixture.

Loop bounds use the integer logarithm $\operatorname{clog}_2$
defined in \cref{app:numerical-primitives}. As above, prime and
integer draws are analytic until the finite implementation in
\cref{cdgs:implementation}.

\begin{lemma}[Counting primitive directions]
\label{cdgs:count-interface}
Given $L$, a positive rational squared radius $r^2$,
rational $0<\eta\le1/10$, integer $G\ge2$, and
$0<\zeta_{\rm pre}<1/4$, an analytic algorithm using SVP returns a
tagged result $\mathsf{GOOD}$ or $\mathsf{DEGENERATE}$.
It makes a deterministically bounded polynomial number of queries.
There is an event $E_{\rm pre}(r)$ of probability at least
$1-\zeta_{\rm pre}$, depending only on preprocessing, such that
on every transcript in this event:
\begin{enumerate}[label=(\roman*)]
 \item A $\mathsf{GOOD}$ result supplies $N>0$, an integer scale $M$,
 a prime $p$, an integer $A$, and a sampling procedure, with
 \[
  (1-\eta)C_L(r)\le N\le(1+\eta)C_L(r),\qquad
  C_L(r)\le M,\quad AM\le p\le2AM.
 \]
 For any subsequent history and any $0<\zeta_{\rm s}<1/4$,
 the procedure makes at most
 \begin{equation}\label{cdgs:direction-cap}
  R_{\rm s}=7A\operatorname{clog}_2(1/\zeta_{\rm s})
 \end{equation}
 queries and returns either a direction in $\mathcal Y_r$ or
 $\mathsf{TIMEOUT}$. The timeout probability is at most
 $\zeta_{\rm s}$. Conditional on success, each direction has
 probability in $[(1-\eta)/C_L(r),(1+\eta)/C_L(r)]$.
 \item A $\mathsf{DEGENERATE}$ result guarantees
 \begin{equation}\label{cdgs:degenerate-condition}
  C_L(r)<\frac{r}{G\lambda_1(L)}.
 \end{equation}
\end{enumerate}
The probability bounds hold for every rule for choosing a shortest
vector based on the current and previous queries. The query caps hold
on every execution.
\end{lemma}

\begin{proof}[Construction and counting guarantee]
An initial SVP query gives $s_1=\lambda_1(L)^2$, if it is not
already available. If $r^2<s_1$, return $\mathsf{DEGENERATE}$.
Otherwise, using rational squared comparisons, compute
\[
 R=\left\lceil\sqrt{r^2/s_1}\right\rceil,\quad
 m=\max\left\{1,\left\lceil\sqrt{r^2/(G^2s_1)}\right\rceil\right\},
 \quad U=(1+2R)^n.
\]
Packing gives $C:=C_L(r)\le U$. Put
\begin{equation}\label{cdgs:count-parameters}
 \begin{split}
 V_0&=\max\{U,m,2\},\qquad
 S=101+G+\eta^{-1}+\operatorname{clog}_2V_0,\\
 A&=\lceil10^4S\rceil,\qquad
 k=\min\{j\ge0:m2^j\ge U\},\qquad K=k+1,\\
 e&=\frac{\eta}{1000A},\qquad
 s=\left\lceil\frac{\operatorname{clog}_2(2K/\zeta_{\rm pre})}
 {2e^2}\right\rceil.
 \end{split}
\end{equation}
Scan $M=m2^k,m2^{k-1},\ldots,m$ in this order.
At each scale generate a prime $p\in[AM,2AM]$ by independent integer
trials with deterministic primality testing. The prime-density estimate
above supplies the geometric tail used by the finite budget. For this fixed $p$, make
$s$ independent kernel lattice queries, recording the indicators
$X=\ind[\lambda_1(L_z)\le r]$. In this analytic interface, coordinates
are exactly uniform in $\mathbb F_p$. Their implemented draws use
\cref{app:numerical-primitives}.
Let $\widehat q$ be their average.
At the first scale satisfying
\begin{equation}\label{cdgs:count-trigger}
 \widehat D:=p\widehat q\ge M/3,
\end{equation}
return $\mathsf{GOOD}$ with $N=\widehat D$ and retain $M,p,A$.
If no scale triggers, return $\mathsf{DEGENERATE}$.

All generated primes satisfy $p\le4AV_0$. Since
$\log A<S$ and $\log V_0\le S\log2$, the chosen constants imply
\begin{equation}\label{cdgs:prime-size-conditions}
 A>G,\qquad 40\log p<80S<A.
\end{equation}
In particular, at any scale $M\ge C$,
$p\lambda_1>r$ and $C\le M<p/(20\log p)$, so
\cref{cdgs:isolation} applies.

Conditional on all previous history and the stage prime, the hit
indicators are independent Bernoulli variables: their values depend
on shortest lengths, not on the oracle's choice of vector.
For $q_p=\Pr[\lambda_1(L_z)\le r]$, Hoeffding's inequality and a
conditional union bound over at most $K$ stages give
\[
 \Pr[E_{\rm pre}(r)]\ge1-\zeta_{\rm pre},\qquad
 E_{\rm pre}(r)=
 \bigcap_{M\text{ reached}}\{|\widehat q-q_{p(M)}|\le e\}.
\]
Fix a transcript in this event. If $C\ge m$, let $M_*$ be the
smallest candidate scale at least $C$, so $C\le M_*<2C$.
All scales preceding $M_*$ are at least $C$. Unless an earlier scale
has triggered, \cref{cdgs:isolation} therefore applies at $M_*$ and
gives
\[
 \widehat D\ge C(1-C/p)-pe
 \ge\frac{M_*}{2}(1-1/A)-\frac{\eta M_*}{500}>M_*/3.
\]
Thus the first trigger occurs at $M\ge M_*\ge C$.
If $C<m$, every candidate is already larger than $C$.
In both cases, a first trigger always lies in the range where the
isolation bounds are valid.

If no scale triggers, $C<m$. Since $C\ge1$ and
$m=\max\{1,\lceil r/(G\lambda_1)\rceil\}$, integrality gives
$C<r/(G\lambda_1)$.
At a trigger, the upper bound $\widehat D\le C+pe$ gives
\begin{equation}\label{cdgs:trigger-lower-count}
 C\ge(1/3-\eta/500)M.
\end{equation}
Consequently,
\[
 |N-C|\le C^2/p+pe
 \le C/A+\eta M/500<0.01\eta C,
\]
using $A\ge10^4\eta^{-1}$. This proves the required count estimate.
There are at most $Ks+1$ queries. The quantities $K,A,s,\log p$ are
polynomial in the input length, $G$, $\eta^{-1}$, and
$\log(1/\zeta_{\rm pre})$, on every preprocessing path.
The sampling assertion follows from the next lemma.
\end{proof}

\begin{lemma}
\label{cdgs:conditional-sampler}
Fix a $\mathsf{GOOD}$ transcript in $E_{\rm pre}(r)$ and retain its
prime $p$ and integer $A$. Independently sample successive
$z\in\mathbb F_p^n$, query SVP on $L_z$, and stop at the first
reply of length at most $r$, returning its canonical direction.
Stop with $\mathsf{TIMEOUT}$ after $R_{\rm s}$ trials as in
\eqref{cdgs:direction-cap}. This procedure has the timeout and
conditional-success guarantees of \cref{cdgs:count-interface},
uniformly over every prior history and every causal oracle.
\end{lemma}
\begin{proof}
Write $C=C_L(r)$,
$\alpha=1/p-(C-1)/p^2$, and $\beta=1/p$.
Condition on any history just before a new trial. Its fresh $z$ is
still uniform, so \cref{cdgs:isolation} gives, for the success event
$S_t$ and its returned direction $Y_t$,
\[
 \alpha\le\Pr[S_t,Y_t=y\mid\text{history}]\le\beta.
\]
Summing and conditioning on $S_t$ gives
\begin{equation}\label{cdgs:conditional-direction}
 \frac{1-(C-1)/p}{C}
 \le\Pr[Y_t=y\mid S_t,\text{history}]
 \le\frac{1}{C(1-(C-1)/p)}.
\end{equation}
Conditional on a hit before the cap, the final direction law is a
convex combination of these laws over the termination time and its
preceding history. Hence the same bounds hold. Since $C/p\le1/A$
and $A\ge10^4\eta^{-1}$, they imply the claimed
$(1\pm\eta)/C$ interval.

By \eqref{cdgs:trigger-lower-count} and $p\le2AM$, every history has
conditional hit probability at least
\[
 C\alpha\ge\frac{1/3-\eta/500}{2A}(1-1/A)>\frac1{7A}.
\]
Iterated conditioning bounds the probability of $R_{\rm s}$ failures
by $e^{-R_{\rm s}/(7A)}\le\zeta_{\rm s}$.
No independence assumption on the oracle's choices is used.
The cap is imposed even when preprocessing was incorrect.
\end{proof}

\subsection{Bounding the radii with too few primitive directions}
\label{cdgs:degenerate-radii}

Finally, \cite[Lemma~2.19]{SD16} gives, for $n\ge2$ and
$r\ge\lambda_2(L)$,
\begin{equation}\label{cdgs:primitive-growth}
 C_L(r)>
 \frac{\sqrt{r^2-\lambda_2(L)^2}}{\lambda_1(L)}
 +\left\lfloor\frac{r-\lambda_2(L)}{\lambda_1(L)}\right\rfloor.
\end{equation}
In both counting results, taking the $L$-primitive cores of the
constructed vectors preserves their distinct real directions and
does not increase their lengths.

\begin{lemma}
\label{cdgs:thin-annulus}
For $n\ge2$ and $G\ge2$, put
$r_*:=\lambda_2(L)/\sqrt{1-G^{-2}}$.
Then
\begin{equation}\label{cdgs:bad-radii}
 \{r\ge\lambda_1:C_L(r)>1, C_L(r)<r/(G\lambda_1)\}
 \subseteq[\lambda_2,r_*).
\end{equation}
In particular, on valid preprocessing transcripts every layer with
radius at least $r_*$ is $\mathsf{GOOD}$.
For the global parameters $Q,G,T,R_{\rm g}$ and $\eta$ of this
reduction, if $\lambda_2\le s\le T$ and $s<r_*$, then
\begin{equation}\label{cdgs:thin-weight}
 r_*\le R_{\rm g},\qquad
 h_\sigma(s)\le e^\eta h_\sigma(r_*).
\end{equation}
\end{lemma}
\begin{proof}
More than one primitive direction in the ball implies $r\ge\lambda_2$.
By \eqref{cdgs:primitive-growth},
$C_L(r)>\sqrt{r^2-\lambda_2^2}/\lambda_1$.
For $r\ge r_*$ this exceeds or equals the degenerate threshold
$r/(G\lambda_1)$, proving \eqref{cdgs:bad-radii} and the assertion
about valid tags.

For the last claim, integrate the logarithmic-derivative bound of
\cref{cdgs:log-derivative} to obtain
\[
 \log\frac{h_\sigma(s)}{h_\sigma(r_*)}
 \le13\max\{1,r_*^2/\sigma^2\}
 \left[-\tfrac12\log(1-G^{-2})\right]
 \le\frac{104Q}{9G^2}<\eta.
\]
Here $s\le T$ implies $r_*\le R_{\rm g}$ and
$r_*^2/\sigma^2\le4Q/3$, while
$-\tfrac12\log(1-G^{-2})\le2/(3G^2)$.
The last inequality follows from $G\ge10^4FQ$ and
$\eta=1/(1000F)$.
\end{proof}

The case $n=1$ uses a one-dimensional sampler directly.
For $n\ge2$, a degenerate layer with $C_L(r)=1$ contains only the
chosen shortest direction and can be represented exactly by its point
mass. A degenerate layer omitting any other direction must lie in
\eqref{cdgs:bad-radii}. All statements use closed balls, so equal length
directions and grid endpoints need no perturbation.

\subsection{Sampling directions by a mixture over radii}
\label{cdgs:mixture}

At each radius $r_i$ run \cref{cdgs:count-interface} with the
parameters in \eqref{eq:cdgs-parameters} and
\[
 \zeta_{\rm pre}=\frac{\kappa}{8(J+1)}.
\]
Reuse the initial shortest length in these calls.
For a $\mathsf{GOOD}$ layer retain its $N_i$ and sampling procedure.
For a $\mathsf{DEGENERATE}$ layer set $N_i=1$ and use the point
mass on $y_0$.
Define
\begin{equation}\label{eq:cdgs-mixture-weights}
 w_i=h_\sigma(r_i)-h_\sigma(r_{i+1})\quad(i<J),\qquad
 w_J=h_\sigma(r_J).
\end{equation}
These weights turn the retained direction samplers into the desired
Gaussian mixture. Give the zero vector weight one and set the direction
failure allowance to $\zeta_{\rm s}=\kappa/8$. After choosing a layer and
drawing its direction, sample a nonzero integer multiple as in
\cref{alg:centered-dgs}. Finite draws use \cref{app:cdgs-numerics},
with zero returned on timeout or a failed check.

\begin{algorithm}[!ht]
\caption{Centered Gaussian sampling by a mixture over radii}
\label{alg:centered-dgs}
\KwIn{A rational lattice $L$, squared width $\sigma^2>0$, accuracy $f\ge1$, and an SVP oracle.}
\KwOut{A vector in $L$, with the accuracy of \cref{thm:cdgs-reduction}.}
Set the parameters of \cref{eq:cdgs-parameters} and the finite budgets of \cref{app:cdgs-numerics}\;
Set $\zeta_{\rm s}\gets\kappa/8$\;
\If{$n=1$}{\Return a scalar Gaussian multiple of the input generator using \cref{app:scalar-sampling}\;}
Obtain a shortest direction $y_0$ by SVP and set $q_0\gets\|y_0\|^2/\sigma^2$\;
\If{$q_0\ge Q$}{\Return $0$\;}
Construct the multiplicative squared-radius grid of \cref{eq:cdgs-grid}\;
\For{$i=0,\ldots,J$}{
  Run the counting interface at $r_i^2$ with $\zeta_{\rm pre}=\kappa/(8(J+1))$\;
  \If{the finite counting procedure times out}{\Return $0$\;}
  \eIf{the returned tag is $\mathsf{DEGENERATE}$}{
    Set $N_i\gets1$ and retain the point mass at $y_0$\;
  }{
    Retain the $\mathsf{GOOD}$ layer's count $N_i$ and capped direction sampler\;
  }
}
Compute the positive weights $w_i$ from \cref{eq:cdgs-mixture-weights}\;
Choose zero or layer $i$ with weights $1,N_0w_0,\ldots,N_Jw_J$\;
\If{zero is chosen}{\Return $0$\;}
Draw $y$ from layer $i$ using its retained procedure and failure allowance $\zeta_{\rm s}$\;
\If{the direction sampler returns $\mathsf{TIMEOUT}$}{\Return $0$\;}
Draw $z$ from $D_{\Z\setminus\{0\},\sigma/\|y\|,0}$ using the scalar sampler\;
\Return $zy$\;
\end{algorithm}

Let $E_{\rm pre}$ be the event that all layer interfaces satisfy
their assertions. The conditional union bound gives
$\Pr[E_{\rm pre}^c]\le\kappa/8$.
Fix a transcript $\omega\in E_{\rm pre}$. In the analytic mixture,
use exact categorical weights and
the complete capped direction sampling transcript conditioned on
success. Its direction marginal $P_{i,\omega}$ is
within $(1\pm\eta)/C_L(r_i)$ on $\mathsf{GOOD}$ layers.
On degenerate layers it is the point mass at $y_0$.
Keeping the complete transcript permits subsequent oracle choices
to depend on all previous queries and replies.
The unnormalized weight of a direction is
\[
 \widetilde h_{\omega}(y)=
 \sum_{i:\|y\|\le r_i}N_iP_{i,\omega}(y)w_i.
\]

\begin{lemma}\label{cdgs:mixture-bound}
Put $u=(1-\eta)^2$, $v=(1+\eta)^2$,
$a=u/((1+\eta)e^\eta)$, and $b=v$.
For every good preprocessing transcript and every direction of length at most $T$,
\[
 a h_\sigma(\|y\|)\le\widetilde h_{\omega}(y)
 \le b h_\sigma(\|y\|).
\]
For $T<\|y\|\le R_{\rm g}$ the upper bound still holds.
\end{lemma}
\begin{proof}
On a good layer the multiplier $N_iP_{i,\omega}(y)$
lies in $[u,v]$. A degenerate layer gives multiplier one to $y_0$
and zero to all other directions.
Thus $y_0$ has total weight in
$[u h_\sigma(\lambda_1),v h_\sigma(\lambda_1)]$.
For any other $y$, let $j(y)$ be its first eligible layer.
The upper bound follows from nonnegativity and telescoping:
$\sum_{i\ge j(y)}w_i=h_\sigma(r_{j(y)})\le h_\sigma(\|y\|)$.
If all eligible layers are good, the same sum and
\cref{cdgs:grid-bound} give the lower bound with factor
$u/(1+\eta)$.

Otherwise an eligible degenerate layer has $C_L(r_i)>1$.
By \cref{cdgs:thin-annulus},
$\lambda_2\le\|y\|\le r_i<r_*=
\lambda_2/\sqrt{1-G^{-2}}$.
Since $\|y\|\le T$, we have $r_*\le R_{\rm g}$.
Let $j_*$ be the first grid index with $r_{j_*}\ge r_*$.
Every layer from $j_*$ onward is good and contains $y$.
The endpoint $j_*=J$ is allowed. Therefore
\[
 \widetilde h_{\omega}(y)
 \ge u\sum_{i=j_*}^Jw_i
 =u h_\sigma(r_{j_*})
 \ge\frac{u}{1+\eta}h_\sigma(r_*)
 \ge a h_\sigma(\|y\|),
\]
using the grid bound and \eqref{cdgs:thin-weight}.
\end{proof}

\subsection{Normalization and the two accuracy parameters}
\label{cdgs:accuracy}

Let
\[
 \mathcal A=\{0\}\cup
 \{zy:y\in\mathcal P(L),\ \|y\|\le T,\ z\in\Z\setminus\{0\}\},
 \qquad \tau=D_{L,\sigma}(\mathcal A^c)<\varepsilon/64,
\]
and let $D_{\mathcal A}$ be the target conditioned on $\mathcal A$.
For a fixed good transcript $\omega$, let $\widehat Q_\omega$ be
the analytic mixture law. Its zero weight is one, and the weight of $zy$ is
\[
 \widetilde h_{\omega}(y)
 \frac{e^{-\pi z^2\|y\|^2/\sigma^2}}{h_\sigma(\|y\|)}.
\]
Every point in $\mathcal A$ has a multiplier in $[a,b]$ relative to its
target unnormalized mass, including zero since $a\le1\le b$.
Thus $Q_\omega=\widehat Q_\omega(\,\cdot\mid\mathcal A)$ satisfies
\begin{equation}\label{eq:cdgs-normalized}
 \frac ab D_{\mathcal A}(x)\le
 Q_\omega(x)\le\frac baD_{\mathcal A}(x).
\end{equation}
Mixing $Q_\omega$ over the good preprocessing transcripts gives
$\overline Q_{\mathcal A}$ with the same bounds. The direction
sampler's guarantee permits every causal choice of shortest vectors.
Moreover,
\[
 \log(b/a)=3\log(1+\eta)-2\log(1-\eta)+\eta
 <7\eta<\log(1+1/F)\le\log(1+1/f).
\]
The unnormalized buffer mass is at most $b\rho_\sigma(L)\tau$,
while the mass in $\mathcal A$ is at least
$a\rho_\sigma(L)(1-\tau)$. Since $\eta\le1/2000$ and
$\tau<1/128$, this gives
\begin{equation}\label{eq:cdgs-buffer-error}
 \TV(\widehat Q_\omega,Q_\omega)
 =\widehat Q_\omega(\mathcal A^c)
 \le\frac ba\frac{\tau}{1-\tau}
 <2\tau<\varepsilon/32=\kappa/8.
\end{equation}
This accounts for the buffer without rejecting its directions.

For now leave prime and uniform integer draws analytic, and denote
the output law with the other finite budgets by $\mu_0$.
On a good preprocessing transcript, couple the finite choices to
$\widehat Q_\omega$. At the direction invocation retain the actual
complete transcript on success. On timeout use a fresh transcript
from its success-conditioned kernel for the reference and charge
probability at most $\zeta_{\rm s}$. This preserves oracle histories
up to the first disagreement. Then compare $\widehat Q_\omega$ to
$Q_\omega$ using \eqref{eq:cdgs-buffer-error}.
On a bad preprocessing transcript the reference uses an independent
sample of $\overline Q_{\mathcal A}$ and charges that branch as a
disagreement. Together with the finite numerical implementation below,
these comparisons give
\[
 \TV(\mu_0,\overline Q_{\mathcal A})
 \le\underbrace{\kappa/8}_{\rm preprocessing}
   +\underbrace{\kappa/8}_{\rm direction\ timeout}
   +\underbrace{\kappa/8}_{\rm buffer}
   +\underbrace{\kappa/8}_{\rm categorical}
   +\underbrace{\kappa/8}_{\rm scalar}
 <\kappa.
\]
Failures return zero and contribute only to this additive error.
Their mass is not included in the multiplicative comparison.

Define a full-support witness by
\[
 P(x)=
 \begin{cases}
 (1-\tau)\overline Q_{\mathcal A}(x),&x\in\mathcal A,\\
 D_{L,\sigma}(x),&x\notin\mathcal A.
 \end{cases}
\]
It is normalized, has the same support as $D_{L,\sigma}$, and
satisfies $D_{L,\sigma}/(b/a)\le P\le(b/a)D_{L,\sigma}$.
Furthermore
$\TV(\overline Q_{\mathcal A},P)=\tau$, so
\begin{equation}\label{eq:cdgs-error-slack}
 \TV(\mu_0,P)<\kappa+\varepsilon/64
 =17\varepsilon/64.
\end{equation}
This proves the accuracy of the analytic interface, with slack for the
finite prime and integer budgets and the randomized oracle implementation.

\subsection{Evaluation of the radius mixture}
\label{app:cdgs-numerics}
Use the parameters of \eqref{eq:cdgs-parameters} and write
\[
 H(q)=2\sum_{k\ge1}e^{-\pi k^2q},\qquad
 w_i=H(q_i)-H(q_{i+1})\ (i<J),\qquad w_J=H(q_J).
\]
The zero vector contributes weight one. This lower bound on total
mass lets us use absolute accuracy for the differences $w_i$,
even when adjacent radii are very close.

For nonnegative rational counts, put
$N_*=\max\{1,N_0,\ldots,N_J\}$.
To achieve error $0<\delta\le1$, evaluate each $H(q_i)$ to absolute error
$\delta/(8(J+1)N_*)$, take successive differences, and clip them
to be nonnegative. The total weighted error is at most $\delta/4$,
and the lower bound one on total mass controls normalization.

For $q\ge1/4$, the Gaussian series converges geometrically.
For smaller $q$, Poisson summation gives
$H(q)=q^{-1/2}(1+H(1/q))-1$.
Thus evaluation has polynomial cost in the grid and count descriptions
and $\log(1/\delta)$, uniformly for small $q$ as well.
Categorical rounding to
$p=\lceil\log_2(4(J+2)/\delta)\rceil$ bits completes a draw within
TV distance $\delta$ under a fixed polynomial budget.

Use error $\kappa/8$ for this categorical draw and for the one
scalar draw from
$D_{\Z\setminus\{0\},\sigma/\|y\|,0}$ supplied by
\cref{app:scalar-sampling}. Their combined error is at most
$\kappa/4=\varepsilon/16$.

\subsection{Query bounds and the randomized SVP implementation}
\label{cdgs:implementation}
Write $N=\operatorname{size}(B,\sigma^2)$ and choose rational
$\ell\le\|B^{-1}\|_F^{-1}$. Initialize every SVP query with a shortest
basis column and retain only shorter nonzero lattice vectors.
The initial answer has length
$\ell\le\widehat\lambda\le\min_j\|b_j\|$ on every execution.

The grid bound in \cref{cdgs:grid-bound} is therefore polynomial
in $N,f$, even with an incorrect initial answer. Packing bounds the
logarithm of the number of directions in the largest ball by
\[
 O\!\left(n\log(1+R_{\rm g}/\ell)\right)=\poly(N,f).
\]
Substitution in \eqref{cdgs:count-parameters} bounds every layer's
candidate scales and trial counts by computable polynomials.
Thus we can fix a polynomial bound
\begin{equation}\label{eq:cdgs-query-cap}
 Q_*=\poly(N,f)
\end{equation}
on all oracle queries before the first call.

The kernel bases in \eqref{cdgs:kernel-basis} have polynomial
encoding length because each query starts from $B$ and uses a prime
of polynomial bit length. The length checks give the same bound
for answers and primitive cores, uniformly over causal oracle
choices. Together with \cref{app:cdgs-numerics}, these bounds give
fixed polynomial work and space outside the oracle.
The prime and integer draws use \cref{app:numerical-primitives}
with total failure at most $\varepsilon/8$. Enforcing these budgets
and returning zero on exhaustion adds this allowance to
\eqref{eq:cdgs-error-slack}.

Replace each exact query by independent executions of
\cref{thm:svp-main}, retaining the shortest candidate, with
\[
 \delta_{\rm or}=\frac{\varepsilon}{8Q_*},\qquad
 r=\left\lceil\log_3(\delta_{\rm or}^{-1})\right\rceil
\]
repetitions. Conditional on any preceding transcript, the amplified
reply fails with probability at most $\delta_{\rm or}$.
Sequential coupling over $Q_*$ queries adds at most
$\varepsilon/8$ to total variation. Hence the complete additive
error is less than
\[
 17\varepsilon/64+\varepsilon/8+\varepsilon/8
 =33\varepsilon/64<\varepsilon,
\]
with the same multiplicative witness as in the analytic proof.
The algorithm remains $(1+1/f,2^{-f})$-close to its target.

Every query has dimension $n$, and the repetition count is
polynomial in $N,f$. Reusing the workspace gives time
$2^{0.4130n+o(n)}$ and space $2^{0.3138n+o(n)}$, up to polynomial
factors in the input and accuracy parameters.

\section{Proofs of the further consequences}
\label{app:consequences}
This appendix gives the sampling argument behind the time--approximation
tradeoff. The two geometric estimates apply at every Gaussian width.
One produces a point near a prescribed shift. The other produces a short
vector outside an arbitrary proper subspace. The latter permits a direct
SMP algorithm in the original dimension. \Cref{app:block-consequences}
applies the rational projected basis construction of
\cref{app:rational-projections} to polynomial-factor approximations.

\subsection{Gaussian mass in a small ball}
Write $B_R=\{x:\|x\|\le R\}$ and use
\eqref{eq:consequence-gaussian-facts}. Markov's inequality at width
$\sqrt\beta s$, followed by the scale comparison, gives
\begin{equation}\label{eq:centered-small-ball-mass}
 \rho_s(L\cap B_{R_0})
 \ge\frac{\tau}{1+\tau}\beta^{n/2}\rho_s(L),
 \qquad R_0^2=(1+\tau)\beta ns^2/(2\pi).
\end{equation}

\begin{lemma}[Gaussian mass near a shifted closest point]
\label{lem:shifted-small-ball}
Let $r=\operatorname{dist}(t,L)$, $s>0$, $0<\beta\le1$, and $\tau>0$.
Set
\[
 R_0^2=\frac{(1+\tau)\beta ns^2}{2\pi},\qquad R^2=r^2+R_0^2.
\]
Then
\begin{equation}\label{eq:shifted-small-ball}
 \rho_s((L-t)\cap B_R)
 \ge\frac{\tau}{2(1+\tau)}\,
       \beta^{n/2}e^{-\pi r^2/s^2}\rho_s(L).
\end{equation}
\end{lemma}
\begin{proof}
Put $C=L\cap B_{R_0}$ and use
\eqref{eq:centered-small-ball-mass}.
Choose a closest lattice point $x_0$ and put $u=x_0-t$. For each
unordered pair $\{x,-x\}$ in $C\setminus\{0\}$, choose a sign such
that $\langle u,\pm x\rangle\le0$. Then
\[
 \|u\pm x\|^2\le r^2+\|x\|^2\le R^2,\qquad
 \rho_s(u\pm x)\ge e^{-\pi r^2/s^2}\rho_s(x).
\]
The selected points belong to $L-t$ and are distinct for distinct pairs.
The point $x=0$ contributes $u$ and satisfies the same weight bound.
Consequently the shifted ball has mass at least
$\tfrac12 e^{-\pi r^2/s^2}\rho_s(C)$, which gives the claim by
\eqref{eq:centered-small-ball-mass}.
\end{proof}

\begin{lemma}[A short sample outside a subspace]
\label{lem:small-ball-escape}
Let $W\subsetneq\R^n$ be a linear subspace and $v\in L\setminus W$.
For $s>0$, $0<\beta\le1$, and $\tau>0$, put
\[
 R_0^2=\frac{(1+\tau)\beta ns^2}{2\pi},\qquad
 R^2=\|v\|^2+R_0^2.
\]
Then
\begin{equation}\label{eq:small-ball-escape}
 \Prb_{X\sim D_{L,s}}[X\notin W,\ \|X\|\le R]
 \ge\frac{\tau}{3(1+\tau)}\,
       \beta^{n/2}e^{-\pi\|v\|^2/s^2}.
\end{equation}
\end{lemma}
\begin{proof}
Again put $C=L\cap B_{R_0}$ and let
\[
 A=\rho_s(C\setminus W),\qquad B=\rho_s(C\cap W),\qquad
 q=e^{-\pi\|v\|^2/s^2},\qquad G=(L\setminus W)\cap B_R.
\]
Since $C\setminus W\subseteq G$, we have $\rho_s(G)\ge A$.
For each pair $\{x,-x\}$ in $C\cap W$, select $v+x$ or $v-x$
with nonpositive cross term. The selected point lies outside $W$,
has squared norm at most $\|v\|^2+\|x\|^2$, and has weight at least
$q\rho_s(x)$. Distinct pairs give distinct points, and the weight
bound also holds for $x=0$, so $\rho_s(G)\ge qB/2$.
The two sources of points can overlap, but
\[
 \rho_s(G)\ge\max\{A,qB/2\}
 \ge\frac{q}{2+q}(A+B)\ge\frac q3\rho_s(C).
\]
Use \eqref{eq:centered-small-ball-mass} and divide by $\rho_s(L)$.
\end{proof}

\subsection{A general sampling reduction}
We use $\Phi$ from \eqref{eq:approximation-phi}. Approximation factors
are effective in the sense stated in \cref{sec:sampling-reductions}.
\begin{proof}[Proof of \cref{prop:sampler-approximation}]
We first treat a rational $\gamma>1$. Write
$\Delta=\gamma^2-1$, $\tau=1/n$, and
\begin{equation}\label{eq:consequence-beta}
 \beta=\min\{1,\Delta\}.
\end{equation}
Suppose a squared scale $r_0^2$ is available with
\begin{equation}\label{eq:consequence-good-scale}
 \ell/(1+1/n)<r_0\le\ell,
\end{equation}
where $\ell$ is the unknown distance or successive minimum relevant to
the application. Choose a positive rational $s$ satisfying
\begin{equation}\label{eq:consequence-width}
 \frac{A_0}{(1+1/n)^2}\le s^2\le A_0,
 \qquad A_0=\frac{2\pi\Delta r_0^2}{(1+1/n)\beta n}.
\end{equation}
These rational squared scales and a lower approximation to $\pi$
give
\begin{equation}\label{eq:consequence-width-bounds}
 R_0^2\le\Delta r_0^2\le\Delta\ell^2,
 \qquad
 \frac{\pi\ell^2}{s^2}
 \le\frac{n\beta}{2\Delta}(1+1/n)^5
 \le\frac{n\beta}{2\Delta}+16.
\end{equation}
Thus discretization costs only an absolute constant in the exponent.
Optimizing the remaining sampling cost gives
\begin{equation}\label{eq:consequence-optimization}
 \min_{0<\beta\le1}
       \left(\frac\beta\Delta+\log\frac1\beta\right)
 =\Phi(\Delta).
\end{equation}
Differentiating in $\beta$ gives \eqref{eq:consequence-beta}.

\paragraph{Finding independent vectors in dimension $n$.}
At stage $i$, keep
$W=\operatorname{span}(v_1,\ldots,v_{i-1})$ fixed during the entire
stage. There exists $v\in L\setminus W$ with
$\|v\|\le\lambda_i(L)$, since $i$ independent vectors attain the
$i$th successive minimum. At a scale satisfying
\eqref{eq:consequence-good-scale} with $\ell=\lambda_i(L)$,
\cref{lem:small-ball-escape} and
\eqref{eq:consequence-width-bounds} show that
\begin{align}
 &\Prb_{X\sim D_{L,s}}
       [X\notin W,\ \|X\|\le\gamma\lambda_i(L)]\nonumber\\
 &\hspace{2cm}\ge\frac{e^{-16}}{3(n+1)}
       \exp\!\left[-\frac n2\Phi(\Delta)\right].
 \label{eq:consequence-smp-hit}
\end{align}
This estimate holds for every history of the preceding stages.

Let $Q$ clear the input denominators and
$U^2=\max_j\|b_j\|^2$. A descending squared scale grid of ratio
$(1+1/n)^2$ covering $[Q^{-2},U^2]$ has polynomial size and
satisfies \eqref{eq:consequence-good-scale} for every $\lambda_i(L)$.
At stage $i$, sample at every scale and retain the shortest vector
outside $W$, initialized by a qualifying input basis vector.
The good scale and \eqref{eq:consequence-smp-hit} give
$\|v_i\|\le\gamma\lambda_i(L)$ once a useful sample appears.

\paragraph{Finding a closest-vector approximation in dimension $n+1$.}
Given $t\in\mathbb Q^n$, first test $t\in L$ by exact rational linear
algebra, and return $t$ if so. Otherwise set
$r=\operatorname{dist}(t,L)>0$. At each candidate width $s$ construct
the full rank rational lattice
\begin{equation}\label{eq:consequence-cvp-embedding}
 \Lambda_s
 =\{(v-kt,ks):v\in L,\ k\in\Z\}
 =\begin{pmatrix}B&-t\\0&s\end{pmatrix}\Z^{n+1}.
\end{equation}
Call the centered sampler on $\Lambda_s$ at width $s$. By shift
maximality,
\begin{equation}\label{eq:consequence-embedding-mass}
 \rho_s(\Lambda_s)
 =\sum_{k\in\Z}e^{-\pi k^2}\rho_s(L-kt)
 \le\rho_1(\Z)\rho_s(L).
\end{equation}
The ideal probability of a sample $(z,s)$ with
$z\in L-t$ and $\|z\|\le R$ is therefore at least
\begin{equation}\label{eq:consequence-cvp-hit}
 \frac{e^{-\pi}}{\rho_1(\Z)}
 \frac{\tau}{2(1+\tau)}\,
 \beta^{n/2}e^{-\pi r^2/s^2}.
\end{equation}
Here \cref{lem:shifted-small-ball} defines
$R^2=r^2+(1+\tau)\beta ns^2/(2\pi)$.
At a scale satisfying \eqref{eq:consequence-good-scale} with
$\ell=r$, \eqref{eq:consequence-width-bounds} gives $R\le\gamma r$.
The bound is on the first block of the sample, which is the
distance to the target.
Letting $Q$ also clear the denominators of $t$, use the same squared
scale grid over $[Q^{-2},\|t\|^2]$.
A sample $(z,ks)$ with $k=\pm1$ yields the candidate $t+kz$.
Retaining the closest verified candidate over all scales, initialized
by zero, gives the approximation when the good scale supplies a hit.

\paragraph{Precision and a predetermined repetition count.}
Since $\rho_1(\Z)<2$, both
\eqref{eq:consequence-smp-hit} and \eqref{eq:consequence-cvp-hit}
are bounded below by
\begin{equation}\label{eq:consequence-hit-lower}
 p_*:=\frac{e^{-24}}{n+1}
          \exp\!\left[-\frac n2\Phi(\Delta)\right]
\end{equation}
at their respective good scales. With $N=n$ for SMP and $N=n+1$
for CVP, choose $f(N)=N^2$. For every good event $E$, the
two-parameter guarantee implies
\begin{equation}\label{eq:consequence-precision}
 \Prb_{\rm actual}(E)
 \ge\frac{\Prb_{\rm ideal}(E)}{1+1/N^2}-2^{-N^2}
 \ge p_*/2
\end{equation}
for large $n$, since $p_*=e^{-O(n)}$ for fixed $\gamma>1$.
Increase the precision in the finitely many remaining ranks so
that $f\ge4$ and $2^{-f}\le p_*/4$.
The pointwise multiplicative guarantee is essential here: it
preserves an event whose probability is exponentially smaller
than the sampler's total variation error.

At every scale use
\begin{equation}\label{eq:consequence-repetitions}
 M=\left\lceil\frac{2}{p_*}\log(6n)\right\rceil
\end{equation}
calls. Standard amplification gives failure at most $1/(6n)$ per
stage and at most $1/6$ over all SMP stages. Multiplying $M$ by
the polynomial number of scales and stages proves
\eqref{eq:sampler-approximation-calls}.

The rational grids and query bases have polynomial encoding length,
so sequential calls give the stated space bound.

\paragraph{Effective constants and growing approximation factors.}
For a fixed nonrational effective $\gamma>1$, choose a rational
$\widehat\gamma\le\gamma$ with
$\widehat\gamma^2-1=(\gamma^2-1)(1-O(1/n))$.
This changes $n\Phi(\gamma^2-1)$ by $O(1)$ and yields at least
the requested approximation quality. All computations above then use
$\widehat\gamma$.

For a rational effective $\gamma(n)=\omega(1)$, run the construction
with $\widehat\gamma(n)=\min\{\gamma(n),n\}$, for $n\ge2$.
Eventually $\widehat\Delta=\widehat\gamma(n)^2-1\ge1$, so
$\beta=1$ and
\[
 \frac n2\Phi(\widehat\Delta)
 =\frac{n}{2(\widehat\gamma(n)^2-1)}=o(n).
\]
The constant loss in \eqref{eq:consequence-width-bounds} is uniform,
and $f(N)=N^2$ still gives \eqref{eq:consequence-precision}.
Rounding an effective factor downward with relative squared error
$O(1/n)$ preserves divergence. Capping at $n$ costs only $O(1/n)$
in $n\Phi$, while keeping the parameter and grids polynomially
encoded. Hence the call count is uniformly
$2^{o(n)}\poly(|I|,n)$.
\end{proof}

\subsection{Block reduction consequences}\label{app:block-consequences}
The rational projected basis interface of
\cref{app:rational-projections} supplies a $2$-approximate SVP
oracle in each block. Exact projected preprocessing uses
\cref{app:suffix-proof}.

For every fixed $c\ge1$, \cite[Corollary~4.4]{ALNS20}, with
oracle factor $2$ and slack $1/n$, reduces $O_c(n^c)$-SVP to
blocks of rank $k=\lfloor n/(c+1)\rfloor$.
For $1/2<c<1$, \cite[Corollary~3.4]{ALNS20} gives the same factor
with $k=\lceil n/(2c)\rceil$.

For HSVP, \cite[Theorem~2.3]{ALNS20} gives
\[
 \|v\|\le(1+1/n)
       (4\gamma_k)^{(n-1)/(2(k-1))}\det(L)^{1/n},
\]
where $\gamma_k\le k$ is Hermite's constant.
Taking $k=\lceil n/(2c)\rceil$ for fixed $c>1/2$ gives
$(n-1)/(2(k-1))=c+O_c(1/n)$ and hence factor $O_c(n^c)$.
This theorem applies for $n>k\ge2$.
At $c=1/2$, direct SVP and
$\lambda_1(L)\le\sqrt n\det(L)^{1/n}$ give the claim.

The polynomial call and encoding bounds of \cite[Section~2.3]{ALNS20}
remain valid under \cref{app:rational-projections}.
Standard amplification and sequential workspace reuse therefore give
total failure at most $1/3$ and time and space
\[
 2^{\svptime k+o(n)}
 \qquad\text{and}\qquad
 2^{\svpspace k+o(n)}.
\]
Substituting the three block sizes proves
\cref{cor:block-consequences}.

\subsection{Short independent vectors}\label{sec:fast-sivp}
\begin{proof}[Proof of \cref{cor:fast-sivp}]
Apply the reduction of \cite[Lemma~7.5]{ADRSD15} using independent
calls to \cref{thm:single-main}, with $\alpha=2$ and $\xi=1$.
The usual basis bounds on $\eta=\eta_{1/10}(L)$ give a polynomial
squared width grid containing $2\eta<\sigma\le4\eta$.
At that width, with probability $1-o(1)$, $n^2$ samples supply
independent vectors of length
$O(\sqrt n\,\sigma)=O(\sqrt{n\log n})\lambda_n(L)$, while their
joint sampling error is
$n^2\exp[-\Omega(n/\log^3(n+2))]=o(1)$.
Retain the verified independent tuple with the smallest maximum
length over all scales, initialized by the input basis.
Sequential sampling preserves the resource exponents.
\end{proof}

\Needspace{20\baselineskip}
\section{Reusable decoding and promise problem reductions}
\label{app:decoding}
This appendix proves the decoding interfaces and the geometric
reductions in \cref{sec:reusable-bdd,sec:decoding-reductions}.
All calls have fixed work, precision, and reply length limits.
Illegal replies are discarded before subsequent arithmetic.

\subsection{Uniform advice from dual Gaussian batches}
\label{app:decoding-advice}
\begin{lemma}[Uniform Gaussian decoding]
\label{lem:drs-mass-upper-bound}
Suppose $s>0$, $0<\varepsilon<1/200$, and
\[
 \rho_{1/s}(L\setminus\{0\})\le\varepsilon=2^{-\Theta(n)}.
\]
For every fixed $0<\theta<1/2$, let
$N=\lceil Cn\log(1/\varepsilon)/\sqrt\varepsilon\rceil$ for a
sufficiently large constant $C$. Then $N$ independent
$D_{L^*,s}$ samples determine, with probability $1-2^{-\Omega(n)}$,
a deterministic function correct simultaneously on all rational
targets at distance at most
\begin{equation}\label{eq:drs-absolute-radius}
 \frac\theta s\sqrt{\frac{\log(1/\varepsilon)}\pi}
\end{equation}
from $L$. Each query takes $N\poly(n,\mathcal B_t)$ work.
A fixed radius margin permits finite arithmetic with the same bound.
\end{lemma}
\begin{proof}
We adapt the periodic Gaussian decoder of
\cite[Theorem~3.1 and Proposition~3.2]{DRS14}.
The two points to verify are that a known mass upper bound determines
the sample count and that one sample event works for all targets.

Normalize to $K=sL$, $W_i=w_i/s\sim D_{K^*,1}$, and put
\[
 f(t)=\rho_1(K+t)/\rho_1(K),\qquad
 f_N(t)=N^{-1}\sum_i\cos(2\pi\langle W_i,t\rangle).
\]
Poisson summation gives $\E[f_N]=f$. On the ball of radius
$R_\theta=\theta\sqrt{\log(1/\varepsilon)/\pi}$, symmetric pairing
gives $f(t)\ge\rho_1(t)\ge\varepsilon^{\theta^2}$.
Completion of squares and shifted theta maximality give
\[
 \Pr_{Y\sim D_{K,1}}[|\langle Y,v\rangle|>u,\ Y\ne0]
 \le\min\{\varepsilon,2e^{-\pi u^2}\}
\]
for every unit $v$.
Put $S=\sqrt{\log(2/\varepsilon)/\pi}$.
The double tail integral in \cite[Lemma~4.2]{DRS14}, split at $S$,
is at most
$C_\theta(1+\log(1/\varepsilon))\varepsilon
e^{2\pi SR_\theta}\|t\|$ throughout the ball.
Since $\varepsilon e^{2\pi SR_\theta}
=\varepsilon^{1-2\theta-o(1)}$, this gives
\begin{equation}\label{eq:drs-ideal-contraction}
 \left\|\frac{\nabla f(t)}{2\pi f(t)}+t\right\|
 \le C_\theta\varepsilon^{1-2\theta-o(1)}\|t\|.
\end{equation}
Only the upper bound on nonzero mass enters these estimates.

Here is the uniform empirical estimate. The moment-generating bound
$\E[e^{u\langle W,v\rangle}]\le e^{u^2/(4\pi)}$ makes the summands of
\[
 v^{\mathsf T}\nabla^2f_N(t)v
 =-\frac{4\pi^2}{N}\sum_i
   \langle W_i,v\rangle^2\cos(2\pi\langle W_i,t\rangle)
\]
uniformly subexponential. Bernstein's inequality at deviation
$\delta=c\varepsilon^{1/4}$ has failure $2e^{-c'N\delta^2}$.
A constant-mesh sphere net costs $e^{O(n)}$ points.
Gaussian norm tails give $\max_i\|W_i\|=O(\sqrt n)$ with failure
$2^{-\Omega(n)}$, so the empirical and expected Hessians are
$O(n^{3/2})$-Lipschitz. A target mesh of spacing
$\delta/(Cn^{3/2})$ has logarithmic size
$O(n\log(1/\varepsilon))$.
Thus $N\varepsilon^{1/2}=\Omega(n\log(1/\varepsilon))$ yields a
uniform Hessian error $O(\varepsilon^{1/4})$.
Both gradients vanish at zero. Integration, together with bounded
cosine concentration on the same mesh, gives
\begin{equation}\label{eq:drs-uniform-empirical}
 |f_N(t)-f(t)|\le C_1\varepsilon^{1/4},\qquad
 \|\nabla f_N(t)-\nabla f(t)\|
 \le C_1\varepsilon^{1/4}\|t\|
\end{equation}
simultaneously throughout the ball.
Since $\theta^2<1/4$, these errors and
\eqref{eq:drs-ideal-contraction} imply the contraction
\[
 \left\|t+\frac{\nabla f_N(t)}{2\pi f_N(t)}\right\|
 \le\frac12\|t\|.
\]
Periodicity gives the same contraction around every lattice point
on this single sample event.

After $O(\log n)$ steps the residual is below $1/(4C\sqrt n)$.
Theta derivatives give expected covariance
$I/(2\pi)+O(\varepsilon\log(1/\varepsilon))$.
Constant-accuracy Bernstein bounds on a sphere net show that
$O(n)$ initial samples contain $n$ independent rows of norm at most
$C\sqrt n$, with failure $2^{-\Omega(n)}$.
Round the current point's inner products with these rows to integers,
solve the resulting system for the unique lattice point underlying the
residual, and verify membership. Rescaling proves
\eqref{eq:drs-absolute-radius}.
\end{proof}

\begin{proof}[Proof of \cref{thm:batch-bdd}]
Put $C_0=A_{\rm KL}/\sqrt{2\pi e}$.
The KL estimate in~\cite[Lemma~6.1, Eq.~(9)]{ADRSD15} gives
\begin{equation}\label{eq:bdd-kl-widths}
 \lambda_1(L)\eta_{1/2}(L^*)\le(C_0+o(1))\sqrt n,\qquad
 \lambda_1(L)\eta_\varepsilon(L^*)
       \le(C_0+o(1))\sqrt n\,\varepsilon^{-1/n}.
\end{equation}
Take $\varepsilon=2^{-2bn}$ and ideal width
$s=(1+\xi)\alpha C_0\sqrt n/\lambda_1(L)$ with small fixed
$\xi>0$. The feasible parameters obey $\alpha>2^{2b}$.
For the full batch this is $3/2>4/3$.
In the general region, if $\alpha\ge4/3$ use $b<\dgsspace$.
For $1<\alpha<4/3$, the function
$1-\alpha^{-2}-\log\alpha$ increases from zero, which implies
$J(\alpha^{-2})<\tfrac12\log_2\alpha$.
Thus both width requirements in \eqref{eq:bdd-kl-widths} hold.

The decoder needs $\poly(n)2^{bn}$ samples, supplied by polynomially
many independent batches. Their total joint TV error is $o(1)$.
Substituting their ideal samples into
\cref{lem:drs-mass-upper-bound}, and taking $\theta$ close to $1/2$
and $\xi$ small, yields every strict radius in
\eqref{eq:batch-bdd-radius}. TV transfers the single global good
advice event to the actual samples.

A polynomial rational grid of ratio $1+1/n$ between basis bounds
contains $\ell\le\lambda_1(L)<(1+1/n)\ell$.
Use $\ell$ in the width formula, with rational rounding inside the
fixed radius margin, and retain the closest legal reply over all
scales. Their fixed budgets preserve the exponents even outside
the sampling promise. On the good scale, one advice event guarantees
correctness for every target at once, so this selection also
preserves the uniform guarantee.

On good advice $f_N\ge\tfrac12\varepsilon^{\theta^2}=2^{-O(n)}$,
so polynomial precision preserves the contraction and the fixed slack
in the final rounding. With fixed iteration and precision bounds,
including failure replies on other scales, every advice defines a
deterministic bounded query function.
\end{proof}

\subsection{Perturbations and larger distance guarantees}
\label{app:decoding-perturbations}
We work with a fixed good advice function on the relevant lattice.
For $\|v\|=d>a>0$, rotate $v$ to $de_1$ and put $q=a/d$.
The event in \eqref{eq:reduction-cap} becomes
$U_1\le-\sqrt{1-q^2}$.
The standard spherical cap estimate~\cite[Lemma~2.1]{BDGL16}
gives probability $\Theta_q(n^{-1/2}q^n)$.
When $q$ approaches one, cap monotonicity supplies a uniform
lower bound from any fixed smaller ratio.
For any other perturbation length $u>0$, the cap threshold is
\[
 \frac{d^2+u^2-a^2}{2du}\ge\sqrt{1-a^2/d^2},
\]
with equality at $u=\sqrt{d^2-a^2}$. Thus this radius maximizes
the probability of hitting the specified ball with one spherical
perturbation. Averaging any direction-independent perturbation
distribution over the error direction gives no larger probability.

\begin{proof}[Proof of \cref{prop:bdd-radius-amplification}]
First suppose $\ell=\lambda_1(L)$ is known. Fix a closest point
$y$ to $t$ and put $d=\|t-y\|$. An initial query handles
$d\le r\ell$. Otherwise put $\eta=r\ell/n^2$ and
$a=r\ell-2\eta$. Use a distance grid with spacing at most
$\eta$ on $[r\ell,R\ell+\eta]$.
Some grid value satisfies $d\le d_i\le d+\eta$.
Query at $t+\sqrt{d_i^2-a^2}U$.
Extend $t-y$ in its own direction to a vector $\widehat v$ of
length $d_i$. If
$\|\widehat v+\sqrt{d_i^2-a^2}U\|\le a$, then the actual
perturbed residual has length at most $a+\eta<r\ell$.
The cap ratio is at least
\[
 \frac{r(1-2/n^2)}{R+r/n^2},
\]
whose $n$th power is $(r/R)^ne^{-O_{r,R}(1/n)}$.
Polynomially many distance guesses and a polynomial multiple of
$(R/r)^n$ trials give any fixed conditional success probability.

For unknown $\lambda_1(L)$, a polynomial rational grid between
basis bounds contains $\ell\le\lambda_1(L)<(1+1/n^3)\ell$.
Enlarging the distance endpoint accordingly and using radius
$r\ell$ costs only $e^{O_{r,R}(1/n)}$ in cap probability.
All scales share the same advice, and retaining the closest
verified reply proves the claim.
\end{proof}

\begin{proof}[Proof of \cref{thm:usvp-bdd-tradeoff}]
Rank one is immediate. Otherwise fix a shortest vector $v=Bz$.
It is primitive, so $z$ is nonzero modulo every prime.
For a fixed prime $p>g$, a uniform nonzero linear function $h$
on $\mathbb F_p^n$ satisfies
\[
 \Pr[h(z)\ne0]=\frac{p^{n-1}(p-1)}{p^n-1}>1-1/p.
\]
On this event its kernel lattice $L'$ contains no parallel multiple
of $v$ shorter than $p\|v\|$. Every nonparallel vector has length
at least $\lambda_2(L)\ge g\|v\|$, so
$\lambda_1(L')\ge g\|v\|$. Solve a modular equation for
$w\in L$ with $h(B^{-1}w)=1$, and enumerate $x_j=jw$, $1\le j<p$.
At $j=h(z)$ the point $y=x_j-v$ belongs to $L'$.
All representatives and perturbation scales use one preprocessing
of this $L'$.

If $gr\ge1$, direct decoding at this representative returns $y$.
For $q=gr<1$, a perturbation of length
$\|v\|\sqrt{1-q^2}$ hits its ball of radius $q\|v\|$
with probability at least $n^{-O(1)}q^n$.
For the unknown $\lambda=\|v\|$ use the preceding rational grid.
At a good lower scale $\ell$ set
$a=q\ell(1-2/n^2)$, $d_i=(1+1/n^3)\ell$, and perturb by
length $\sqrt{d_i^2-a^2}$.
The extended true residual differs by $O(\ell/n^3)$.
Thus the extra slack preserves the same main cap exponent.
At $gr=1$, direct decoding uses the closed promise. Shrinking
the geometric radius by $1-O(1/n^2)$ and adding small perturbations
also gives polynomial overhead for a strict-distance interface.

For each reply $y'\in L'$, keep the nonzero residual $x_j-y'$.
The perturbation is not subtracted from this residual. Every retained
candidate belongs to $L$, and at a hit it equals $v$.
Keeping the shortest candidate, initialized by a basis vector, gives
constant success from the sublattice, advice, and cap events.
The argument applies also to $g\le2$, including $g=1$.
The kernel basis and representatives have polynomial encoding length,
and standard amplification gives success $2/3$.
\end{proof}

Generate spherical perturbations by normalizing Gaussian vectors.
Standard tail and small-norm bounds permit a distance coupling
within $r\ell/(4n^2)$, with total error $\zeta$ over $Q$ queries,
using polynomial precision in the input and $\log(Q/\zeta)$.
Shrinking the hit radius by $r\ell/n^2$ preserves its exponent
and keeps every coupled target inside the decoding promise.

\subsection{The GapSVP conflict test on a finite grid}
\label{app:decoding-gap}
\begin{proof}[Proof of \cref{thm:gapsvp-bdd-tradeoff}]
Fix the advice, iteration limits, precision, and exceptional returns.
Replace every illegal reply by $0\in L$, obtaining a complete
deterministic function $F$ on the rational points in $B[0,1)^n$.
Write $R=r\gamma d$. For a test point $e$, compute exactly
$e=x+y$ with $x\in B[0,1)^n$ and $y\in L$, and check
$F(x)=-y$. Declare YES if any check fails and NO otherwise.

In the NO case good advice passes every point with $\|e\|\le R$,
since $\|x-(-y)\|\le r\lambda_1(L)$ and $r<1/2$.
The only errors are bad advice and the global implementation error.
In the YES case fix $0\ne v\in L$ with $\|v\|\le d$.
Whenever $e,e+v$ both lie in the ball, their representatives are
equal but they require the distinct answers $-y,-y-v$.
For every fixed $F$, including bad advice, the failure indicators obey
$I(e)+I(e+v)\ge1$.

In the continuous geometry, put
$A=B(0,R)\cap(B(0,R)-v)$. Integration bounds the failure fraction
below by $\operatorname{vol}(A)/(2\operatorname{vol}(B(0,R)))$.
With axis along $v$, a cylinder about the perpendicular bisector
has half-length $\delta=(R-d/2)/n$ and cross-section radius
$b_0=\sqrt{R^2-(d/2+\delta)^2}$.
It lies in the overlap. Since $R/d=r\gamma>1/2$ is fixed,
\[
 \frac{\operatorname{vol}(A)}{\operatorname{vol}(B(0,R))}
 \ge\frac{2\delta V_{n-1}b_0^{n-1}}{V_nR^n}
 \ge n^{-O(1)}\left(1-\frac1{4r^2\gamma^2}\right)^{n/2}.
\]
Only the finite counting version below is used for actual queries.

Let $Q_B$ be the product of the input denominators and take
$h=(Q_B2^k)^{-1}$, so $L\subset h\Z^n$.
On $\Omega_h=h\Z^n\cap B(0,R)$, translation by $v\in L$
preserves the grid exactly. Summing the indicator conflict gives
\begin{equation}\label{eq:gap-finite-conflict}
 \Pr_{e\sim\operatorname{Unif}(\Omega_h)}[I(e)=1]
 \ge\frac{|\Omega_h\cap(\Omega_h-v)|}{2|\Omega_h|}.
\end{equation}
Grid cubes of radius $\delta_h=\sqrt n h/2$ sandwich ball counts
between the volumes at radii $R\pm\delta_h$. The overlap count
has error at most two boundary layers, of relative volume
$O(n\delta_h/R)$. Choose $k=O(n+\mathcal B+\log Q)$ so this
error is at most one tenth of the preceding overlap lower bound.
A rational radius approximation with relative error $2^{-Cn}$
preserves that estimate and the closed NO radius bound.
The finite conflict probability is therefore at least
$n^{-O(1)}2^{-c(r,\gamma)n}$.

Efficient grid ball sampling avoids rejection from an enclosing cube.
Draw continuously from $B(0,R+\delta_h)$, round to the nearest
grid point, and retain the point precisely when it lies in
$\Omega_h$. Every accepted grid cube lies entirely in the expanded
ball and has volume $h^n$. Thus the conditional reference law is
exactly uniform on $\Omega_h$, and its acceptance probability is
$1-O(n\delta_h/R)$.
The continuous ball draw is
$(R+\delta_h)U_0^{1/n}Z/\|Z\|$, with independent uniform
$U_0\in[0,1]$ and standard Gaussian $Z$.
Certified finite approximations control both ball boundaries and
rounding cell boundaries. A coordinate error much smaller than
$h\zeta/(Q\poly(n))$ makes total rounding disagreement at most
$\zeta$, using the bounded coordinate density and a union bound
over its grid boundary strips. Gaussian tails and the small-norm
cutoff are handled as for perturbations.
Cap each rejection run at $O(\log Q+\log(1/\zeta))$ attempts.
All generation overflows then cost at most another $\zeta$.
The finite program is coupled to a discrete reference distribution,
so \eqref{eq:gap-finite-conflict} remains exact.

Take $Q=\poly(n)2^{c(r,\gamma)n}$ trials with a sufficient
polynomial factor. On YES, conditional on any fixed advice, the
miss probability is at most $1/100$ plus implementation error.
On NO it is bounded by the single bad advice probability and
implementation error. Constant independent whole run repetitions
and majority vote give success at least $2/3$.
Only the advice, current test point, decomposition, and counters
are stored, giving the stated $P+QD$ time and space.
\end{proof}

When \cref{thm:batch-bdd} supplies the advice, one may instead
construct it at the known scale $\gamma d$ on a GapSVP input.
Only NO requires legal Gaussian widths. On YES every bounded
preprocessing result still defines a fixed function, and the same
conflict proof applies. Keeping the query function deterministic
is essential for both formulations.

\section*{AI methodology}
\phantomsection
\label{sec:ai-methodology}
\addcontentsline{toc}{section}{AI methodology}
Building on our earlier work~\cite{GFH26a,GFH26b}, the authors used Codex
(GPT-5.6-sol, GPT-6.1-sol, and GPT-6-astra) to explore provable
algorithms for SVP and DGS below the $2^{n/2}$ barrier.
We initially aimed to make BDGL sieving provable and proposed
tracking the first two counting moments under repeated differencing
as a dynamical system. With assistance from GPT-5.6-sol in algorithm
design and mathematical analysis, we obtained an SVP algorithm with
running time below $2^{n/2}$ on September~10, 2026.
We subsequently named the framework Gaussian difference dynamics
(GDD), refined the algorithms, and developed the single sample DGS
and SVP results in \cref{thm:single-main,thm:svp-main} and their
consequences.

After reading Doulgerakis and Laarhoven~\cite{DL26}, we recognized
that their rejection sampling method could extend the single sample
construction to Gaussian batches (\cref{thm:batch-main}). We worked with GPT-6-astra to
design the batch algorithm and develop its proof, then worked with
GPT-6.1-sol to organize and draft the manuscript.
We thoroughly checked and revised the entire paper and remain
responsible for all mathematical claims, proofs, and references.

\clearpage
\phantomsection

\end{document}